\documentclass[11pt,twoside,singlespace]{report}
\usepackage{pdflscape}

\usepackage[square,numbers,sectionbib]{natbib}

\usepackage[dvipsnames]{xcolor}
\usepackage{fancyhdr}
\usepackage{lscape}

\usepackage{url}
\usepackage[a-1b]{pdfx}
\usepackage{hyperref}
\hypersetup{colorlinks=true,linkcolor=blue,allcolors=blue}

\usepackage{bm}
\usepackage{amssymb}
\usepackage{amsmath}
\usepackage{verbatim}
\usepackage{empheq}
\usepackage[most]{tcolorbox}%
\usepackage{mathtools}
\usepackage{mathrsfs}
\usepackage{upgreek}
\usepackage[makeroom]{cancel}

\usepackage{tcolorbox}
\usepackage{bold-extra}
\usepackage{enumitem}

\usepackage{array}
\usepackage{colortbl}
\usepackage{tabularx}
\usepackage{longtable}
\usepackage{multicol}
\usepackage{booktabs}
\usepackage{supertabular}
\makeatletter
\@ifpackagelater{supertabular}{2020/02/03}
	{\newcommand\STspace{\kern\dimexpr2\fontdimen2\font-2\fontdimen4\font\relax}}
	{\newcommand\STspace{\space}}
\makeatother
\usepackage{ctable}

\usepackage{graphicx}
\usepackage{placeins}
\usepackage{caption}
\usepackage{subcaption}
\usepackage{float}
\usepackage{pgfplots}
\usepackage{wrapfig}

\usepackage{etoolbox}

\usepackage{tocloft}

\usepackage[hang,flushmargin]{footmisc}

\usepackage{textcomp}

\usepackage[titletoc,toc,page]{appendix}

\noappendicestocpagenum

\usepackage[a4paper,
	bindingoffset=0.0in,
	left=0.9in,
	right=0.9in,
	top=1in,
	bottom=1in,
	footskip=0.5in]{geometry}
\makeatletter
\patchcmd{\@makechapterhead}{\vspace*{50\p@}}{}{}{}%
\patchcmd{\@makeschapterhead}{\vspace*{50\p@}}{}{}{}%
\makeatother

\definecolor{codegreen}{rgb}{0,0.6,0}
\definecolor{codegray}{rgb}{0.5,0.5,0.5}
\definecolor{codepurple}{rgb}{0.58,0,0.82}
\definecolor{backcolour}{rgb}{0.95,0.95,0.92}
\lstdefinestyle{mystyle}{
	backgroundcolor=\color{backcolour},
	commentstyle=\color{codegreen},
	keywordstyle=\color{magenta},
	numberstyle=\tiny\color{codegray},
	stringstyle=\color{codepurple},
	basicstyle=\ttfamily\footnotesize,
	breakatwhitespace=false,
	breaklines=true,
	captionpos=b,
	keepspaces=true,
	numbers=left,
	numbersep=5pt,
	showspaces=false,
	showstringspaces=false,
	showtabs=false,
	showlines=true,
	escapechar=@,
	tabsize=4,
}
\let\origthelstnumber\thelstnumber
\makeatletter
\def\lst@PlaceNumber{\ifnum\value{lstnumber}=0\else
	\llap{\normalfont\lst@numberstyle{\thelstnumber}\kern\lst@numbersep}\fi}
\newcommand*\Suppressnumber{%
	\lst@AddToHook{OnNewLine}{%
		\let\thelstnumber\relax%
		\advance\c@lstnumber-\@ne\relax%
	}%
}

\newcommand*\Reactivatenumber{%
	\lst@AddToHook{OnNewLine}{%
		\let\thelstnumber\origthelstnumber%
		\advance\c@lstnumber\@ne\relax}%
}
\makeatother

\pgfmathdeclarefunction{gauss}{2}{%
	\pgfmathparse{1/(#2*sqrt(2*pi))*exp(-((x-#1)^2)/(2*#2^2))}%
}

\DeclareFontFamily{U}{wncy}{}
\DeclareFontShape{U}{wncy}{m}{n}{<->wncyr10}{}
\DeclareSymbolFont{mcy}{U}{wncy}{m}{n}
\DeclareMathSymbol{\Sh}{\mathord}{mcy}{"58}

\newcommand{\vm}{{\vspace{-2mm}}}

\newcommand{\qqq}{\qquad \qquad}

\newcommand{\FM}{F_\text{Ms}}
\newcommand{\hatg}{\hat{g}_{\mathbf{k},s}}
\newcommand{\hatvarphi}{\hat{\varphi}_{\mathbf{k}}}
\newcommand{\zetaB}{\zeta_\text{B}}
\newcommand{\thetaB}{\theta_\text{B}}
\newcommand{\dv}{\text{d}^3v}
\newcommand{\ir}{\mathrm{i}\hspace*{0.2mm}}
\newcommand{\irr}{\mathrm{i}\hspace*{0.5mm}}
\newcommand{\bolddeltaf}{$\boldmath{\boldsymbol{\delta}\hspace{-0.5mm}  f\hspace{-0.5mm} }$ }
\newcommand{\deltaf}{$\delta\hspace*{-0.4mm} f$}
\newcommand{\deltafm}{\delta\hspace*{-0.4mm} f}
\newcommand{\nablam}{\boldsymbol{\nabla}\hspace*{-0.2mm}}
\newcommand{\nablamm}{\boldsymbol{\nabla}\hspace*{-0.3mm}}
\newcommand{\nablammm}{\boldsymbol{\nabla}\hspace*{-0.4mm}}
\newcommand{\nablaperp}{\boldsymbol{\nabla}\hspace*{-1mm}_\perp}

\newcommand\myprime{\mkern0.5mu\raise0.6ex\hbox{$\scriptstyle\prime$}}
\newcommand{\hatb}{\boldsymbol{\hat{b}}}
\newcommand{\wminus}{\textcolor{white}{$-$}}
\newcommand{\wzero}{\textcolor{white}{$0$}}

\graphicspath{{figures/}}

\newtcbox{\mymath}[1][]{%
	nobeforeafter, math upper, tcbox raise base,
	enhanced, colframe=blue!30!black,
	colback=blue!30, boxrule=1pt,
	#1}

\def\lineup{\def\0{\hbox{\phantom{\footnotesize\rm 0}}}%
	\def\m{\hbox{$\phantom{-}$}}%
	\def\-{\llap{$-$}}}

\newcommand{\ms}{\noalign{\vspace{3\p@ plus2\p@ minus1\p@}}}

\usepackage{makeidx}

\usepackage{pifont}

\definecolor{black}{rgb}{0.0, 0.0, 0.0}
\definecolor{blue}{rgb}{0.0, 0.0, 1.0}

\makeindex

\begin{document}

\pagenumbering{roman}

\begin{center}
\thispagestyle{empty}

\textcolor{white}{Go down}
\vspace{10mm}

{\Huge \textbf{Simulations of turbulent heat and \\[3mm] \!\!\!\!particle transport in stellarator plasmas\!\!\!\!\!\! }}

\vspace{10mm}
{\LARGE by}

\vspace{10mm}
{\huge Hanne Thienpondt}

\vspace{25mm}
{\Large A dissertation submitted in partial fulfillment of the \\[1.5mm]  requirements for the degree of Doctor of Philosophy in}

\vspace{6mm}
{\Large Plasma Physics and Nuclear Fusion}

\vspace{25mm}
{\Large Universidad Carlos III de Madrid}

\vspace{20mm}
{\Large Advisors:}

\vspace{3mm}
{\Large Dr. Jos\'e Manuel Garc\'ia Rega\~na} \\[1.5mm]
{\Large Prof. Iv\'an Calvo Rubio}

\vspace{10mm}
{\Large Tutor:}

\vspace{3mm}
{\Large Prof. Luis Ra\'ul S\'anchez Fern\'andez}

\vspace{25mm}
{\Large \textcolor{black}{June 2025}}
\end{center}

\newpage
\thispagestyle{empty}

\textcolor{white}{Go down}
\vspace*{\fill}
\begin{center}
	{\large This thesis is distributed under license \\[0mm]
	``Creative Commons \textbf{Attribution -- Non Commercial -- Non Derivatives}''.}
\end{center}

\begin{figure}[h!]
	\centering
	\includegraphics[width=0.22\linewidth] {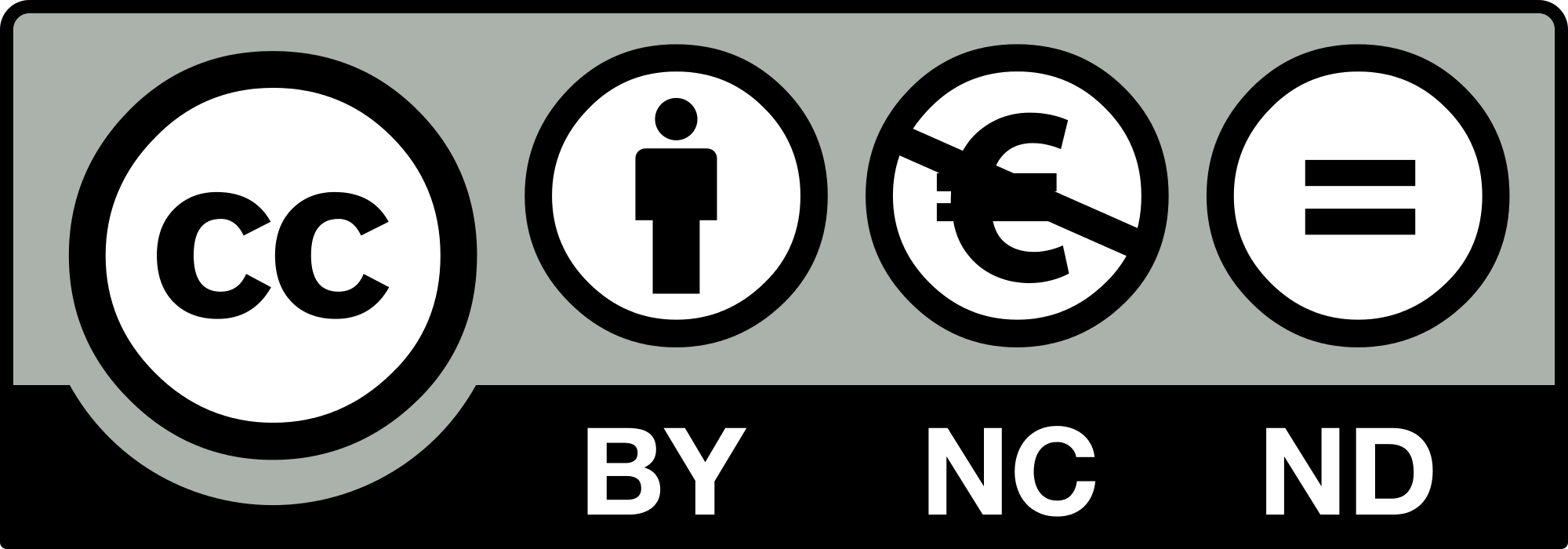}
\end{figure}
\FloatBarrier

\newpage
\begin{center}
	\section*{Acknowledgments}
\end{center}

\vspace{2mm}
\noindent
First and foremost, I would like to express my deepest gratitude to my thesis supervisors, Prof.~Iv\'an Calvo and Dr.~J.~M.~Garc\'ia-Rega\~na, for their unwavering support, invaluable guidance, and never-ending patience throughout the course of this work. Their mentorship has shaped not only the content of this thesis, but also my growth as a researcher. I am truly fortunate to have had the opportunity to learn from their expertise and dedication.

\vspace{3mm}
\noindent
Most of this work has been carried out at the National Fusion Laboratory, CIEMAT (Spain), and I am grateful for the institutional support and collaborative environment provided throughout my PhD. I also acknowledge the use of computational resources: simulations were performed on the Marconi supercomputer at CINECA (Italy) and on the Xula supercomputer located at CIEMAT. Their availability was essential to the completion of this work.

\vspace{3mm}
\noindent
I would also like to thank my family---my parents, my grandparents, and my aunts---for their constant encouragement and unconditional support over the years. Their belief in me has been a steady source of strength and motivation through the challenges of this journey, and I am deeply grateful for their presence in my life.

\vspace{3mm}
\noindent
Special thanks go to Georgia Acton, a great friend, colleague, and the best rubber duck one could ask for. Her insight, kindness, and humor made both science and life more manageable. I am also grateful to Paul Mulholland for his friendship and thoughtful collaboration during our time working together.

\vspace{3mm}
\noindent
To the Salsa group of Los Tucusitos Del Son and the many dancers of ``La Ermita'', thank you for being my family here in Madrid and for bringing joy, rhythm, and community into my life. I am equally thankful to all my friends in Madrid and abroad whose presence, whether near or far, has supported me in countless ways during these years.

\newcommand{\HTetal}{\textbf{H.~Thienpondt} et al}
\newcommand{\HT}{\textbf{H.~Thienpondt}}
\newcommand{\JM}{J.~M.~Garc\'ia-Rega\~na}
\newcommand{\IC}{I.~Calvo}

\newcommand{\paperDOIExtra}[9]{
	\hspace{-11mm}\STspace%
	\begin{supertabular}{p{0.3cm}p{15.7cm}}
		#1 & #2 (#3) #4 \ \textit{#5}, #6. DOI: \href{http://doi.org/#7}{#7} \\[2mm]
		& #8 \\[2mm]
		& #9 \\[2mm]
	\end{supertabular}
}

\newpage
\begin{center}
	\section*{Published and submitted content}
\end{center}
\label{page:content}

\noindent
The content of the following publications, appropriately modified, is partly or fully included in \mbox{this thesis.}

\vspace{4mm}
\subsection*{\underline{\smash{Journal Papers}}}

\vspace{-2mm}
\textcolor{white}{.}

\paperDOIExtra{\cite{thienpondt2024influence}}{\HT, \JM, \IC, G.~Acton, and M.~Barnes}{2024}{Influence of the density gradient on turbulent heat transport at ion-scales: an inter-machine study with the gyrokinetic code \texttt{stella}.}{Nuclear Fusion}{65.1, p.\,016062} {10.1088/1741-4326/ad9ab9}
{Simulations with the gyrokinetic code \texttt{stella} and the analysis of the results was carried out by H.~Thienpondt. The paper was written by H.~Thienpondt, with major revisions by the PhD advisors, J.~M.~Garc\'ia-Rega\~na and Iv\'an Calvo. G.~Acton and M.~Barnes provided support for the \texttt{stella} code throughout the course of this work.}
{The material from this article is modified and fully included in this thesis under license \href{https://creativecommons.org/licenses/by/4.0/}{CC BY 4.0} and it is not singled out with typographic means and references. Material from this article is included in chapter \ref{chap:heat}.}

\paperDOIExtra{\cite{thienpondt2023prevention}}{\HT, \JM, \IC, J.~A.~Alonso, J.~L.~Velasco et al.}{2023}{Prevention of core particle depletion in stellarators by turbulence.}{Physical Review Research}{5.2, p.\,L022053}{10.1103/PhysRevResearch.5.L022053}
{H.~Thienpondt carried out the main work of the paper, namely obtaining the turbulent transport results through simulations using the \texttt{stella} code. The paper was written by H.~Thienpondt, with major revisions by the PhD advisors, J.~M.~Garc\'ia-Rega\~na and Iv\'an Calvo. J.~L.~Velasco contributed by performing the neoclassical calculations with \texttt{KNOSOS} and \texttt{DKES}. I.~Calvo and J.~A.~Alonso developed the neutral source model introduced in the paper, and J.~A.~Alonso applied it to the analyzed discharge. K.~Brunner, O.~Ford, G.~Fuchert, J.~Knauer, E.~Pasch, and L.~Van\'o are responsible for the diagnostics of the W7-X stellarator, which provided the information of the profiles. M.~Barnes provided support for the \texttt{stella} code throughout the course of this work and A.~Gonz\'alez-Jerez carried out the implementation of the boundary conditions that the \texttt{stella} code considered in the simulations. }
{The material from this article is modified and fully included in this thesis under license \href{https://creativecommons.org/licenses/by/4.0/}{CC BY 4.0}, and it is not singled out with typographic means and references. Material from this article is included in chapter \ref{chap:particles}.}

\paperDOIExtra{\cite{alonso2024density}}{J.~A.~Alonso, D.~Alegre, J.~Alonso, ..., \HT\ et al.}{2024}{Density profiles in stellarators: An overview of particle transport, fueling and profile shaping studies at TJ-II.}{Nuclear Fusion}{64.11, p.\,112018}{10.1088/1741-4326/ad67ef}
{H.~Thienpondt carried out the numerical simulations with the gyrokinetic code \texttt{stella} that are presented in figures 1 and 2 of this article.}
{Figures 1 and 2 from this article are modified and fully included in this thesis under license \href{https://creativecommons.org/licenses/by/4.0/}{CC BY 4.0}, and they are not singled out with typographic means and references. Material from this article is included in chapter \ref{chap:particles}.}

\newcounter{mycounter}
\newcommand\showmycounter{\stepcounter{mycounter}\themycounter}

\newcommand{\Parra}{F.~I.~Parra}
\newcommand{\WX}{{\fontfamily{qtm}\selectfont W7-X}}
\newcommand{\TitleParticleTransport}{Turbulent transport avoids core particle depletion in stellarators}
\newcommand{\TitleHeatTransport}{Turbulent heat flux versus density gradient: an inter-machine study with the gyrokinetic code \stellaExtra}
\newcommand{\three}{{\fontfamily{qtm}\selectfont 3}}
\newcommand{\six}{{\fontfamily{qtm}\selectfont 6}}
\newcommand{\stellaExtra}{{\fontfamily{lmtt}\selectfont stella}}

\newcommand{\researchvisit}[3]{
	\hspace{-4mm}\STspace%
	\begin{supertabular}{p{1.5cm}p{16cm}}
		\textsc{#1} & #2  \\
		& #3 \\
	\end{supertabular}
}

\newcommand{\course}[4]{
	\hspace{-4mm}\STspace%
	\begin{supertabular}{p{1.5cm}p{16cm}}
		\textsc{#1} & #2  \\
		& #3 \\
		& #4 \\
	\end{supertabular}
}

\newcommand{\paper}[4]{
	\hspace*{-8mm}\STspace%
	\begin{supertabular}{p{0.05cm}p{15.7cm}}
		\vspace*{-2mm}\textsc{\tiny\ding{108}} & #1 (#2) #3 \ \textit{#4}  \\[2mm]
	\end{supertabular}
}
\newcommand{\paperDOI}[5]{
	\hspace*{-8mm}\STspace%
	\begin{supertabular}{p{0.05cm}p{15.7cm}}
		\vspace*{-2mm}\textsc{\tiny\ding{108}} & #1 (#2) #3 \ \textit{#4}. DOI: \href{http://doi.org/#5}{#5} \\[2mm]
	\end{supertabular}
}
\newcommand{\paperDOIVolume}[6]{
\hspace*{-8mm}\STspace%
\begin{supertabular}{p{0.05cm}p{15.7cm}}
	\vspace*{-2mm}\textsc{\tiny\ding{108}} & #1 (#2) #3 \ \textit{#4}, #5. \mbox{DOI: \href{http://doi.org/#6}{#6}} \\[2mm]
\end{supertabular}
}
\newcommand{\paperDOIVolumeTwo}[6]{
\hspace*{-8mm}\STspace%
\begin{supertabular}{p{0.05cm}p{15.7cm}}
	\vspace*{-2mm}\textsc{\tiny\ding{108}} & #1 (#2) #3 \ \textit{#4}, #5. \\[0mm]
	& \mbox{DOI: \href{http://doi.org/#6}{#6}} \\[2mm]
\end{supertabular}
}
\newcommand{\talk}[7]{
	\hspace*{-8mm}\STspace%
	\begin{supertabular}{p{0.05cm}p{15.7cm}}
		\vspace*{-2mm}\textsc{\tiny\ding{108}} & #1. #2. \\
		&  #3. #4, #5. \href{#7}{#6} \\[2mm]
	\end{supertabular}
}
\newcommand{\talkAWARD}[8]{
	\hspace*{-8mm}\STspace%
	\begin{supertabular}{p{0.05cm}p{15.7cm}}
		\vspace*{-2mm}\textsc{\tiny\ding{108}} & #1. #2.  \\
		& #3. #4, #5. \href{#7}{#6} \\[-2mm]
		& {\textsc{#8}} \\[2mm]
	\end{supertabular}
}
\newcommand{\conference}[5]{
	\hspace*{-8mm}\STspace%
	\begin{supertabular}{p{0.05cm}p{15.7cm}}
		\vspace*{-2mm}\textsc{\tiny\ding{108}} & #1. #2, #3. \href{#5}{#4} \\[2mm]
	\end{supertabular}
}
\newcommand{\thesis}[6]{
	\hspace*{-8mm}\STspace%
	\begin{supertabular}{p{0.05cm}p{15.7cm}}
		\vspace*{-2mm}\textsc{\tiny\ding{108}} & #1. #2. #3. #4. \href{#6}{#5} \\[2mm]
	\end{supertabular}
}

\newpage
\begin{center}
	\section*{Other Research Merits}
\end{center}

\noindent
In addition to the core research presented in this dissertation, the PhD work has contributed to a number of collaborative scientific efforts. This includes significant contributions to the development and maintenance of the \texttt{stella} and \texttt{stellapy} codes, which have supported not only this thesis but also the wider fusion research community. Moreover, this section highlights co-authored contributions (not included in the section ``published and submitted content'') in which H.~\mbox{Thienpondt} has been involved, as well as conferences and workshops where the results from this thesis have been presented or disseminated. These contributions reflect the broader scientific engagement and collaborative context of this doctoral research.

\vspace{8mm}
\noindent
\textbf{\underline{\smash{\large Journal papers}}}
\vspace{5mm}

\paperDOIVolume{D.~Carralero, T.~Estrada, \JM, E.~S\'anchez, ..., \HT\ et al}{2025}{First experimental observation of zonal flows in the optimized stellarator Wendelstein 7-X.}{Physical Review Research}{7.2, p.\,L022009}{10.1103/PhysRevResearch.7.L022009}

\paperDOIVolume{J.~L.~Velasco, I.~Calvo, ..., \HT\ et al}{2024}{Piecewise omnigenous stellarators.}{Physical Review Letters}{133.18, p.\,185101}{10.1103/PhysRevLett.133.185101}

\paperDOIVolume{\JM, \IC, E.~S\'anchez, \HT, J.~L. Velasco and J.~A.~Capit\'an}{2024}{Reduced electrostatic turbulence in the quasi-isodynamic stellarator configuration CIEMAT-QI4.}{Nuclear Fusion}{65.1, p.\,016036}{10.1088/1741-4326/ad86cb}

\paperDOIVolumeTwo{G.~Acton, M.~Barnes, S.~Newton and \HT}{2024}{Optimisation of Gyrokinetic Microstability Using Adjoint Methods.}{Journal of Plasma Physics}{90.4, p.\,905900406}{10.1017/S0022377824000709}

\paperDOIVolumeTwo{\JM, \IC, \Parra, and \HT}{2024}{Reduction or Enhancement of Stellarator Turbulence by Impurities.}{Physical Review Letters}{133.10, p.\,105101}{10.1103/PhysRevLett.133.105101}

\paperDOIVolume{J.~A.~Alonso, D.~Alegre, J.~Alonso, ..., \HT\ et al}{2024}{Density profiles in stellarators: an overview of particle transport, fueling and profile shaping studies at TJ-II.}{Nuclear Fusion}{64.11, p.\,112018}{10.1088/1741-4326/ad67ef}

\paperDOIVolume{O.~Grulke, C.~Albert, J.~A.~Alcuson Belloso, ..., \HT\ et al}{2024}{Overview of the first Wendelstein 7-X long pulse campaign with fully water-cooled plasma facing components.}{Nuclear Fusion}{64.11, p.\,112002}{10.1088/1741-4326/ad2f4d}

\paperDOIVolume{T.~S.~Pedersen, ..., \HT\ et al}{2022}{Experimental confirmation of efficient island divertor operation and successful neoclassical transport optimization in Wendelstein {\fontfamily{qtm}\selectfont 7-X}.}{Nuclear Fusion}{62.4, p.\,042022}{10.1088/1741-4326/ac2cf5}

\paperDOIVolume{C.~Hidalgo, E. Ascas\'ibar, D.~Alegre, A.~Alonso, ..., \HT\ et al\  }{2022}{\  Overview of the TJ-II stellarator research programme towards model validation in fusion plasmas.}{\mbox{Nuclear Fusion}}{62.4, p.\,042025}{10.1088/1741-4326/ac2ca1}

\paperDOIVolume{\JM, M.~Barnes, ..., \HT\ et al}{2021}{Turbulent transport of impurities in {\fontfamily{qtm}\selectfont 3D} devices.}{Nuclear Fusion}{61.11, p.\,116019}{10.1088/1741-4326/ac1d84}

\paperDOIVolume{G.~Van Oost, P.~Beyer, A.~Devitre, ..., and \HT}{2021}{The European Master of Science in Nuclear Fusion and Engineering Physics (FUSION-EP): 15 years of experience.}{\mbox{European Journal of Physics}}{42.2, p.\,024002}{10.1088/1361-6404/abc6e5}

\noindent
\textbf{\underline{\smash{\large Invited talks}}}
\vspace{3mm}

\talk{\HTetal}{\TitleParticleTransport}{{\fontfamily{qtm}\selectfont 49$^\text{th}$} European Conference on Plasma Physics (EPS)}{03/07/2023}{France}{Programme}{https://eps2023.github.io/EPS-2023-program.html}

\talk{\HTetal}{\TitleHeatTransport. \ \  \textit{Received the most popular talk award}}{{\fontfamily{qtm}\selectfont 2$^\text{nd}$} Fusion HPC Workshop}{02/12/2021 -- 03/12/2021}{Virtual}{Programme}{https://www.bsc.es/es/printpdf/news/events/2nd-fusion-hpc-workshop}{}

\vspace{6mm}
\noindent
\textbf{\underline{\smash{\large Oral contributions}}}
\vspace{3mm}

\talk{\HTetal}{\TitleHeatTransport}{{\fontfamily{qtm}\selectfont 23$^\text{rd}$} International Stellarator-Heliotron Workshop (ISHW)}{20/06/2022}{Poland}{Abstract}{https://www.ifpilm.pl/images/ishw2021/Oral_presentations/Hanne_Thienpondt.pdf}

\vspace{6mm}
\noindent
\textbf{\underline{\smash{\large Workshops}}}
\vspace{3mm}

\talk{\HTetal}{\TitleHeatTransport. \ \  \textit{Received the Best Student Talk Prize}}{{\fontfamily{qtm}\selectfont 14$^\text{th}$} Plasma Kinetics Working Meeting}{24/07/2023 -- 28/07/2023}{Austria}{Programme}{https://www-thphys.physics.ox.ac.uk/research/plasma/wpi/workshop2023.html}

\vspace{6mm}
\noindent
\textbf{\underline{\smash{\large Seminars}}}
\vspace{3mm}

\talk{\HTetal}{Gyrokinetic turbulence in stellarators: 	An overview of Fundamentals}{Oxford Plasma Theory Seminar}{16/07/2024}{University of Oxford, UK} {Programme}{https://www-thphys.physics.ox.ac.uk/research/plasma/seminarsT24.html}

\talk{\HTetal}{Gyrokinetic turbulence in stellarators: 	An overview of Fundamentals}{FUSION-EP Summer Event}{11/07/2024}{Universidad Complutense, Madrid} { }{ }

\talk{\HTetal}{\TitleParticleTransport}{Oxford Plasma Theory Seminar}{06/09/2022}{University of Oxford, UK} {Programme }{https://www-thphys.physics.ox.ac.uk/research/plasma/seminarsM22.html}

\talk{\HTetal}{\TitleHeatTransport}{FUSION-EP Summer Event}{13/07/2021}{University of Ghent, Belgium} { }{ }

\vspace{6mm}
\noindent
\textbf{\underline{\smash{\large Other talks}}}
\vspace{3mm}

\talk{\HTetal}{\TitleParticleTransport}{Weekly \WX\ physics meetings}{26/06/2023}{Virtual, IPP Greifswald}{ Slides}{https://indico.euro-fusion.org/event/2691/}

\talk{\HTetal}{\TitleHeatTransport}{CIEMAT Fusion student seminars}{28/04/2023}{Virtual, CIEMAT}{ }{ }

\talk{\HTetal}{\TitleParticleTransport}{Weekly CIEMAT physics meetings}{14/11/2022}{Virtual, CIEMAT}{ }{ }

\talk{\HTetal}{\TitleParticleTransport}{Turbulence Topical Group of \WX}{19/05/2022}{Virtual, IPP Greifswald}{Slides}{https://indico.euro-fusion.org/event/2097/}

\talk{\HTetal}{\TitleParticleTransport}{CIEMAT Fusion student seminars}{13/05/2022}{Virtual, CIEMAT}{ }{ }

\talk{\HTetal}{\TitleHeatTransport}{Weekly CIEMAT physics meetings}{13/09/2021}{Virtual, CIEMAT}{ }{ }

\talk{\HTetal}{\TitleHeatTransport}{Turbulence Topical Group of \WX}{01/07/2021}{Virtual, IPP Greifswald}{Slides}{https://indico.euro-fusion.org/event/1127/}

\vspace{6mm}
\noindent
\underline{\smash{\textbf{\large Posters}}}
\vspace{3mm}

\talk{\HTetal}{\TitleHeatTransport}{Fusenet's 8-bit PhD Event}{22/11/2021 -- 23/11/2021}{Virtual}{ }{ }

\talk{\HTetal}{\TitleHeatTransport}{{\fontfamily{qtm}\selectfont 19$^\text{th}$} European Fusion Theory Conference (EFTC)}{11/10/2021 -- 15/10/2021}{Virtual}{Poster } {https://indico.cern.ch/event/934747/contributions/4516409/}

\vspace{5mm}
\noindent
\underline{\smash{\textbf{\large Codes}}}
\vspace{3mm}

\newcommand{\codes}[1]{
	\hspace*{-8mm}\STspace%
	\begin{supertabular}{p{0.05cm}p{15.7cm}}
		\vspace*{-2mm}\textsc{\tiny\ding{108}} & #1 \\[2mm]
	\end{supertabular}
}

\codes{The \texttt{stella} Fortran code. During the course of this PhD work, H.~Thienpondt has made numerous contributions to the development and maintenance of the gyrokinetic code \texttt{stella}. In collaboration with G.~Acton, a major restructuring and clean-up of the code was carried out to improve usability, readability, and robustness. H.~Thienpondt also played a key role in merging several important code branches, including the full-flux-surface and electromagnetic extensions of \texttt{stella}. To enhance reliability and reproducibility, an automated testing infrastructure was implemented to verify code validity with each commit to the GitHub repository. Additionally, the build system was reorganized to simplify compilation and improve maintainability. Throughout the PhD, various bugs were identified and resolved, and several minor enhancements, as well as novel diagnostics, were introduced to support ongoing development. The \texttt{stella} code is open source and can be found at \href{https://github.com/stellaGK/stella}{https://github.com/stellaGK/stella}.}

\codes{The \texttt{stellapy} Python code. As part of this PhD work, H.~Thienpondt has developed a dedicated Python diagnostics package, \texttt{stellapy}, designed to analyze and visualize the outputs of the gyrokinetic code \texttt{stella}. The \texttt{stellapy} package has been developed in collaboration with J.~M.~Garc\'ia-Rega\~na and A.~Gonz\'alez-Jerez. This tool enables users to extract and examine a wide range of physical quantities either via the command line or within Python scripts. These include, but are not limited to, heat and particle fluxes, linear growth rates, parallel mode structures, phase shift data, velocity-space distribution functions, real-space potential fluctuations, as well as density and temperature fluctuations. In addition to standard diagnostics, several utility tools have been implemented. These include scripts for determining the optimal flux-tube length based on the selected parallel boundary condition, as well as routines for selecting appropriate input parameters to achieve a desired numerical resolution. Furthermore, diagnostic tools have been added to visualize and analyze \texttt{VMEC} output files that define the magnetic equilibrium configurations. The \texttt{stellapy} code is open source and can be found at \href{https://github.com/stellaGK/stella/tree/master/POST_PROCESSING/stellapy}{https://github.com/stellaGK/stella/tree/master/POST\_PROCESSING/stellapy}.}

\newpage
\chapter*{Abstract}

\vspace{0mm}
\noindent

Achieving sustained energy production in magnetic confinement fusion reactors requires a deep understanding of turbulent transport, especially concerning its role on the confinement of heat and particles. This thesis investigates turbulent transport in different fusion devices, with a particular emphasis on stellarator configurations. The research is structured around two central questions: how do density gradients affect the turbulent transport of heat, and can turbulence help retain particles in the core of fusion plasmas. By means of extensive numerical simulations with the $\deltafm$ gyrokinetic code \texttt{stella}, the complex interplay between microinstabilities and plasma profiles is explored across a range of fusion devices. Specifically, the \mbox{Wendelstein 7-X} (W7-X), Large Helical Device (LHD), TJ-II, and National Compact Stellarator Experiment (NCSX) stellarators are included---which represent helias, heliotron, heliac and quasi-axisymmetric configurations, respectively---as well as the ASDEX Upgrade (AUG) tokamak and the tokamak geometry \mbox{of the Cyclone Base Case (CBC).}

In the first part of the thesis, the impact of density gradients on turbulent heat transport is analyzed through a comprehensive inter-machine study. Linear and nonlinear simulations are performed across a wide range of parameters to explore the effects of density gradients, temperature gradients, and kinetic electrons. A key finding is the strong reduction of the ion heat flux in NCSX and W7-X with increasing density gradients, attributed to the stabilization of ion-temperature-gradient-driven (ITG) modes without significantly destabilizing density-gradient-driven modes, such as trapped-electron modes (TEM) and universal instabilities. This beneficial effect is absent in tokamaks, where the heat flux is found to increase with the density gradient. This contrast highlights the role of magnetic geometry on turbulence. Moreover, the behavior of ion heat fluxes in stellarators does not align with that of linear growth rates, if only the fastest-growing mode is taken into account. Therefore, the results underscore the importance of considering  nonlinear effects. Additionally, this study provides physical insights into the microinstabilities, emphasizing the dominance of trapped-electron modes in CBC, AUG, TJ-II, LHD and NCSX, while passing-particle-driven universal instabilities contribute \mbox{significantly in W7-X.}

In the second part, the thesis addresses the long-standing discrepancy between neoclassical predictions and experimental observations of core density profiles in stellarators. Namely, in reactor-relevant plasmas, neoclassical transport drives an outward particle flux in the core of large stellarators and predicts strongly hollow density profiles. However, this theoretical prediction is contradicted by experiments. In particular, in W7-X, the first large optimized stellarator, flat or weakly peaked density profiles are generally measured, indicating that neoclassical theory is not sufficient and that an inward contribution to the particle flux is missing in the core. Through extensive nonlinear simulations and a detailed particle balance analysis, this work demonstrates that turbulence provides a robust inward particle flux, driven by the simultaneous presence of finite ion and electron temperature gradients, that can account for the observed profiles. Specifically, the sign (direction) of the turbulent particle flux matches the sign of the discrepancy between the experimental and neoclassical particle fluxes across the entire plasma radius in W7-X, confirming its key role in core density retention. These results significantly advance the understanding of \mbox{particle transport in stellarators.}

Together, this dissertation deepens our understanding of turbulence-driven transport and highlights the critical role of magnetic geometry and plasma profiles in regulating confinement. Moreover, it provides essential insights for the design and optimization of future fusion reactors, advancing the predictive capabilities necessary to achieve robust, turbulence-resistant operation scenarios.

\vspace{4mm}
\noindent
\textit{Keywords:} Gyrokinetics, Turbulence, Heat transport, Particle transport, Stellarator, Tokamak, Magnetic confinement fusion, Microinstabilities, ITG, TEM, Universal instability

\newpage
{\hypersetup{linkcolor=black}
\let\clearpage\relax
\setcounter{tocdepth}{2}
\tableofcontents}

\fancyhf{}
\lhead[\thepage]{\leftmark}
\rhead[\leftmark]{\thepage}
\renewcommand{\chaptermark}[1]{\markboth{\sc{\chaptername\ \thechapter.\ #1}}{}}
\setlength{\headsep}{6pt}
\setlength{\parskip}{0pt plus 0pt minus 0pt}

\pagenumbering{arabic}

\part{Introduction}

\chapter{Magnetic Confinement Fusion, Microinstabilities and Plasma Turbulence}
\label{chap:intro}

The pursuit of sustainable, clean energy has become one of the most pressing challenges of the 21$^\text{st}$ century, driven by rapidly increasing global energy demands, and the urgent need to mitigate climate change by reducing greenhouse gas emissions. Traditional energy sources like fossil fuels, which currently dominate global energy production, are major contributors to carbon dioxide emissions and are finite, posing both environmental and resource limitations. This has intensified the search for energy sources that can deliver high output, reliability, and a low environmental impact. Among the various options, nuclear fusion has emerged as a promising pathway toward achieving a sustainable energy future.

\section{Nuclear fusion}
\vspace{-2mm}

Fusion, the process that powers the stars, generates energy by fusing light atomic nuclei under extreme temperatures and pressures. Among the different possible reactions, the fusion of two hydrogen isotopes, deuterium (D) and tritium (T), is mainly considered for future fusion reactors because its cross-section (a measure of the probability of the reaction occurring) is higher than for other reactions at relatively low temperatures. During the deuterium-tritium fusion reaction,
\begin{equation}
	\text{D} + \text{T} \ \rightarrow \  ^4\text{He} \, (3.5\,\text{MeV}) + \text{n}\,(14.1\,\text{MeV})
\end{equation}
a helium nucleus (He) is produced, accompanied by the release of a neutron (n).
During this reaction, part of the mass of the reactants is converted into kinetic energy (17.6\,MeV).

Nuclear fusion has certain advantages compared to fossil fuels, renewable energy sources and its nuclear counterpart---fission. First of all, the fusion fuel is sustainable: deuterium and tritium are extracted from seawater and lithium respectively, both of which are abundant. Additionally, nuclear fusion does not emit greenhouse gases like carbon dioxide into the atmosphere, and energy production does not depend on the weather like most renewable energy sources. Finally, unlike fission, fusion produces minimal long-lived radioactive waste and has an inherently lower risk of catastrophic failure, making it a highly attractive option for future energy systems.

To fuse together two hydrogen nuclei, the Coulomb barrier needs to be overcome, which prevents charged nuclei to fuse under normal circumstances. When the nuclei are within 10$^{-15}$\,m of each other, the strong attractive nuclear force, that binds protons and neutrons together in the nucleus, becomes dominant and pulls the particles together, causing them to fuse. However, even on a perfect collision course, the two nuclei must have an extremely high velocity to get sufficiently close to each other. In order for the nuclei to reach these velocities, the fusion fuel must be heated to extremely high temperatures of the order of hundreds of million degrees Celsius.

\index{Plasma}
At the temperatures required for nuclear fusion, the atoms that make up the fusion fuel are completely ionized, and the nuclei and electrons form a globally neutral gas of charged particles, called \textit{plasma}. This hot plasma cannot be confined by any physical barrier, as it would rapidly destroy or melt any solid material. To achieve plasma confinement, three main methods exist: \mbox{gravitational}, inertial and magnetic. In stars, such as our Sun, confinement is achieved through the immense gravitational field generated by their massive size. This gravitational force compresses and heats the plasma, creating conditions where fusion occurs naturally, which releases vast amounts of \mbox{energy}. Unfortunately, replicating gravitational confinement on Earth is not feasible. Instead, two other confinement methods have been devised to create and control plasma on Earth.
Inertial confinement involves compressing small pellets of hydrogen isotopes to extremely high densities by means of high-powered lasers or particle beams. When the pellet is compressed rapidly enough, it reaches the temperature and pressure conditions necessary for fusion. The extreme compression and heating cause fusion reactions to occur briefly before the pellet disintegrates, releasing energy in a controlled burst.
The third approach is \textit{magnetic confinement fusion}, which relies on powerful magnetic fields that prevent the plasma from reaching the reactor walls on excessively short time scales, thereby protecting them from melting under the intense heat. Magnetic confinement fusion, which is the main focus of this dissertation, is the leading method for achieving controlled nuclear fusion and is currently the most promising approach for generating useful energy from fusion.
\index{Magnetic confinement fusion}

\section{Magnetic confinement fusion: tokamaks and stellarators}
\label{sec:intro:MCF}

\begin{figure}[h!]
	\centering
	\includegraphics[trim={0mm 0mm 0mm 0mm}, clip]{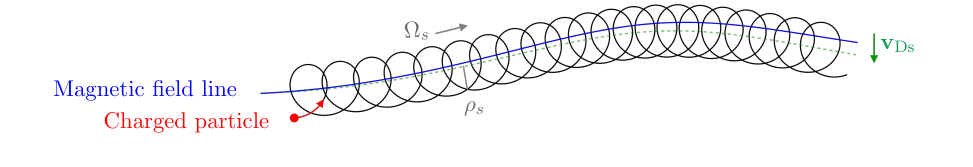}
	\captionsetup{singlelinecheck=false, margin=0.0cm}\vspace{-11.5mm}
	\caption{In the presence of an inhomogeneous magnetic field charged particles (red) gyrate around magnetic field lines (blue) with a gyrofrequency $\Omega_s$ and gyroradius $\rho_s$, while slowly drifting across the magnetic field lines with a velocity $\mathbf{v}_\text{Ds}$ (green) if the Larmor radius $\rho_s$ is small.} \vspace{-0mm}
	\label{fig:gyration}
\end{figure}

In magnetic confinement fusion, the charged particles which make up the plasma are confined by strong magnetic fields. These charged particles (ions and electrons) have a specific mass $m_s$, temperature $T_s$ (in energy units), charge number $Z_s$ and thermal velocity \mbox{$v_{\text{th},s} = \sqrt{2T_s/m_s}$}, where the subindex refers to the particle species, with $s=e$ for electrons and $s=i$ for ions. In the presence of a magnetic field, they experience a force called the \textit{Lorentz force}, given by,
\begin{equation}
	\mathbf{F} = Z_s e\,(\mathbf{v} \times \mathbf{B}),
\end{equation}
with $e$ the proton charge, $\mathbf{v}$ the velocity of the particle and $\mathbf{B}$ the magnetic field.	If the velocity of the charged particles has a component perpendicular to the magnetic field, $v_\perp$, and if the magnetic field is uniform and all other forces are absent, then the Lorentz force causes a circular motion of the particles around the magnetic field lines, called the \textit{gyromotion} or \textit{gyration}. The radius of this circular motion, known as the \textit{gyroradius} or \textit{Larmor radius}, is given by $\rho_s = v_{\text{th},s}/|\Omega_s|$, with $\Omega_s = Z_s e B/m_s$ the \textit{gyrofrequency} and $B = |\mathbf{B}|$ the magnetic field strength. The stronger the magnetic field, the tighter this gyration becomes, keeping the particles closer to the field lines. Moreover, note that ions gyrate in slightly larger circles and in the opposite direction to electrons, in the plane perpendicular to the magnetic field. On the other hand, if the velocity of the charged particles has a component parallel to the magnetic field, $v_\parallel$, then the particles move parallel to the magnetic field lines. As a result of both velocity components, the particle's trajectory forms a helix, as illustrated by the black line in figure \ref{fig:gyration}.
\index{Gyromotion or gyration}
\index{Gyroradius or Larmor radius}
\index{Gyrofrequency}

\vspace{0mm}

Parallel to the magnetic field lines, charged particles are able to move freely. In a simple cylindrical or linear configuration, magnetic field lines are open, leading to plasma escaping through the ends. Therefore, the magnetic field lines have to be arranged in closed loops or they have to trace out a closed surface to prevent end losses. However, in doing so, the magnetic field is no longer homogeneous, since the magnetic field lines are bent and the magnetic field strength typically varies spatially. As a result, both the the curvature of the magnetic field lines ($\boldsymbol{\kappa} = \hatb \cdot \nablam \hatb$ with $\smash{\hatb} = \mathbf{B}/B$ the unit vector along the magnetic field) and the gradient of the magnetic field strength ($\nablammm B$) affect the motion of the charged particles.
In strongly magnetized plasmas, where the Larmor radius is much smaller than the characteristic macroscopic length scale of the system ($\rho_s/L \ll 1$), charged particles are affected by the inhomogeneity of the magnetic field in a perturbative manner. This allows their behavior to be analyzed using perturbation theory, by expanding the particle dynamics in terms of a small parameter $\rho_{*s} = \rho_s/L$. Under these assumptions, it can be shown that  charged particles slowly drift across field lines with a drift \mbox{velocity $\mathbf{v}_\text{Ds}$}, defined as,
	\begin{equation}
		\label{eq:MagneticDriftsIntro}
		\mathbf{v}_{\text{D}s} = \mathbf{v}_{\boldsymbol{\kappa},s} + \mathbf{v}_{\nablammm B, s} = \frac{m_s v_\parallel^2}{Z_s e}  \frac{\mathbf{B} \times  \boldsymbol{\kappa}}{B^2} +  \frac{1}{2} \frac{m_s v_\perp^2}{Z_s e} \, \frac{\mathbf{B} \times  \nablammm B}{B^3}.
	\end{equation}
As a result, the motion of charged particles can be decomposed in the \textit{gyromotion}, which is the circular motion around magnetic field lines, and the \textit{guiding center motion}, which is the average trajectory of the particle's center of gyration, illustrated by \mbox{the green dashed line in figure \ref{fig:gyration}.}
\index{Gyromotion or gyration}
\index{Guiding center motion}

\begin{figure}[b!]
	\vspace{-4mm}
	\centering
	\hspace*{-4mm}
	\includegraphics{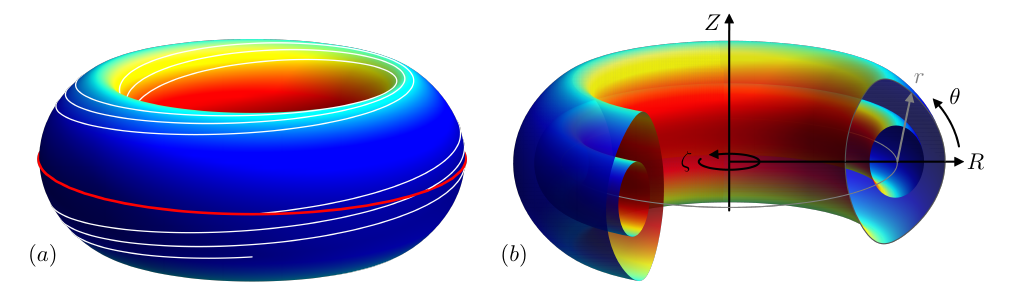}
	\captionsetup{singlelinecheck=false,  margin=0cm}\vspace{-6.5mm}
	\caption{(a) Toroidal magnetic field consisting of circular  (red) or twisted (white) magnetic field lines.  (b) Nested magnetic flux surfaces of a toroidal magnetic field illustrating the cylindrical coordinates $(R,Z,\zeta)$, with $\zeta$ the toroidal angle. Additionally, the poloidal angle $\theta$ and radial coordinate $r$ are indicated. The magnetic field strength is represented by the color scale (red is maximum, blue is minium). Note that shaped (non-circular) magnetic coils are required to generate the toroidal magnetic field in this figure. } \vspace{-4mm}
	\label{fig:AUGIntroduction}
\end{figure}

Naively, one could think that, in order to confine particles, it is sufficient to bend magnetic field lines into circles and to construct a toroidal or donut-shaped magnetic field, as illustrated in figure \ref{fig:AUGIntroduction}a, where the red magnetic field line closes onto itself. To generate this magnetic field, which is purely toroidal, circular magnetic coils can be arranged side by side to form a torus. This \mbox{configuration} naturally creates an inhomogeneous magnetic field that is stronger at the center of the torus (inboard side) than on the outside of the torus (outboard side), as highlighted by the color scale in figure \ref{fig:AUGIntroduction}a. Both the curvature vector and $\nablammm B$ vector point towards the center of the red circle, while the magnetic field vector $\mathbf{B}$ is tangent to the magnetic field lines. For ions ($Z_s>0$) the magnetic drifts (eq.~\ref{eq:MagneticDriftsIntro}) always point downwards if the magnetic \mbox{field is directed clockwise} when viewed from above. As a result, when considering circular magnetic field lines, the ions will slowly drift downwards and the electrons will slowly drift upwards, and the plasma is not confined.

To mitigate this loss, a twist must be introduced into the magnetic field lines, allowing them to explore the entire poloidal cross-section of the torus, as illustrated by the white magnetic field lines in figure \ref{fig:AUGIntroduction}a. This is achieved by adding a poloidal component to the magnetic field. A measure of the twisting of magnetic field lines is the \textit{safety factor}, $q = \psi'_t/\psi'_p$, or \textit{rotational transform}, $\iota = \psi'_p/\psi'_t$. Here $2\pi\psi_p$ and $2\pi\psi_t$ represent the poloidal and toroidal \textit{magnetic fluxes}, which quantify how much magnetic field passes through a given poloidal or toroidal surface, respectively, and primes denote derivatives with respect to the radial coordinate. These field lines trace out closed surfaces, known as \textit{magnetic flux surfaces}, which are assumed to be nested. At the very center of these nested surfaces lies the innermost magnetic flux surface, which collapses to a single line known as the \textit{magnetic axis} (represented by the gray circle in figure \ref{fig:AUGIntroduction}b). The outermost layer of these nested surfaces is known as the \textit{last closed flux surface} (LCFS), which defines the effective boundary of the confined plasma region. Beyond the LCFS, magnetic flux surfaces are no longer closed. These magnetic flux surfaces are parameterized by a radial coordinate $r$ \mbox{or flux-surface label $\psi$.}
\index{Safety factor}
\index{Rotational transform}
\index{Magnetic flux surfaces}
\index{Magnetic flux}
\index{Magnetic axis}
\index{Last closed flux surface (LCFS)}

\begin{figure}[h!]
	\centering
	\includegraphics{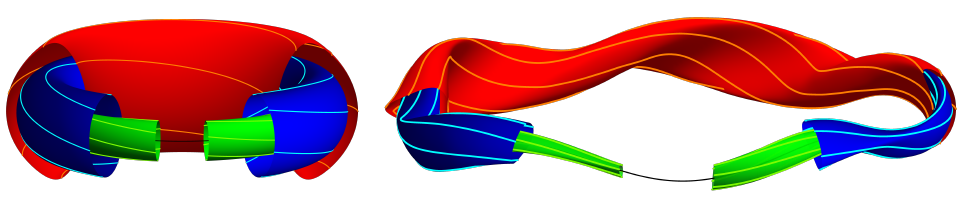}
	\captionsetup{singlelinecheck=false,  margin=0cm}\vspace{-8mm}
	\caption{Magnetic field lines on three different nested magnetic flux surfaces for two specific designs of magnetic confinement fusion devices: a generic tokamak (left) and the W7-X stellarator (right).}
	\label{fig:NestedFFSAUGandW7X}
\end{figure}

According to Spitzer \cite{spitzer1958stellarator}, there are three ways to twist the magnetic field: (i) creating a
poloidal magnetic field by driving a toroidal current in the plasma; (ii) rotating the poloidal cross-section of stretched flux surfaces around the torus; (iii) making the magnetic axis non-planar. While \textit{tokamaks} use the first approach, as illustrated in figure \ref{fig:AUGIntroduction}, \textit{stellarators} usually rely on the latter two methods \mbox{(see figure \ref{fig:NestedFFSAUGandW7X})}. Because tokamaks use a large toroidal current to generate the poloidal field required for confinement, it makes them susceptible to current-driven instabilities and challenges steady-state operation.
In contrast, stellarators are non-axisymmetric and typically rely on intricately shaped, three-dimensional magnetic field coils to twist the magnetic field and confine the plasma without requiring a large plasma current. As a result, stellarators are inherently free from current-driven \mbox{instabilities and can operate in a steady state.}

\vspace{1mm}
\FloatBarrier
\vspace{1mm}

In order to characterize the plasma confinement in more detail, it is essential to categorize the charged particles in the plasma into two distinct groups: \textit{passing particles} and \textit{trapped particles}. Passing particles are those that travel primarily along magnetic field lines, completing full toroidal and poloidal loops around the device without becoming confined to a specific region. In contrast, trapped particles oscillate back and forth between two points in the magnetic field. This trapping occurs due to the conservation of the particle's magnetic moment, $\mu = v_\perp^2/2B$, defined per unit of mass---a principle that holds for both trapped and passing particles. As a particle moves into regions of higher magnetic field strength (e.g.~on the inboard side of the torus in figure \ref{fig:AUGIntroduction}), its perpendicular velocity increases in order to preserve its magnetic moment. This increase of the perpendicular velocity occurs at the expense of its parallel velocity, to ensure that the particle's energy remains unchanged. If the parallel velocity becomes zero, the particle reverses direction, creating a bounce motion between two reflection points, known as \textit{bounce points} or \textit{mirror points}. Trapped particles are thus confined to regions where the magnetic field strength reaches a minimum, called \textit{magnetic wells}, corresponding to the outboard side of the torus in figure \ref{fig:AUGIntroduction}. Seeing that trapped particles are confined to a specific region of the fusion device, they are more prone to be influenced by localized magnetic inhomogeneities. As a result, trapped particles typically have worse confinement properties than passing particles.
\index{Trapped and passing particles}
\index{Bounce points or mirror points}
\index{Magnetic wells}

In an axisymmetric magnetic field with twisted field lines, passing particles which explore the entire poloidal cross-section, will spend half of their time in the upper half of the device and the other half in the lower half of the device. As a result, their radial drift due to the curvature and $\nablammm B$ effects (eq.~\ref{eq:MagneticDriftsIntro}) alternates: in a clockwise configuration ions drift radially inwards in the upper half of the device and radially outwards in the lower half of the device, while electrons drift in the opposite direction. On average, these opposing radial drifts cancel out over a complete poloidal cycle, confining passing particles to a specific flux surface in the absence of collisions. Moreover, one can prove that in axisymmetric configurations the radial drift vanishes on average also for trapped particles.
In contrast, non-axisymmetric devices like stellarators lack toroidal symmetry, leading to more complex particle trajectories and drift patterns. This asymmetry can disrupt the cancellation of orbit-averaged radial drifts, resulting in net radial transport. Note that passing particles are well confined, but, in a generic stellarator, trapped particles are not \cite{helander2014theory}. To address this, modern stellarator designs use optimization techniques to improve the plasma confinement by minimizing radial drifts. Achieving comparable confinement to tokamaks is an ongoing area of innovation in stellarator design.
Moreover, there are many additional factors which can influence the plasma confinement, such as collisions, magnetic inhomogeneities, particle drifts, turbulent fluctuations, etc., leading to \mbox{radial transport of particles and heat.}

\begin{figure}[h!]
	\centering
	\includegraphics[trim={0mm 10mm 0mm 0mm}, clip] {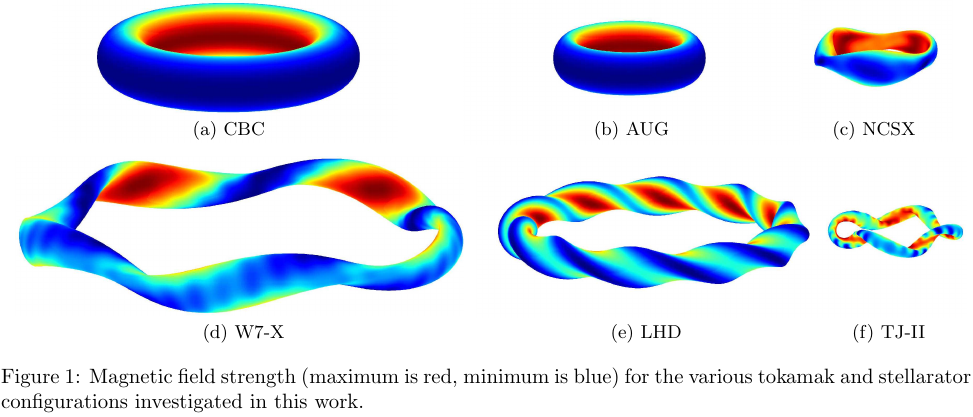}
	\captionsetup{singlelinecheck=false,  margin=0cm}\vspace{-7mm}
	\caption{Magnetic field strength (maximum is red, minimum is blue) for the various tokamak and stellarator configurations investigated in this work.} \vspace{-0mm}
	\label{fig:fluxsurfacesintro}
\end{figure}

In this work, the ASDEX Upgrade tokamak (AUG) and the tokamak geometry of the Cyclone Base Case (CBC) \cite{dimits2000simulation} are investigated, as well as various stellarators such as Wendelstein 7-X \mbox{(W7-X)}, the Large Helical device (LHD), the National Compact Stellarator Experiment (NCSX) and TJ-II. The tokamak geometry of CBC is defined by specifying a set of Miller parameters that are used to construct a local solution to the Grad--Shafranov equation \cite{miller1998noncircular}, whereas the magnetic equilibria of the other 5 devices are generated by the three-dimensional MHD equilibium code \mbox{\texttt{VMEC} \cite{hirshman1983steepest}.} The magnetic field strength on the last closed flux surfaces is illustrated in \mbox{figure \ref{fig:fluxsurfacesintro}.}

\vspace{4mm}
\vspace{0mm}

\section{Classical, Neoclassical and Turbulent transport}

To achieve magnetic confinement fusion, it is crucial to understand and control the transport of particles and heat. There are three primary mechanisms responsible for the loss of particles and heat: classical, neoclassical, and turbulent transport.
Collisions between particles can enable them to cross magnetic field lines, allowing, through multiple interactions, for a particle to eventually escape. This mechanism of particle transport---and the associated heat and energy transfer---across magnetic field lines is referred to as \textit{classical transport}.
While classical transport on its own is typically negligible in fusion devices, its impact is greatly intensified by the drifts resulting from the inhomogeneity of the magnetic field (curvature and $\nablammm B$ drifts). The transport due to the combined effect of collisions and drifts is \mbox{known as \textit{neoclassical transport}}.
\index{Classical transport}
\index{Neoclassical transport}

Finally, \textit{plasma turbulence} can lead to enhanced transport of particles and energy across magnetic surfaces. Specifically, small-scale perturbations of the plasma can be driven unstable by the free energy present in the temperature and density gradients of the plasma.
These instabilities generate fluctuations in the electric and magnetic fields, leading to a fluctuating $\mathbf{E} \times \mathbf{B}$ drift. The resulting cross-field motion is usually the dominant transport mechanism in tokamaks, since neoclassical transport is generally much smaller thanks to their axisymmetry. In contrast, stellarators, which lack axisymmetry, tend to experience significantly larger neoclassical transport. Nonetheless, in neoclassically optimized stellarators, neoclassical transport can be reduced to the point where  turbulent transport becomes the dominant transport mechanism. Therefore, turbulent transport ultimately becomes the limiting factor for confinement in both tokamaks and stellarators. As a result, understanding and mitigating turbulent transport is the \mbox{main focus of this dissertation.}
\index{Turbulent transport}

\section{Modeling of electrostatic microinstabilities in fusion plasmas}
\label{sec:intro:micro}

Magnetically confined plasmas may become unstable to perturbations with wavelengths comparable to the ion or electron gyroradius, $\rho_i$ or $\rho_e$, and frequencies smaller than the cyclotron frequency, $\Omega_i$ or $\Omega_e$. Due to their size, these instabilities are called \textit{microinstabilities} \cite{doyle2007plasma}, which encompasses a large variety of destabilizing mechanisms.
The primary concern regarding microinstabilities lies in the nonlinear or saturated state, as the fluctuating electric and magnetic fields give rise to a fluctuating $\mathbf{E} \times \mathbf{B}$ drift that generates turbulent transport across magnetic surfaces, which can significantly increase the particle and energy losses from the confined plasma volume.
\index{Microinstabilities}

To predict the contribution of turbulent transport, numerical codes are developed that are capable of simulating microturbulence. An accurate treatment of microinstabilities requires a kinetic model describing the evolution of the distribution function of each species in six dimensional phase space. The problem is simplified by averaging the trajectories of particles over the fast gyromotion around magnetic field lines, thus dropping the dependence on the gyrophase---which is the angle of the circular motion around magnetic field lines---and thereby reducing the phase space from six to five dimensions. This reduction forms the basis of \textit{gyrokinetic theory}, described in detail in chapter \ref{chap:gyrokinetics}, which enables the study of microinstabilities and the turbulence they drive. To further reduce the necessary computational power required to carry out a simulation, only a reduced simulation domain corresponding to a narrow plasma volume aligned with a magnetic field line is considered, known as the \textit{flux-tube approach} \cite{beer1995field}, which is discussed in detail in \mbox{chapter \ref{chap:fluxtube}.}  Significant contributions to the understanding of microinstabilities have been made possible thanks to this approximation. Finally, in chapter \ref{chap:stella} it is explained how the flux-tube gyrokinetic equations are implemented in the gyrokinetic code \texttt{stella} \cite{stella}.

When a perturbation---such as a small fluctuation in density or temperature---occurs in a plasma, it typically propagates through the plasma in the form of a wave. For example, just as ordinary sound waves travel through a gas by transmitting vibrations via collisions between neutral particles, \textit{ion acoustic waves} propagate through a plasma due to electrostatic interactions between charged ions.
\index{Ion acoustic waves}
If no net energy is exchanged between the wave and plasma, the wave remains purely oscillatory. However, if the wave absorbs energy from the plasma, its amplitude can grow and it can develop into an instability, conversely, if the wave loses \mbox{energy to the plasma, it becomes damped.}

Any periodic motion of the plasma can be decomposed using Fourier analysis into a superposition of sinusoidal oscillations, each with a distinct frequency $\omega$ and wavevector $\mathbf{k}$, of the form $\exp(\ir \mathbf{k} \cdot \mathbf{r} - \ir \omega t)$, with $\mathbf{r}$ the particle's position. Each individual sine or cosine wave---defined by its specific frequency, amplitude, phase and wavevector---is called a \textit{mode}. The overall periodic behavior of the plasma is thus the sum of these modes, which can be described by gyrokinetic theory. The complex frequency $\omega = \omega_r + \ir\gamma$ consists of a real part $\omega_r$, representing the \textit{oscillation frequency}, and an imaginary part $\gamma$, which denotes the \textit{growth rate}. A positive growth rate indicates that the mode is unstable and grows over time, conversely, the mode is damped if the growth rate is negative. Generally, the frequencies of these modes need to be determined numerically due to the complexity of the system. Nonetheless, in simplified magnetic geometries or considering specific assumptions, these modes can be studied analytically. The mathematical equation that describes the relation between the frequency and the wavevector of a plasma wave (or mode) is the so-called \textit{dispersion relation}, $\omega(\mathbf{k})$, which characterizes the waves that can exist in the plasma.

For a plasma perturbation (or mode) to grow into an instability, there must be a source of free energy and a mechanism to extract and utilize that energy. The microinstabilities responsible for driving turbulence in plasmas can be categorized based on the type of free energy they access, the mechanisms by which they extract it, and the magnetic geometry of the system. While there are various ways to excite instabilities in plasmas, the turbulent fluctuations observed in magnetically confined plasmas are believed to be driven primarily by the free energy stored in the density and temperature gradients. One pathway for a mode to absorb this energy is through a \textit{feedback loop} involving the initial perturbation, the charge separation caused by the relative magnetic drifts between ions and electrons and the resulting $\mathbf{E}\times\mathbf{B}$ drift. This $\mathbf{E}\times\mathbf{B}$ drift can reinforce the original perturbation, which amplifies the perturbation, giving rise to an instability. If instead, the $\mathbf{E}\times\mathbf{B}$ drift opposes the original perturbation, the perturbation is damped. Another pathway for the free energy stored in the plasma gradients to be tapped is through \mbox{\textit{wave-particle resonant interactions}}, where plasma waves resonate with particle motions, enabling a sustained energy exchange between the waves and the plasma. Resonance occurs when the natural motion of particles---such as the bounce motion or drift motion---approximately matches the frequency or phase velocity of a wave. Under these conditions, the particles can interact with the wave over an extended time, allowing efficient transfer of energy and momentum.  If energy is transferred from the plasma to the wave, the perturbation is amplified, leading to the growth of small-scale instabilities. These microinstabilities, in turn, can cause turbulent transport of heat and particles, degrading \mbox{plasma confinement.}

An important factor to consider when studying microinstabilities is the gyromotion of charged particles perpendicular to the magnetic field (figure \ref{fig:gyration}). Although gyrokinetic theory averages over this fast gyromotion---eliminating the dependence on the gyrophase---the resulting equations still retain its influence. Specifically, perturbations with a small scale perpendicular to the magnetic field (large $k_\perp\rho_s$) are typically damped because the particles ``average out'' the perturbation over their gyro-orbit. This \textit{gyro-averaging} effectively smooths out the effect of the perturbation on the particle's motion. As a result, the plasma responds less effectively to the perturbation. The lack of a coherent response from the particles reduces the energy transfer between the perturbation and the plasma, inhibiting the growth of the perturbation. Therefore, these so-called \textit{finite-Larmor-radius (FLR) effects} can act as a damping mechanism, \mbox{suppressing small-scale instabilities.}

Moreover, it is important to point out that \textit{magnetic shear}---which is a measure of how the twist of magnetic field lines changes between flux surfaces---can have a significant effect on microinstabilities. As a perturbation moves along a field line, it is tilted in the plane perpendicular to the magnetic field due to magnetic shear. As a result, the perpendicular scale of the perturbation ($k_\perp$) changes as the perturbation propagates along a field line. If the magnetic shear is very strong, it is capable of shearing the perturbation into smaller scale fluctuations perpendicular to the magnetic field which are consequently damped by the gyro-average or FLR effects. Additionally, the distance between flux surfaces and  between field lines can vary throughout the device. Consequently, $k_\perp$ depends not only on magnetic shear but also on flux-surface and field-line compression.

In the absence of collisions and magnetic perturbations, turbulent transport is thought to be primarily driven by \textit{ion-temperature-gradient-driven modes} (ITG), \textit{electron-temperature-gradient-driven modes} (ETG), and \textit{trapped-electron modes} (TEM). Additionally, other instabilities, such as the \textit{passing-particle-driven universal instability}, generally play a subdominant role compared to these modes. However, they can become dominant in configurations that are especially stable against trapped-electron modes or temperature-gradient-driven modes. The following sections provide an overview of these analytical instabilities---considering peaked density and temperature profiles, i.e.~$dn_s/dr<0$ and $dT_s/dr<0$---which are accurate only under specific limits and assumptions. In realistic plasmas, where the kinetic response of both ions and electrons is considered, alongside the full plasma profiles, turbulence arises from a complex interplay of stabilizing and destabilizing effects, making it challenging to attribute the resulting behavior to specific, analytically tractable instabilities. Furthermore, the analytic instabilities can transition into one another through continuous changes of the plasma parameters \cite{kammerer2008exceptional}. Therefore, under realistic conditions, the identification of the instabilities should be done very carefully or be \mbox{avoided altogether.}

\begin{figure}[b!]
	\centering
	\vspace{2mm}
	\includegraphics[trim={0mm 30mm 0mm 5mm}, clip] {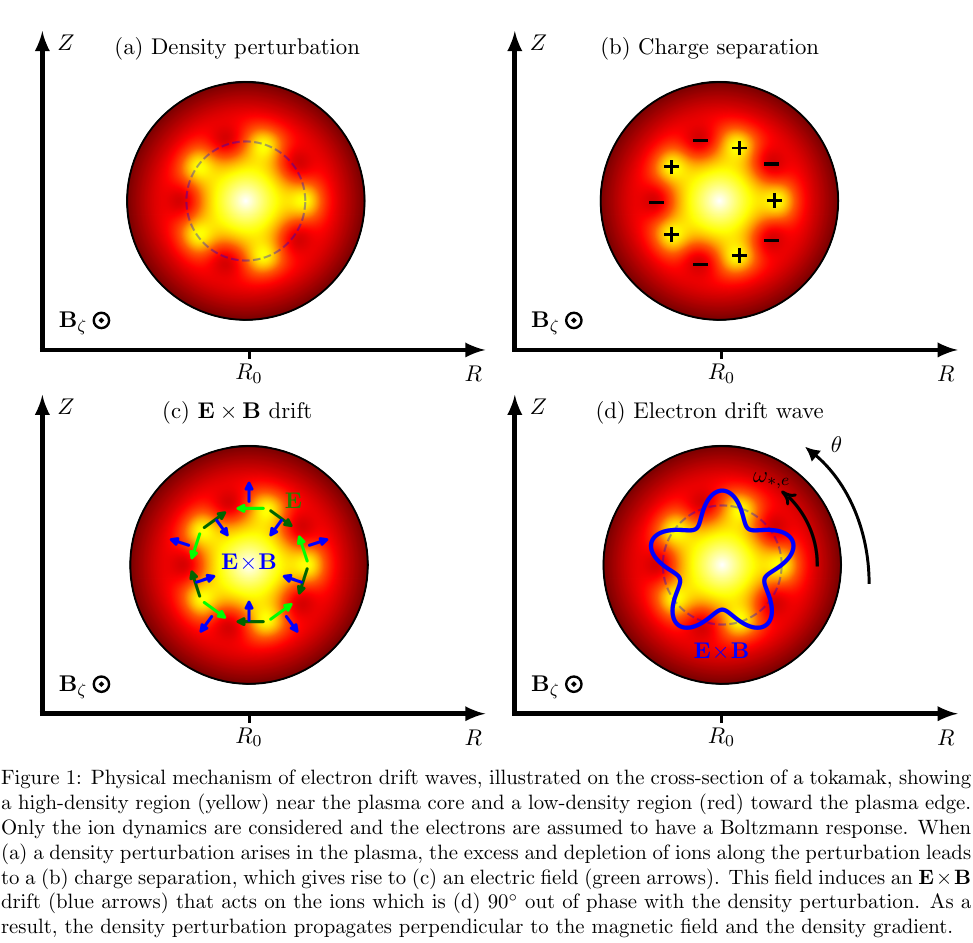}
	\vspace{-8mm}
	\caption{Physical mechanism of electron drift waves, illustrated on the cross-section of a tokamak, showing a high-density region (yellow) near the plasma core and a low-density region (red) toward the plasma edge. Only the ion dynamics are considered and the electrons are assumed to have a Boltzmann response. When (a) a density perturbation arises in the plasma, the excess and depletion of ions along the perturbation leads to a (b) charge separation, which gives rise to (c) an electric field (green arrows). This field induces an $\mathbf{E} \times \mathbf{B}$ drift (blue arrows) that acts on the ions which is (d) 90$^\circ$ out of phase with the density perturbation. As a result, the density perturbation propagates perpendicular to the magnetic field and the density gradient.} \vspace{-2mm}
	\label{fig:driftwave}
\end{figure}

\subsection{Electron drift wave}

In general, microinstabilities in plasmas are associated with drift waves that are driven unstable under specific conditions \cite{horton1999drift, chen2016introduction, goldston1995introduction}. The fundamental driving force behind the drift wave instability is the pressure gradient, which gives rise to collective oscillations of the plasma, known as \textit{drift waves}. These waves propagate perpendicular to both the magnetic field and the pressure gradients, with a phase velocity corresponding to the \textit{diamagnetic drift velocity}, defined as,
\begin{equation}
	\label{eq:omegastarsv}
	\mathbf{v}_{*s} = - \frac{\nablamm p \times \mathbf{B}}{Z_s e n_s B^2}.
\end{equation}
Specifically, density or temperature perturbations characterized by a wavevector $\mathbf{k}_\perp$ perpendicular to the magnetic field, propagate through the plasma at the diamagnetic drift velocity as drift waves. This allows us to define the \textit{diamagnetic frequency} $\omega_{*s}$ related to a density gradient as,
\begin{equation}
	\label{eq:omegastars}
	\omega_{*s} = \mathbf{k}_\perp \cdot \mathbf{v}_{*s} = \frac{T_s k_\alpha}{Z_s e \psi'_t} \frac{n'_s}{n_s},
\end{equation}
where primes denote derivatives with respect to the radial coordinate, $\alpha$ is a field-line label and $k_\alpha$ characterizes the spatial scale of the perturbation across field lines---i.e.~in the so-called \textit{binormal direction}, which is perpendicular to the magnetic field and perpendicular to the radial direction. Similarly, we define the diamagnetic frequency associated with a temperature gradient as,
\begin{equation}
	\label{eq:omegastars2}
	\eta_s \omega_{*s} = \frac{k_\alpha}{Z_s e \psi'_t} \, T'_s,
\end{equation}
with $\eta_s$ the ratio of the temperature and density gradient, and $L_{n_s}$ and $L_{T_s}$ the density and temperature \textit{gradient scale lengths}, defined through $1/L_{n_s} = - (\ln n_s)'$ and $1/L_{T_s} = - (\ln T_s)'$, i.e.
\index{Gradient scale lengths $L_{n_s}$ and $L_{T_s}$}
\begin{equation}
	\label{eq:etai}
	\eta_s = \frac{(\ln T_s)'}{(\ln n_s)'} = \frac{1}{T_s} \frac{dT_s}{dr} \Big( \frac{1}{n_s} \frac{dn_s}{dr} \Big)^{-1} \!\! = \frac{L_{n_s}}{L_{T_s}}.
\end{equation}
When studying microinstabilities, the diamagnetic frequencies associated with the density gradient and temperature gradient typically appear together as a function of the particle's energy as,
\begin{equation}
	\label{eq:omegstarT}
	\omega_{*s}^T = \omega_{*s} \Big[ 1 + \eta_s \Big( \frac{m_s v^2}{T_s} - \frac{3}{2} \Big) \Big].
\end{equation}

The physical mechanism of the \textit{electron drift wave} is depicted in figure \ref{fig:driftwave}. For simplicity only a finite density gradient is considered. The analysis focuses solely on the non-adiabatic response of the ions, while electrons are assumed to follow a \textit{Boltzmann response}, because they are free to move along the magnetic field lines and their response time is much faster than that of ions. As a result, electrons can quickly adjust to maintain quasineutrality in response to fluctuations of the ion density. Mathematically, the electron density perturbation $\delta n_e$ approximately follows the potential fluctuations $\varphi$ as, \mbox{$\delta n_e/n_0 \approx e\varphi/T_e$}.
When a density perturbation arises in the plasma, the resulting excess or depletion of ions along the perturbation leads to a charge separation, which gives rise to an electric field $\mathbf{E} = - \nablam \varphi$ (indicated by the green arrows in figure \ref{fig:driftwave}c). This electric field, in turn, produces an $\mathbf{E} \times \mathbf{B}$ drift (blue arrows) that acts on the perturbation of the ions. In the absence of a phase shift, the density and potential perturbations are exactly 90$^\circ$ out of phase, as illustrated in figure \ref{fig:driftwave}d, and the plasma wave associated with the perturbation is purely oscillatory.   Consequently, the density perturbation propagates perpendicular to the magnetic field and the density gradient, with the electron diamagnetic drift frequency $\omega_{*e}$ (eq.~\ref{eq:omegastars}).

However, if the density and potential perturbations are not exactly 90$^\circ$ out of phase, the resulting $\mathbf{E} \times \mathbf{B}$ drift can either reinforce or oppose the perturbation. If the drift is directed outward where the plasma has already been shifted outward, and inward where it has been shifted inward, the perturbation grows, leading to instability. Conversely, if the drift opposes the displacement, the perturbation is damped. Therefore, the phase shift between the density and potential perturbations determines whether the drift wave remains oscillatory, becomes unstable,  or is damped.

To summarize, \textit{electron drift waves} correspond to collective oscillations of the plasma occurring at the electron diamagnetic drift frequency $\omega_{*e}$.  We typically assume that the ion and electron density profiles are equal, in this case the ion diamagnetic frequency is given by $\omega_{*i} = -(T_i/T_e)\,\omega_{*e}$. As a result, plasma waves characterized by a frequency $\omega_{*i}$ propagate in the opposite direction as electron drift waves. Moreover, note that a temperature gradient gives rise to similar drift waves.

\subsection{Temperature-gradient-driven modes: ITG and ETG modes}
\label{sec:ITG}

When an ion temperature gradient exists in the plasma, the free energy stored in this gradient can drive both drift waves and ion acoustic waves unstable. The resulting instabilities are the so-called \textit{ion-temperature-gradient-driven modes} or ITG modes \cite{coppi1967instabilities, biglari1989toroidal, plunk2014collisionless}. The theory of the ITG mode is developed considering the response of ions to an ion temperature gradient, while the electrons are modeled with a Boltzmann response. By applying the gyrokinetic theory, an expression for the dispersion relation of ITG modes can be derived. This dispersion relation shows that the complex frequency $\omega$ of a plasma wave or mode, depends on both the perpendicular and parallel wavenumbers, $k_\perp$ and $k_\parallel$, as well as on a complex interplay of various physical effects.

First of all, the complex frequency $\omega$ of the mode depends on the velocity-dependent ion diamagnetic frequency $\omega^T_{*i}$ (eq.~\ref{eq:omegstarT}). Consequently, the mode is influenced by the ion density gradient and  ion temperature gradient, which  appear explicitly in $\omega_{*i}$ and $\eta_i\omega_{*i}$, respectively. The ITG mode is also known as the $\eta_i$ mode, since it is generally destabilized by the ion temperature gradient and stabilized by the density gradient, making $\eta_i$ the relevant stability parameter. Additionally, the toroidal geometry of the magnetic field---i.e.~the curvature $\boldsymbol{\kappa}$ of the field lines and the magnetic field inhomogeneity $\nablammm B$---can play a significant role, which is reflected by the dependence of $\omega(\mathbf{k})$ on the magnetic drift frequency $\omega_{\text{D}i} = \mathbf{k}_\perp \cdot \mathbf{v}_{\text{D}i}$ (eq.~\ref{eq:MagneticDriftsIntro}). The modes also depend on the ion transit frequency $\omega_{\parallel i} = k_\parallel v_{\text{th},i}$, which characterizes how fast ions move along magnetic field lines due to their thermal motion. Finally, the ion-to-electron temperature ratio $T_i/T_e$, magnetic shear, flux surface compression $\nablam r$, field line compression $\nablam \alpha$, and others, can also impact the mode behavior.

The stability of the ITG mode clearly depends on a wide range of factors. To analyze these microinstabilities analytically, the theoretical framework is simplified. Specifically, the ITG mode is typically split up into two branches: the slab branch and the toroidal branch.
If the magnetic drift frequency is negligible ($\omega_{\text{D}i}\ll\omega$) the \textit{slab branch} is obtained \cite{coppi1967instabilities, cowley1991considerations}. This is valid, for example, in a slab magnetic geometry where a uniform and straight magnetic field is considered, hence $\boldsymbol{\kappa}=0$ and $\nablammm B=0$, which yields $\omega_{\text{D}i}=0$. If the thermal velocity $v_{\text{th},i}$ of the ions becomes comparable to the phase velocity of the waves, i.e.~if $\omega \approx k_\parallel v_{\text{th},i}$, the wave-particle interactions lead to \textit{Landau damping} or \textit{Landau resonance}---named after its discoverer L.\,D.\,Landau \cite{landau1946vibrations}---which can either stabilize of destabilize the mode depending on the direction of the energy transfer. Alternatively, if the wave-particle resonances can be ignored, i.e.~if $k_\parallel v_{\text{th},i} \ll \omega$, the slab ITG mode can be seen as an ion acoustic wave, characterized by a frequency $\omega_{\parallel i} = k_\parallel v_{\text{th},i}$, which is driven unstable by the ion temperature gradient through a feedback loop between the initial perturbation, a charge separation and the resulting $\mathbf{E} \times \mathbf{B}$ drift.

On the other hand, if the parallel drift of the ions is negligible ($k_\parallel v_{\text{th},i} \ll \omega$), then the toroidal geometry of the magnetic field, captured by the magnetic drift frequency $\omega_{\text{D}i}$, plays a dominant role in driving the ITG mode unstable, corresponding to the \textit{toroidal branch}.  Similar to the slab ITG mode, a toroidal ITG mode can be stabilized or destabilized by wave-particles resonances if the phase velocity of the mode becomes comparable to the ion magnetic drift, i.e.~if $\omega \approx \omega_{\text{D}i}$. Alternatively, in the non-resonant limit ($\omega_{\text{D}i}\ll \omega$) the toroidal ITG mode is driven unstable by a feedback loop  between the initial perturbation, a charge separation and the resulting \mbox{$\mathbf{E} \times \mathbf{B}$} drift. Moreover, Biglari et al.~has shown in Ref.~\cite{biglari1989toroidal} that instability requires $\eta_i \omega_{*i} \omega_{\text{D}i} > 0$ in the non-resonant limit.
Considering the usual orientation of the plasma gradients ($\eta_i>0$) this means that only in regions of the fusion device where the ion magnetic drift, $\omega_{\text{D}i}$, is aligned with the ion diamagnetic drift, $\omega_{*i}$, a perturbation is amplified, leading to an unstable non-resonant toroidal ITG mode. This instability requirement related to the magnetic drift frequency defines the so-called \textit{``bad-curvature'' regions}. In a low-$\beta$ magnetohydrodynamic equilibrium, with $\beta$ the ratio of the plasma pressure to the magnetic pressure, we can approximate $\boldsymbol{\kappa} \approx (1/B) \nablammm_\perp B$. As a result, the bad-curvature regions, corresponding to $\eta_i \omega_{*i} \omega_{\text{D}i} > 0$, are defined as,
\begin{equation}
	\label{eq:badCurvatureIntro}
	\frac{k_\alpha T'_i}{\psi'_t} (\mathbf{B} \times \nablammm B) \cdot \mathbf{k}_\perp > 0. \vspace{-1mm}
\end{equation}
The bad-curvature regions thus depend on the magnetic curvature and magnetic field inhomogeneity, despite the fact that the name only refers to the curvature. Moreover, seeing that the magnetic drift frequency depends on the perpendicular wavenumber $\mathbf{k}_\perp$, the bad-curvature regions depend on the scale of the perturbation perpendicular to the magnetic field. The perpendicular scale $k_\perp$ can be influenced by magnetic shear, flux surface compression $\nablam r$ and field line compression $\nablam \alpha$. Therefore, despite the name, the bad-curvature regions are not a purely geometric quantity, since they also depend on the perpendicular characteristics of the mode. Finally, it is important to note that the perpendicular compression of ions due to $\omega_{\text{D}i}$ has a stabilizing influence on the instability. As a result, if the ion magnetic drift is too large, it will stabilize the toroidal ITG mode \cite{biglari1989toroidal}.

\begin{figure}[b!]
	\centering
	\vspace{-5mm}
	\includegraphics[clip, width=0.99\linewidth, trim={0mm 36mm 0mm 2mm}] {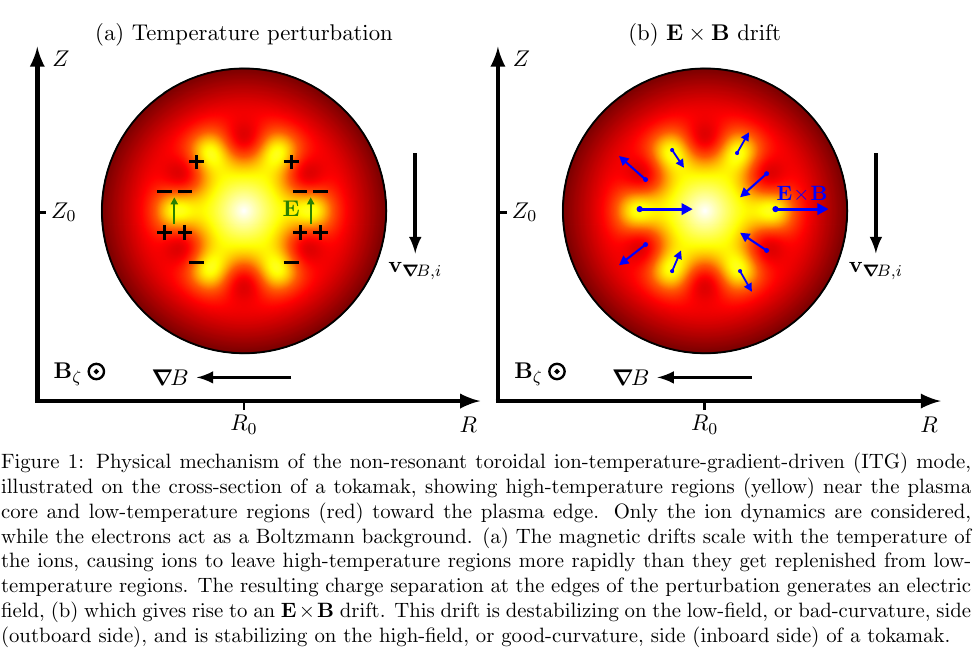}  \vspace{-2mm}
	\caption{Physical mechanism of the non-resonant toroidal ion-temperature-gradient-driven (ITG) mode, illustrated on the cross-section of a tokamak, showing high-temperature regions (yellow) near the plasma core and low-temperature regions (red) toward the plasma edge. Only the ion dynamics are considered, while the electrons act as a Boltzmann background. (a) The magnetic drifts scale with the temperature of the ions, causing ions to leave high-temperature regions more rapidly than they get replenished from low-temperature regions.  The resulting charge separation at the edges of the perturbation generates an electric field, (b) which gives rise to an $\mathbf{E} \times \mathbf{B}$ drift. This drift is destabilizing on the low-field, or bad-curvature, side (outboard side), and is stabilizing on the high-field, or good-curvature, side (inboard side) of a tokamak.}  \vspace{-3mm}
	\label{fig:perturbation}
\end{figure}

In tokamaks, the bad-curvature regions are found on the low-field (or outboard) side of the device. In contrast, in stellarators, the bad-curvature regions are not necessarily confined to the outboard side, nor do they have to overlap with minima of the magnetic field strength. The location of these regions in stellarators is thus less straight-forward than in tokamaks due to their \mbox{complex 3D geometry.} Note that in tokamaks and stellarators, both the slab drive and the toroidal drive are typically present, resulting in mixed slab-toroidal ITG modes. Nonetheless, the toroidal ITG mode generally dominates over the slab ITG mode at \mbox{sufficiently large $k_\alpha$ \cite{plunk2014collisionless}.}

The physical mechanism that drives perturbations of the ion temperature unstable due to a feedback loop driven by the ion magnetic drift is illustrated in figure \ref{fig:perturbation}. To study this instability, note that the curvature and $\nablammm B$ drifts (eq.~\ref{eq:MagneticDriftsIntro}) depend on the parallel and perpendicular kinetic energy of the particles, respectively, and the drifts thus scale with the particle's temperature. Moreover,  the $\boldsymbol{\kappa}$ and $\nablammm B$ vectors point toward the center of the torus $(R=0)$ in a tokamak with circular poloidal cross-sections. As a result, the combined $\nablammm B$-curvature drift of the ions is directed downwards in a clockwise magnetic field, as illustrated in figure \ref{fig:perturbation}. In the presence of a temperature perturbation, there are regions with higher temperatures, in which ions have larger magnetic drift velocities than those located in regions with lower temperatures. As a result, ions leave regions of high temperature more rapidly than they are replenished from the regions of low temperature, which results in a charge separation.  This charge separation generates an electric field between the boundaries of the perturbation, which in turn gives rise to an $\mathbf{E} \times \mathbf{B}$ drift, as indicated by the blue arrows in figure \ref{fig:perturbation}b. On the low-field or outboard side (corresponding to the right side in figure \ref{fig:perturbation}b) the $\mathbf{E} \times \mathbf{B}$ drift is directed outward where the plasma has already been shifted outward, enhancing the perturbation. In contrast, on the high-field or inboard side, the $\mathbf{E} \times \mathbf{B}$ drift counteracts the temperature perturbation, which damps the perturbation.
For the toroidal ITG mode to be unstable, the magnetic drift thus needs to be in the same direction as the ion diamagnetic drift when considering $\eta_i > 0$, i.e.~instability requires $\eta_i \omega_{*i} \omega_{\text{D}i} > 0$.
Temperature perturbations with $\eta_i > 0$ are thus destabilized on the low-field (or outboard) side of a tokamak while they are stabilized on the high-field (or inboard) side. At the top and bottom of the device, where the perturbation is parallel to the magnetic drifts, the \mbox{perturbation remains unaffected}.

In summary, ITG modes are destabilized and stabilized by a large number of complex mechanisms. Similar to ITG modes, \textit{electron-temperature-gradient-driven modes} or ETG modes are driven by the temperature gradient of the electrons. In this case, the driving mechanism relies purely on the electron dynamics, while the ions are treated with a Boltzmann response, since the ions act as a nearly stationary background compared to the \mbox{fast electron response.}

\subsection{Trapped-electron modes (TEM)}
\label{sec:TEM}

In toroidal magnetic confinement systems, such as tokamaks or stellarators, a fraction of the electrons becomes trapped in regions of the magnetic field where the field strength is weakest. These trapped electrons bounce back and forth along field lines within a specific magnetic well. The effect of trapped electrons on plasma perturbations can be studied, as they drive so-called \textit{trapped-electron modes} (TEM) \cite{adam1975destabilization, helander2013collisionless} unstable, whereas the contribution of passing electrons is typically negligible. Specifically, these modes are driven unstable by magnetically trapped electrons in the presence of an electron density gradient or electron temperature gradient.

Recall that the ITG mode discussed in the previous section, only relies on the ion dynamics. If we now take into account the electron dynamics, and we avoid strong Landau damping \mbox{$(k_\parallel v_{\text{th},i}\ll \omega \ll k_\parallel v_{\text{th},e})$} and we assume that the ion magnetic drift frequency is much smaller than
the diamagnetic frequency $(\omega_{\text{D}i} \ll \omega_{*i})$ then we can not excite slab nor toroidal ITG modes. By applying the gyrokinetic theory under these assumptions, a dispersion relation is obtained which describes a different kind of modes, relying on the existence of trapped electrons. This dispersion relation reveals that the complex frequency $\omega$ of these modes depends on the velocity-dependent ion and electron diamagnetic frequencies $\omega_{*i}^T$ and $\omega_{*e}^T$ (eq.~\ref{eq:omegstarT}). Consequently, the mode is influenced by the density and temperature gradients of both the ion and electron populations.
Moreover, trapped-electron modes depend on the \textit{bounce-averaged electron magnetic  drift frequency} or \textit{precession frequency} $\overline{\omega}_{\text{D}e}$, which is the magnetic drift that an electron trapped in a magnetic well experiences on average along its orbit. The overbar thus denotes a bounce-average for trapped particles.
Note that trapped particles in a tokamak do not exhibit a net radial drift over a complete bounce orbit, since charged particles are, on average, confined to a specific magnetic flux surface in the absence of collisions. However, trapped particles do experience a net drift in the toroidal direction, known as the \textit{precession drift} of trapped particles, associated to the \textit{precession frequency} $\overline{\omega}_{\text{D}e}$. In contrast, in stellarators, trapped particles can experience a net radial drift on top of their precession; nonetheless, this drift is reduced in neoclassically optimized stellarators.

The mode needs to travel in the same  direction as the trapped precessing electrons to allow for a resonant power transfer between the drift wave and the trapped electrons, \mbox{i.e.~$\omega_r \overline{\omega}_{\text{D}e}>0$ \cite{helander2013collisionless}.} Recall that wave-particle interactions can damp or amplify the waves depending on the direction of energy flow. For the resonant power transfer to drive the mode unstable, i.e.~for the precessing electrons to transfer power into the wave, a resonance between the velocity-dependent electron diamagnetic drift and the bounce-averaged magnetic drift of the trapped electrons is required, i.e.~$\omega_{*e}^T \overline{\omega}_{\text{D}e}>0$, at least in some parts of phase space \cite{helander2013collisionless}.  In particular, this instability grows if the trapped electrons feed energy into the wave, which comes from the plasma gradients. The density gradient drives the so-called $\nablam n$-driven TEM, whereas the temperature gradient drives \mbox{the $\nablam T$-driven TEM.} Based on these requirements, it is clear that trapped-electron modes are driven unstable in regions where trapped electrons are present and where the curvature is unfavorable, i.e.~where $\omega_{*e}^T \overline{\omega}_{\text{D}e}>0$, which implies that trapped electrons precess in the same direction as electron drift waves, and therefore have the potential of destabilizing them. Note that the velocity-dependent diamagnetic frequency has the same sign as $\omega_{*e}$ regardless of the particle's energy if $0 \leq \eta_e\leq 2/3$. In this case the bad-curvature requirement reduces to $\omega_{*e} \overline{\omega}_{\text{D}e}>0$.

Trapped-electron modes can thus be stabilized or destabilized by a resonance between the phase velocity of the mode and the bounce-averaged magnetic drift of the electrons, i.e.~if $\omega \approx \overline{\omega}_{\text{D}e}$. In order for the resonant power transfer to drive the trapped-electron modes unstable we require $\omega_{*e}^T \overline{\omega}_{\text{D}e} > 0$. Additionally, trapped-electron modes can be driven unstable through a feedback loop between the initial perturbation, a charge separation and the resulting $\mathbf{E} \times \mathbf{B}$ drift, since the trapped-electron mode persist in the non-resonant limit $\overline{\omega}_{\text{D}e} \ll \omega$ \cite{coppi1974fluid, proll2014trapped}. In this case these non-resonant, or fluid-like, trapped-electron modes are driven unstable by the electron precession drift in the bad-curvature region, through a similar physical mechanism as illustrated for the ITG mode in figure \ref{fig:perturbation}, with the difference that the precession drift lies in the toroidal direction.
\hfill

\newpage
Similar to ITG modes, which are destabilized by the local bad-curvature, trapped-electron modes are thus driven unstable by the bounce-averaged bad-curvature. Moreover, TEMs reside in regions where electrons are trapped, which is inside the magnetic wells. In tokamaks, the bad-curvature regions usually overlap with magnetic wells, making TEMs particularly problematic. Specifically, $\omega_{*e} \overline{\omega}_{\text{D}e} > 0$ on the outboard side of a typical tokamak due to this overlap. Since trapped electrons precess in the same direction as electron drift waves, the trapped electrons can destabilize the drift waves. In stellarators, these regions do not necessarily overlap to the same extent. In particular, separating the trapping and bad-curvature regions can have a stabilizing influence on TEMs.
If the trapping is perfectly separated from bad-curvature regions, one would obtain $\omega_{*e} \overline{\omega}_{\text{D}e} < 0$ for all orbits. This is satisfied in \textit{maximum-$J$ devices}, where the parallel adiabatic invariant $J$, which corresponds to the integral of a trapped particle's parallel momentum $mv_\parallel$ over a full bounce orbit, is constant on magnetic flux surfaces and decreases from the magnetic axis to the last closed flux surface ($\partial J/ \partial r < 0$).
Realistic devices can only approach this \mbox{maximum-$J$} condition, which may offer enhanced stability against trapped-electron modes.

Both ions and electrons can be trapped in magnetic minima, leading to two types of instabilities: trapped-electron modes (TEM) and \textit{trapped-ion modes} (TIM). However, the TIM does not play a notable role on turbulent transport \cite{cohen1976non}. The trapped-electron mode discussed in this section is the classical \textit{electron-driven trapped-electron mode}. Nonetheless, trapped-electron modes can be driven unstable by ions, which are referred to as \textit{ion-driven trapped-electron modes} \mbox{(iTEM) \cite{plunk2017collisionless}.} While the iTEM is typically subdominant to the classical TEM, it can become important in devices that are especially stable against electron-driven trapped-electron modes, such as devices that approach the maximum-J condition. The ion-driven trapped-electron modes appear in the dispersion relation when analyzing trapped-electron modes considering $\omega_{\text{D}i}\neq 0$. In this case, it is shown that ions have a destabilizing influence on trapped-electron modes if $\omega_{*i} \omega_{\text{D}i} > 0$ when considering $\eta_i > 0$. The trapped electrons do not provide free energy to the mode but they are required, because they reduce the phase velocity of the mode sufficiently to allow for resonance with the ion magnetic drift. These modes can propagate in both the ion and electron diamagnetic direction, depending on the balance between the ion diamagnetic drift and the trapped-electron diamagnetic drift \cite{plunk2017collisionless}.

In the limit $\omega_{\text{D}i} = 0$ we rule out the possibility of driving toroidal ITG modes as well as ion-driven trapped-electron modes (iTEM). Moreover, if we neglect parallel Landau damping for the ions ($k_\parallel v_{\text{th},i} = 0$) we avoid the excitation of slab ITG modes. As mentioned before, in this case the turbulence is typically dominated by trapped-electron modes. However, in configurations that are particularly stable against trapped-electron modes, such as configurations approaching the maximum-$J$ condition, we need to take into account the contribution of passing electrons. This contribution has been neglected in the theoretical framework of the trapped-electron mode, since the trapped-electron contribution typically overwhelms the passing-electron contribution. If the passing-electron contribution becomes dominant, it gives rise to the so-called \textit{passing-electron driven universal instability} \cite{helander2015universal, costello2023universal}, driven by a parallel Landau resonance with the passing electrons ($\omega \approx k_\parallel v_{\text{th},e}$) or a feedback loop associated with the thermal motion of the electrons.  These instabilities can be dominant at low $k_\perp$ which are the non-linearly relevant scales for heat transport. In the steep-density-gradient limit ($\omega_{*e}\gg \omega_{\text{D}e}$) the frequency branch $\omega(\mathbf{k})$ of the passing-electron-driven universal instability  starts with frequencies in the electron diamagnetic direction for small $k_\perp$, it crosses $\omega = 0$ for some intermediate value of $k_\perp$, and it continues with frequencies in the ion diamagnetic direction for large $k_\perp$ \cite{helander2015universal}.

\newpage
\section{Scope of this thesis}

To achieve net energy in magnetic confinement fusion reactors, it is important to understand, predict and to some extent control heat and particle transport, and in particular transport driven by turbulence. This doctoral thesis addresses, via numerical simulations, the complex challenges of turbulent transport in magnetic confinement fusion, focusing specifically on the impact of the density gradient on heat transport in chapter \ref{chap:heat} and examining turbulent particle transport within the core of fusion devices in chapter \ref{chap:particles}. A comprehensive analysis of turbulent transport mechanisms in both tokamaks and stellarators is provided, contributing valuable insights towards improving heat and particle confinement in fusion reactors. This work ultimately advances our understanding of how specific magnetic geometries and plasma profiles influence confinement, thereby supporting the design of more efficient, turbulence-resistant fusion reactors.

The theoretical framework for this study, including the gyrokinetic theory and the flux-tube approximation, are discussed in chapters \ref{chap:gyrokinetics} and \ref{chap:fluxtube}. The gyrokinetic code \texttt{stella} \cite{stella} is utilized to evolve the turbulent component of the distribution function and the electrostatic potential, in order to study the effects of the magnetic geometry and plasma parameters on the turbulent heat and particle transport, as detailed in chapter \ref{chap:stella}. The magnetic equilibria of the devices considered in this work are shown in figure \ref{fig:fluxsurfacesintro} and will be discussed in detail in  \mbox{sections \ref{chap:equilibria} and \ref{sec:MC}.}

\subsection{Inter-machine comparison of the influence of the density gradient on electrostatic gyrokinetic stability and turbulent heat transport}

Experimental studies have shown that steep density profiles can reduce turbulent heat transport \cite{komori2006overview, bozhenkov2020high, 2022-Cortes-Enhanced}, likely due to the stabilization of ion-temperature-gradient (ITG) and electron-temperature-gradient (ETG) modes by large density gradients. However, when density gradients increase, modes driven by these gradients, such as trapped-electron modes, can become destabilized. Chapter \ref{chap:heat} therefore explores the range of density gradients that stabilizes temperature-gradient-driven modes without significantly destabilizing density-gradient-driven turbulence.

While many studies have investigated turbulence and transport in individual fusion devices, the results are often scattered throughout the literature and are based on differing sets of plasma parameters, making direct comparisons difficult. To address this, the present work conducts a comprehensive comparative study across a range of fusion devices,  including different families of stellarators. Specifically, the Wendelstein 7-X (W7-X) stellarator, the Large Helical Device (LHD), the TJ-II stellarator and the National Compact Stellarator Experiment (NCSX) are considered, which represent helias, heliotron, heliac and quasi-axisymmetric configurations, respectively. Additionally, the Axially Symmetric Divertor Experiment or ASDEX Upgrade (AUG) tokamak and the tokamak geometry of the Cyclone Base Case (CBC) \cite{dimits2000simulation} are included for comparison.

On the other hand, stability and transport studies in fusion plasmas are often confined to a limited range of parameters, particularly in the case of turbulent transport. Furthermore, common approximations---such as assuming an adiabatic electron response or neglecting the electron temperature gradient when treating electrons kinetically---question the accuracy of these results. To address these limitations, both linear and nonlinear simulations are performed in this work across a wide range of parameters to explore the effects of density gradients, temperature gradients, and kinetic electrons. It is shown that the density gradient has a beneficial effect on heat transport in several devices. Specifically, a strong reduction in ion heat flux with increasing density gradients is found in NCSX and W7-X due to the stabilization of temperature-gradient-driven modes without significantly destabilizing density-gradient-driven modes.  Notably, the behavior of ion heat fluxes in stellarators does not align with that of linear growth rates, if only the fastest-growing mode is taken into account. Additionally, this study provides physical insights into the microinstabilities, emphasizing the dominance of trapped-electron modes in CBC, AUG, TJ-II, LHD and NCSX, while both trapped-electron modes and \mbox{passing-particle-driven universal instabilities contribute significantly in W7-X.}

\subsection{Turbulent particle transport at the plasma core}

Neoclassical transport theory predicts significant outward particle fluxes in the plasma core of large stellarators, which should yield hollow density profiles. However, this theoretical prediction is contradicted by experiments. In particular, in Wendelstein 7-X, the first large optimized stellarator, flat or weakly peaked density profiles are generally measured, indicating that neoclassical theory is
not sufficient and that an inward contribution to the particle flux is missing in the core. Therefore, in chapter \ref{chap:particles} it is investigated whether turbulent particle transport can account for the discrepancy in the observed core density profiles.

Simulations of turbulent particle flux are performed for the W7-X and TJ-II stellarators, to examine how turbulence contributes to particle retention in the core of stellarators, potentially counteracting the neoclassical outward flux. The results suggest that turbulent transport plays a significant role in maintaining the experimentally observed density profiles.  Specifically, a one-dimensional particle balance analysis is performed for a W7-X discharge, where the sign of the turbulent particle flux agrees, over the entire plasma radius, with the sign of the difference between the experimental and the neoclassical particle fluxes. In particular, the simulations predict a sign change of the turbulent flux, accounting for the fact that the neoclassical flux is too large at the core and too small at the edge of the plasma to explain
the experimental particle flux. Moreover, it is shown that the existence of an inward turbulent particle flux is very robust and is found over broad regions of the scanned parameter space.

\addtocontents{toc}{\protect\vspace*{5mm}}
\part{Modeling electrostatic turbulence with the gyrokinetic \bolddeltaf code \texttt{stella}}

\chapter{Electrostatic \bolddeltaf gyrokinetics for stellarators}
\label{chap:gyrokinetics}

\vspace{2mm}

Modeling plasma turbulence is a critical endeavor in the study of fusion devices. The primary goal is to accurately predict, for a given set of plasma profiles, the levels of heat, momentum, and particle loss due to turbulence. This turbulence---largely caused by microscopic instabilities within the plasma---poses a significant challenge to achieving efficient and sustainable fusion. Plasma turbulence is inherently complex, driven by the interactions of charged particles moving through a magnetic field, as discussed in sections \ref{sec:intro:MCF} and \ref{sec:intro:micro}. To describe this turbulence accurately, one would need to solve the six-dimensional Vlasov equations coupled to the Maxwell's equations.

This six-dimensional kinetic description of plasma turbulence can be simplified by exploiting the distinctive characteristics of microturbulence in strongly magnetized plasmas, i.e.~when \mbox{$\rho_{*s} = \rho_s/L \ll 1$}. Specifically, microturbulence evolves on timescales that are significantly faster than the equilibrium changes in the plasma, yet much slower than the rapid oscillatory motion of particles due to their gyromotion around magnetic field lines. This disparity in timescales, combined with the fact that turbulent structures are often very elongated along magnetic field lines while having perpendicular scales comparable to the gyroradius, allows to reduce the complexity of the system significantly. \textit{Gyrokinetic theory} exploits these orderings by  systematically expanding the equations of motion in $\rho_{*s}$ while averaging over the fastest degree of motion, namely the gyrophase. This procedure removes the dependence on the gyration angle, effectively reducing the dimensionality of the problem from six to five. The resulting gyrokinetic equations describe the dynamics of charged rings---particles that appear as rings due to their fast circular motion---as they stream along magnetic field lines and slowly drift across them. The derivation of the gyrokinetic equations has been extensively covered elsewhere, see e.g.~Refs.~\cite{catto1978linearized, frieman1982nonlinear, abel2008linearized, calvo2012long, calvo2013stellarators, davies2022kinetic}. Specifically, the full derivation of the electrostatic gyrokinetic equations to high order in $\rho_{*s}$ for general magnetic geometry, along with all associated assumptions, is detailed by Parra and Calvo in Ref.~\cite{parra2011phase}. For the purposes of this work, only a summary of the essential aspects is provided.

Gyrokinetic theory provides a powerful framework to model microturbulence in strongly magnetized plasmas by systematically reducing the full kinetic equations. While this reduction significantly simplifies the description of plasma dynamics, the resulting gyrokinetic equations remain analytically intractable due to their inherent nonlinearity and high dimensionality. Consequently, they must be solved numerically, which is made possible by the development of advanced computational codes and high-performance computing centers. In this work, a relatively recent code is employed, the \texttt{stella} gyrokinetic code \cite{stella}, designed to tackle the numerical solution of the gyrokinetic equations within general magnetic geometries. Thus, the code is capable of handling not only axisymmetric devices like tokamaks but also the complex 3D geometries of stellarators.

In this chapter, the main gyrokinetic equations and the assumptions that underpin them are summarized, setting the stage for their implementation within the \texttt{stella} code, which will be discussed in chapter \ref{chap:stella}. The gyrokinetic equations, that are introduced in section \ref{sec:gyrokineticEquations}, are derived by expanding the fundamental equations of motion (section \ref{sec:kinetic-description}) in terms of the gyrokinetic ordering parameter $\rho_{*s}$ (section \ref{sec:gyrokineticOrderings}), and averaging over the gyrophase. To facilitate gyroaveraging, it is essential to define appropriate coordinates for the particle position and velocity (section \ref{sec:guiding-center-coordinates}).

\newpage
\section{Kinetic description of plasma}
\label{sec:kinetic-description}

The starting point of the gyrokinetic theory is the kinetic description of a plasma.
The kinetic theory of plasmas describes the motion of charged particles---electrons and ions---in the presence of electromagnetic fields.  The particles are characterized by their position $\mathbf{r}$ and \mbox{velocity $\mathbf{v}$} at each time instant $t$, for which the microscopic information is averaged out to obtain statistical, kinetic equations. The \textit{distribution function} $f_s(\mathbf{r}, \mathbf{v}, t)$\index{Distribution function!Distribution function $f_s$} of species $s$ is defined such that,
\begin{equation}
	\label{eq:fs}
	f_s(\mathbf{r}, \mathbf{v}, t) \,\, \text{d}^3r \,\, \dv = \text{d}\hspace{-0.2mm}N(\mathbf{r}, \mathbf{v}, t)
\end{equation}
is the number of particles in the infinitesimal phase-space volume element $\text{d}V = \text{d}^3r \,\, \dv$. The evolution of the distribution function then follows from the conservation of phase-space density,
\begin{equation}
\label{eq:VlasovEquation}
\frac{\partial f_s}{\partial t}  + \mathbf{v} \cdot \nablam f_s + \frac{Z_s e}{m_s} \Big( {\mathbf{E}} + \mathbf{v} \times {\mathbf{B}} \Big) \cdot \frac{\partial f_s}{\partial \mathbf{v}} = 0,
\end{equation}
which is the so-called \textit{Vlasov equation}, with ${\mathbf{E}}$ the electric field, ${\mathbf{B}}$ the magnetic field, $Z_s$ the charge number, $e$ the proton charge and $m_s$ the mass of particle $s$. Introducing a collision operator, alongside other source terms, on the right-hand side of equation (\ref{eq:VlasovEquation}) yields the \textit{Fokker-Planck kinetic equation}. In this work, a collisionless plasma without source terms is considered. The electric and magnetic fields obey the Maxwell’s equations, given by,

\vspace{-2mm}
\hspace{-0.8cm}
\begin{minipage}{0.45\textwidth}
	\begin{align}
		\label{eq:GaussLaw}
		\qquad\qquad \nabla \cdot {\mathbf{E}} & =  \frac{{\varrho}}{\varepsilon_0}  \\[2mm]
		\label{eq:MagneticGaussLaw}
		\qquad\qquad \nabla \cdot {\mathbf{B}} & = 0
	\end{align}
\end{minipage}
\begin{minipage}{0.55\textwidth}
	\begin{align}
		\label{eq:FaradaysInductionLaw}
		\qquad\nabla \times {\mathbf{E}}   & = - \frac{\partial {\mathbf{B}}}{\partial t}  \\[0mm]
		\label{eq:AmpereMaxwellLaw}
		\qquad\nabla \times {\mathbf{B}} & = \mu_0 \, {\mathbf{j}} + \frac{1}{c^2} \frac{\partial {\mathbf{E}}}{\partial t},
	\end{align}
\end{minipage}
\vspace{1mm}

\noindent
with $\varepsilon_0$ the vacuum permittivity, $\mu_0$ the vacuum permeability, and $c$ the speed of light, related through $\varepsilon_0\mu_0 =  1/c^2$. Plasma effects enter the Maxwell equations through the charge density ${\varrho}$ and current density ${\mathbf{j}}$, produced by the response of a plasma to electromagnetic fields, defined as,
\begin{align}
\label{eq:charge}
{\varrho} & = \sum_s Z_s e \int f_s(\mathbf{r}, \mathbf{v}, t) \  \dv, \\[-2mm]
\label{eq:current}
{\mathbf{j}} & = \sum_s Z_s e \int f_s(\mathbf{r}, \mathbf{v}, t) \  \mathbf{v} \ \dv .
\end{align}
\vspace{-3mm}

\noindent
Equations (\ref{eq:VlasovEquation})--(\ref{eq:current}) are the fundamental equations that provide a complete kinetic description of a collisionless plasma, called the \textit{Vlasov-Maxwell plasma model}. This system of equations is further simplified by neglecting the Debye-scale and relativistic effects. The Debye length is defined as $\smash{\lambda_{\text{De}} = \sqrt{\varepsilon_0 T_e/n_e e^2}}$ and the thermal velocity as $\smash{v_{\text{th,s}} = \sqrt{2T_s /m_s}}$, with $T_s$ and $n_s$ the mean temperature and density of species $s$. These assumptions are implemented by the orderings  $k^2_\perp \lambda^2_{\text{De}} \ll 1 $ and ${v^2_{\text{th,s}}}/{c^2} \ll 1$, with $2\pi/k_\perp$ the characteristic scale of the turbulent fluctuations perpendicular to the magnetic field. As a result, Gauss's law (eq.\ \ref{eq:GaussLaw}) is replaced by the \textit{quasineutrality condition},
\begin{equation}
\label{eq:SimplifiedQuasiNeutralityEquation}
{\varrho} = 0.
\end{equation}
Moreover, the displacement current is neglected in Amp\`ere's law (eq.\ \ref{eq:AmpereMaxwellLaw}),
\begin{equation}
\label{eq:SimplifiedAmperesLaw}
\nabla \times {\mathbf{B}} = \mu_0\,{\mathbf{j}}.
\end{equation}
In this work, the effects of electromagnetic perturbations are neglected, by considering a constant magnetic field, i.e.~$\mathbf{B} = \mathbf{B}_0$. Consequently, the electric field can be derived from an electrostatic \mbox{potential ${\varphi}$} and the magnetic field can be derived from a vector potential ${\mathbf{A}}$, through,
\begin{align}
\label{eq:SimplifiedElectrostaticPotential}
{\mathbf{E}} & = - \nablam {\varphi} (\mathbf{r},t), \\[2mm]
\label{eq:SimplifiedVectorPotential}
{\mathbf{B}} & = \nablam \times {\mathbf{A}} (\mathbf{r}).
\end{align}
To summarize, the kinetic description of a plasma---in the absence of collisions and sources, and neglecting the Debye-scale effects, relativistic effects and electromagnetic perturbations---is governed by the Vlasov equation (eq.~\ref{eq:VlasovEquation}) coupled to the four simplified Maxwell's equations (eqs.~\ref{eq:SimplifiedQuasiNeutralityEquation}--\ref{eq:SimplifiedVectorPotential}).

\section{Guiding-center coordinates}
\label{sec:guiding-center-coordinates}
\index{Guiding-centre coordinates}
\index{Coordinate systems!Guiding-centre coordinates}

\begin{figure}[h!]
	\centering
	\includegraphics{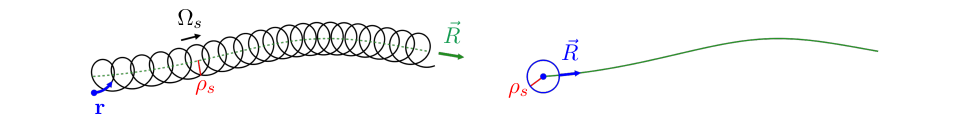}
	\captionsetup{singlelinecheck=false,  margin=0.0cm}\vspace{-8mm}
	\caption{The motion $\mathbf{r}$ of a charged particle in a magnetic field  $\mathbf{B}_0$, with a gyrofrequency $\Omega_s  = Z_s e B/m_s$ and gyroradius $\rho_s= v_{\text{th},s}/|\Omega_s|$ (left) and the motion $\mathbf{R}$ of an equivalent charged ring with radius $\rho_s$ (right). }
	\label{fig:6Dto5D}
\end{figure}

This kinetic description of a plasma is in essence a 6-dimensional description in phase space since it describes the motion of charged particles, denoted by $\mathbf{r}$ in figure \ref{fig:6Dto5D}, as they gyrate around magnetic field lines---with a \textit{gyroradius} $\rho_s$\index{Guiding-centre coordinates!Gyroradius $\rho_s$}  and \textit{gyrofrequency} $\Omega_s$\index{Guiding-centre coordinates!Gyrofrequency $\Omega_s$}---and slowly drift across it (see section \ref{sec:intro:MCF}). The {motion of the guiding center} is denoted by $\mathbf{R}$, and illustrated by the green line on the left of figure \ref{fig:6Dto5D}. Gyrokinetic theory reduces this 6-dimensional problem to a 5-dimensional system, by averaging out the fast gyromotion, and thus eliminating the \textit{gyroangle} $\vartheta$\index{Gyroangle} from the system of equations. As a result, gyrokinetics is often called the kinetics of charged rings, as it describes how charged rings, shown on the right of figure \ref{fig:6Dto5D}, move along the guiding-center trajectory.

\vspace{3mm}
\FloatBarrier
\vspace{3mm}

\noindent
The motion of a charged particle in the presence of a magnetic field $\mathbf{B}_0(\mathbf{r})$ and an electrostatic potential $\varphi(\mathbf{r},t)$ is thus a superposition of the fast gyromotion and the slow guiding-center motion. Microturbulence is characterized by low-frequency fluctuations with $\omega < \Omega_s$, where $\omega$ is a characteristic frequency of the fluctuations. Since the fluctuations have low frequencies, all dynamics of interest are considered slow compared to the particle gyration about a mean magnetic field. This justifies the decomposition of the particle motion into its fast gyromotion and slow guiding-center motion. The actual particle position is therefore given by,
\begin{equation}
	\label{eq:r}
	\mathbf{r} = \mathbf{R} + \boldsymbol{\rho}_s(\mathbf{R}, \mu, \vartheta) + \dots,
\end{equation}
with $\mathbf{R}$ the \textit{guiding-center position}\index{Guiding-centre coordinates!Guiding-center position $\mathbf{R}$} and $\boldsymbol{\rho}_s$ the \textit{gyroradius vector} or \textit{Larmor radius vector}\index{Guiding-centre coordinates!Gyroradius vector $\boldsymbol{\rho}_s$},
\begin{equation}
	\boldsymbol{\rho}_s(\mathbf{R}, \mu, \vartheta)  = \Omega^{-1}_s \, \mathbf{\hatb} \times \mathbf{v} = \frac{\sqrt{2\mu B}}{\Omega_s} \Big( \cos \vartheta \ \mathbf{\hat{e}}_2(\mathbf{R}) - \sin \vartheta \  \mathbf{\hat{e}}_1(\mathbf{R}) \Big).
\end{equation}
Here $\{ \mathbf{\hat{e}}_1, \mathbf{\hat{e}}_1, \mathbf{\hatb} \}$ is an orthonormal set satisfying $\smash{\mathbf{\hat{e}}_1(\mathbf{R}) \times \mathbf{\hat{e}}_2(\mathbf{R}) = \mathbf{\hatb}(\mathbf{R})}$.  The dots ``$\dots$'' in equation (\ref{eq:r}) stand for higher-order corrections in the $\rho_{*s}$ expansion, as explained in the next section.

\section{Orderings and assumptions in gyrokinetic theory}
\label{sec:gyrokineticOrderings}

Numerous orderings and assumptions are imposed to simplify the Vlasov equation (eq.~\ref{eq:VlasovEquation}). As previously mentioned, we restrict our attention to plasmas with electrostatic fluctuations, and we eliminate the gyroangle $\vartheta$ as a phase-space variable. The reduction to 5 dimensional phase-space is justified if the characteristic frequency $\omega$ of the turbulence is much smaller than the gyrofrequency $(\omega \ll \Omega_s)$, if the plasma is strongly magnetized so that the gyroradius is small compared to a characteristic length associated with the equilibrium $(\rho_s \ll L)$, and if the fluctuating electrostatic potential is small compared to the characteristic energy of the particles $(e_s \varphi \ll T_s)$. Additionally, fluctuations are assumed to be very elongated along magnetic field lines while only having short spatial scales perpendicular to the field lines $(k_\parallel \ll k_\perp)$. Here, $k_\parallel$ and $k_\perp$ are typical parallel and perpendicular wavenumbers of the fluctuations.
These orderings can be summarized as,
\begin{equation}
\label{eq:GyrokineticOrderings}
\frac{\omega}{\Omega_s} \sim \frac{\rho_s}{L} \sim \frac{e \varphi}{T_s} \sim \frac{k_\parallel}{k_\perp}  \sim \rho_{*s} \ll 1.
\end{equation}
Moreover, the \deltaf-approximation is employed, which assumes that there is a separation of space-time scales between the plasma equilibrium and the turbulent fluctuations. This makes it possible to decompose the particle distribution function $f_s$ (eq.~\ref{eq:fs}), into an equilibrium component $F_s$, and a fluctuating component $\deltafm_s$, i.e.~$f_s = F_s + \deltafm_s$.  The perturbed distribution  is assumed to be much smaller than the background distribution $(\deltafm_s \ll F_s)$, and the background is assumed to vary on the length scale of the system size $(\nablam F_s \sim 1/L)$.
On the other hand, the electrostatic potential is assumed to only consist of a fluctuating component, which varies on the same length and time scales as the microturbulence: it is assumed to fluctuate with some characteristic frequency of the turbulent fluctuations $(\partial_t \varphi \sim \omega)$, its perpendicular gradient is of the order of the ion gyroradius $(\nablaperp \varphi \sim 1/\rho_i)$, while the parallel gradient is of the order of the size of the device $(\hatb \cdot \nablam \varphi \sim 1/L)$.

Note that the turbulence can be affected by the magnitude and gradients of an equilibrium flow, which gives rise to the Coriolis and centrifugal drifts. Moreover, perpendicular and parallel shear flows, also known as  $\mathbf{E}\!\times\!\mathbf{B}$ shear, can suppress or excite turbulence. These effects are neglected in this work by considering the
low-flow low-Mach limit \cite{parra2011phase, abel2013multiscale}.

\section{Gyrokinetic equations}
\label{sec:gyrokineticEquations}

The gyrokinetic equations are obtained by systematically expanding equations (\ref{eq:VlasovEquation}), (\ref{eq:SimplifiedQuasiNeutralityEquation}) and (\ref{eq:SimplifiedAmperesLaw}) in the gyrokinetic ordering parameter $\rho_{*s}$, and averaging the equations over the gyrophase $\vartheta$ \cite{catto1978linearized, davies2022kinetic, parra2011phase}. The lowest-order, electrostatic gyrokinetic equation for the gyro-averaged perturbed distribution function, denoted as $g_s(\mathbf{R},v_\parallel,\mu,t) = \langle \deltafm_s \rangle_\mathbf{R}$, is given by,
\begin{equation}
	\label{eq:GyrokineticEquation}
	\begin{split}
		\frac{\partial g_s}{\partial t}
		+ v_\parallel \hatb \cdot \nablam g_s
		+ v_\parallel  \frac{Z_s e }{T_s}\,   \hatb \cdot \nablam \langle \varphi \rangle_\mathbf{R} \, F_{\text{M}s}
		- \mu \, \hatb \cdot \nablammm B \, \frac{\partial g_s}{\partial v_\parallel} \qquad
		&\\
		+ \,\mathbf{v}_{\text{D}s} \cdot \nablam g_s
		+ \frac{Z_s e}{T_s} \,\mathbf{v}_{\text{D}s} \cdot   \nablam \langle \varphi \rangle_\mathbf{R} \, F_{\text{M}s}
		+ \langle \mathbf{v}_\text{E} \rangle_\mathbf{R} \cdot \nablam g_s
		+ \langle \mathbf{v}_\text{E} \rangle_\mathbf{R} \cdot \nablam|_\mathcal{E} F_{\text{M}s} & = 0,
	\end{split}
\end{equation}
where, unless noted otherwise, derivatives are taken holding $\mathbf{R}$, $v_\parallel$ and $\mu$ constant, whereas $\nablam|_\mathcal{E}$ is performed at constant energy. This is the gyrokinetic equation implemented (in its flux-tube version) in \texttt{stella} by  Barnes et al., corresponding to equation (2) in Ref.~\cite{stella}. Here $\mathbf{v}_{\text{D}s}$ is the \textit{magnetic drift velocity}\index{Drift velocities!Magnetic drift $\mathbf{v}_{\text{D}s}$} (eq.~\ref{eq:MagneticDriftsIntro}), $\mathbf{v}_\text{E}$ is the lowest order, fluctuating \textit{$\mathbf{E} \!\times\! \mathbf{B}$ drift velocity}\index{Drift velocities!$\mathbf{E} \!\times\! \mathbf{B}$ drift $\mathbf{v}_\text{E}$}, given by,
\begin{equation}
	\label{eq:EcrossBDrift}
	\mathbf{v}_\text{E} = \frac{1}{B^2} \, \Big( \mathbf{B} \times \nablam \varphi \Big),
\end{equation}
and $F_{\text{M}s}$ is a Maxwellian distribution in the kinetic energy $\mathcal{E}_s = m_s v_s^2/2 = m_s v_\parallel^2/2 + m_s \mu B$, i.e.
\begin{equation}
	\label{eq:FMs}
	F_{\text{M}s} = n_s\, \Big( \frac{m_s}{2\pi T_s} \Big)^{3/2} \! \exp \Big( - \frac{\mathcal{E}_s}{T_s} \Big).
\end{equation}
The average over the gyrophase $\vartheta$, or \textit{gyroaverage}, is defined as,
\begin{equation}
	\label{eq:gyroaverage}
	\langle \,\cdot\, \rangle_\mathbf{R} = \frac{1}{2\pi} \oint \, (\,\cdot\,) \ d\vartheta,
\end{equation}
where the integral is performed holding $\mathbf{R}$, $v_\parallel$, and $\mu$ constant. Note that the turbulent electrostatic potential $\varphi(\mathbf{r},t)$ is given as a function of the particle position. Therefore, it needs to be transformed to guiding-center coordinates before the gyroaverage can be performed.
In the gyrokinetic equation, both $g_s(\mathbf{R},v_\parallel,\mu,t)$ and $\varphi(\mathbf{r},t)$ are unknown. The system is closed through the lowest-order turbulent terms of the \textit{quasineutrality condition} (\ref{eq:SimplifiedQuasiNeutralityEquation}),
\index{Quasineutrality condition!Quasineutrality in real space}
\begin{equation}
	\label{eq:quasineutrality}
	\sum_s Z_s e \,\delta n_s =  \sum_s Z_s e
	\int \,
	\Big( g_s - \frac{Z_s e}{T_s} \big( \varphi - \langle \varphi \rangle_\mathbf{R} \big)\, F_{\text{M}s} \Big) \ \dv = 0.
\end{equation}
where the integral is performed holding $\mathbf{r}$, $v_\parallel$, and $\mu$ constant.

Note that each term in the gyrokinetic equation (\ref{eq:GyrokineticEquation}) can be described based on their physical significance. The terms containing $\hatb \cdot \nablam$ include the effects of \textit{parallel streaming}\index{Terms in the gyrokinetic equation!Parallel streaming}, and those with $\hatb \cdot \nablammm B$ include the effects of the \textit{magnetic mirror}\index{Terms in the gyrokinetic equation!Magnetic mirror}. The influence of the \textit{plasma gradients}, such as the temperature and density gradients\index{Terms in the gyrokinetic equation!Plasma gradients}\index{Terms in the gyrokinetic equation!Plasma gradients}, are included in the $\nablam|_\mathcal{E} F_{\text{M}s}$-term, which can be seen explicitly in equation (\ref{eq:GradientOfF0s}). Lastly, seeing that the  $\langle \mathbf{v}_\text{E} \rangle_\mathbf{R} \cdot \nablam g_s$ term depends on both the turbulent distribution function $g_s(\mathbf{R},v_\parallel,\mu,t)$ and the fluctuating electrostatic potential $\varphi(\mathbf{r},t)$, this term thus contains the nonlinear effects of the system\index{Terms in the gyrokinetic equation!Nonlinear term}.

\section{Adiabatic Boltzmann response}

The gyrokinetic system of equations (\ref{eq:GyrokineticEquation} and \ref{eq:quasineutrality}) can describe electrostatic fluctuations with full kinetic electron effects and an arbitrary number of kinetic ion species.
However, solving these equations for two kinetic species with significantly different masses can be computationally expensive. The computation time can be reduced by using a \textit{Boltzmann (adiabatic) response} instead.
To lowest order, the turbulent component of the distribution function \cite{davies2022kinetic, abel2013multiscale} is given by,
\begin{equation}
	\label{eq:deltaf}
	\deltafm_{s} = - \frac{Z_s e}{T_s} \ \varphi (\mathbf{r}, t)\, F_{\text{M}s} + h_s(\mathbf{R}, v_\parallel, \mu, t).
\end{equation}
The first term is gyrophase-dependent and represents the adiabatic, or Boltzmann, response\index{Boltzmann response} to the electrostatic field. In contrast, $h_s$ corresponds to the non-adiabatic component \mbox{of $\deltafm_{s}$}. Physically, $h_s$ characterizes the \textit{distribution of Larmor rings}\index{Distribution function!Distribution of Larmor rings $h_s$}\index{Distribution function!Non-adiabatic part $h_s$}, and encapsulates the essential kinetic response of the plasma to microturbulence.
To model electron-Larmor-scale turbulence, kinetic electrons are considered while the ions can be treated adiabatically. That is to say, the non-adiabatic part of the distribution function of the ions, $h_i$, is assumed to be negligible. As a result, the guiding-center distribution function of the ions, $g_i(\mathbf{R},v_\parallel,\mu,t) = \langle \deltafm_i \rangle_\mathbf{R}$, reduces to a purely adiabatic part, i.e. \index{Boltzmann response}\index{Adiabatic response}\index{Modfiied Boltzmann response}
\begin{equation}
	g_i = h_i - \frac{Z_i e }{T_i} \, \langle \varphi \rangle_\mathbf{R}\, F_{\text{M}i} \approx - \frac{Z_ie}{T_i} \, \langle \varphi \rangle_\mathbf{R}\, F_{\text{M}i}.
\end{equation}
The perturbation of the ion density in the quasineutrality condition is then replaced by,
\begin{equation}
	\delta n_i = \int \,
	\Big( g_i - \frac{Z_i e}{T_i} \big( \varphi - \langle \varphi \rangle_\mathbf{R} \big)\, F_{\text{M}i} \Big) \ \dv \approx  - \frac{Z_i e}{T_i} \, \varphi \! \int \! F_{\text{M}i} \, \dv = - \frac{Z_i e n_i}{T_i} \, \varphi.
\end{equation}
On the other hand, to model ion-Larmor-scale turbulence, kinetic ions and adiabatic electrons can be considered. In this case, $\varphi$ needs to be replaced by $(\varphi - \overline{\varphi})$, with $\overline{\varphi}$ the flux-surface average \mbox{of $\varphi$}, in order to allow for the excitation of sheared poloidal $\mathbf{E} \times \mathbf{B}$ flows, also known as zonal \mbox{flows \cite{beer1995field, hammett1993developments}.} Therefore, the so-called \textit{modified Boltzmann response} for the electrons, is given by,
\begin{equation}
	\delta n_e = \frac{en_e}{T_e} (\varphi - \overline{\varphi}).
\end{equation}

\addtocontents{toc}{\protect\newpage}

\chapter{Flux-tube approach to gyrokinetics}
\label{chap:fluxtube}

In this chapter, advantage is taken of the fact that turbulence driven by microinstabilities in strongly magnetized plasmas is generally very elongated along magnetic field lines while only having short spatial scales perpendicular to the field lines, which is implied by the ordering in equation (\ref{eq:GyrokineticOrderings}). To exploit this scale separation, the turbulent distribution function $g_s(\mathbf{R},v_\parallel,\mu,t)$ and the fluctuating electrostatic potential $\varphi(\mathbf{r},t)$ are Fourier expanded in the fast variation in $\psi$ and $\alpha$ in section \ref{sec:fluxtube:approximation}, with $\psi$ a flux-surface label and $\alpha$ a field-line label, which is in essence the flux-tube approximation.
The gyrokinetic equation and the quasineutrality condition are then rewritten in function of the wavenumbers $k_\psi$ and $k_\alpha$ in \mbox{section \ref{sec:fluxtube:gyrokineticequation}}. We define straight-field-line coordinates $(\psi,\theta_s,\zeta_s)$ in section \ref{sec:straight-field-line-coords} and field-aligned coordinates $(\psi,\alpha,z)$ in \mbox{section \ref{sec:field-aligned-coordinates}}. As a consequence of describing gyrokinetics by means of field-aligned coordinates, the simulation domain---which is called a flux tube---becomes sheared, which is explained in detail in section \ref{sec:fluxtube:shearing}. In deriving the gyrokinetic equations, the rapid gyromotion of particles around the magnetic field lines has been averaged out. Nonetheless, the effects of the rapid gyromotion of the particles on the plasma behavior is retained through the so-called finite-Larmor-radius (FLR) effects, discussed in \mbox{section \ref{sec:FLR}}. Finally, section \ref{sec:parallelBC} explains how the simulation domain is connected onto itself in the parallel direction through the application of parallel boundary conditions.

\vspace{2mm}
\section{Straight-field-line flux coordinates $(\psi,\theta_s,\zeta_s)$}
\label{sec:straight-field-line-coords}

\begin{figure}[h!]
	\centering\vspace{2mm}
	\begin{subfigure}{\linewidth}
		\includegraphics{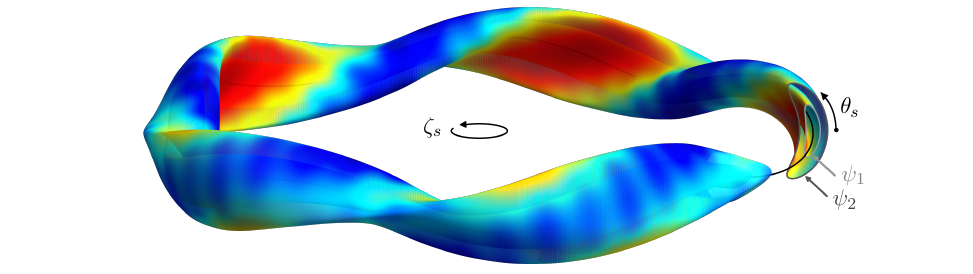}
	\end{subfigure}
	\captionsetup{singlelinecheck=false,  margin=0cm}\vspace{-5.5mm}
	\caption{Straight-field-line flux coordinates $(\psi,\theta_s,\zeta_s)$ illustrated on the magnetic flux surfaces of W7-X. The magnetic field strength is represented by the color scale (maximum is red, minimum is blue).} \vspace{-2mm}
	\label{fig:straight-field-line-coords}
\end{figure}

In order to provide a simple description of field lines and magnetic flux surfaces, a set of \textit{straight-field-line flux coordinates}\index{Coordinate systems!Straight-field-line flux coordinates} $(\psi,\theta_s,\zeta_s)$ is defined \cite{haeseleer2012flux}, which are depicted in figure \ref{fig:straight-field-line-coords}. The magnetic equilibria are assumed to consist of nested flux surfaces. As a result, any quantity which is constant on a flux surface, can serve as a \textit{flux-surface label}\index{Coordinates!Flux-surface label} $\psi$  (gray contours in \mbox{figure \ref{fig:straight-field-line-coords}}). Typically, the choice $\psi = \psi_p$ is made in the tokamak literature, while $\psi = \psi_t$ is more popular for stellarators, with $\psi_p$ and $\psi_t$ the poloidal and toroidal flux divided by $2\pi$, respectively. Radial locations within magnetic equilibria are typically described using the \textit{effective minor radius} $r = a\sqrt{\psi_t/\psi_{t,\text{LCFS}}}$, which is normalized to the minor radius $a$ of the device, with $\psi_{t,\text{LCFS}}$ the value of $\psi_t$ at the last closed flux surface  (LCFS). Using this radial coordinate, $r/a=0$ represents the magnetic axis, while  $r/a=1$ denotes the last closed flux surface. Note that the effective minor radius can be used as a flux-surface label, offering a more intuitive alternative to the magnetic flux itself.

\vspace{0mm}
\FloatBarrier

\begin{figure}[h!]
	\centering
	\includegraphics[trim={0mm 15mm 0mm 0mm},clip] {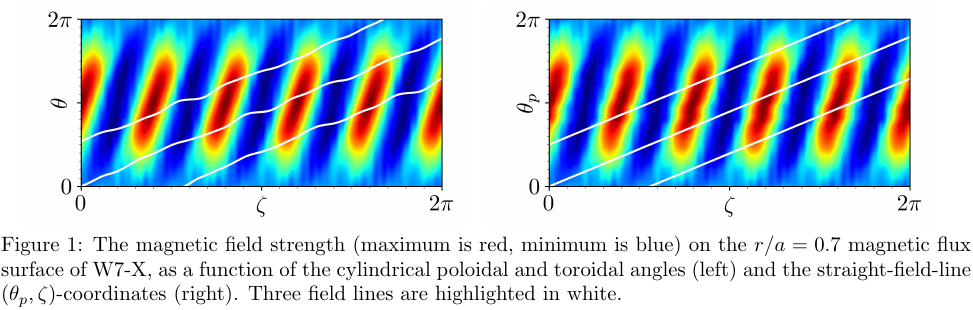}
	\captionsetup{singlelinecheck=false,  margin=0.0cm}
	\vspace{-7mm}
	\caption{The magnetic field strength (maximum is red, minimum is blue) on the $r/a=0.7$ magnetic flux surface of W7-X, as a function of the original poloidal and toroidal angles (left) and the straight-field-line $(\theta_p,\zeta)$-coordinates (right). Three field lines are highlighted in white. }
	\vspace{1mm}
	\label{fig:straightFieldLine}
\end{figure}

\newpage
The $(\theta_s, \zeta_s)$-coordinates are obtained by modifying an initial set of poloidal and toroidal angles in such a way that field lines appear straight in these coordinates. The coordinate system in which the field lines are straight is not unique. Specifically, for axisymmetric geometries which are based on a set of Miller parameters, \texttt{stella} 	uses Boozer coordinates \cite{boozer1981plasma} whereas for \texttt{VMEC} equilibria, the \texttt{PEST} coordinates $(\theta_p,\zeta)$ are employed \cite{grimm1983ideal}.
In figure \ref{fig:straightFieldLine}, a magnetic flux surface at $r/a=0.7$, generated by \texttt{VMEC}, is represented in the original $(\theta,\zeta)$-coordinates (left) and in straight-field-line $(\theta_p,\zeta)$-coordinates (right). It is clear that the field lines which lie on top of these magnetic flux surfaces, highlighted in white, are straight in $(\theta_p,\zeta)$-coordinates.

\FloatBarrier

\begin{figure}[!b]
	\centering
	\includegraphics[trim={0mm 22mm 0mm 0mm},clip] {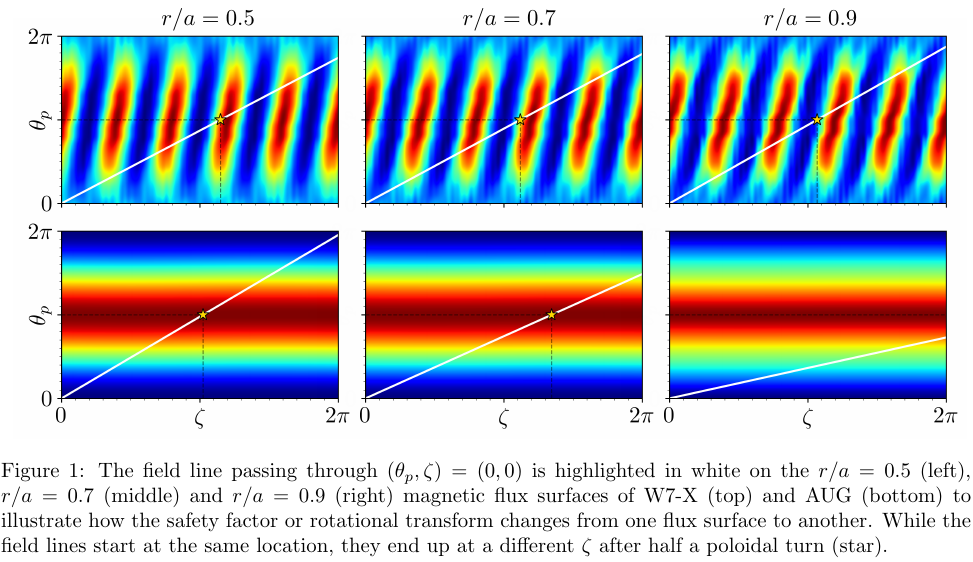}
	\captionsetup{singlelinecheck = false, margin=0cm}
	\vspace{-6.5mm}
	\caption{The field line passing through $(\theta_p, \zeta)=(0,0)$ is highlighted in white on the $r/a = 0.5$ (left), $r/a = 0.7$ (middle) and $r/a = 0.9$ (right) magnetic flux surfaces of W7-X (top) and AUG (bottom) to illustrate how the safety factor, or rotational transform, changes from one flux surface to another. While the field lines start at the same location, they end up at a different $\zeta$ after half a poloidal turn (star). }  \vspace{-5mm}
	\label{fig:MagneticShear}
\end{figure}

Note that magnetic field lines on different magnetic flux surfaces have a different tilt, which is illustrated in figure \ref{fig:MagneticShear} for the Wendelstein 7-X stellarator (W7-X) and the ASDEX Upgrade tokamak (AUG). The rate at which the safety factor $q = \psi'_t/\psi'_p$, or the rotational transform \mbox{$\iota = \psi'_p/\psi'_t$}, changes from flux surface to flux surface, is given \mbox{by the \textit{global magnetic shear}, defined as,}
\index{Magnetic shear!Global magnetic shear}
\begin{equation}
	\label{eq:shat}
	\hat{s} = \frac{r}{q} \Big(\frac{dq}{dr}\Big) = -\frac{r}{\iota} \Big(\frac{d\iota}{dr}\Big).
\end{equation}
It is clear from figure \ref{fig:MagneticShear} that in W7-X (top row) the rotational transform changes only slightly from $r/a = 0.5$ to $r/a = 0.9$, specifically, $\iota = 0.87 \rightarrow 0.94$. Therefore, W7-X is a so-called \textit{low-shear device}. For comparison, in other devices such as AUG, the rotational transform changes strongly from $r/a = 0.5$ to $r/a = 0.9$, namely, $\iota = 0.97 \rightarrow 0.36$. Consequently, AUG is a \textit{high-shear device}. Magnetic shear can play an important role on turbulence, as will be explained in section \ref{sec:fluxtube:shearing}.
\index{Magnetic shear!High-shear device}
\index{Magnetic shear!Low-shear device}

\vspace{2mm}
\section{Field-aligned flux coordinates $(\psi,\alpha,z)$} \vspace{-1mm}
\label{sec:field-aligned-coordinates}

\begin{figure}[!h]
	\centering
	\includegraphics{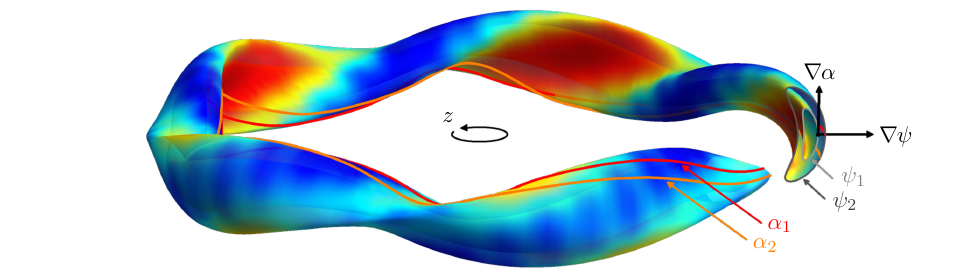}
	\captionsetup{singlelinecheck = false, margin=0cm}
	\vspace{-6.5mm}
	\caption{Field-aligned flux coordinates $(\psi,\alpha,z)$ illustrated on the magnetic flux surfaces of W7-X, with $\psi$ a flux-surface label, $\alpha$ a field-line label and $z$ a coordinate along magnetic field lines. The magnetic field strength is represented by the color scale, with red being its maximum and blue its minimum.}
	\label{fig:W7XGlobalCoordinates}
\end{figure}

A set of \textit{field-aligned flux coordinates} $(\psi,\alpha,z)$ simplifies the expression of the magnetic field $\mathbf{B}$ and the gyrokinetic equation (\ref{eq:GyrokineticEquation}). These $(\psi,\alpha,z)$-coordinates are illustrated in \mbox{figure \ref{fig:W7XGlobalCoordinates}}.  First, a flux-surface label $\psi$ is defined (gray contours in \mbox{figure \ref{fig:W7XGlobalCoordinates}}) which acts as a radial coordinate. Next, a coordinate $z$ is introduced which measures distances along magnetic field lines. Generally, $z=\theta_s$ is chosen for tokamaks, because in axisymmetric devices the magnetic field is independent of the toroidal angle, while $z=\zeta_s$ is typically used for stellarators. Finally, a \textit{field-line label} $\alpha$ is defined  which labels the field lines (red and orange lines in figure \ref{fig:W7XGlobalCoordinates}) on a given magnetic flux surface.
\index{Coordinate systems!Field-aligned flux coordinates}
\index{Coordinates!Field-line label}

\vspace{5mm}
\FloatBarrier
\vspace{3mm}

\noindent
In flux coordinates, the magnetic field can be expressed in \textit{Clebsch form}\index{Clebsch form} \mbox{(see section 6.1 in \cite{haeseleer2012flux}) as,}
\begin{equation}
	\label{eq:B}
	\mathbf{B} = \psi'_t \, \nablam \psi \times \nablam \alpha,
\end{equation}
with $\alpha =  \theta_s  - \iota \zeta_s$. Moreover, for the choice $\psi=\psi_t$ of the flux-surface label, this reduces to $\mathbf{B} = \nablam \psi_t \times \nablam \alpha$. In the gyrokinetic code \texttt{stella}, the stellarator conventions, $\alpha =  \theta_s  - \iota \zeta_s$ and $\mathbf{B} = \psi'_t \, \nablam \psi \times \nablam \alpha$ are employed when reading the magnetic geometry from a \texttt{VMEC} equilibrium. If instead the Miller formalism is used to construct the magnetic geometry, \texttt{stella} defines $\alpha = \zeta_s - q \theta_s$ for which the magnetic field is given by $\mathbf{B} = \psi'_p \nablam \alpha \times \nablam \psi$. For the choice $\psi=\psi_p$ the expression of the magnetic field simplifies to  $\mathbf{B} = \nablam \alpha \times \nablam \psi_p$, which is the Clebsch form typically used in tokamak literature. The coordinates and expressions implemented in \texttt{stella} when defining the magnetic geometry based on a set of Miller parameters are defined in Appendix \ref{chap:app:tok_stell}.
Specifically, the Clebsch form of the magnetic field for different choices of the \mbox{$(\psi,\alpha)$-coordinates are listed in Appendix \ref{app:clebsch}.}

\vspace{1mm}
\section{Flux tube approximation}
\label{sec:fluxtube:approximation}

Turbulence driven by microinstabilities in strongly magnetized plasmas is generally characterized by being very elongated along magnetic field lines while only having short spatial scales perpendicular to the field lines ($k_\parallel \ll k_\perp$). Instead of modeling the entire device, one can thus simplify the problem by solving the gyrokinetic equations in a small tube surrounding a specific field \mbox{line $\alpha_0$}, located on a magnetic flux surface $\psi_0$, which is called a \textit{flux tube}\index{Flux tube approximation!Flux tube}.
In the local flux-tube approximation, quantities are approximated by first order Taylor expansions in $(\psi,\alpha)$ around the center of the flux tube, e.g.~$\iota(\psi) \approx \iota_0 + (d\iota/d\psi) (\psi - \psi_0)$ and $n(\psi) \approx n_0 + (dn/d\psi) (\psi - \psi_0)$, with $(d\iota/d\psi)$ and $(dn/d\psi)$ evaluated at $(\alpha_0,\psi_0)$. As a result, the magnetic geometry and plasma parameters within the simulation domain depend solely on their values along the selected magnetic \mbox{field line $\alpha_0$}, and their derivatives across it. Quantities evaluated along $(\alpha_0,\psi_0)$ are denoted by an index ``0''.

\begin{figure}[h!]
	\centering
	\includegraphics{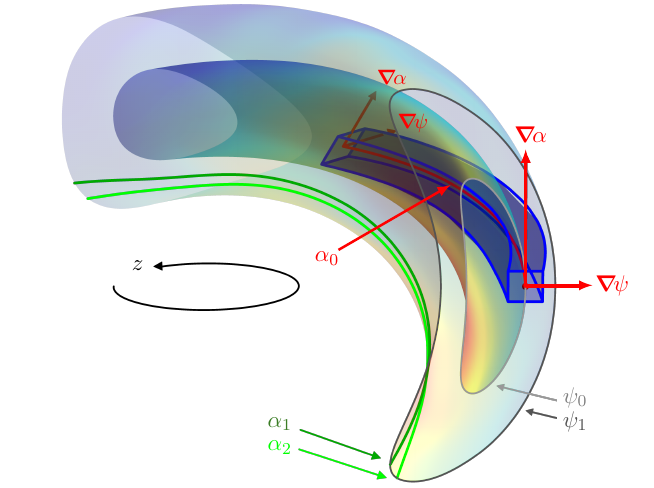}
	\captionsetup{singlelinecheck = false, margin=0cm} \vspace{0mm}
	\caption{Half a period of the W7-X stellarator illustrating the field-aligned $(\psi,\alpha,z)$-coordinates, with $\psi$ a flux-surface label, $\alpha$ a field-line label and $z$ a coordinate along magnetic field lines. The last closed flux surface and the flux surface located at mid-radius are shown, highlighting two field lines in green. The flux tube surrounding the $\alpha_0=0$ field line at mid-radius is shown in blue, covering a quarter of a period.}
	\label{fig:W7XFluxtube}
\end{figure}

The simulation domain---namely, the flux tube---is illustrated in blue in figure \ref{fig:W7XFluxtube}. The chosen magnetic field line $\alpha_0$, located on the magnetic flux surface $\psi_0$, is highlighted in red. The flux tube is constructed by placing small parallelograms around the chosen magnetic field line, aligned with the $\nablam\psi$ and $\nablam\alpha$ directions, with a width $(L_\psi, L_\alpha)$. The field-line label $\alpha = \theta_s - \iota(\psi)\zeta_s$ involves the rotational transform $\iota(\psi)$, which changes with the radial coordinate $\psi$. This radial variation introduces global magnetic shear  as defined by equation (\ref{eq:shat}). As a result, although the $\nablam\psi$ and $\nablam\alpha$ directions may be chosen to be perpendicular at the origin in $z=\zeta_s$, the flux tube cross-section becomes increasingly sheared along the magnetic field line if the global magnetic shear $\hat{s}$ is large. This causes the initial rectangular cross-section to gradually distort into a parallelogram, reflecting the geometric twist inherent to the magnetic field structure.
This flux tube allows us to model turbulence driven by a set of modes whose wavenumbers in the radial and binormal directions are multiples of the fundamental modes defined by the extent of our domain---that is, modes with \textit{radial wavenumbers} $k_\psi=2\pi j/L_\psi$ and \textit{binormal wavenumbers} $k_\alpha = 2\pi k/L_\alpha$, \mbox{where $j$ and $k$ are integers.}
\index{Wavenumbers!Radial wavenumber $k_\psi$}
\index{Wavenumbers!Binormal wavenumber $k_\alpha$}

\vspace{2mm}
\FloatBarrier
\vspace{3mm}

\noindent
In order to simulate turbulence within the cross-section $(L_\psi,L_\alpha)$ of the flux tube, the statistical properties of the fluctuations at $\psi_0-L_\psi/2$  are assumed to be the same as at \mbox{$\psi_0+L_\psi/2$}. Similarly, the fluctuations are assumed to be statistically identical at $\alpha_0-L_\alpha/2$ and \mbox{$\alpha_0+L_\alpha/2$}.  If the simulation box-width $L_\psi$ is larger than the radial correlation length of the fluctuations, and if $L_\alpha$ is larger than the binormal correlation length, then the turbulence should be insensitive to the \textit{perpendicular boundary conditions}, and they can be chosen to be periodic \cite{beer1995field}. As a result, turbulent eddies are effectively modeled to be entering and exiting the perpendicular simulation domain. Seeing that the $(\psi,\alpha,z)$-coordinates are already aligned with the magnetic field, statistical periodicity in the radial and binormal directions is easily enforced through,
\index{Boundary conditions!Perpendicular boundary conditions}
\begin{align}
	\label{eq:fluxtube:parallelbc1}
	\varphi(\psi_0-L_\psi/2, \alpha, z, t) & = \varphi(\psi_0+L_\psi/2, \alpha, z,t) \\[2mm]
	\varphi(\psi, \alpha_0-L_\alpha/2, z, t) & = \varphi(\psi, \alpha_0+L_\alpha/2, z,t)
	\label{eq:fluxtube:parallelbc2}
\end{align}

\newpage
\noindent
Therefore, the turbulent distribution function $g_s$ and the fluctuating electrostatic \mbox{potential $\varphi$}
are \textit{Fourier expanded}\index{Flux tube approximation!Fourier expansion} in the fast variation in $\psi$ and $\alpha$, around the chosen field line $(\psi_0,\alpha_0)$, i.e.
\begin{align}
	\label{eq:gFourier}
	g_s(\psi,\alpha,z,v_\parallel,\mu,t) & = \sum_{k_\psi,k_\alpha} \hat{g}_s(\psi_0,\alpha_0,z,k_\psi,k_\alpha,v_\parallel,\mu,t) \,  \exp \Big(\ir k_\psi (\psi-\psi_0)+\ir k_\alpha (\alpha-\alpha_0) \Big), \\
	\label{eq:phiFourier}
	\varphi(\psi,\alpha,z,t) & = \sum_{k_\psi,k_\alpha} \hat{\varphi}(\psi_0,\alpha_0,z,k_\psi,k_\alpha,t) \,  \exp \Big(\ir k_\psi (\psi-\psi_0)+ \ir k_\alpha (\alpha-\alpha_0) \Big).
	\\[-8mm] \notag
\end{align}
which intrinsically provides periodicity in the $(\psi,\alpha)$-directions. The Fourier coefficients, $\hat{g}_s(\psi_0,\alpha_0,$ $z,k_\psi,k_\alpha,v_\parallel,\mu,t)$ and $\hat{\varphi}(\psi_0,\alpha_0,z,k_\psi,k_\alpha,t)$ will be denoted by $\hatg$ and $\hatvarphi$ for brevity.

\section{Gyrokinetic equations in the flux-tube formalism}
\label{sec:fluxtube:gyrokineticequation}
\vspace{-1mm}

To derive the gyrokinetic equation (\ref{eq:GyrokineticEquation}) in the flux-tube approximation, the equation must be Fourier transformed, expressing all terms in guiding-center coordinates $\mathbf{R}$. This yields the \textit{gyrokinetic equation} for the mode $\hat{g}_s(\mathbf{R}, k_\psi, k_\alpha, v_\parallel, \mu,t)$, i.e.
\index{Gyrokinetic equation!Gyrokinetic equation in Fourier space}
\begin{equation}
	\begin{split}
		\label{eq:FourierGyrokineticEquation}
		\partial_t \hatg + (v_\parallel \hatb \cdot \nablammm z \, \partial_z - \mu \hatb \cdot \nablammm B \, \partial_{v_\parallel}) \, \hatg + \irr \mathbf{k}_\perp \! \cdot \mathbf{v}_\text{Ds} \, \hatg \qquad\qquad\qquad \\[2mm]
		+\,\frac{Z_s e v_\parallel}{T_s} \, \hatb \cdot \nablammm z \,\partial_z [\hatvarphi \, J_0(k_\perp\rho_s)]\, \FM + \frac{\ir Z_se}{T_s} \,\mathbf{v}_\text{Ds} \cdot \mathbf{k}_\perp \, \hatvarphi  \, J_0(k_\perp \rho_s)\, \FM \qquad\\[2mm]
		- \, \frac{\ir k_\alpha}{\psi_t'} \, \hatvarphi  \, J_0(k_\perp \rho_s) \, \Big[ \frac{n'_s}{n_s} + \frac{T'_s}{T_s} \Big( \frac{m_sv_\parallel^2/2 + \mu m_s B}{T_s} - \frac{3}{2} \Big) \Big]\, \FM \qquad\qquad\\[2mm]
		- \, B^{-1} \!\!\!\!\!\!\! \sum_{k_\psi^a,k_\alpha^a,k_\psi^b,k_\alpha^b} \!\!\!\!\!\! (\mathbf{k}_\perp^a  \times \mathbf{k}_\perp^b ) \cdot \hatb\ \hat{\varphi}  (k_\psi^a,k_\alpha^a) \ J_0 (k_\perp^a\rho_s)\ \hat{g}_s(k_\psi^b, k_\alpha^b) \Big|_{\mathbf{k}_\perp^a + \mathbf{k}_\perp^b = \mathbf{k}_\perp} = 0, \\
	\end{split}
\end{equation}
with $\mathbf{k}_\perp = k_\psi \nablam \psi + k_\alpha \nablam \alpha$ the perpendicular wavenumber and $J_0(k_\perp \rho_s)$ the zeroth-order Bessel function of the first kind, appearing as a result of the gyroaverages $\langle \varphi \rangle_\mathbf{R}$ in equation (\ref{eq:GyrokineticEquation}).
\index{Wavenumbers!Perpendicular wavenumber $\mathbf{k}_\perp$}
To close the system, the \textit{quasineutrality condition} (\ref{eq:quasineutrality}) is Fourier transformed, which gives,
\index{Quasineutrality condition!Quasineutrality in Fourier space}
\begin{equation}
	\label{eq:FFTQuasiNeautrality}
	\sum_s Z_s e \!
	\int\limits_{-\infty}^{\infty} \! dv_\parallel
	\int\limits_{0}^{\infty} \! d\mu   \
	\hatg \, J_0(k_\perp\rho_s) - \sum_s  \frac{Z_s^2 e^2 n_s}{2\pi B T_s} \, \hatvarphi \Big( 1 - \Gamma_0 \Big( \frac{k_\perp^2\rho_s^2}{2} \Big) \Big) = 0
\end{equation}
where all quantities are evaluated at constant particle position $\mathbf{r}$ and $\Gamma_0(y) = \exp(-y)\, I_0(y)$ with $I_0$ the zeroth-order modified Bessel function of the first kind.

Equations (\ref{eq:FourierGyrokineticEquation}) and (\ref{eq:FFTQuasiNeautrality}) are the gyrokinetic equation and quasineutrality condition implemented in \texttt{stella} by Barnes et al., corresponding to equations (19) and (25) in Ref.~\cite{stella}. Note that in the flux-tube approximation, the gyrokinetic equations have no explicit dependence on neither $\psi$ nor $\alpha$, and they are therefore solved spectrally in those coordinates. This allows one to solve the gyrokinetic equations independently for each Fourier mode when the nonlinear term---which is the last term on the left-hand-side of equation (\ref{eq:FourierGyrokineticEquation})---is neglected, as interactions between different Fourier modes with perpendicular wavenumbers $\mathbf{k}_\perp^a$ and $\mathbf{k}_\perp^b$ occur \mbox{solely through this term.}

\section{Shearing of the flux tube cross-section}
\label{sec:fluxtube:shearing}
\vspace{-1mm}

\index{Magnetic shear!Shearing of the flux-tube}
While each term in the Fourier series (eqs.~\ref{eq:gFourier}-\ref{eq:phiFourier}) is a plane wave in $(\psi,\alpha)$-coordinates, the wavefronts in real space can be very distorted. The most pronounced distortion typically arises from magnetic shear. The effect of magnetic shear on turbulent fluctuations and on the perpendicular cross-section of the flux tube is explained in section \ref{eq:shearingRealSpace}. Next, it is examined how the global magnetic shear affects the perpendicular  scale $k_\perp$ of the turbulent fluctuations in section \ref{sec:shearingFourierSpace}.

\subsection{Shearing of the perpendicular cross-section and turbulent fluctuations}
\label{eq:shearingRealSpace}

\begin{figure}[b!]
	\centering
	\includegraphics[trim={0mm 17mm 0mm 0mm},clip]{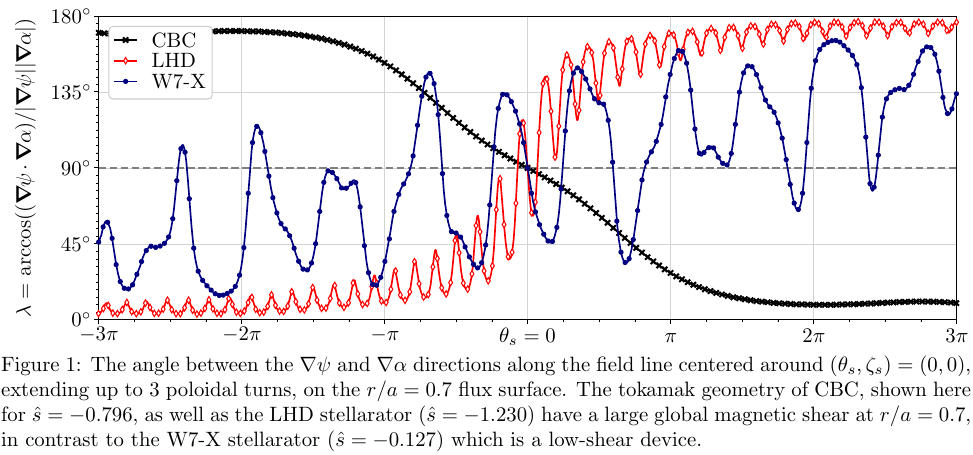}
	\captionsetup{singlelinecheck=false,  margin=0.0cm} \vspace{-7mm}
	\caption{The angle between the $\nablam\psi$ and $\nablam\alpha$ directions along the field line centered around $(\theta_s,\zeta_s) = (0,0)$, extending up to 3 poloidal turns, on the $r/a=0.7$ flux surface. The tokamak geometry of CBC ($\hat{s} = 0.796$) as well as the LHD stellarator $(\hat{s} = -1.230)$ have a large global magnetic shear at $r/a=0.7$, in contrast to the W7-X stellarator $(\hat{s} = -0.127)$ which is a low-shear device.  }
	\label{fig:lambda}
\end{figure}

In the flux-tube formalism, the magnetic geometry is determined entirely by its values along the chosen magnetic field line $\alpha_0$ and by derivatives across it, such as that of the rotational transform, which introduces the global magnetic shear $\hat{s}$ (eq.~\ref{eq:shat}). Therefore, field lines on different flux surfaces have a different rotational transform (see section \ref{sec:straight-field-line-coords}), which means that the field lines on the $\psi_0 - L_\psi/2$ flux surface are tilted differently with respect to the field lines on the $\psi_0 + L_\psi/2$ flux surface. As a result, the cross-section of the flux tube is twisted as one moves along the field line.
This can be seen by Taylor expanding the rotational transform, $\iota(\psi) \approx \iota_0 + (d\iota/d\psi)_0\, (\psi-\psi_0)$, with $(d\iota/d\psi)_0 \! = \!- \hat{s}_0\iota_0/r_0$ evaluated at $(\alpha_0, \psi_0)$. Substituting $\iota(\psi)$ into the field-line label gives,
\begin{align}
	\label{eq:alphaq}
	\alpha & = \theta_s - \iota(\psi)\,\zeta_s \approx \theta_s - \iota_0\zeta_s  + \zeta_s (\hat{s}_0\iota_0/r_0)  (\psi - \psi_0).
\end{align}
The third term has a secular dependence on the parallel location ($z=\zeta_s$) which scales with the global magnetic shear. To analyze how magnetic shear deforms the flux-tube cross-section perpendicular to the magnetic field lines, we examine the angle $\lambda$ between the $\nablam\psi$ and $\nablam\alpha$ directions,
\index{Magnetic shear!Angle $\lambda$ between the $(\psi,\alpha)$ directions}
\index{Coordinates!Angle $\lambda$ between the $(\psi,\alpha)$-directions}
\index{Magnetic shear!Local magnetic shear}
\begin{equation}
	\label{eq:coslambda}
	\cos \lambda = \frac{\boldsymbol{\nablam}\psi  \cdot \boldsymbol{\nablam}\alpha}{|\boldsymbol{\nablam}\psi|\,|\boldsymbol{\nablam}\alpha|} \approx \frac{\boldsymbol{\nablam}\psi  \cdot (\boldsymbol{\nablam}\theta_s - \iota_0 \boldsymbol{\nablam}\zeta_s)} {|\boldsymbol{\nablam}\psi|\,|\boldsymbol{\nablam}\alpha|} + \frac{|\boldsymbol{\nablam}\psi|}{|\boldsymbol{\nablam}\alpha|} \Big( \frac{\hat{s}_0\iota_0}{r_0} \Big) \, \zeta_s.
\end{equation}
The second term continuously increases in magnitude with the parallel location ($z=\zeta_s$) and thus increasingly shears the flux-tube cross-section due to the \textit{global magnetic shear}, $\hat{s}$, which plays a big role in high-shear devices. On the other hand, in low-shear devices, the first term, which generally changes its sign rapidly along $z$, plays a more prominent role.
\index{Magnetic shear!Global magnetic shear}
If the first term is large, then the flux tube is typically sheared and unsheared multiple times. This is illustrated in \mbox{figure \ref{fig:lambda}}, where the angle between the $\nablam\psi$ and $\nablam\alpha$ directions is depicted along the $\alpha=0$ field line. Note that the perpendicular cross-section of the flux tube is generally chosen such that it is rectangular $(\lambda=90^\circ)$ at the midpoint of the field line $(z=0)$. The tokamak geometry of CBC has a large global magnetic shear on the considered flux surface. As a result, the flux-tube cross-section becomes extremely sheared after one poloidal turn. In contrast, in the low-shear stellarator W7-X the global magnetic shear is very small, and is responsible for the slight increase that $|\lambda(z)|$ experiences on average. However, the effect of the local magnetic shear is much more prominent, which shears and unshears the cross-section continuously. As a result, there are multiple locations along the flux tube where the $\nablam\psi$ and $\nablam\alpha$ directions are perpendicular $(\lambda=90^\circ)$. Finally, for a high-shear stellarator such as LHD, both effects contribute significantly to the flux-tube shearing.

\begin{figure}[h!]
	\centering
	\vspace{-2mm}
	\includegraphics[trim={0mm 20mm 0mm 0mm},clip] {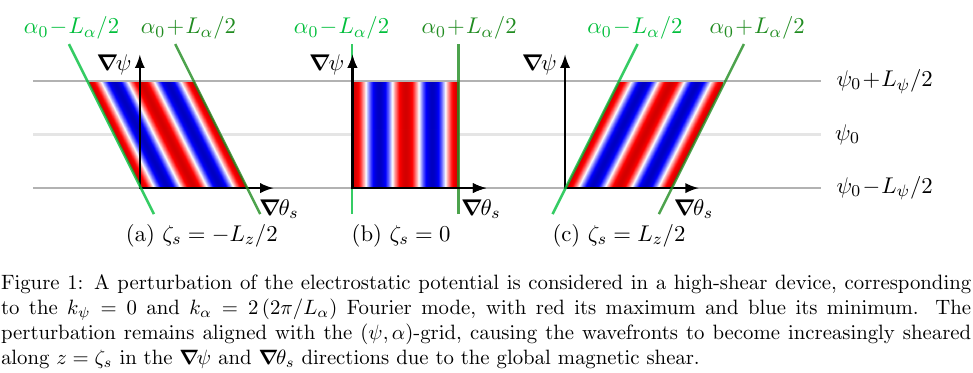}
	\captionsetup{singlelinecheck=false, margin=0.0cm} \vspace{-7.5mm}
	\caption{A perturbation of the electrostatic potential is considered in a high-shear device, corresponding to the $k_\psi=0$ and $k_\alpha = 2k_{\alpha,\text{min}}$ Fourier mode, with $k_{\alpha,\text{min}} = 2\pi/L_\alpha$, where red represents the maximum potential and blue the minimum. The perturbation remains aligned with the $(\psi, \alpha)$-grid, causing the wavefronts to become increasingly sheared along $z = \zeta_s$ in the $\nablam \psi$ and $\nablam \theta_s$ directions due to \mbox{the global magnetic shear.}}  \vspace{-3mm}
	\label{fig:parallelogramFourier}
\end{figure}

Charged particles move rapidly along magnetic field lines while slowly drifting across them (section \ref{sec:intro:MCF}). As a result, perturbations of the plasma tend to stick to magnetic field lines, causing them to align with the $(\psi, \alpha)$-grid, which is illustrated in figure \ref{fig:parallelogramFourier}. A perturbation of the electrostatic potential $\varphi(\mathbf{r},t)$ is introduced in the flux tube corresponding to the \mbox{$(k_\psi, k_\alpha) = (0, 2 k_{\alpha,\text{min}})$}  Fourier mode, with $k_{\alpha,\text{min}} = 2\pi/L_\alpha$, where red represents the maximum potential and blue the minimum. The $\alpha_0 - L_\alpha/2$ (light green) and $\alpha_0 + L_\alpha/2$ (dark green) field lines, as well as the $\psi_0 - L_\psi/2$ (bottom line)  and $\psi_0 + L_\psi/2$ (top line) flux surfaces are highlighted, defining the perpendicular boundaries of the flux tube. It can be seen that the wavefronts remain aligned with the $(\psi,\alpha)$-grid at every $\zeta_s$-location. Consequently, the wavefronts in \mbox{real space become sheared.}

\FloatBarrier

\subsection{Effect of global magnetic shear on the perpendicular scale of turbulence}
\label{sec:shearingFourierSpace}

\begin{figure}[h!]
	\centering
	\vspace{-2mm}
	\includegraphics[trim={0mm 10mm 0mm 0mm},clip] {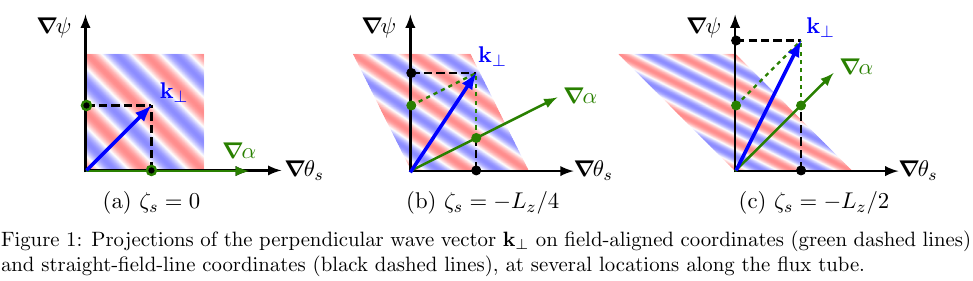}
	\captionsetup{singlelinecheck=false, margin=0.0cm} \vspace{-7.5mm}
	\caption{Projections of the perpendicular wavevector $\mathbf{k}_\perp$ on field-aligned coordinates (green dashed lines) and straight-field-line coordinates (black dashed lines), at several locations along the flux tube.}  \vspace{-6mm}
	\label{fig:parallelogramFourierKpsiKalpha}
\end{figure}

\noindent
It is essential to understand how the global magnetic shear affects the perpendicular scale of turbulent perturbations at different locations along the flux tube. Using equation (\ref{eq:alphaq}), which reads \mbox{$\alpha = \theta_s - \iota_0\zeta_s + \zeta_s (\hat{s}_0\iota_0/r_0) (\psi - \psi_0)$}, the perpendicular wavevector can be expressed as,
\begin{equation}
	\label{eq:kperp2}
	\mathbf{k}_\perp = k_\psi \nablam \psi + k_\alpha \nablam \alpha = \Big( k_\psi +  k_\alpha \frac{\hat{s}_0\iota_0}{r_0} \, \zeta_s \Big) \nablam \psi + k_\alpha ( \nablam\theta_s - \iota_0\nablam\zeta_s ).
\end{equation}
Therefore, the Fourier modes are plane waves along the $\nablam\psi$ and $\nablam\alpha$ directions in field-aligned coordinates, corresponding to constant $(k_\psi,k_\alpha)$ wavenumbers.
This is illustrated in figure~\ref{fig:parallelogramFourierKpsiKalpha}, which shows how the perpendicular wavevector of an electrostatic potential perturbation evolves along the flux tube due to the global magnetic shear.  In this example a Fourier mode with \mbox{$k_\psi = 2k_{\psi,\text{min}}$} and \mbox{$k_\alpha = 2k_{\alpha,\text{min}}$} is considered. Moreover, to isolate the effect of $\hat{s}$, a magnetic geometry is considered where only the global magnetic shear determines the flux tube shearing---that is, only the second term in equation (\ref{eq:coslambda}) needs to be considered. The perpendicular wavevector $\mathbf{k}_\perp$ is orientated perpendicular to the wavefronts. The projection of $\mathbf{k}_\perp$ onto the field-aligned coordinates (green dashed lines) reveals constant wavenumbers $(k_\psi,k_\alpha)$ along $\nablam\psi$ and $\nablam\alpha$, since both the perturbation and coordinates are sheared at the same rate. As a result of this shearing, turbulent eddies become elongated, and the magnitude of the perpendicular wavevector increases.

\FloatBarrier

\section{Finite-Larmor-radius (FLR) effects within a flux tube}
\label{sec:FLR}

\begin{figure}[h!]
	\centering
	\includegraphics[trim={0mm 10mm 0mm 0mm},clip] {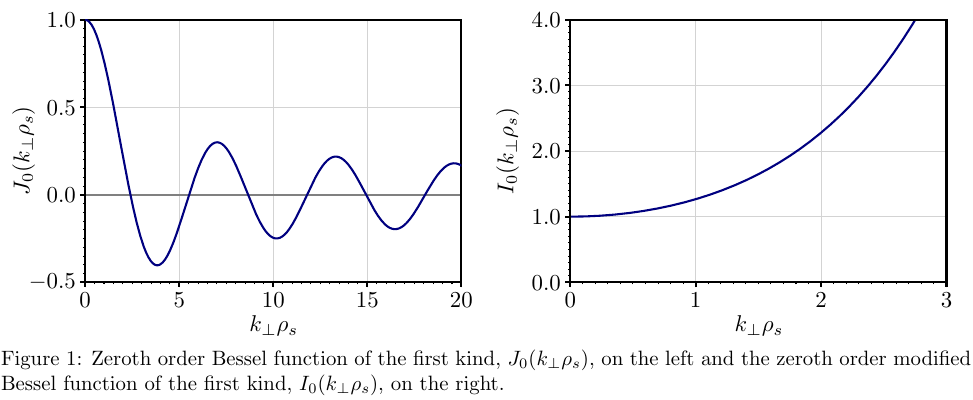}
	\captionsetup{singlelinecheck=false, margin=0cm} \vspace{-6.5mm}
	\caption{Zeroth order Bessel function of the first kind, $J_0(k_\perp \rho_s)$, on the left and the zeroth order modified Bessel function of the first kind, $I_0(k_\perp \rho_s)$, on the right.} \vspace{2mm}
	\label{fig:ZerothOrderBesselFunction}
\end{figure}

In deriving the gyrokinetic equations (\ref{eq:GyrokineticEquation}) and (\ref{eq:quasineutrality}), the rapid gyromotion of particles around magnetic field lines has been averaged out, as illustrated in figure \ref{fig:6Dto5D}. Nonetheless, the effects of the rapid gyromotion of the particles on the plasma behavior is retained by modifying the electrostatic potential and distribution function with $J_0(k_\perp \rho_s)$ in equations (\ref{eq:FourierGyrokineticEquation}) and (\ref{eq:FFTQuasiNeautrality}). This factor $J_0(k_\perp \rho_s)$ captures the dependence of the rapid gyromotion on the Larmor radius $\rho_s$.  The \textit{finite-Larmor-radius (FLR) effects} thus modify the interaction between particles and electromagnetic fields and play an important role in accurately describing the plasma dynamics. The dependence of $J_0(k_\perp \rho_s)$---and of $I_0(k_\perp \rho_s)$---on $k_\perp \rho_s$ is shown in figure \ref{fig:ZerothOrderBesselFunction}.

\vspace{3mm}
\FloatBarrier
\vspace{0mm}

\noindent
For large-scale fluctuations $(k_\perp \rho_s \ll 1)$, the zeroth order Bessel function is given by \mbox{$J_0(k_\perp \rho_s) \approx 1$.} Therefore, the dynamics of particles in the presence of large-scale fluctuations are largely unaffected by the gyro-average. In contrast, for small-scale fluctuations $(k_\perp \rho_s \gg 1)$, the zeroth order Bessel function approaches zero, meaning that the response of the particles to these small-scale fluctuations is significantly reduced. Physically, this reduction occurs because, as the particles gyrate in a circular motion around  magnetic field lines, the small-scale fluctuations effectively average out over a full gyro-orbit. Consequently, small-scale fluctuations are damped due to the gyro-average or finite-Larmor-radius (FLR) effect, as the particles cannot respond effectively to them. Note that this FLR-damping is much stronger for ions than for electrons since $\rho_i = \sqrt{m_e/m_i} \, \rho_e$. Specifically, when analyzing ion-scale turbulence, i.e.~when $k_\perp\rho_i \sim 1$, we find that $k_\perp\rho_e \ll 1$ and $J_0(k_\perp \rho_e) \approx 1$.

\begin{figure}[h!]
	\centering
	\vspace{1mm}
	\includegraphics{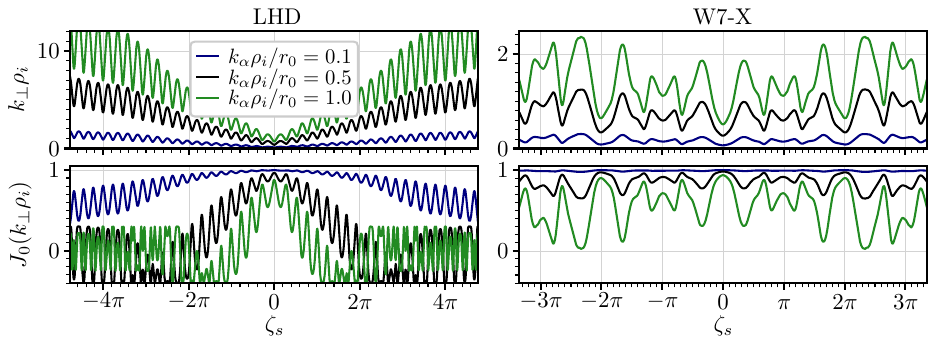}
	\captionsetup{singlelinecheck=false, margin=0cm}
	\vspace{-2mm}\caption{The perpendicular wavenumber (top row) and the zeroth order Bessel function of the first kind (bottom row) for different modes along the $\alpha=0$ field line in LHD (left) and W7-X (right) for $k_r\rho_i = 0$.}
	\vspace{1mm}
	\label{fig:BesselAlpha}
\end{figure}

To analyze the finite-Larmor-radius effects, it is essential to examine how $J_0(k_\perp\rho_s)$ varies within the flux tube. The value of $J_0(k_\perp\rho_s)$ depends on the magnitude of the perpendicular wavenumber,
\begin{equation}
	\label{eq:kperp}
	k_\perp = \sqrt{k_r^2 \,|\nablam r|^2 + 2\,k_r k_\alpha\nablam r \cdot \nablam \alpha + k_\alpha^2 |\nablam \alpha |^2},
\end{equation}
which changes along the field line. The flux-surface label is chosen to be the effective minor radius, i.e.~$\psi=r$. Consequently, an instability with wavenumbers $({k}_r, {k}_\alpha)$ tends to be localized in regions along the field line where $k_\perp \rho_s(z)$ is minimum. Specifically, modes with a vanishing radial wavenumber, $k_r=0$, have a perpendicular wavenumber $k_\perp = |k_\alpha\!\nablam \alpha|$. For such modes $J_0(k_\perp\rho_s)$ can approach zero along the field line if the $(\hat{s}_0\iota_0/r_0) \, \zeta_s \boldsymbol{\nablam} r$ term in $\nablam\alpha$ (eq.~\ref{eq:alphaq}) becomes sufficiently large, which occurs in high-shear devices.
This is illustrated for LHD on the left of figure \ref{fig:BesselAlpha}. It can be seen that modes with $k_r=0$ would likely be localized around the center of the flux tube, where $k_\perp\rho_s$ is small (top row) and hence $J_0(k_\perp\rho_s) \approx 1$ (bottom row). On the other hand, $k_\perp\rho_s$ becomes very large towards the ends of the flux tube, where the modes are damped as $J_0(k_\perp\rho_s) \rightarrow 0$.
This behavior of damping modes far away from the center of the flux tube is due to the global magnetic shear. Physically, perturbations are sheared into smaller scale fluctuations perpendicular to the magnetic field, which are consequently suppressed by the FLR effects. In devices with low magnetic shear, such as W7-X, the local shear dominates the behavior, as can be seen on the right side of figure \ref{fig:BesselAlpha}. For W7-X, there are multiple locations along the field line where the modes remain undamped because $J_0(k_\perp\rho_s) \approx 1$, whereas for locations where $J_0(k_\perp\rho_s) \approx 0$ the modes are suppressed. The flux tube would have to be extended significantly before the damping effect of the  global magnetic shear can be observed in W7-X.

\FloatBarrier

\begin{figure}[h!]
	\centering
	\vspace{1mm}
	\includegraphics{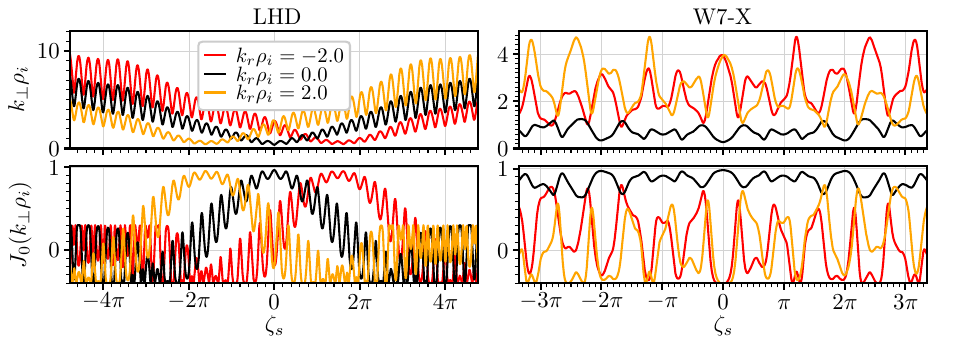}
	\captionsetup{singlelinecheck=false, margin=0cm}
	\vspace{-2mm}\caption{The perpendicular wavenumber (top row) and the zeroth order Bessel function of the first kind (bottom row) for different modes along the field line in LHD (left) and W7-X (right) for $k_\alpha\rho_i/r_0 = 0.5$.}
	\vspace{-1mm}
	\label{fig:BesselPsi}
\end{figure}

The effect of considering a finite radial wavenumber $(k_r \neq 0)$ is illustrated in figure \ref{fig:BesselPsi}. Modes with $k_r=0$ and a sufficiently large $k_\alpha$ would be localized at the center of the flux tube in LHD due its large global magnetic shear (figure \ref{fig:BesselAlpha}). As the radial wavenumber is increased, the location where the mode remains undamped is shifted along the field line.
From equation (\ref{eq:kperp2}) it follows that the magnitude of the perpendicular wavenumber is the smallest if $k_r + k_\alpha (\hat{s}_0\iota_0/r_0)\,\zeta_s = 0$, when the last term of eq.~(\ref{eq:coslambda}) dominates. At $\zeta_s = -(k_r/k_\alpha)(r_0/\hat{s}_0\iota_0)$, the mode is the least damped. This position is referred to as the \textit{ballooning angle}, as it marks the location where the instability is most likely to grow---or ``balloon''---most strongly in a high-shear device.
\index{Magnetic shear!Ballooning angle}
Specifically, modes with $k_r>0$ (orange) are excited towards $\zeta_s<0$ in LHD, whereas modes with $k_r<0$ (red) are excited towards $\zeta_s>0$. The direction of this translation depends on the sign of  $(r_0/\hat{s}_0\iota_0)$. On the other hand, in W7-X, the most prominent effect of considering a finite radial wavenumber, is that the magnitude of  $k_\perp$ increases, and modes are damped more strongly along the field line, which would make them more localized.

\vspace{1mm}
\FloatBarrier
\vspace{1mm}

\begin{figure}[h!]
	\centering
	\includegraphics{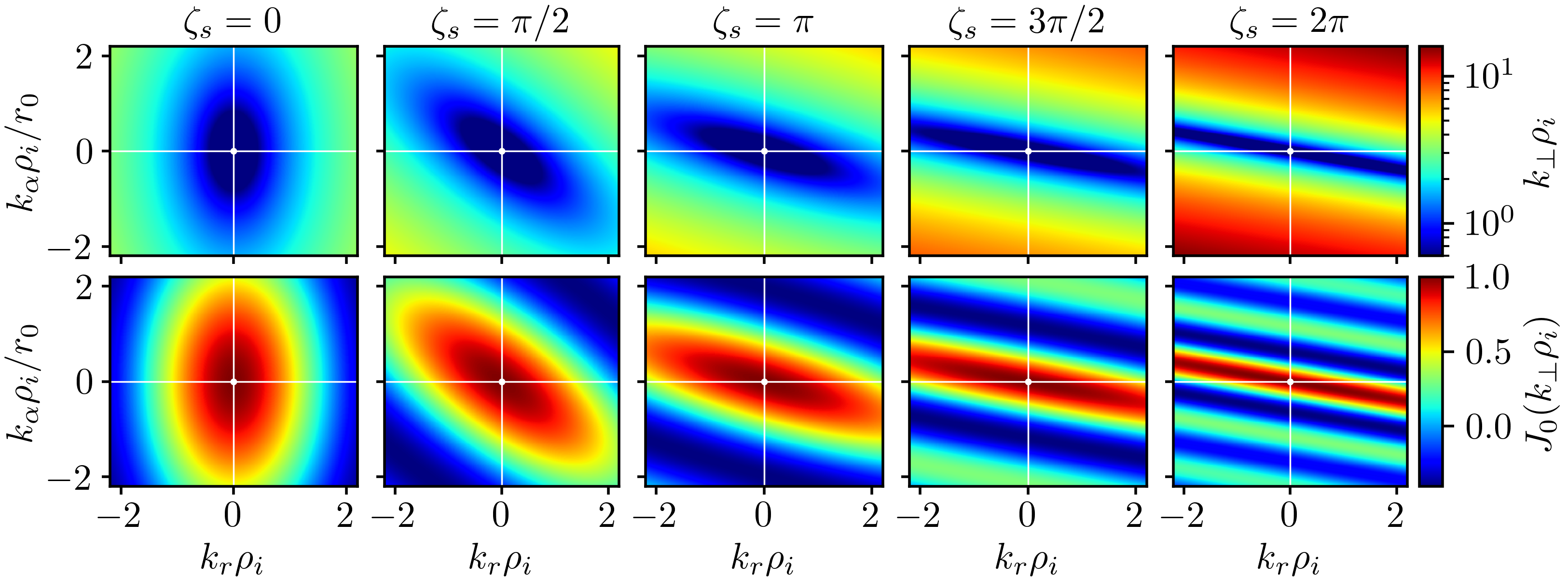}
	\captionsetup{singlelinecheck=false, margin=0cm}
	\vspace{-7mm}\caption{The perpendicular wavenumber represented on a logarithmic scale (top row) and the zeroth order Bessel function of the first kind (bottom row) at 5 locations along the field line for $r/a=0.7$ in LHD. } \vspace{-1mm}
	\label{fig:BesselLHDFixedZPoints}
\end{figure}

\newpage
Instead of considering a fixed wavenumber, a fixed location along the field line can be considered to examine $J_0(k_\perp\rho_s)$. If the strongest instability drives are found at a certain location $z=\zeta_s$ along the field line, then the most unstable mode, is most likely the $(k_r, k_\alpha)$-mode that minimizes $k_\perp \rho_s(k_r, k_\alpha)$. This is illustrated on the $(k_r, k_\alpha)$-plane in figure \ref{fig:BesselLHDFixedZPoints} for LHD, by looking at the perpendicular wavenumber (top row) and the zeroth order Bessel function (bottom row) at fixed positions along the field line. If an instability, or mode, is excited with a large perpendicular wavenumber (corresponding to the reddish regions in the top row) it will experience strong damping as $J_0 \approx 0$ (indicated by the bluish regions in the bottom row). In contrast, modes which have a small perpendicular wavenumber at a certain parallel location (bluish regions in the top row) will remain largely undamped since $J_0 \approx 1$ (reddish regions in the bottom row).
Moreover, the effect of the large global magnetic shear of LHD can be clearly observed in \mbox{figure \ref{fig:BesselLHDFixedZPoints}}, since the minima of the perpendicular wavenumber are found where $k_r + k_\alpha (\hat{s}_0\iota_0/r_0)\,\zeta_s = 0$, since the flux tube cross-section is chosen to be perpendicular at $\zeta_s=0$. Specifically, the minima are found for $k_r=0$ at $\zeta_s=0$, whereas they are found along a diagonal in the $(k_r, k_\alpha)$-plane as $|\zeta_s|$ increases.

\vspace{-1mm}
\FloatBarrier

\begin{figure}[h!]
	\centering
	\includegraphics{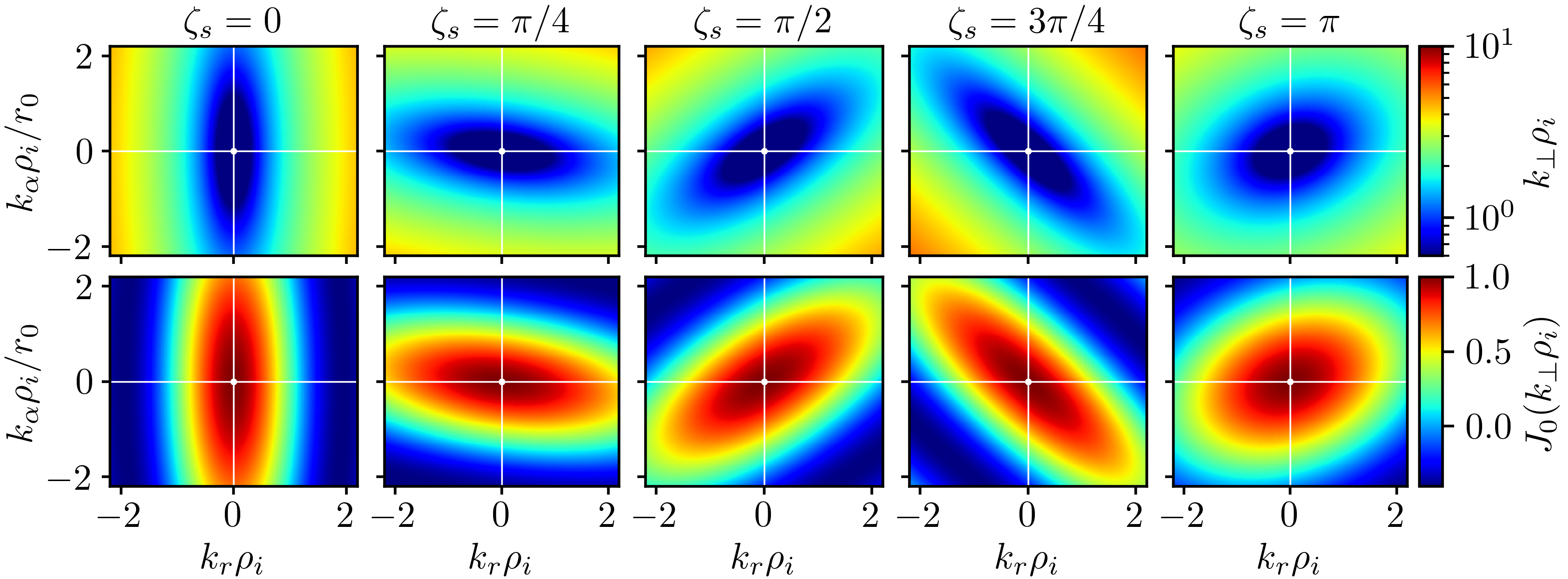}
	\captionsetup{singlelinecheck=false, margin=0cm}
	\vspace{-7.5mm}\caption{The perpendicular wavenumber represented on a logarithmic scale (top row) and the zeroth order Bessel function of the first kind (bottom row) at 5 locations along the field line for $r/a=0.7$ in W7-X.} \vspace{-5mm}
	\label{fig:BesselW7XFixedZPoints}
\end{figure}

\noindent
For a low-shear device like W7-X, shown in figure \ref{fig:BesselW7XFixedZPoints}, very different behavior emerges.
In this case, the global magnetic shear is very small, so the magnitude of the perpendicular wavenumber is primarily governed by local shear effects, captured by the variation of $\nablam \psi$, $\nablam \theta_s$, and $\nablam \zeta_s$ along the field line (eq.~\ref{eq:kperp2}). As a result, the flux tube cross-section undergoes alternating shearing and unshearing along the field line. The smallest perpendicular wavenumbers are found along $k_r=0$ at $\zeta_s=0$, where the flux tube cross-section is chosen to be perpendicular. However, at other parallel locations they tend to be found along various diagonals.

\vspace{-1mm}
\FloatBarrier

\newpage
\section{Parallel boundary conditions}
\label{sec:parallelBC}

Periodic boundary conditions are imposed along $\psi$ and $\alpha$ by Fourier expanding the fluctuating quantities in the fast variation in $\psi$ and $\alpha$ (eqs.~\ref{eq:gFourier}-\ref{eq:phiFourier}). On the other hand, \textit{parallel boundary conditions}, i.e.~boundary conditions pertaining to the parallel coordinate $z$, need to be defined.
\index{Parallel boundary conditions}
The most common approach for such boundary conditions is to use the so-called ``twist-and-shift'' condition, derived by Beer et al.~in Ref.~\cite{beer1995field}, which are summarized in section \ref{sec:twistandshift}. Alternatively, a new set of orthogonal $(u,v)$-coordinates can be defined to apply \textit{stellarator-symmetric boundary conditions} as derived by Martin et al.~in Ref.~\cite{martin2018parallel}, and implemented in \texttt{stella} by Gonzalez for the simulations in Ref.~\cite{gonzalez2024electrostatic}. These boundary conditions are outlined in section \ref{sec:StellaratorSymmetricBC}.
\index{Twist-and-shift boundary conditions}
\index{Stellarator-symmetric boundary conditions}

\subsection{Twist-and-shift parallel boundary conditions}
\label{sec:twistandshift}

\begin{figure}[h!]
	\centering
	\includegraphics[trim={0mm 0mm 0mm 12mm},clip] {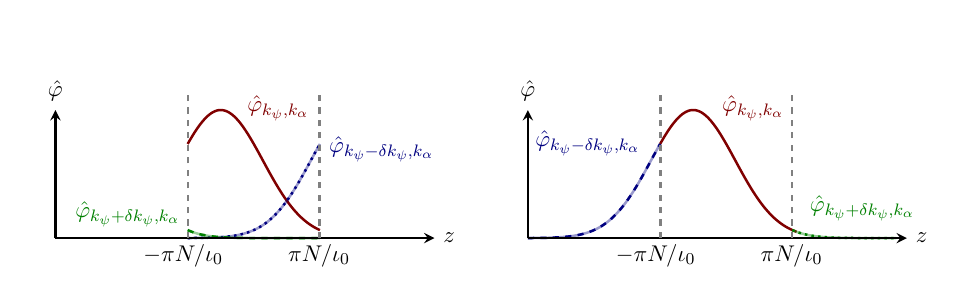}
	\captionsetup{singlelinecheck=false, margin=0cm}
	\vspace{-13mm}\caption{Boundary condition for $	\hat{\varphi}_{k_\psi,k_\alpha}(z)$ along the parallel coordinate, represented on the $z$-domain (left) and on an extended $z$-domain (right). Different $k_\psi$ modes with the same $k_\alpha$ are connected at $z=-\pi N/\iota_0$ and $z=\pi N/\iota_0$. On an extended $z$-grid, the Fourier coefficients form a continuous chain along $z$. \vspace{1mm} }
	\label{fig:TwistAndShiftGaussians}
\end{figure}

To apply the standard twist-and-shift boundary conditions, it is assumed that fluctuations are statistically identical at two locations with the same poloidal angle (but different toroidal angle). This is valid in axisymmetric devices, however it is formally incorrect in stellarators. These boundary conditions are derived by Beer et al.~in Ref.~\cite{beer1995field}, and are summarized here. The starting point is the Fourier expansion in straight-field-line coordinates obtained in by substituting $\alpha = \theta_s - \iota(\psi)\zeta_s$ and $\iota(\psi) = \iota_0 - (\hat{s}_0\iota_0/r_0)\, (\psi-\psi_0)$ into equation (\ref{eq:phiFourier}) which gives,
\index{Parallel boundary conditions:Standard twist-and-shift parallel boundary conditions}
\begin{align}
	\label{eq:varphikpsikalpha3}
	\varphi(\psi,\alpha,z,t)
	& = \sum_{k_\psi,k_\alpha} \hat{\varphi}_{k_\psi,k_\alpha} (z, t) \, \exp \Bigg( \ir \Big( k_\psi + k_\alpha \frac{\hat{s}_0\iota_0}{r_0} \zeta_s \Big) (\psi-\psi_0) +  \ir k_\alpha (\theta_s - \iota_0\zeta_s ) \Bigg).
\end{align}
Fluctuating quantities are assumed to be periodic after $N$ poloidal turns, with $N$ \mbox{an integer number.} Using stellarator conventions, we assume periodicity in $z=\zeta_s$ with a period $2\pi N/\iota_0$ while holding $(\psi,\theta_s)$ fixed. This implies that when a particular $(k'_\psi,k'_\alpha)$ Fourier mode gets to the \mbox{end of the} $z$-domain $(\zeta_s = \pi N/\iota_0)$, the mode connects exactly to another Fourier mode $(k_\psi,k_\alpha)$ at the beginning of the the $z$-domain $(\zeta_s = -\pi N/\iota_0)$. Substituting this condition into equation (\ref{eq:varphikpsikalpha3}) gives,
\begin{equation}
	\label{eq:parallelBC}
	\begin{split}
		\hat{\varphi}_{k_\psi,k_\alpha}\Big( - \frac{\pi N}{\iota_0}, t \Big) \,  \exp  \! \Bigg(\ir \Big(k_\psi \,-\, k_\alpha \frac{\hat{s}_0\iota_0}{r_0}  \frac{\pi N}{\iota_0}    \Big) (\psi-\psi_0) + \ir k_\alpha \Big(\theta_s \,+\, \iota_0  \frac{\pi N}{\iota_0} \Big) \Bigg) \\
		=
		\hat{\varphi}_{k'_\psi,k'_\alpha} \Big( \frac{\pi N}{\iota_0}, t \Big) \,  \exp \! \Bigg(\ir \Big(k'_\psi + k'_\alpha \frac{\hat{s}_0\iota_0}{r_0}  \frac{\pi N}{\iota_0} \Big)  (\psi-\psi_0)  + \ir k'_\alpha \Big(\theta_s - \iota_0  \frac{\pi N}{\iota_0}  \Big) \! \Bigg)
	\end{split}
\end{equation}
For this equation to hold for all field lines, the factors in front of $\theta_s$ should match. Therefore, the binormal wavenumber remains unchanged, $k'_\alpha=k_\alpha$. Moreover, for equality to hold for all flux surfaces, the factors in front of $\psi$ should match, thus we must have,
\begin{equation}
	\label{eq:kpsi}
	\delta k_\psi = k'_\psi - k_\psi = -k_\alpha \, \frac{\hat{s}_0}{r_0} \, 2\pi N.
\end{equation}
Recall that the wavenumbers $(k_\psi, k_\alpha)$ in the Fourier expansions (eqs.~\ref{eq:gFourier} and \ref{eq:phiFourier}) are discretized. Specifically, \mbox{$k_\psi=2\pi j/L_\psi$} and $k_\alpha = 2\pi k/L_\alpha$, where $j$ and $k$ are integers. To ensure that $k'_\psi$ belongs to the wavenumber grid, and assuming that the $k_\psi$ grid consists of integer multiples of some \mbox{$k_{\psi,\text{min}} = 2\pi/L_\psi$}, then the difference $k'_\psi-k_\psi$ should be an integer multiple of $k_{\psi,\text{min}}$, i.e.%
\begin{equation}
	\delta k_\psi = k'_\psi - k_\psi  =  J\, k_{\psi,\text{min}}
\end{equation}
In particular, equation (\ref{eq:kpsi}) must be true for the smallest nonzero $k_{\alpha,\text{min}}=2\pi/L_\alpha$, which gives,
\begin{equation}
	\label{eq:kalphamin}
	-k_{\alpha,\text{min}} \, \frac{\hat{s}_0}{r_0} \, 2\pi N = J\, k_{\psi,\text{min}}
\end{equation}
This condition determines the radial box size $L_\psi$ (or equivalently the step size $k_{\psi, \text{min}}$) of the flux tube for a given binormal box size $L_\alpha$. Therefore, it imposes a quantization on the perpendicular domain's aspect ratio. This corresponds to an upper limit on the smallest radial wavenumber $k_{\psi,\text{min}}$ that has to be included in our $(k_\psi,k_\alpha)$-grid. Specifically, for $J=1$ the largest  $k_{\psi,\text{min}}$ is obtained, whereas for $J>1$ it becomes smaller. It is clear that for low-shear devices ($|\hat{s}_0\iota_0/r_0| \ll 1)$ we are forced to take a very large radial box size $L_\psi$, equivalent to a very small $k_{\psi,\text{min}}$. As a result, many grid points along $k_\psi$ need to be included in order to include the relevant scales at $k_\perp\rho_i \approx 1$. Using this quantization, the parallel periodicity condition \mbox{imposed in equation (\ref{eq:parallelBC}) gives,~}
\begin{align}
	\label{eq:bcphaseshift}
	\hat{\varphi}_{k_\psi,k_\alpha}(-\pi N/\iota_0, t) \, \exp \Big( \ir  k_\alpha  2\pi N \Big) = \hat{\varphi}_{k_\psi+\delta k_\psi,k_\alpha}(\pi N/\iota_0, t),
\end{align}
with $C_{k_\alpha} = \exp ( i  k_\alpha 2\pi N )$ a phase-factor which recenters the flux tube around $\alpha_0 + 2\pi N$ as a perturbation is connected across the $z$-boundary. The phase factor is equal to one if $N$ is an integer and $L_\alpha=2\pi$. This means that on a rational surface, i.e.~if $N/\iota_0$ is an integer, field lines connect exactly back onto themselves after $N/\iota_0$ toroidal turns.
This phase factor is important in full-flux-surface simulations, to ensure that field lines are connected properly. In contrast, in flux-tube simulations, the binormal wavenumber $k_\alpha$ is assumed to be sufficiently small in order to approximate the phase factor $C_{k_\alpha}$ to be 1. Therefore, in the flux-tube approximation, we assume,
\begin{equation}
	C_{k_\alpha} = \exp ( \ir  k_\alpha  2\pi N ) \approx 1.
\end{equation}
For modes with $k_\alpha=0$, we obtain $C_{k_\alpha} = 1$ and $\delta k_\psi = 0$. Therefore, the $k_\alpha=0$ modes connect onto themselves and their periodicity condition simplifies to \textit{self-periodicity}. Note that the field-line averaged components of the $k_\alpha=0$ modes are referred to as the \textit{zonal modes}, which have no variation in the radial direction nor in the parallel direction. The self-periodic condition states,
\begin{equation}
	\label{eq:self-periodicity}
	\hat{\varphi}_{k_\psi,0}(\zeta_s+2\pi N/\iota_0,t) = \hat{\varphi}_{k_\psi,0}(\zeta_s,t).
\end{equation}
The periodicity condition on the mode amplitudes in equation (\ref{eq:bcphaseshift}) states that the value of a coefficient $\hat{\varphi}_{k_\psi,k_\alpha}$ at one end of the flux tube is specified by the value of another coefficient, with the same $k_\alpha$ but a different $k_\psi$, at the other end of the flux tube. This is represented graphically in figure \ref{fig:TwistAndShiftGaussians}, where the Fourier mode $\hat{\varphi}_{k_\psi,k_\alpha}(z)$ is taken to be a Gaussian on an extended $z$-grid. The flux tube domain is limited to $z \in [-\pi N/\iota_0, \pi N/\iota_0]$. Therefore, to capture the complete $\hat{\varphi}_{k_\psi,k_\alpha}(z)$ mode within this limited $z$-domain,  it is connected to the $\hat{\varphi}_{k_\psi + \delta k_\psi,k_\alpha}(z)$ mode at the end of the domain, while it connects to the $\hat{\varphi}_{k_\psi - \delta k_\psi,k_\alpha}(z)$ mode at the start of the domain.

\vspace{-2mm}
\FloatBarrier

\noindent
The twist-and-shift boundary condition is commonly applied to stellarators, even though they are formally incorrect in stellarators. In this case, fluctuating quantities are assumed to be periodic in $\zeta_s$ with a period $2\pi P$, with $P$ a rational number. Typically $P$ is the inverse of the number of field periods, for instance, $P=1/5$ for W7-X. Alternatively, one could set $P=1$ to simulate the entire toroidal domain, or choose $P=\text{integer}/ \text{(number of field periods)}$ to simulate an intermediate domain size. The standard twist-and-shift parallel boundary condition is derived using tokamak conventions in Appendix \ref{app:twistandshift} while a graphic derivation can be found in Appendix \ref{app:visualTwistAndShift}.

\subsection{Stellarator symmetric boundary conditions}
\label{sec:StellaratorSymmetricBC}

In a stellarator, two locations with the same poloidal angle (but different toroidal angle) are typically not geometrically equivalent, making the validity of the standard twist-and-shift parallel boundary conditions derived in the previous section questionable. Fortunately, all modern stellarators are designed to possess so-called \textit{stellarator symmetry} \cite{dewar1998stellarator}, for which geometrically equivalent locations can be found along field lines centered around  \textit{stellarator-symmetric points}. Specifically, two unique stellarator-symmetric field lines exist, the first one is centered around the start of each field period, \mbox{$(\theta_s,\zeta_s) = (0,2\pi m/N_\text{nfp})$}, and the second one is centered around the middle of each field period, $(\theta_s,\zeta_s) = (0,\pi/N_\text{nfp}+2\pi m/N_\text{nfp})$, with $m$ an integer and $N_\text{nfp}$ the number of field periods of the device.  These stellarator-symmetric field lines are characterized by,
\begin{align}
	\label{eq:stellaratorSymmetry1}
	|B|_{z_-} & = \ |B|_{z_+}, \\
	|\nablam \psi|_{z_-} & = \ |\nablam \psi|_{z_+}, \\
	|\nablam \alpha|_{z_-} & = \ |\nablam \alpha|_{z_+},
	\label{eq:stellaratorSymmetry3}
\end{align}
where $z_\pm$ indicates the two flux tube ends.
The length of the flux tube needs to be several times larger than the parallel correlation length, in order to assume that the fluctuations at the end points are statistically identical, in which case parallel boundary conditions can be applied at these geometrically equivalent points. To formulate these boundary conditions, the Fourier expansion is written as a function of the integrated local magnetic shear, which is performed by Martin et al.~\mbox{in \cite{martin2018parallel}}, and is summarized here. First, an orthonormal coordinate system is constructed with,
\begin{equation}
	\boldsymbol{\hat{b}} = \frac{\mathbf{B}}{|\mathbf{B}|},
	\qqq
	\mathbf{\hat{e}}_u = \frac{\nablam \psi}{|\nablam \psi|},
	\qqq
	\mathbf{\hat{e}}_v = \frac{\boldsymbol{\hat{b}} \times \nablam \psi}{|\nablam \psi|}.
\end{equation}
As a result, the flux-tube cross-section can be parametrized using the following $(u,v)$ coordinates,
\begin{align}
	u & = (\mathbf{r} - \mathbf{r}_0) \cdot \mathbf{\hat{e}}_u = \frac{(\psi-\psi_0)}{|\nablam \psi |}, \\[1mm]
	v & = (\mathbf{r} - \mathbf{r}_0) \cdot \mathbf{\hat{e}}_v = - \frac{\psi'_t}{B\,|\nablam\psi|}  \, \Big( |\nablam\psi|^2\,(\alpha-\alpha_0) - (\nablam\psi \cdot \nablam\alpha) \,(\psi-\psi_0) \Big),
\end{align}
with $\mathbf{B} = \psi'_t \, \nablam \psi \times \nablam \alpha$ (eq.~\ref{eq:B}). Rearranging these equations gives expressions for $(\psi-\psi_0)$ and $(\alpha-\alpha_0)$ which can be substituted into equation (\ref{eq:phiFourier}), to obtain,
\begin{align}
	\label{eq:FourierLocalShear}
	\varphi(\psi,\alpha,z,t)
	& = \sum_{k_\psi,k_\alpha} \! \hat{\varphi}_{k_\psi,k_\alpha} (z, t) \,  \exp \Bigg(\ir \, |\nablam\psi| \, \Big( k_\psi + k_\alpha \, \frac{\nablam\psi \cdot \nablam\alpha}{|\nablam \psi|^2} \Big) \, u + \ir \, \Big( k_\alpha \, \frac{B}{\psi'_t\, |\nablam\psi |} \Big)\, v  \Bigg).
\end{align}
In $(u,v)$-coordinates, the local magnetic shear, captured within \mbox{$\nablam \psi \cdot \nablam \alpha = |\nablam\psi| \, |\nablam\alpha| \, \cos\lambda$}, enters the Fourier expansion explicitly, whereas in field-aligned $(\psi, \alpha)$-coordinates this effect is contained within the deformation of the flux-tube cross-section along the field line.
Similarly as in section \ref{sec:twistandshift}, when a particular $(k'_\psi,k'_\alpha)$ Fourier mode gets to the end of the $z$-domain $(z=L_z/2)$, the mode needs to connect exactly to another Fourier mode $(k_\psi,k_\alpha)$ at the beginning of the $z$-domain $(z=-L_z/2)$. Therefore, the coefficients multiplying $u$ and $v$ in equation (\ref{eq:FourierLocalShear}) must match at the ends of the flux tube, which yields,
\begin{align}
	\Bigg[ k_\alpha \, \frac{1}{\psi'_t} \frac{B}{|\nablam \psi|} \Bigg]_{z_+} & = \ \Bigg[ k_\alpha \,\frac{1}{\psi'_t} \frac{B}{|\nablam \psi|} \Bigg]_{z_-}, \\
	\Bigg[ \, |\nablam \psi| \Big(k_\psi +  k_\alpha \, \frac{\nablam\psi \cdot \nablam\alpha}{|\nablam \psi|^2}\Big) \Bigg]_{z_+} & = \ \Bigg[ \, |\nablam \psi| \Big(k_\psi +  k_\alpha \, \frac{\nablam\psi \cdot \nablam\alpha}{|\nablam \psi|^2}\Big) \Bigg]_{z_-}.
	\label{eq:kpsizcondition}
\end{align}
For stellarator-symmetric field lines, which satisfy equations (\ref{eq:stellaratorSymmetry1}--\ref{eq:stellaratorSymmetry3}), the binormal wavenumber remains unchanged across the $z$-boundary. On the other hand, equation (\ref{eq:kpsizcondition}) is simplified by noting that $(\nablam \psi \cdot \nablam\alpha)$ is an odd function along the field line, i.e.~\mbox{$[\nablam \psi \cdot \nablam\alpha]_{z_-} = - [\nablam \psi \cdot \nablam\alpha]_{z_+}$}. Therefore, the radial wavenumber is shifted across \mbox{the $z$-boundary by,}
\begin{equation}
	\label{eq:deltakpsisym}
	\delta k_\psi = [k_\psi]_{z_+} - [k_\psi]_{z_-} =  \frac{k_\alpha}{|\nablam \psi|^2} \Big( [\nablam\psi \cdot \nablam\alpha]_{z_-} -  [\nablam\psi \cdot \nablam\alpha]_{z_+}  \Big) =   \frac{2  k_\alpha}{|\nablam \psi|^2} \, [\nablam\psi \cdot \nablam\alpha]_{z_-}.
\end{equation}
If $k'_\psi = [k_\psi]_{z_+}$ belongs to the wavenumber grid, and assuming the $k_\psi$ grid consists of integer multiples of some $k_{\psi,\text{min}}$, then the difference $\delta k_\psi$ should be an integer multiple of $k_{\psi,\text{min}}$. In particular this must be true for the smallest nonzero $k_{\alpha,\text{min}}$, which gives,
\begin{equation}
	\label{eq:kalphaminlocal}
	k_{\alpha,\text{min}} \,\frac{ 2 \, [\nablam \psi \cdot  \nablam \alpha]_{z_-} }{|\nablam \psi|^2}  =  J \, k_{\psi,\text{min}}
\end{equation}
Note that the stellarator-symmetric boundary condition (eq.~\ref{eq:kalphaminlocal}) reduces to the standard twist-and-shift boundary condition (eq.~\ref{eq:kalphamin}) in axisymmetric devices when the flux tube extends up to $N$ poloidal turns $(L_z = 2\pi N/\iota_0)$. To see this, note that the dot product $\nablam \psi \cdot \nablam \alpha$ \mbox{is given by,}
\begin{equation}
	\nablam \psi \cdot \nablam \alpha = \nablam \psi \cdot \nablam \theta_s - \iota_0 \nablam \psi \cdot \nablam \zeta_s + \zeta_s (\hat{s}_0\iota_0/r_0) \, | \nablam \psi|^2
\end{equation}
Axisymmetry implies that any field line centered around $\theta_s = 0$ satisfies,
\begin{align}
	|B|_{z_-} & = \ |B|_{z_+}, \\[1mm]
	|\nablam \psi \cdot \nablam \zeta_s |_{z_-} & = \ |\nablam \psi \cdot \nablam \zeta_s |_{z_+}, \\[1mm]
	|\nablam \psi \cdot \nablam \theta_s |_{z_-} & = \ |\nablam \psi \cdot \nablam \theta_s |_{z_+}.
\end{align}
In this case, a Fourier mode undergoes a shift in $k_\psi$  as it crosses the $z$-boundary, given by,
\begin{equation}
	\delta k_\psi =  \frac{k_\alpha}{|\nablam \psi|^2} \Big( [\nablam\psi \cdot \nablam\alpha]_{z_-} -  [\nablam\psi \cdot \nablam\alpha]_{z_+}  \Big) = - k_\alpha \, \frac{\hat{s}_0}{r_0} \, 2\pi N,
\end{equation}
which corresponds to the shift obtained with the standard twist-and-shift boundary condition (eq.~\ref{eq:kpsi}). Therefore, the stellarator-symmetric boundary conditions can be applied if the two ends of the flux tube correspond to stellarator-symmetric locations or if the flux tube ends are separated by an integer number of poloidal turns in an axisymmetric device.

Similar to equation (\ref{eq:kalphamin}), which is derived for the standard twist-and-shift boundary condition,  equation (\ref{eq:kalphaminlocal}) imposes a  quantization on the perpendicular domain's aspect ratio in real space. However, this condition allows for significantly more flexibility. To resolve fluctuations up to $k_{\psi,\text{max}}$ in a low-shear device, i.e.~when $|\hat{s}_0\iota_0/r_0| \ll 1$, the standard twist-and-shift-boundary condition forces one to include a large number of radial wavenumbers $(N_{k_\psi}\gg 1)$ since $k_{\psi,\text{min}}$, given by equation (\ref{eq:kalphamin}), can be very small. Alternatively, very long flux tubes need to be employed $(N>1)$ to increase the step size $k_{\psi,\text{min}}$. Both these options are computationally expensive. In contrast, the stellarator-symmetric boundary conditions do not depend on the global magnetic shear. Instead, the length of the flux tube $L_z$ can be chosen freely to obtain an optimal value of $\nablam \psi \cdot  \nablam \alpha/|\nablam \psi|^2$ at $z=\pm L_z/2$. For example, a flux tube length can be chosen so that \mbox{$2 \, |\nablam \psi \cdot  \nablam \alpha| / |\nablam \psi|^2 = J$} at the flux tube ends, which means that the perpendicular cross-section has $L_\alpha=L_\psi$, corresponding to $k_{\alpha,\text{min}} = k_{\psi,\text{min}}$. Alternatively, if the flux-tube cross-section is chosen to be perpendicular at $z=\pm L_z/2$, then $[\nablam \psi \cdot  \nablam \alpha]_{z_\pm}\! = 0$ and $k_{\psi,\text{max}}$ can be chosen freely. This enforces self-periodicity for each Fourier mode (given by eq.~(\ref{eq:self-periodicity}) if one assumes $C_{k_{\alpha}} = 1$). Finally, the magnetic drifts ($\nablammm B$-drift and curvature drift)  can be made continuous across the $z$-boundary. To achieve this, one needs $\mathbf{B} \times \nablammm B \cdot \nablam \psi = 0$ at $z=\pm L_z/2$. However, these locations do no necessarily coincide with good choices for $\nablam \psi \cdot  \nablam \alpha$. In this work, we choose $L_z$ such that $L_\alpha=L_\psi$ is satisfied.

\subsection{Self-periodic boundary conditions}
\label{sec:selfperiodic}

If the global magnetic shear is extremely small, then the cross-section is only deformed by the first term in equation  (\ref{eq:coslambda}). In an axisymmetric device, $\nablam \psi \cdot \nablam \theta_s$ and $\nablam \psi \cdot \nablam \zeta_s$ are the same at the forward and backward end of the flux tube.
As a result, the two perpendicular cross-sections at the ends of the flux tube can be connected directly onto each other, provided that the distance between the two flux tube ends is multiple times larger than the parallel correlation length.  Approximating $\hat{s}_0\iota_0/r_0 \approx 0$, the parallel \mbox{boundary condition (eq.~\ref{eq:parallelBC}) simplifies to}
\begin{equation}
	\begin{split}
		\hat{\varphi}_{k_\psi,k_\alpha} (-L_z/2, t)  \, \exp  \Big(\ir k_\psi (\psi-\psi_0) + \ir k_\alpha \theta_s \Big) \, e^{\ir k_\alpha \iota_0L_z}  \\[2mm]
		= \hat{\varphi}_{k'_\psi,k'_\alpha} (L_z/2, t) \,  \exp  \Big(\ir  k'_\psi (\psi-\psi_0)  + \ir k'_\alpha \theta_s  \Big). \,\,\,\,
	\end{split}
\end{equation}
For this equation to hold for all field lines and flux surfaces, the factors in front of $\theta_s$ and $\psi$ should match, respectively. Therefore, both the binormal wavenumber and the radial wavenumber remain unchanged, i.e.~$k'_\alpha=k_\alpha$ and $k'_\psi=k_\psi$. If the global magnetic shear is taken to be zero in an axisymmetric device, the boundary condition thus simplifies to \textit{self-periodicity}, given by,
\begin{equation}
	\label{eq:self-periodicity}
	\hat{\varphi}_{k_\psi,k_\alpha}(-L_z/2,t) \exp (\ir k_\alpha \iota_0 L_z ) = \hat{\varphi}_{k_\psi,k_\alpha}(L_z/2,t)
\end{equation}
This is typically referred to as \textit{forced periodicity} since the global magnetic shear is generally finite, making these boundary conditions formally incorrect. However, when applying the stellarator-symmetric boundary conditions, \textit{exact periodicity} is achieved if the flux tube ends are chosen to correspond to a location where the cross-section is perpendicular, i.e.~where $[\nablam \psi \cdot  \nablam \alpha]_{z_\pm}\! = 0$. In both forced and exact periodicity the step size $k_{\psi,\text{min}}$ can be chosen freely.

\begin{figure}[b!]
	\centering
	\vspace{1mm}
	\includegraphics[width=0.99\textwidth,trim={0mm 0mm 0mm 79mm},clip] {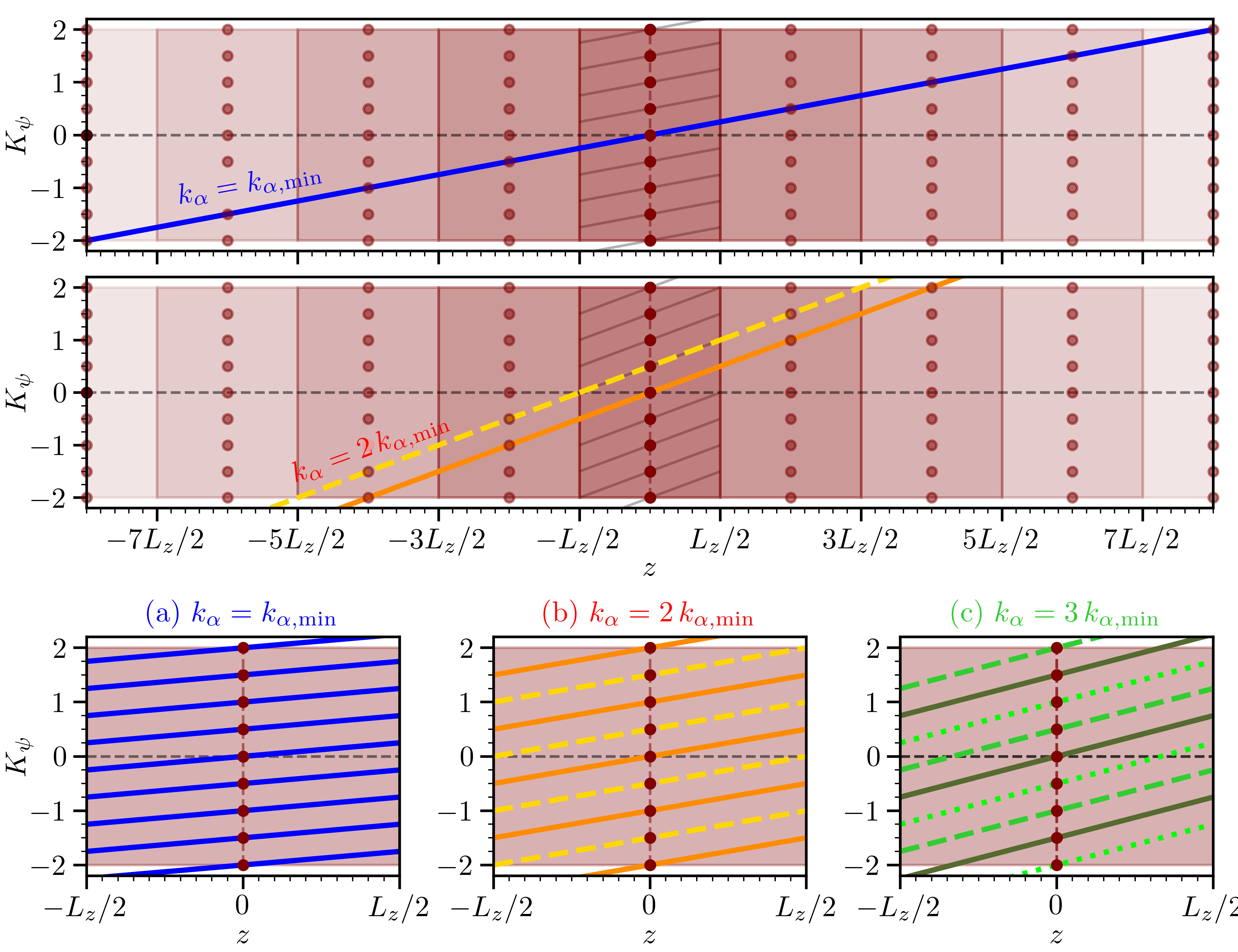}
	\captionsetup{singlelinecheck=false,  margin=0.0cm} \vspace{-2mm}
	\caption{Evolution of $K_\psi = k_\psi +  k_\alpha (\hat{s}_0\iota_0/r_0)\,z$ along the parallel coordinate, for $\smash{k_\alpha = k_{\alpha, \text{min}}}$ (left), $\smash{k_\alpha = 2k_{\alpha, \text{min}}}$ (middle) and $\smash{k_\alpha = 3k_{\alpha, \text{min}}}$ (right). The discrete $k_\psi$-grid, with  $\smash{N_{k_\psi}}=9$ and $|k_{\psi, \text{max}}| = 2$ is illustrated by the markers at the middle of each $L_z$-domain. In this example $k_{\alpha, \text{min}} =  k_{\psi, \text{min}} = 0.5$,  $J=1$ and $\hat{s}_0\iota_0/r_0 = 1/L_z$ which satisfies equation (\ref{eq:kalphamin}). Each color represents a chain of connected $k_\psi$ modes. }
	\label{fig:zgrid}
\end{figure}

\subsection{Connections of $(k_\psi,k_\alpha)$-modes along the extended $z$-grid}
\label{sec:connections}

\noindent
To visualize the connections between the Fourier modes across the $z$-boundary, the perpendicular components of the wavevector are examined in this section. The factor in front of $\nablam \psi$ in \mbox{equation (\ref{eq:kperp2})} corresponds to the \textit{effective radial wavenumber} $K_\psi$, given by,
\begin{align}
	\label{eq:Kpsi}
	K_\psi = k_\psi +  k_\alpha \frac{\hat{s}_0\iota_0}{r_0} \,z
\end{align}
While $k_\alpha$ remains constant along the flux tube, perturbations are sheared radially according to \mbox{$K_\psi = k_\psi +  k_\alpha (\hat{s}_0\iota_0/r_0) \,z$}. Seeing that the wavenumbers $k_\psi=2\pi j/L_\psi$ and $k_\alpha=2\pi k/L_\alpha$ in the Fourier expansions (eqs.~\ref{eq:gFourier}--\ref{eq:phiFourier}) are discretized, they form a discrete wavenumber grid. Of course, computer simulations can not retain an infinite number of $k_\psi$ and $k_\alpha$ modes. Instead, enough $k_\psi$ and $k_\alpha$ modes are kept to be able to resolve up to a desired value of $k_\perp\rho_i$, above which the Fourier coefficients $\hat{\varphi}(k_\psi,k_\alpha,z,t)$ and $\hat{g}(k_\psi,k_\alpha,z,v_\parallel,\mu,t)$ are assumed to vanish due to FLR effects \mbox{(section \ref{sec:FLR}).}
Specifically, $N_{k_\psi}$ and $N_{k_\alpha}$  Fourier modes are considered in the radial and binormal directions. The evolution of the effective radial wavenumber $K_\psi(k_\psi, k_\alpha, z)$ along the parallel coordinate $z = \zeta_s$ is shown in figure \ref{fig:zgrid}, considering $\smash{k_\alpha = k_{\alpha, \text{min}}}$ (left), $\smash{k_\alpha = 2k_{\alpha, \text{min}}}$ (middle) and $\smash{k_\alpha = 3k_{\alpha, \text{min}}}$ (right). The discrete $k_\psi$-grid is illustrated by the markers at the middle of the $L_z$-domain, since $K_\psi$ coincides with the discrete wavenumber grid at $z=0$.
For each $(k_\psi, k_\alpha)$-mode in the Fourier expansion, a $K_\psi$-line can be drawn, where $k_\psi$ determines which grid point the line passes through at $z=0$; and the binomial wavenumber, $k_\alpha$, and the global magnetic shear, $\hat{s}_0\iota_0/r_0$, determine the slope of the line.

\begin{figure}[b!]
	\centering
	\vspace{-3mm}
	\begin{subfigure}{\linewidth}
		\centering
		\includegraphics[width=0.99\textwidth,trim={0mm 48mm 0mm 0mm},clip] {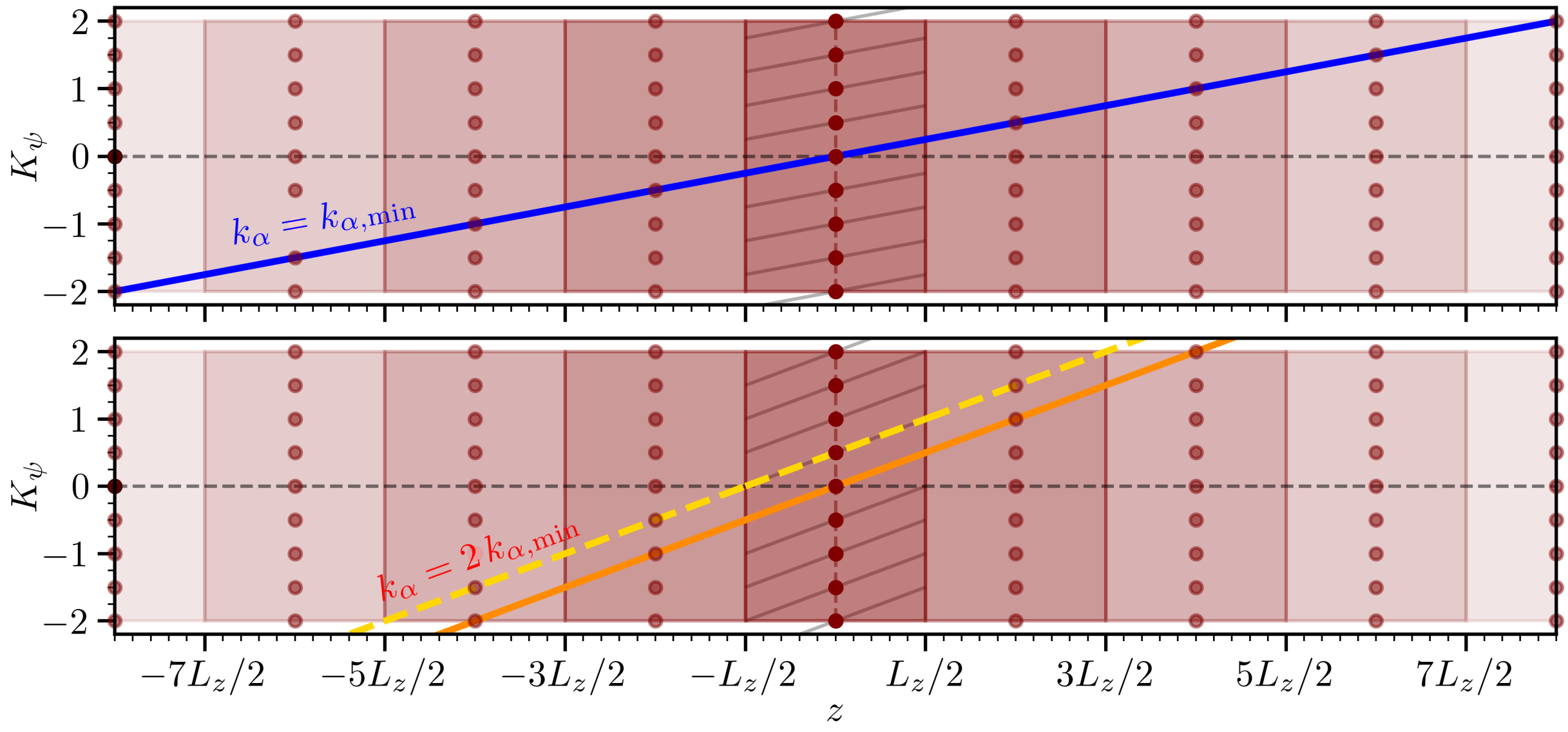}
	\end{subfigure}
	\captionsetup{singlelinecheck=false,  margin=0.0cm}\vspace{-8mm}
	\caption{Evolution of $K_\psi = k_\psi + k_\alpha\,(\hat{s}_0\iota_0/r_0)\,z$ along the parallel coordinate, for $\smash{k_\alpha = k_{\alpha, \text{min}}}$ (top row) and $\smash{k_\alpha = 2k_{\alpha, \text{min}}}$ (bottom row). The discrete $k_\psi$-grid, with  $\smash{N_{k_\psi}}=9$ and $|k_{\psi, \text{max}}| = 2$ is illustrated by the markers at the middle of each $L_z$-domain, and the extent of each $L_z$-domain is highlighted in red. In this example $k_{\alpha, \text{min}} =  k_{\psi, \text{min}} = 0.5$,  $J=1$ and $\hat{s}_0\iota_0/r_0 = 1/L_z$ which satisfies equation (\ref{eq:kalphamin}). }  \vspace{-1mm}
	\label{fig:extendedzgrid}
\end{figure}

Fluctuating quantities are assumed to be periodic in $z$ with a period  $z = L_z$. To apply the standard twist-and-shift boundary condition, a  $(k_\psi,k_\alpha)$-mode is connected to another $(k_\psi + k_\alpha (\hat{s}_0\iota_0/r_0) L_z,k_\alpha)$-mode across the \mbox{$z$-boundary}, with $L_z=2\pi N/\iota_0$ (section \ref{sec:twistandshift}). To ensure that \mbox{$k'_\psi = k_\psi + k_\alpha (\hat{s}_0\iota_0/r_0) L_z$} belongs to the discrete wavenumber grid, $k_\alpha (\hat{s}_0\iota_0/r_0) L_z$ has to be an integer multiple of the step size $k_{\psi, \text{min}}$ of the grid.
In other words, the change of $K_\psi$ over a single $L_z$-domain has to be equal to an integer multiple of the step size $k_{\psi, \text{min}}$.
The standard twist-and-shift boundary condition then effectively connects a $K_\psi(k_\psi, k_\alpha, z)$ line at the end of the $L_z$-domain to a $K_\psi(k_\psi + k_\alpha (\hat{s}_0\iota_0/r_0) L_z, k_\alpha, z)$ line at the beginning of the \mbox{$L_z$-domain}, which are separated by $Jk$ grid points at $z=0$, with $k=k_\alpha/k_{\alpha, \text{min}}$.
In \mbox{figure \ref{fig:zgrid}a}, $k_\alpha = k_{\alpha, \text{min}}$ and $J=1$ are considered. As a result, each Fourier coefficient $\hat{\varphi}(k_\psi,k_{\alpha, \text{min}},z)$ is connected at the end of the flux tube to the next Fourier coefficient $\hat{\varphi}(k_\psi+k_{\psi,\text{min}},k_{\alpha, \text{min}},z)$ at the beginning of the flux tube. Eventually a big chain is formed by the connected Fourier modes, and their effective radial wavenumber is depicted by the blue line.
For $k_\alpha = 2k_{\alpha, \text{min}}$ (figure \ref{fig:zgrid}b) the slope of $K_\psi(k_\psi,k_\alpha,z)$ is twice as large. Consequently, $K_\psi(k_\psi,k_\alpha,z)$ only connects every second $k_\psi$-point at the middle of the $L_z$-domain (orange solid line). To ensure that all $k_\psi$ values are included, a second chain of Fourier modes needs to be considered, whose $K_\psi$ passes through $k_{\psi, \text{min}}$ at $z=0$ (yellow dashed line). This reasoning can be followed for every $k_\alpha$ value, which makes it clear that $Jk$ distinct chains of Fourier modes need to be included for a specific $k_\alpha$. Of course, the number of eigenmodes is limited to the number of $k_\psi$ grid points, hence \mbox{$N_\text{chains} = \text{minimum}(Jk, N_{k_\psi})$}, and for $N_\text{chains} = N_{k_\psi}$ each mode is unconnected and forms its own chain consisting of a single $L_z$-segment.

Instead of representing these chains of Fourier modes on a single $L_z$-domain, they can be represented on an \textit{extended $z$-grid}, which has been demonstrated in figure \ref{fig:TwistAndShiftGaussians}. In essence, each Fourier mode $\hat{\varphi}(k_\psi,k_\alpha,z)$ is shifted along $z$ by an integer multiple of $L_z$ so that they connect directly to the next Fourier mode $\hat{\varphi}(k_\psi + k_\alpha (\hat{s}_0\iota_0/r_0) L_z,k_\alpha,z)$, at the $z$-point that they have in common. This chain of connected Fourier modes can now be seen as an \textit{eigenmode} of the system on the extended $z$-domain.
The effective radial wavenumber of these chains of Fourier modes is shown in figure \ref{fig:extendedzgrid}, where the top row corresponds to the $k_\alpha = k_{\alpha, \text{min}}$ and $J=1$ case discussed in \mbox{figure \ref{fig:zgrid}a}. On a single $L_z$-domain there are $N_{k_\psi}$ lines of $K_\psi(k_\psi,k_\alpha,z)$ for each value of $k_\alpha$, since the effective radial wavenumber can be represented for each Fourier mode (gray lines). On the extended $z$-grid, only \mbox{$N_\text{chains}$} lines are needed, corresponding to the chains of Fourier modes.

Ultimately, $K_\psi$ reaches the final grid point, after which the mode gets sheared out of the discrete radial wavenumber grid, since only a finite number of $k_\psi$-points are considered. Therefore, turbulent eddies with $|K_\psi| > k_{\psi, \text{max}}$ are lost. This approximation is valid if $k_{\psi, \text{max}}$ is taken to be large enough so that small-scale fluctuations with $|K_\psi| > k_{\psi, \text{max}}$ would be suppressed by finite-Larmor-radius effects (section \ref{sec:FLR}).
Therefore, the beginning and end of each chain of connected $L_z$-segments, remains unconnected. This can be seen in \mbox{figure \ref{fig:extendedzgrid}} where the radial wavenumber grid is defined by $N_{k_\psi} = 9$ and $k_{\psi, \text{max}} = 2$. The $(k_\psi,k_\alpha) = (0, k_{\alpha, \text{min}})$ mode (blue line) remains confined to the $k_\psi$-grid for $z \in [-4L_z, 4L_z]$, whereas the $(0, 2k_{\alpha, \text{min}})$ mode (orange line) remains confined for $z \in [-2L_z, 2L_z]$. Essentially, limiting the number of radial wavenumbers, thus limits the length of the chains of Fourier modes.
Note that this is only valid if the standard twist-and-shift boundary conditions or stellarator-symmetric boundary conditions are used with a finite $\delta k_\psi$. In the case of self-periodic boundary conditions, the step size $k_{\psi,\text{min}}$ is chosen freely, and each $(k_\psi,k_\alpha)$ Fourier mode forms its own chain which is connected onto itself.

\begin{figure}[h!]
	\centering
	\vspace{-2mm}
	\includegraphics{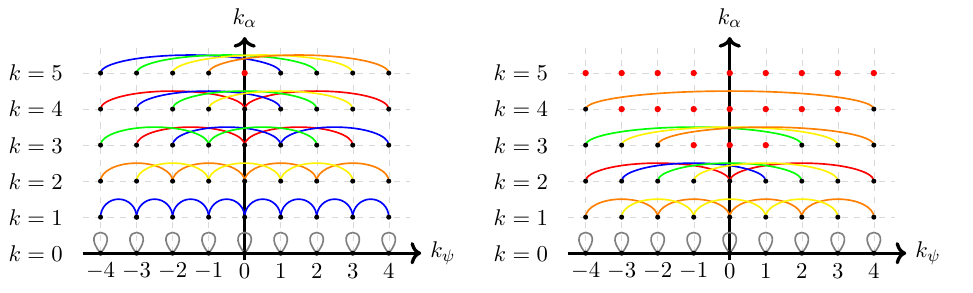}
	\captionsetup{singlelinecheck=false,  margin=0.0cm} \vspace{-8mm}
	\caption{Connections between different $(k_\psi, k_\alpha)$-grid points, for $N_{k_\psi} = 9$ and considering $J=1$ (left) and $J=2$ (right). Grid points without connections are highlighted with a red marker.}
	\vspace{-8mm}
	\label{fig:connections}
\end{figure}

Alternatively, the connections between the different $(k_\psi,k_\alpha)$ modes linking $z=-L_z/2$ to \mbox{$z=L_z/2$} can be represented on the $(k_\psi,k_\alpha)$-grid, which is shown by the colored arcs in \mbox{figure \ref{fig:connections}.} The $k_\alpha=0$ modes are characterized by $K_\psi=k_\psi$ (eq.~\ref{eq:Kpsi}). Therefore, the $(k_\psi,0)$ modes are connected onto themselves, demonstrated by the gray loops (bottom row). For $k_\alpha=k_{\alpha, \text{min}}$ and $J=1$ (left) each $k_\psi$ point is connected to its neighbor by the blue connections, while for $J=2$ (right) every second $k_\psi$ grid point is connected for $k_\alpha=k_{\alpha, \text{min}}$.
It is important to note that the central radial wavenumber, $k_\psi=0$, is sheared out of the wavenumber grid before reaching the end of the $L_z$-domain for $k=5$ if $J=1$ (left) and for $k=3$ if $J=2$ (right). Therefore, it has no $k_\psi$ grid point to connect to, and this eigenmode thus consists of a single $L_z$ segment, which can be problematic.
Specifically, for each $k_\alpha$, there is a location $\zeta_s = - (k_\psi/k_\alpha) (\hat{s}_0\iota_0/r_0)$ along the flux tube---i.e.~the ballooning angle---for which $K_\psi=0$, where a certain Fourier mode $(k_\psi,k_\alpha)$ is effectively unsheared by the global magnetic shear. At this location, the perpendicular wavenumber $k_\perp\rho_s$ obtains a minimum along $z$, provided that the local magnetic shear is small. Therefore, it is important that enough radial wavenumbers $k_\psi$ are considered, to ensure that $\zeta_s = - (k_\psi/k_\alpha) (\hat{s}_0\iota_0/r_0)$ covers the entire $L_z$-domain for all the relevant $k_\alpha$ wavenumbers. If this is not satisfied, instabilities would be artificially damped in a part of the flux tube since they are always considered to be sheared, which decreases their perpendicular scales and \mbox{increases the FLR damping (section \ref{sec:FLR}).}

\chapter[Implementation of $\boldsymbol{\delta}\hspace{-0.2mm}{f}$ gyrokinetic equations in the code \texttt{stella}]{Implementation of $\boldsymbol{\delta}\hspace{-0.8mm}\bm{f}$ gyrokinetic equations in the code \texttt{stella}}
\label{chap:stella}

Solving the 5D gyrokinetic equations in stellarators can be challenging due to their complex geometry. Moreover, the high dimensionality of the problem and its intrinsic nonlinear character requires the use of a large number of grid points in 5D phase-space. Therefore, gyrokinetic codes require state-of-the-art simulation and numerical techniques and must run on highly parallelized computers. Various numerical schemes have been developed, each with its own advantages and drawbacks. The numerical scheme employed in \texttt{stella} \cite{stella} is distinguished from other gyrokinetic codes through its use of an operator-split, implicit-explicit time advance scheme. The \texttt{stella} code has been extensively benchmarked for stellarator geometry \cite{gonzalez2022electrostatic, sanchez2021gyrokinetic} and has been applied to the study of turbulent heat transport \cite{thienpondt2024influence, velasco2024piecewise, garcia2024reduced}, turbulent particle transport \cite{thienpondt2023prevention, alonso2024density} and turbulent impurity transport \cite{garcia2021turbulent, garcia2021turbulent2, garcia2024reduction}. Moreover, \texttt{stella} has been used to develop optimization methods \cite{acton2024optimisation} and in the validation of gyrokinetic results against experimental measurements \mbox{in W7-X \cite{gonzalez2024electrostatic, carralero2025first}.}

By using the local approximation, microturbulence is simulated within a small region of the fusion device called a flux tube (see chapter \ref{chap:fluxtube}).
In this chapter, the gyrokinetic equation (\ref{eq:FourierGyrokineticEquation}) and quasineutrality condition (\ref{eq:FFTQuasiNeautrality}) are rewritten in local coordinates $(k_x,k_y,z,v_\parallel,\mu)$ which are defined in \mbox{section \ref{sec:localcoordinates}.} In section \ref{sec:normalizedquantities}, the variables are normalized such that they are of order unity, which are used in section \ref{sec:normalizedGyrokineticEquation} to normalize the system of equations, ensuring that \texttt{stella} evolves dimensionless quantities. To numerically evolve turbulent fluctuations within a flux tube, the domain in $(k_x, k_y, z, v_\parallel, \mu)$ must be discretized, and the magnetic geometry and plasma parameters within the simulation domain---i.e.~within the flux tube---have to be provided to the code through a \texttt{VMEC} \cite{hirshman1983steepest} output file and a \texttt{stella} input file, respectively, which is explained in section \ref{sec:inputs}. Finally, the diagnostics needed to analyze the turbulence driven by microinstabilities are discussed in \mbox{section \ref{sec:stella:diag}.}
The most important input parameters used for the \texttt{stella} code, as well as the most important output variables for diagnosing turbulence, are defined \mbox{in this chapter using \texttt{teletype} font.}

\section{Local coordinates $(k_x, k_y, z, v_\parallel, \mu)$}
\label{sec:localcoordinates}

\index{Local coordinates $(x, y, z, v_\parallel, \mu)$}
To describe turbulent fluctuations within the simulation domain---i.e.~within a flux tube---a set of local $(x, y, z, v_\parallel, \mu)$-coordinates are defined in \texttt{stella}. The velocity coordinates consist of the speed along the magnetic field, $\smash{v_\parallel = \hatb \cdot \mathbf{v}}$, and the magnetic moment per unit of mass, $\smash{\mu = v^2_\perp /2B}$. The coordinate $z$ measures distance along the magnetic field lines. Currently two options are available in \texttt{stella} \cite{stella} to define the magnetic geometry. The first option is to use a magnetic equilibrium generated by the three-dimensional MHD equilibium code \texttt{VMEC} \cite{hirshman1983steepest}. The second option, applicable only to axisymmetric magnetic field configurations, involves specifying a set of Miller parameters that are used to construct a local solution to the Grad--Shafranov equation \cite{miller1998noncircular}. By default, $z=\thetaB$ is used when defining the magnetic geometry based on a set of Miller parameters, with $\thetaB$ the Boozer poloidal angle, whereas $z=\zeta$ is used when reading \texttt{VMEC} equilibria, with $\zeta$ the cylindrical toroidal angle. Nonetheless, it is possible to perform \texttt{stella} simulations using $z = \ell$, with $\ell$ the arc length of the field line, which is typically used in the simulations presented in this work.

In the flux-tube approximation, quantities are approximated by first order Taylor expansions in $(\psi, \alpha)$, expanded around the center of the flux tube. Consequently, the effective minor radius  is approximated by $r \approx r_0 + (dr/d\psi)_0 (\psi - \psi_0)$ within the flux tube, with $(dr/d\psi)_0$ evaluated at $(\psi_0, \alpha_0)$. This leads to the definition of the local radial coordinate $x = (dr/d\psi)_0 (\psi - \psi_0)$, corresponding to $x \approx r-r_0$. Similarly, the local binormal coordinate is defined as a scaled version of $\alpha$ such that $dy/d\alpha = r_0$, i.e.~$y=r_0(\alpha-\alpha_0)$.  Various options for the  $(x, y)$-coordinates have been implemented in \texttt{stella}, depending on the choice of the flux-surface label $\psi$. In this work, we employ $\psi=r$ with $r = a\sqrt{\psi_t/\psi_{t,\text{LCFS}}}$ for simplicity, which gives $x=r-r_0$. This convention has been implemented in the \texttt{stella} code as a part of this PhD work. In summary,
\begin{align}
	\label{eq:stella:rxy}
	x & = r - r_0 \\[2mm]
	y & = r_0 (\alpha  - \alpha_0)
\end{align}
with $\psi=r$ and $\alpha=\theta_s - \iota\zeta_s$ are the conventions used in this dissertation.
Note that Barnes et al. has implemented $\psi=-\psi_t$ in Ref.~\cite{stella} giving $dr/d\psi= - \text{sgn}(\psi_t) /(r_0B_r)$ with $B_r = 2\,|\psi_{t,\text{LCFS}}|/a^2$ for \texttt{VMEC} equilibria; and $\psi=\psi_p$ giving $dr/d\psi=q_0/(r_0B_r)$ with $B_r$ a user-speciﬁed reference magnetic field strength for magnetic geometries based on a set of Miller parameters.

\section{Normalized quantities}
\label{sec:normalizedquantities}

In order to evolve the gyrokinetic equations, they need to be normalized first.
The temperature $T_{1}$, density $n_{1}$ and mass $m_{1}$ of the first species, which is usually taken to be those of the main ions, are used as the reference $T_r$, $n_r$ and $m_r$.  Quantities with an index $s = \{1, 2, \dots \}$ denote the actual quantities of each species, whereas the index $r$ denotes the reference quantities. Note that species 1 refers to the one defined in the \texttt{species\_parameters\_1} namelist specified in the \texttt{stella} input file (see Appendix \ref{app:defaultInputVariables}). In the normalized equations, only factors $T_s/T_r$, $n_s/n_r$ and $m_s/m_r$ appear, therefore
these reference quantities are completely arbitrary, and they can be set to any desired value after the simulations have been run, by correctly denormalizing the variables.

The reference temperature and mass are used to calculate the reference thermal velocity $\smash{v_{\text{th},r} = \sqrt{2T_r/m_r}}$. Other gyrokinetic codes are typically normalized with respect to the ion sound speed $\smash{c_r = \sqrt{T_r/m_r}}$, therefore, in \texttt{stella} factors of $\sqrt{2}$ appear that other gyrokinetic codes usually do not have. The velocity-space coordinate normalizations are species-dependent, creating factors of $\sqrt{T_s/T_r}$ and $\sqrt{m_s/m_r}$. When comparing normalized quantities between codes, these factors have to be taken into account.
A list of all reference quantities used by \texttt{stella} is given in table \ref{table:reference_quantities}.
Based on these reference quantities, all quantities appearing in the gyrokinetic equation can be normalized, and the normalized quantities are denoted with a tilde. An overview of these normalizations is given in table \ref{table:normalization}.

\vspace{3mm}
\begin{table}[h!]
	\captionsetup{singlelinecheck = false, margin=0cm}
	\caption[Reference quantities used for the normalization]{Reference quantities used for the normalization in \texttt{stella}. Here index 1 refers to species 1. The reference length $a$ and magnetic field $B_r$ depend \mbox{on the implemented geometry.}}
	\vm
	\centering
	\begin{tabular}{|cc cc l c|}
		\hline \rowcolor[HTML]{C0C0C0} \qqq\qquad &\quad& \qqq\qqq\quad &\quad\quad& \qqq\qqq\qqq\qqq & \qqq\qqq\ \\[-3.5mm]
		\rowcolor[HTML]{C0C0C0}
		\textbf{Symbol} && \textbf{Definition} && \textbf{Reference quantity} & \textbf{Unit} \\[0.5mm]
		\hline &&&&&\\[-3mm]
		$T_r$ && $T_{1}$ && Temperature & J  \\[0.5mm]
		$n_r$ && $n_{1}$ && Density & m$^{-3}$ \\[0.5mm]
		$m_r$ && $m_{1}$ && Mass & kg \\[0.5mm]
		$q_r$ && $|e|$ && Elementary charge & C \\[0.5mm]
		$v_{\text{th},r}$ && $\sqrt{2T_r/m_r}$ && Thermal velocity & m/s \\[0.5mm]
		$\Omega_r$ && $q_r B_r/m_r$ && Gyrofrequency  & 1/s \\[0.5mm]
		$\rho_r$ && $v_{\text{th},r}/\Omega_r$ && Gyroradius & m \\[0.5mm]
		$a$ && Depends on geometry && Length & m \\[0.5mm]
		$B_r$ && Depends on geometry && Magnetic field & T \\[0.5mm]\hline
	\end{tabular}
	\vspace{2mm}
	\label{table:reference_quantities}
\end{table}
\FloatBarrier

The reference magnetic field $B_r$ and reference length $a$ depend on the implemented geometry.  Specifically, when using a \texttt{VMEC} equilibrium $B_r = 2\,|\psi_{t,\text{LCFS}}|/a^2$ and $a$ corresponds to the minor radius. On the other hand, if the equilibrium is defined through a set of Miller parameters, then $a$ and $B_r$ are not explicitly used and their values can be imposed afterwards by correctly denormalizing the variables. In this case, $B_r$ is a user-specified reference magnetic field strength, and $a$ is the half-diameter of the plasma volume at the height of the magnetic axis.

\vspace*{2mm}
\begin{table}[h!]
	\captionsetup{singlelinecheck = false, margin=0cm}
	\caption{Normalized quantities used in \texttt{stella} are denoted by a tilde, and the parallel coordinate is normalized with respect to $L_z=\text{max}(z) - \text{min}(z)$ and $L_\text{mid} = [\text{max}(z) + \text{min}(z)]/2$ to ensure that $\tilde{z} \in [-\pi,\pi]$ if \texttt{nperiod = 1} is set in the input file (see Appendix \ref{app:stella:grids}). Note that the radial location in \texttt{stella} is set in the input file through \texttt{torflux} which is defined as $s = \psi_{t}/\psi_{t,\text{LCFS}} = (r/a)^2$. }
	\centering
	\vspace{-2mm}
	\begin{tabular}{|cc cc l c|}
		\hline \rowcolor[HTML]{C0C0C0} \qqq\qquad &\quad& \qqq\qqq\quad &\quad\quad& \qqq\qqq\qqq\qqq\qquad & \qqq\qqq\ \\[-3.5mm]
		\rowcolor[HTML]{C0C0C0}
		\textbf{Symbol} && \textbf{Definition} && \textbf{Normalized quantity} &\textbf{\texttt{stella} variable}  \\[0.5mm] \hline &&&&& \\[-3.5mm]
		$\tilde{T}_{s}$ && $T_{s}/T_r$ && Temperature & \texttt{temp} \\[0.5mm]
		$\tilde{n}_{s}$ && $n_{s}/n_r$ && Density & \texttt{dens} \\[0.5mm]
		$\tilde{m}_{s}$ && $m_{s}/m_r$ && Mass & \texttt{mass} \\[0.5mm]
		$\tilde{v}_{\text{th},s}$ && $v_{\text{th},s}/v_{\text{th},r}$ && Thermal velocity & - \\[0.5mm]
		\hline &&&&& \\[-3mm]
		$s$ && $\psi_{t}/\psi_{t,\text{LCFS}}$ && Normalized toroidal flux & \texttt{torflux} \\[0.5mm]
		$\tilde{r}$  && $r/a$  && Radial coordinate & \texttt{rho}, \texttt{rhotor} \\[0.5mm]
		$\tilde{\psi}_{t}$  && ${\psi_{t}}/{a^2 B_r}$  && Magnetic flux & - \\[0.5mm]
		$\tilde{\alpha}$  && $\alpha$  && Field-line label & - \\[0.5mm]
		\hline &&&&&\\[-3mm]
		$\tilde{t}$  && $(v_{\text{th},r}/a)\,t$ && Time & \texttt{code\_time} \\[0.5mm]
		$\tilde{x} $ && $ x/\rho_r$ && Radial perp. coordinate  & -   \\[0.5mm]
		$\tilde{y} $ && $ y/\rho_r$ && Poloidal perp. coordinate & -   \\[0.5mm]
		$\tilde{z} $ && $ 2\pi (z -L_\text{mid})/L_z$ && Parallel coordinate & \texttt{zed} \\[0.5mm]
		$\tilde{\nablam}$  && $a \nablam$   && Nabla operator& - \\[0.5mm]
		\hline &&&&& \\[-3mm]
		$\tilde{k}_{x}$ && $k_x \rho_r$ && Radial perp.\ wavenumber & \texttt{akx} \\[0.5mm]
		$\tilde{k}_{y}$ && $k_y \rho_r$ && Binormal perp.\ wavenumber & \texttt{aky} \\[0.5mm]
		\hline &&&&& \\[-3mm]
		$\tilde{v}_{\parallel}$ && $v_\parallel/v_{\text{th},s}$ && Parallel velocity & \texttt{vpa} \\[0.5mm]
		$\tilde{\mu}$ && $\mu B_r/v_{\text{th},s}^2 $ && Magnetic moment & \texttt{mu} \\[0.5mm]
		\hline &&&&& \\[-3mm]
		$\tilde{F}_{\text{M}s}$ && $(v_{\text{th},s}^3/n_s)\,F_{\text{M}s}$ && Equilibrium dist.\ function & - \\[0.5mm]
		$\tilde{g}_{\mathbf{k}, s}$  && $\!\!\!\!\!\! (a/\rho_r) \,(1/F_{0s})\, e^{-\tilde{v}^2_s}\,\hat{g}_{\mathbf{k}, s}\!\!\!\!\!\!\!\!\!$  &&  Guiding-center dist.\ function & \texttt{gold, gnew}\\[0.5mm]
		$\tilde{\varphi}_\mathbf{k}$   && $(a /  \rho_r) \, (q_r/T_r) \, \hat{\varphi}_\mathbf{k}$ && Perturbed electrostatic potential \!\!\!\!\!\! & \texttt{phi} \\[0.5mm]
		\hline
	\end{tabular}
	\label{table:normalization}
\end{table}
\FloatBarrier

\section{Normalized gyrokinetic equations}
\label{sec:normalizedGyrokineticEquation}

The normalized gyrokinetic equation for Fourier component $\mathbf{k}$ is obtained by multiplying equation (\ref{eq:FourierGyrokineticEquation}) with the normalizing factor $(a^2/\rho_r v_{\text{th},r}) \exp (-\tilde{v}^2)/\FM$, which yields,
\begin{equation}
\begin{split}
\label{eq:NormalizedGyrokineticEquation}
\frac{\partial \tilde{g}_{\mathbf{k},s}}{\partial \tilde{t}}
+
\tilde{v}_{\text{th},s} \tilde{v}_{\parallel}\, \hatb \cdot \tilde{\nablam} \tilde{z} \, \Big( \frac{\partial \tilde{g}_{\mathbf{k},s}}{\partial \tilde{z}}
+
\frac{Z_s}{\tilde{T}_s} \frac{\partial J_0(a_{\mathbf{k},s}) \tilde{\varphi}_\mathbf{k}}{\partial \tilde{z}} \, e^{-\tilde{v}^2_s} \Big)
-
\tilde{v}_{\text{th},s}
\tilde{\mu} \, \hatb \cdot \tilde{\nablammm} \tilde{B} \,\frac{\partial \tilde{g}_{\mathbf{k},s}}{\partial \tilde{v}_\parallel} \qquad \\
+ \,
\ir \omega_{d,\mathbf{k},s} \Big(  \tilde{g}_{\mathbf{k},s}
+
\frac{Z_s}{\tilde{T}_s} J_0(a_{\mathbf{k},s}) \tilde{\varphi}_\mathbf{k} e^{-\tilde{v}^2_s} \Big)
+ \ir \omega_{*,\mathbf{k},s}
J_0(a_{\mathbf{k},s}) \tilde{\varphi}_\mathbf{k}
+
\mathcal{N}_{\mathbf{k},s}
= 0, \qqq
\end{split}
\end{equation}

\noindent
where $a_{\mathbf{k},s} = k_\perp \rho_s$, and the nonlinear term $\mathcal{N}_{\mathbf{k},s}$ is given by,
\begin{equation}
\label{eq:NormalizedNonlinearTerm}
\mathcal{N}_{\mathbf{k},s}\! = -\frac{1}{\psi'_t}\frac{B_r}{2} \frac{dy}{d\alpha} \frac{dx}{d\psi} \mathcal{F}_\mathbf{k}\Big[ \mathcal{F}_\mathbf{k}^{-1}\! [\,\ir \tilde{k}_y J_0(a_{\mathbf{k},s}) \tilde{\varphi}_\mathbf{k}] \mathcal{F}_\mathbf{k}^{-1}\! [\,\ir \tilde{k}_x \hat{g}_{\mathbf{k},s}] \!-\! \mathcal{F}_\mathbf{k}^{-1}\! [\,\ir \tilde{k}_x J_0(a_{\mathbf{k},s}) \tilde{\varphi}_\mathbf{k}] \mathcal{F}_\mathbf{k}^{-1}\! [\,\ir \tilde{k}_y \hat{g}_{\mathbf{k},s}] \Big].
\end{equation}
The normalized drift frequency $\omega_{d,\mathbf{k},s}$ and diamagnetic frequency $\omega_{*,\mathbf{k},s}$ are given by
\begin{align}
	\label{eq:NormalizedOmegaD}
	\omega_{d,\mathbf{k},s} & = \frac{\tilde{T}_s}{Z_s \tilde{B}} \, \hatb \times \Big(\tilde{v}_\parallel^2 \tilde{\boldsymbol{\kappa}} + \tilde{\mu} \tilde{\nablammm} \tilde{B} \Big)  \cdot (\tilde{k}_y \nablam y + \tilde{k}_x\nablam x) \\[2mm]
	\label{eq:NormalizedOmegaStar}
	\omega_{*,\mathbf{k},s} & =  - \frac{1}{\psi'_t} \, \frac{\tilde{k}_y}{2} aB_r \frac{dy}{d\alpha} e^{-\tilde{v}^2_s} \frac{d\ln F_\text{Ms}}{d\psi}%
\end{align}
with $\tilde{\boldsymbol{\kappa}} = \hatb \cdot \tilde{\nablam} \hatb$.
The normalized Fourier-transformed quasineutrality condition is obtained by multiplying equation (\ref{eq:FFTQuasiNeautrality}) with the normalizing factor $(a/n_r\rho_r)$, which yields,
\begin{equation}
\label{eq:NormalizedQuasiNeutrality}
\sum_s Z_s \Big(\frac{\delta n_{\mathbf{k},s}}{n_r} \frac{a}{\rho_r} \Big) =
\sum_s Z_s \tilde{n}_s \Bigg( \frac{2\tilde{B}}{\sqrt{\pi}} \int\limits_{-\infty}^\infty \!\! d\tilde{v}_\parallel \int\limits_0^\infty \!\! d\tilde{\mu} \, J_0(a_{\mathbf{k},s}) \, \tilde{g}_{\mathbf{k},s} + \frac{Z_s}{\tilde{T}_s} (\Gamma_0(b_{\mathbf{k},s}) - 1) \, \tilde{\varphi}_\mathbf{k}  \Bigg) = 0,
\end{equation}
where $b_{\mathbf{k},s}=k_\perp^2\rho_s^2/2$.
Equations (\ref{eq:NormalizedGyrokineticEquation}) and (\ref{eq:NormalizedQuasiNeutrality}) are the normalized gyrokinetic equation and normalized quasineutrality condition in the flux-tube approximation implemented in \texttt{stella} by Barnes et al., corresponding to equations (19) and (25) in Ref.~\cite{stella}.

\section{Key input parameters for a \texttt{stella} simulation}
\label{sec:inputs}

In \texttt{stella}, the gyrokinetic system of equations (eqs.~\ref{eq:NormalizedGyrokineticEquation} and \ref{eq:NormalizedQuasiNeutrality}) is used to evolve an initial perturbation of the electrostatic potential in time, for any number of kinetic species. The default input parameters used for a \texttt{stella} simulation are listed in Appendix \ref{app:defaultInputVariables}, and a few example input files, corresponding to simulations presented in this thesis, are given in Appendix \ref{app:usedinputfiles}. Several important input parameters are briefly highlighted in this section.

To numerically evolve a perturbation of the electrostatic field within a flux tube, the simulation domain in the $(k_x, k_y, z, v_\parallel, \mu)$ dimensions must be discretized, and a balance has to be found between the simulation accuracy and the computational cost. The number of grid points, the step size between the grid points and the minimum (or maximum) values are specified in the input file, as detailed in Appendix \ref{app:stella:grids}. Typically, a resolution of $(N_x, N_y, N_z, N_\mu, N_{v_\parallel})=(91,91,49,12,48)$ grid points is used, corresponding to setting $(\texttt{nx}, \texttt{ny}, \texttt{nzed}, \texttt{nmu}, \texttt{nvgrid}) =(91,91,48,12,24)$ in the input file. Next, the magnetic geometry within the flux tube is set, either by reading it from  a \texttt{VMEC} equilibrium when \texttt{geo\_option="vmec"}, or by computing it using a set of Miller parameters when \texttt{geo\_option="miller"}. In either case, there are only eight independent geometrical quantities appearing in the gyrokinetic equation, which are listed in Appendix \ref{sec:geometryicQuantities}. Finally, the plasma profiles must be provided to \texttt{stella}, which explicitely enter the gyrokinetic equation through the diamagnetic frequency (see eq.~\ref{app:eq:omegastar}).
The equilibrium plasma profiles $n(\psi)$ and $T(\psi)$ are approximated within the flux tube by Taylor expanding them around $\psi = \psi_0$ yielding,
\begin{align}
	n_s(\psi) & \approx n_s(\psi_0) + (\psi-\psi_0) \, \frac{dn_s}{d\psi} \Big|_{\psi_0}  \\[1mm]
	T_s(\psi) & \approx T_s(\psi_0) + (\psi-\psi_0) \, \frac{dT_s}{d\psi} \Big|_{\psi_0}
\end{align}
To include the radial variation of the plasma profiles, the normalized
density and temperature gradients are specified in the input file  for each particle species as,
\begin{align}
	\label{eq:fprim}
	\texttt{fprim} =  - \frac{a}{n_s} \frac{d n_s}{d x} \Big|_{x_0}
	\qquad \text{and} \qquad
	\texttt{tprim} = - \frac{a}{T_s} \frac{d T_s}{d x} \Big|_{x_0}.
\end{align}
For the implemented geometry options in \texttt{stella}, $dr/dx = 1$. Therefore, the input variables \texttt{fprim} and \texttt{tprim} correspond to the density and temperature \textit{gradient scale lengths}, defined as,
\index{Gradient scale length $a/L_{ns}$ and $a/T_{ns}$}
\begin{align}
	\frac{a}{L_{ns}} = - \frac{a}{n_s} \frac{d n_s}{d r} \Big|_{r_0}
	\qquad\text{and}\qquad
	\frac{a}{L_{Ts}} = - \frac{a}{T_s} \frac{d T_s}{d r} \Big|_{r_0}.
\end{align}

\vspace{8mm}

\section{Diagnostics}
\label{sec:stella:diag}

The \texttt{stella} code simulates microturbulence inside a flux tube by evolving a perturbation of the electrostatic potential in time. This electrostatic perturbation $\varphi(x,y,z)$ modifies the turbulent part of the distribution function, $\deltafm_s(x,y,z,\mu,v_\parallel)$, of the kinetic species in the plasma, which represent the deviation from a Maxwellian distribution function due to turbulence. The gyro-average of the perturbed distribution function is given by $g_s(x,y,z,\mu,v_\parallel)$. In \texttt{stella}, the Fourier components $\hat{\varphi}(k_x,k_y,z)$ and  $\hat{g}_s(k_x,k_y,z,\mu,v_\parallel)$ are evolved. The electrostatic potential and distribution function in real space are defined in section \ref{eq:diagnostics:realspace}. In a linear simulation, a perturbation of the electrostatic potential will grow indefinitely if it is driven unstable by an instability. As a result, the normalized growth rate $\tilde{\gamma} = (a/v_{\text{th},r})\, \Im\,[{\omega}]$ and normalized frequency $\tilde{\omega} = (a/v_{\text{th},r})\, \Re\,[{\omega}]$ can be computed, with ${\omega}$ the  complex frequency of the mode, given in detail in section \ref{sec:diagnostocs:gamma}.   Another important quantity to examine are the density fluctuations as a result of the perturbed electrostatic potential, as well as the particle and heat fluxes in the radial direction, which are defined in section \ref{sec:diagnostics:moments}. A summary of the normalized quantities calculated within \texttt{stella} to diagnose the turbulence is given in table \ref{tab:diagnostics}. Specifically, the fluxes are normalized with respect \mbox{to the Gyro-Bohm fluxes, which are given in table \ref{tab:gyrobohm}.}

\vspace{4mm}
\begin{table}[h!]
	\captionsetup{singlelinecheck = false, margin=0cm}
	\caption{Normalized quantities used in \texttt{stella}, related to the implemented diagnostics.}
	\label{tab:diagnostics}
	\centering
	\vspace{-2mm}
	\begin{tabular}{|cc cc l c|}
		\hline \rowcolor[HTML]{C0C0C0} \qqq&\qquad& \qquad&\qquad\quad& \qqq\qqq\qqq\qqq&\qqq\qqq \\[-3.5mm]
		\rowcolor[HTML]{C0C0C0}
		\textbf{Symbol} && \textbf{Definition} && \textbf{Normalized quantity} &\textbf{\texttt{stella} variable}  \\[1mm] \hline &&&&& \\[-3.5mm]
		$|\tilde{\varphi}|^2$   && $(a /  \rho_r)^2 \, (q_r/T_r)^2 \, |\varphi|^2$ && Perturbed electrostatic potential \!\!\!\!\!\! & \texttt{phi2} \\[1mm]
		$|\tilde{g}_s|^2$  && $\!\!\!\!\!\! (a/\rho_r)^2 \,(1/\FM)^2\, e^{-2\tilde{v}^2_s}\,|{g}_s|^2\!\!\!\!\!\!\!\!\!$  &&  Guiding-center dist.\ function & \texttt{g2}\\[1mm]
		\hline &&&&&\\[-3mm]
		$\tilde{\omega}$ && $ (a/v_{\text{th},r})\ \Re\,[\omega] $ && Mode angular frequency & - \\[1mm]
		$\tilde{\gamma}$ && $ (a/v_{\text{th},r})\ \Im\,[\omega] $ && Mode growth rate & - \\[1mm]
	    \hline &&&&&\\[-3mm]
	    $\delta \tilde{n}_s$ && $(a/\rho_r)(1/n_r)\, \delta n_s$ && Density fluctuations & \texttt{dens} \\[1mm]
	    $\delta \tilde{T}_s$ && $(a/\rho_r)(1/T_r)\, \delta T_s$ && Temperature fluctuations & \texttt{temp} \\[1mm]
		\hline &&&&&\\[-3mm]
		$\tilde{\Gamma}_{s}$ && $\Gamma_s/\Gamma_{\text{gB},r}$ && Particle flux density & \texttt{pflux} \\[1mm]
		$\tilde{Q}_{s}$      && $Q_s/Q_{\text{gB},r}$           && Heat flux density & \texttt{qflux} \\[1mm]
		$\tilde{\Pi}_{s}$    && $\Pi_s/\Pi_{\text{gB},r}$       && Angular momentum flux & \texttt{vflux} \\[1mm] \hline
	\end{tabular}
\end{table}
\FloatBarrier

\vspace{4mm}
\begin{table}[h!]
	\captionsetup{singlelinecheck = false, margin=0cm}
	\caption[Reference quantities used for the normalization]{Gyro-Bohm fluxes. Note that the units of angular momentum are kg\,m$^2$/s, and the units of a flux are 1/(m$^2$s), hence the units of the angular momentum flux are kg/s$^2$. }
	\label{tab:gyrobohm}
	\vm
	\centering
	\begin{tabular}{|cc cc l c|}
		\hline \rowcolor[HTML]{C0C0C0} \qqq&\qquad& \qquad&\qquad\quad& \qqq\qqq\qqq\qqq&\qqq\qqq \\[-3.5mm]
		\rowcolor[HTML]{C0C0C0}
		\textbf{Symbol} && \textbf{Definition} && \textbf{Reference quantity} & \textbf{Unit} \\[1mm]
		\hline &&&&&\\[-3mm]
		$\Gamma_{\text{gB},r}$ && $(\rho_r/a)^2 \,n_r \, v_{\text{th},r}$ && Gyro-Bohm Particle flux density & 1/(m$^2$s) \\[2mm]
		$Q_{\text{gB},r}$ && $(\rho_r/a)^2 \,n_r \, T_r\,  v_{\text{th},r} $ && Gyro-Bohm Heat flux density & J/(m$^2$s) \\[2mm]
		$\Pi_{\text{gB},r}$ && $(\rho_r/a)^2 \,n_r \, a\, m_r\, v_{\text{th},r}^2$ && Gyro-Bohm Angular Momentum flux & kg/s$^2$ \\[2mm]\hline
	\end{tabular}
\end{table}
\vspace*{2mm}

\subsection{Perturbed electrostatic potential in real space}
\label{eq:diagnostics:realspace}

The turbulent distribution $g_s$ and electrostatic potential $\varphi$ are Fourier expanded in the fast variation in $x$ and $y$. These quantities are obtained in real space through an inverse Fourier transformation,
\begin{align}
	\tilde{\varphi}(x,y,z) & = \frac{1}{N_xN_y} \sum_{k_x,k_y} \tilde{\Phi}_\mathbf{k}(z) \, e^{ik_x x+ik_y y}  = \sum_{k_x,k_y} \tilde{\varphi}_\mathbf{k}(z) \, e^{ik_x x+ik_y y},
\end{align}
where $\tilde{\Phi}_\mathbf{k}(z)$ are the normalized Fourier components according to the conventional definitions, while in \texttt{stella} the Fourier components are defined as $\smash{\tilde{\varphi}_\mathbf{k}(z) = \tilde{\Phi}_\mathbf{k}(z)/N_xN_y}$. To perform the Fourier transformation, the $(k_x,k_y)$ grids in reciprocal space are padded with zeros until they have a length $(N_x, N_y)$ to avoid aliasing errors (see Appendix \ref{eq:discretekxkygrid}). The Parseval theorem states that,
\begin{align}
	\sum_{x,y} |\tilde{\varphi}(x,y,z)|^2  & = \frac{1}{N_x N_y} \sum_{k_x,k_y} |\tilde{\Phi}_\mathbf{k}(z)|^2 = N_x N_y \sum_{k_x,k_y} |\tilde{\varphi}_\mathbf{k}(z)|^2.
\end{align}
Therefore, the squared electrostatic potential as a function of the parallel coordinate and averaged over the perpendicular cross-section is given by the sum over the \texttt{stella} \mbox{Fourier components, i.e.}
\begin{equation}
		|\tilde{\varphi}(z)|^2 = \frac{1}{N_x N_y} \sum_{x,y} |\tilde{\varphi}(x,y,z)|^2 = \frac{1}{(N_x N_y)^2} \sum_{k_x,k_y} | \tilde{\Phi}_\mathbf{k}(z)|^2 = \!\! \sum_{k_x,k_y} |\tilde{\varphi}_\mathbf{k}(z)|^2.
\end{equation}

\subsection{Growth rate and frequency of a $(k_x,k_y)$-mode }
\label{sec:diagnostocs:gamma}

In linear simulations, each chain of Fourier modes---consisting of different $(k_x,k_y)$-modes with the same $k_y$ but different $k_x$ values, linked through the parallel boundary condition (see section \ref{sec:connections})---evolves independently when neglecting the nonlinear term (eq.~\ref{eq:NormalizedNonlinearTerm}) in the gyrokinetic equation. If the Fourier mode is driven unstable, then the electrostatic potential will grow exponentially in time. The field-line average of the Fourier coefficient of the electrostatic potential can be fit to,
\begin{equation}
	\int \tilde{\varphi}_\mathbf{k}(z,t) \, \mathcal{J}\,dz \,\,\propto\,\, \exp \Big( -i {\omega} t \Big) = \exp \Big( -i (\tilde{\omega} + i\tilde{\gamma}) t \Big),
\end{equation}
with ${\omega} = \texttt{omega}$ the complex frequency, which is available in the output file, and $\mathcal{J}$ the Jacobian from Cartesian coordinates to the $(y,x,z)$ \texttt{stella} coordinates (eq.~\ref{eq:jac}). The magnitude of the electrostatic potential then grows in time with a normalized growth rate $\tilde{\gamma}$ and oscillates in time with a normalized \mbox{frequency  $\tilde{\omega}$.}
Therefore, the oscillation frequency is the real part of the complex frequency ${\omega}$ and the growth rate is the imaginary part.

\subsection{Moments of the distribution function}
\label{sec:diagnostics:moments}

The density fluctuations resulting from the perturbed distribution function are given by,
\begin{equation}
	\delta n_s =   \int  \deltafm_{s} \ \text{d}\mathbf{v}.
\end{equation}
Considering the Fourier expansions of $g_s = \langle \deltafm_s \rangle_\mathbf{R}$ and $\varphi$, and using equation (\ref{eq:deltaf}) gives the following expression for the normalized density fluctuations, as they are \mbox{calculated in \texttt{stella},}
\begin{equation}
	\label{eq:normalizedDensityFluctuations}
	\delta \tilde{n}_{s} = \sum_{k_x,k_y} \, \tilde{n}_s\,\frac{2 \tilde{B}}{\sqrt{\pi}}  \int\displaylimits_{0}^{\infty} \!\! d\tilde{\mu} \! 	\int\displaylimits_{-\infty}^{\infty} \!\!\!  d\tilde{v}_\parallel \ \Big( \tilde{g}_{\mathbf{k},s} \, J_0(k_\perp\rho_s) +  \frac{\tilde{Z}_s}{\tilde{T}_s} \, \tilde{\varphi}_\mathbf{k}  \Big( J_0(k_\perp\rho_s)^2 - 1 \Big) e^{-\mathcal{E}_s/T_s} \Big).
\end{equation}

\noindent
Additionally, we are interested in the turbulent particle and heat fluxes that occur in the radial direction within the plasma.  The velocity which affects the perturbations in the radial direction is the fluctuating $\mathbf{E}\!\times\!\mathbf{B}$ drift velocity (eq.~\ref{eq:EcrossBDrift}), given by,
\begin{equation}
	\label{eq:EcrossBDriftNoGyroaverage}
	\mathbf{v}_\text{Es} = (1/B^2)\,\mathbf{B} \times \nablam \varphi.
\end{equation}
The turbulent particle flux across the flux tube in the radial direction, is defined as,
\begin{equation}
	\label{eq:Gammas}
	\Gamma_s = \frac{1}{\langle | \nablam \psi |\rangle_z} \, \Big\langle \, \int  \deltafm_{s} \ (\mathbf{v}_\text{Es} \cdot \nablam \psi)  \ \text{d}\mathbf{v} \  \Big\rangle_z,
\end{equation}
with $\langle \, \cdot \, \rangle_z$ the field line average. The flux tube is taken to be small in the perpendicular $(\psi,\alpha)$-directions, hence the average flux within a flux tube reduces to the \mbox{field-line average, defined as,}
\begin{equation}
	\label{eq:fieldlineaverage}
	\langle \, \cdot \,  \rangle_z =  \int \, \cdot \,  \, \mathcal{J} \, dz \ \Big/ \int  \mathcal{J} \, dz.
\end{equation}
Substituting equations (\ref{eq:deltaf}) and (\ref{eq:EcrossBDriftNoGyroaverage}) into \mbox{equation (\ref{eq:Gammas}}), Fourier expanding the turbulent distribution function and perturbed electrostatic potential, and normalizing according to tables \ref{table:reference_quantities}--\ref{tab:diagnostics}, gives an expression for the radial component of the field-line averaged particle flux,
\begin{equation}
	\label{eq:pflux}
	\Re[\tilde{\Gamma}_s]
	=  - \sum_{\mathbf{k}}  \frac{\tilde{n}_s}{2}   \Big( \frac{\rho_r}{a} \frac{d\tilde{y}}{d\alpha} \Big)  \Big(  \frac{dr}{d\psi} \frac{aB_r}{\psi'_t} \Big)
	\frac{k_y  \rho_r}{\langle | \tilde{\nablam} \tilde{r} |\rangle_z} \Bigg\langle  \!
	\,\text{Im} \Bigg[ \,  \tilde{\varphi}^*_{\mathbf{k}} \, \frac{2\tilde{B}}{\sqrt{\pi}}
	\int\displaylimits_{0}^{\infty} \!\! d\tilde{\mu} \!
	\int\displaylimits_{-\infty}^{\infty} \!\!\!  d\tilde{v}_\parallel \
	\tilde{g}_{s,\mathbf{k}}\, J_0(k_\perp\rho_s) \Bigg]  \Bigg\rangle_z.
\end{equation}

\noindent
Similarly, the radial component of the turbulent heat flux is defined as,
\begin{equation}
	Q_s = \frac{1}{\langle | \nablam \psi |\rangle_z} \, \Big\langle \, \int  \deltafm_{s} \ (\mathbf{v}_\text{Es} \cdot \nablam \psi)  \ \text{d}\mathbf{v} \ \frac{m_s v^2}{2}  \Big\rangle_z,
\end{equation}
Using equations (\ref{eq:deltaf}) and (\ref{eq:EcrossBDriftNoGyroaverage}), Fourier expanding the turbulent distribution function and the perturbed electrostatic potential, and normalizing the variables gives,
\begin{equation}
	\label{eq:Q}
	\!\Re[\tilde{Q}_s]
	=  -\! \sum_{\mathbf{k}}  \frac{\tilde{n}_s\tilde{T}_s}{2}   \Big( \frac{\rho_r}{a} \frac{d\tilde{y}}{d\alpha} \Big)  \Big(  \frac{dr}{d\psi} \frac{aB_r}{\psi'_t} \Big)
	\frac{k_y  \rho_r}{\langle | \tilde{\nablam} \tilde{r} |\rangle_z} \Bigg\langle  \!
	\text{Im} \Bigg[ \,  \tilde{\varphi}^*_{\mathbf{k}} \, \frac{2\tilde{B}}{\sqrt{\pi}}
	\int\displaylimits_{0}^{\infty} \!\! d\tilde{\mu} \!
	\int\displaylimits_{-\infty}^{\infty} \!\!\!  d\tilde{v}_\parallel \
	\tilde{g}_{s,\mathbf{k}}\, J_0(k_\perp\rho_s) \, \tilde{v}^2 \Bigg]  \Bigg\rangle_{\!\!z }
\end{equation}
Note that these expressions are derived using the stellarator conventions, $\mathbf{B} = \psi'_t \, \nablam \psi \times \nablam \alpha$ with $\alpha =  \theta_s  - \iota \zeta_s$ which are implemented in \texttt{stella} when reading the magnetic geometry from a \texttt{VMEC} equilibrium. If instead Miller parameters are used to define the magnetic geometry, \texttt{stella} employs the tokamak conventions $\mathbf{B} = \psi'_p \nablam \alpha \times \nablam \psi$ with $\alpha = \zeta_s - q \theta_s$. Therefore, to convert the expressions in this section from stellarator to tokamak conventions, it is sufficient to substitute $\psi'_t$ by $-\psi'_p$.

\part{Turbulent Heat and Particle transport in stellarators}

\chapter{Turbulent heat transport in stellarators: an inter-machine study on the role of the density gradient}
\label{chap:heat}
\vspace{4mm}

\vspace{2mm}
Experimental observations in both tokamaks \cite{1985-Sengoku-Improvement, wolfe1986effect, 1986-Milora-Confinement, kaufmann1988pellet, 1993-Alladio-MHD, tubbing1991h} and \mbox{stellarators \cite{komori2006overview, bozhenkov2020high, 2022-Cortes-Enhanced, stroth1998high}} indicate that peaked density profiles result in a reduction of turbulent heat transport. This reduction is believed to be related to the stabilization of the ion-temperature-gradient (ITG) and electron-temperature-gradient (ETG) driven instabilities by large density gradients. However, an increasing density gradient eventually destabilizes density-gradient-driven modes such as trapped-electron modes (TEM) and passing-particle-driven universal instabilities. Hence, the question
arises whether there exists a specific range of density
gradients wherein temperature-gradient-driven modes
are stabilized, without significantly destabilizing the
density-gradient-driven modes.

In tokamaks, trapped-electron modes are particularly problematic because the majority of the trapped electrons are located in the bad curvature region on the low field side of the device. To explain the observed high-confinement phases in pellet-enhanced performance (PEP) discharges, it is not sufficient to consider density peaking alone. Additional mechanisms that can mitigate turbulence must also be taken into account, such as $E\times B$ shearing, reversed magnetic shear due to large bootstrap current, and high-collisionality which can detrap electrons \cite{hugon1992shear, smeulders1995survey, maget1999drift, baylor2000improved, romanelli2004effects}. In contrast, in stellarators, the bulk of trapped electrons may be separated from the bad curvature regions, which reduces the impact of trapped-electron modes. Additionally, configurations approaching the \mbox{maximum-$J$} property have been shown to have reduced TEM turbulence  \cite{helander2013collisionless, helander2015advances, proll2022turbulence, garcia2024reduced}. As a result, the classical electron-driven trapped-electron mode may be significantly less detrimental in stellarators, with other instabilities driven by density gradients, such as ion-driven trapped-electron modes (iTEM) and passing-particle-driven universal instabilities, playing a more prominent role. Moreover, turbulence in stellarators has been shown to be reduced by ambipolar electric fields \cite{xanthopoulos2020turbulence, fu2021global}, $E\times B$ shearing, zonal flow shearing, energy cascades to dissipative scales \cite{xanthopoulos2011zonal, faber2015gyrokinetic, ishizawa2017multi, zocco2024electron} and energy transfer through damped modes \cite{faber2018stellarator, mckinney2019comparison}, which is beyond the scope of this work.

In this chapter, the effect of the density gradient on turbulent heat transport is isolated to explore how density peaking influences turbulent heat transport and how this impact varies across different magnetic configurations. The $\deltafm$-gyrokinetic code \texttt{stella} \cite{stella} is used to carry out a comprehensive comparative study of different devices, including different families of stellarators, specifically, the Wendelstein 7-X  (W7-X) stellarator, the Large Helical Device (LHD), the \mbox{TJ-II} stellarator and the National Compact Stellarator Experiment (NCSX), which represent helias, heliotron, heliac and quasi-axisymmetric configurations, respectively. Additionally, the Axially Symmetric Divertor Experiment or ASDEX Upgrade (AUG) tokamak and the tokamak geometry of the Cyclone Base Case (CBC) \cite{dimits2000simulation} are included for comparison. The results presented in this chapter closely follow those recently published in Ref.~\cite{thienpondt2024influence},  as indicated in the ``Published and submitted content'' declaration on page~\pageref{page:content}.

Previous cross-device comparisons of the effect of the density gradient were limited to linear results \cite{rewoldt2005comparison, proll2013collisionless},  or only consider a few points in parameter space for nonlinear simulations \cite{helander2015advances, proll2022turbulence, garcia2024reduced,  xanthopoulos2020turbulence, faber2015gyrokinetic, zocco2024electron, mckinney2019comparison}. Specifically, it has been shown that W7-X benefits from reduced ITG turbulence and low TEM-driven transport in the presence of a density gradient \cite{proll2022turbulence, garcia2024reduced}, however, the influence of the electron temperature gradient was neglected in \cite{proll2022turbulence}. On the other hand, the heat transport has been investigated before and after the injection of cryogenic hydrogen pellets in a  \mbox{W7-X} discharge, where it is argued that a decrease of turbulence in the presence of peaked density profiles is either due to a transition from ITG to iTEM-driven turbulence \cite{xanthopoulos2020turbulence}, or due to a more efficient zonal-flow stabilization in the post-pellet phase \cite{zocco2024electron}. In contrast to previous studies, this work includes both linear and nonlinear results across a broad range of parameters, examining the impact of electrons with and without a temperature gradient, as well as analyzing the velocity dependence of the turbulent component of the distribution function.

This chapter is divided into the following sections.  First, section \ref{chap:equilibria} introduces and analyzes
the different magnetic configurations, and in section \ref{sec:localization}
the localization of the microinstabilities is inferred
from the regions of bad curvature and magnetic
wells.
In \mbox{section \ref{sec:linear}} a comprehensive linear stability analysis of the six considered devices is carried out. The growth rates and frequencies of the microinstabilities are discussed in section \ref{sec:gammavskxky}, for specific values of the profile gradients and considering a wide range of radial and binormal wavenumbers. Section \ref{sec:linearlineallkys} explores the evolution of the instabilities with increasing density gradients, revealing that the most unstable mode transitions continuously from an ion-temperature-gradient driven mode to a predominantly density-gradient-driven mode in CBC, AUG, TJ-II and LHD. In contrast, NCSX and W7-X display distinct frequency branches, indicating that different microinstabilities can be distinguished from one another. In section \ref{sec:mostunstable} the most unstable mode for the considered binormal wavenumbers is identified, showing that the linear growth rates of the fastest-growing modes increase with the density gradient in CBC, AUG, TJ-II and LHD---and to a lesser extent in NCSX---whereas a modest reduction of the growth rates is found for a small range of density gradients in \mbox{W7-X}. The contributions of the specific plasma gradients are isolated in \mbox{section \ref{sec:linearlineITGTEM}}, which reveals that in NCSX and \mbox{W7-X} there is a range of density gradients for which the ion-temperature-gradient-driven modes are stabilized, without driving the density-gradient-driven modes particularly unstable yet. To conclude the linear analysis, in section \ref{sec:gammavsfprimtiprim} the ion temperature and density gradient are scanned across a wide range of parameters.

Section \ref{sec:nonlinear} is the most important part of this chapter, as it investigates the effect of the density gradient on the turbulent heat transport by means of nonlinear gyrokinetic simulations. Section \ref{sec:heat} highlights the distinct behaviors of the ion heat flux in different devices. Specifically, in the tokamaks, the ion heat flux increases strongly with the density gradient, whereas in LHD it remains largely unaffected, and in TJ-II a substantial reduction of the ion heat flux occurs for small density gradients.
In contrast, both NCSX and W7-X exhibit a strong reduction of the ion heat flux across a wide range of density gradients, due to the stabilization of temperature-gradient-driven modes, without significantly destabilizing  density-gradient-driven modes, as demonstrated in \mbox{section \ref{sec:nonlinearITGTEM}}. The influence of a finite electron temperature gradient is discussed in \mbox{section \ref{sec:nonlinearWithALTe}}. It is worth noting that in stellarators the ion heat flux shows no correlation with the growth rates of the fastest-growing modes obtained from the linear analysis.
Finally, the distribution function is examined in detail in section \ref{sec:distributon} to characterize the microinstabilities that are driving the turbulence by means of novel diagnostics developed in \texttt{stella} as a part of this PhD thesis. Specifically, the velocity dependence of the perturbed distribution function is examined to quantify the importance of the ion and electron dynamics, and to determine whether the turbulence is primarily driven by the trapped or passing electrons. It is found that at high density gradients, trapped-electron modes predominately drive the heat transport in CBC, AUG, TJ-II, LHD and NCSX, whereas in W7-X, the passing-particle-driven universal instability contributes significantly to the turbulence.

The simulations presented in this chapter are electrostatic, collisionless, consider ion-Larmor scales and include both kinetic hydrogen nuclei and electrons, with equal densities ($n_i=n_e$) and temperatures ($T_i=T_e$). The ion and electron density gradients are assumed to be equal ($a/L_{n_i} =  a/L_{n_e}$) with $a/\hspace{-0.1mm}L_{n_s} = -(a/n_s)\hspace{0.2mm}dn_s/dr$ the normalized density gradient. Moreover, a fixed ion temperature gradient of $a/L_{T_i} = 3.0$ is considered, while exploring both vanishing ($a/L_{T_e} = 0.0$) and finite ($a/L_{T_e} = 3.0$) electron temperature gradients. The density gradient is varied across a broad range of parameters,  ranging from \mbox{$a/L_n = 0.0$} to $a/L_n = 4.0$. These parameters were chosen with the high-performance pellet-injection discharges of W7-X in mind, which are characterized by ($a/L_n = 0.8$ and $a/L_{T_i} = 2.5$) in the pre-pellet phase and ($a/L_n = 3.0$ and $a/L_{T_i} = 4.5$) in the post-pellet phase at $r/a = 0.63$ \cite{xanthopoulos2020turbulence, zocco2024electron}. Given that each device operates under different conditions, the considered gradients may be somewhat extreme for LHD and \mbox{TJ-II} for example. Nonetheless, examining the impact of density peaking in these devices under such gradients is valuable, as they represent distinct stellarator designs and they provide a clear contrast with the results from the tokamak configurations. It is important to note that this work focuses on investigating the effects of 	density peaking on turbulent transport across different reactor designs, rather than on direct comparisons with experimental data.

\vspace{2mm}
\section{Magnetic configurations}
\label{chap:equilibria}

\vspace{1mm}
To address the effect of the magnetic geometry on
linear growth rates and nonlinear heat transport,
this study incorporates the tokamak geometry of the Cyclone Base Case (CBC), which has been used extensively to benchmark gyrokinetic \mbox{codes \cite{dimits2000simulation}},  as well as the ASDEX Upgrade (AUG) tokamak, constructed at IPP (Germany), which commenced operations \mbox{in 1991}.
On the other hand, the \mbox{W7-X}, TJ-II, LHD and NCSX stellarators are included. \mbox{Wendelstein 7-X} is a low-shear five-field-period helias configuration, that has been in operation at IPP (Germany) since 2015, for which the standard EIM-003 \mbox{configuration \cite{Geiger_2015}} is used. Another low-shear device is the four-field-period heliac configuration \mbox{TJ-II}, which has been in operation since 1997 at CIEMAT (Spain). The standard configuration (100\_44\_64) is considered. NIFS (Japan) houses the Large Helical Device,  a ten-field-period heliotron configuration, generating plasmas since 1998, for which the standard configuration with a major radius of \mbox{$R_0=3.73$} is used. Lastly, NCSX is a three-field-period quasi-symmetric stellarator, in particular featuring toroidal quasi-symmetry, also known as quasi-axisymmetry, developed at PPPL (United States). The magnetic configurations of AUG, \mbox{TJ-II}, LHD, NCSX and \mbox{W7-X} are generated by the three-dimensional MHD equilibium code \mbox{\texttt{VMEC} \cite{hirshman1983steepest}}, whereas the considered flux surface of CBC is defined by a set of Miller parameters \cite{miller1998noncircular}, explicitly \mbox{given in \cite{stella}} and in Appendix \ref{app:usedinputfiles}. The configurations are taken to have $\beta = 0$, with $\beta$ the ratio of the \mbox{plasma pressure to the magnetic pressure.}

\begin{figure}[b!]
	\centering
	\includegraphics[clip,  trim={0mm 10mm 0mm -1mm}] {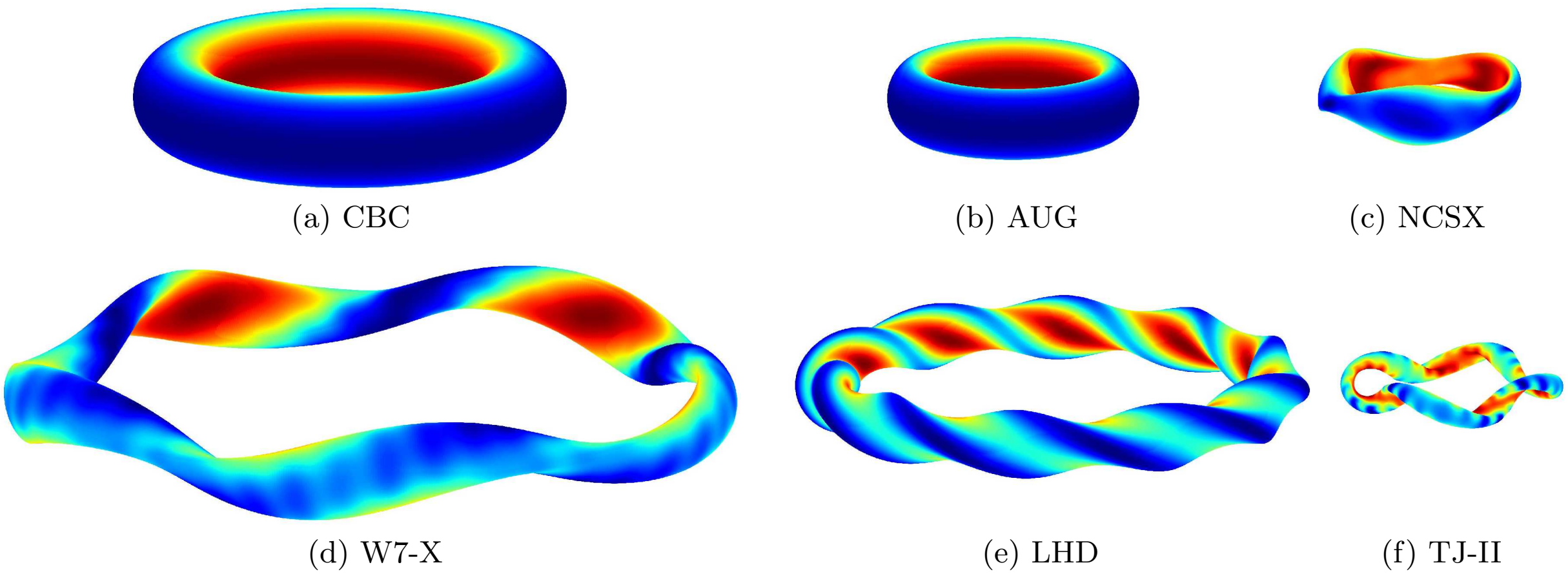} \vspace{-6mm}
	\caption{Magnetic field strength---minimum is blue, maximum is red---on the flux surface $r/a = 0.7$ for the six considered devices. The same scale is used in order to compare the relative size of the reactors.}
	\vspace{9mm}
	\label{fig:fluxsurfaces}
\end{figure}

\newpage
The simulations are performed at the radial
location $r/a = 0.7$, for which the flux surfaces are
shown in figure \ref{fig:fluxsurfaces}.
The magnetic field strength is minimum (blue) on the outboard side and maximum (red) on the inboard side of the tokamaks, as can be seen in figures \ref{fig:fluxsurfaces}a and \ref{fig:fluxsurfaces}b. This trend does not necessarily hold in stellarators, where a minimum of the magnetic field can also be found on the inboard side, and vice versa.  An overview of the geometric characteristics of these flux surfaces is given in \mbox{table \ref{tab:geometrySixDevices}}.
For tokamaks, the strongest turbulence is expected to be found at the outboard midplane, therefore gyrokinetic studies generally focus on the $\alpha_0=0$ field line centered at \mbox{$(\theta,\zeta) = (0,0)$}. The $(\theta,\zeta) = (0,0)$ points lays on the bean-shaped cross-section of \mbox{W7-X} (see figure \ref{fig:W7XFluxtube}), and on the outboard midplane of the tokamaks. Note that the magnetic geometry and plasma parameters included in the simulations depend solely on their values and radial derivatives along the selected magnetic \mbox{field line $\alpha_0$}. Seeing that each field line in a stellarator has a unique magnetic geometry, different field lines will simulate different levels of \mbox{turbulence \cite{sanchez2021gyrokinetic, gonzalez2022electrostatic}}.

Although in stellarators the turbulence may be strongest elsewhere on the flux surface, this study is limited to the $\alpha_0=0$ field line, since it is stellarator-symmetric \cite{dewar1998stellarator}, allowing us to utilize the generalized twist-and-shift boundary \mbox{conditions \cite{martin2018parallel}} discussed in section \ref{sec:StellaratorSymmetricBC}, which are particularly advantageous for low-shear devices such as TJ-II and W7-X.
The field line is extended up to three poloidal turns for linear simulations, and up to approximately one poloidal turn for nonlinear simulations, where the exact length (see table \ref{tab:geometrySixDevices}) is chosen to \mbox{ensure that $\Delta k_x = \Delta k_y$}.

\begin{table}[h!]
	\vspace{4mm}
	\caption{\label{tab:geometrySixDevices} Geometric quantities, with $a$ the minor radius, $R_0$ the major radius, $\iota_0$ the rotational transform and $\hat{s}_0 = - (r_0/\iota_0)(d\iota/dr)_0$ the global magnetic shear at $r/a = 0.7$. \mbox{$N_{f}$ is the} number of field periods of the device and $N_{\rm pol}$ is the length of the flux tube, given in number of poloidal turns, considered in the nonlinear simulations performed in section \ref{sec:nonlinear}. The mirror ratio $(B_\text{max}/B_\text{min})$ is calculated along the $\alpha_0$ field line.} \vspace{-3mm}
	\lineup
	\centering
	\begin{tabular}{@{}cccccccc}
		\specialrule{.12em}{.1em}{.1em} &&&&&&& \\[-4mm]
		Device & $\0a$\,(m) & $\0R_0$\,(m)  & $\m\0\hat{s}_0$ & \m\0$\iota_0$ & $B_\text{max}/B_\text{min}$ & $N_{f}$ & $N_{\rm pol}$\ \\[1mm] \hline &&&&&&& \\[-4mm]
		\hspace{16mm} & \hspace{15mm}  & \hspace{15mm} &
		\hspace{15mm} & \hspace{15mm} & \hspace{15mm} & \hspace{15mm} & \hspace{15mm} \\[-3.5mm]
		CBC   & \m 1.00 & \m 2.78 & \m 0.796 & \m 0.714 & 1.67 & \01 & 1.00 \\[0.5mm]
		AUG   & \m 0.63 & \m 1.62 & \m 1.587 & \m 0.744 & 1.57 & \01 & 1.00 \\[0.5mm]
		NCSX  & \m 0.32 & \m 1.44 & $-$0.703 & \m 0.541 & 1.22 & \03 & 0.95 \\[0.5mm]
		TJ-II & \m 0.19 & \m 1.50 & $-$0.057 & $-$1.592 & 1.55 & \04 & 1.19 \\[0.5mm]
		LHD   & \m 0.60 & \m 3.73 & $-$1.230 & $-$0.628 & 1.51 & 10  & 1.00 \\[0.5mm]
		W7-X & \m 0.51 & \m 5.51 & $-$0.127 & \m 0.894 & 1.24 & \05 & 1.11 \\[0.5mm]
		\specialrule{.12em}{.1em}{.1em}
	\end{tabular}
	\vspace{2.5mm}
\end{table}

\vspace{3mm}
The resolutions used for the simulations performed in this chapter are given in table \ref{tab:resolutionSixDevices}, where the choices are justified  by performing rigorous convergence checks of the nonlinear heat fluxes, for the different devices and gradients considered in this work, which are discussed \mbox{in Appendix \ref{sec:resolution}.} A small amount of hyper-viscosity, with $D=0.1$, was applied in the simulations presented in this work. The form of the hyper-viscosity used in \texttt{stella} is given by equation (53) in \cite{stella}.

\begin{table}[b!]
	\centering
	\vspace{0mm}
	\caption{\label{tab:resolutionSixDevices} Resolution used for the linear and nonlinear simulations included in this chapter. For nonlinear simulations, approximately 1 poloidal turn is used to ensure that $\Delta k_x = \Delta k_y$, the exact length of the flux tube is given in table \ref{tab:geometrySixDevices}. For LHD, $N_z=97$ grid points are used in nonlinear simulations to account for its more detailed magnetic field structure. The Fourier grid corresponds to a computational domain of $L_x = L_y = 94\rho_i$ with $N_x = N_y = 91$ grid points in real space, here $\rho_i$ is the ion-Larmor radius.} \vspace{-2mm}
	\lineup

	\begin{tabular}{@{}ccccccccccc}
		\specialrule{.12em}{.1em}{.1em} \\[-4mm]
		& $N_{\rm pol}$ & $N_z$ & $k_x\rho_i$ & $k_y\rho_i$ & $N_{k_x}$ & $N_{k_y}$ & $v_\parallel/v_{{\rm th},s}$ & $v_\perp/v_{{\rm th},s}$ & \!$N_{v_\parallel}$\! & \!$N_{v_\perp}$\!  \\[1mm] \hline\\[-4mm]
		\qquad&\qquad&\qquad&
		\qquad\quad\;\;\;\; &\qquad\quad\;\;\;\; &\qquad&\qquad&
		\qquad\quad\;\;\; &\qquad\quad\;\;\; &\quad&\quad\\[-4mm]
		Linear & 3 & 513  & 0 & \![0.125, 2.00]\! & 0 & $>$20 & \![-3.0, 3.0]\! & \![0.0, 3.0]\! & \!128\! & 24   \\[1mm]
		Nonlinear & $\sim$1 & 49  & \![0.067, 2.00]\! & \![0.067, 2.00]\! & 61 & 31  & \![-3.0, 3.0]\! & \![0.0, 3.0]\! & 48& 12    \\[1mm]
		\specialrule{.12em}{.1em}{.1em}
	\end{tabular}
	\vspace{0mm}
\end{table}

\begin{figure}[h!]
	\centering
	\vspace{2mm}
	\vspace{2mm}
	\vspace{2mm}
	\includegraphics[trim={0mm  2mm 0mm -1mm}] {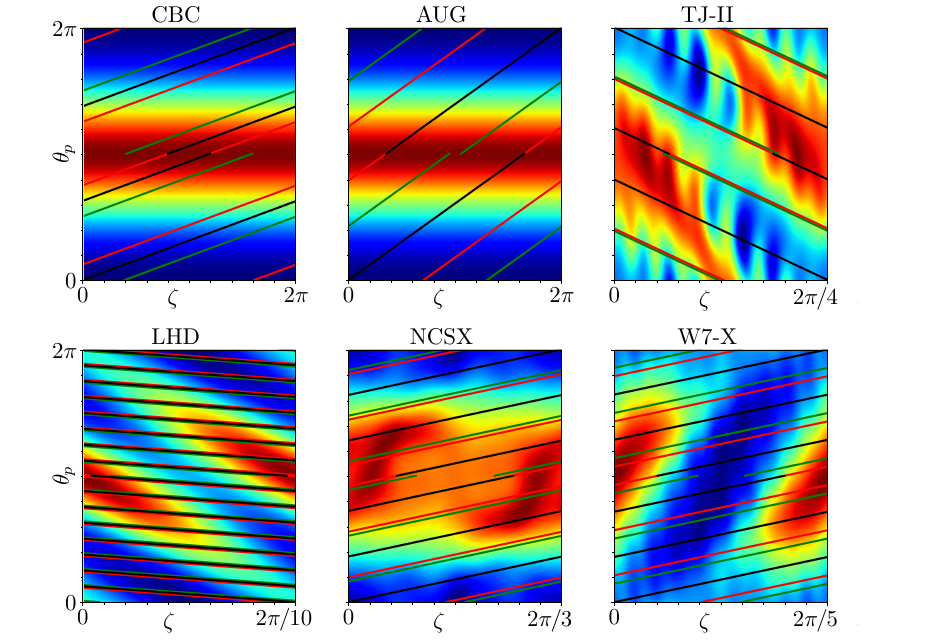}
	\caption{Coverage of the flux surface $r/a=0.7$ by a field line that extends up to one (black), two (red) and three (green) poloidal turns.  Only one field period is shown for each device. \vspace{1mm}}
	\vspace{2mm}
	\vspace{2mm}
	\label{fig:coverage}
\end{figure}

\newpage
It is important to be aware of how densely the flux surface is covered by the simulated field line, which is shown in figure \ref{fig:coverage} for field lines extending up to one (black), two (red) and three (green) poloidal turns, centered at \mbox{$(\theta,\zeta) = (0,0)$}. Since tokamaks are axisymmetric, each poloidal turn is identical, whereas extending the field line for more than one poloidal turn samples new areas of the flux surface for the standard configuration of \mbox{W7-X}. In contrast, for LHD, each additional turn overlaps almost exactly with the previous turn due to the specific value of the rotational transform on the $r/a=0.7$ flux surface. Therefore, extending the flux tube for multiple poloidal turns does not necessarily cover additional area of the flux surface. Nonetheless, it can be important to increase the length of the flux tube to capture modes that extend beyond the first poloidal turn of the device \cite{parisi2020toroidal}, or if the parallel correlation length is longer than the length covered by one poloidal turn \cite{beer1995field, faber2018stellarator,  ball2020eliminating}. In figures \ref{fig:linear_phi_vs_z}, \ref{fig:linear_phi_vs_z2} and \ref{fig:qfluxvsz} it can be seen that the potential and heat fluxes diminish considerably at the ends, indicating that a single poloidal turn should be sufficient.

\begin{figure}[b!]
	\centering
	\vspace{-3mm}
	\includegraphics[width=0.99\linewidth, trim={0mm 10mm 0mm 2mm}, clip] {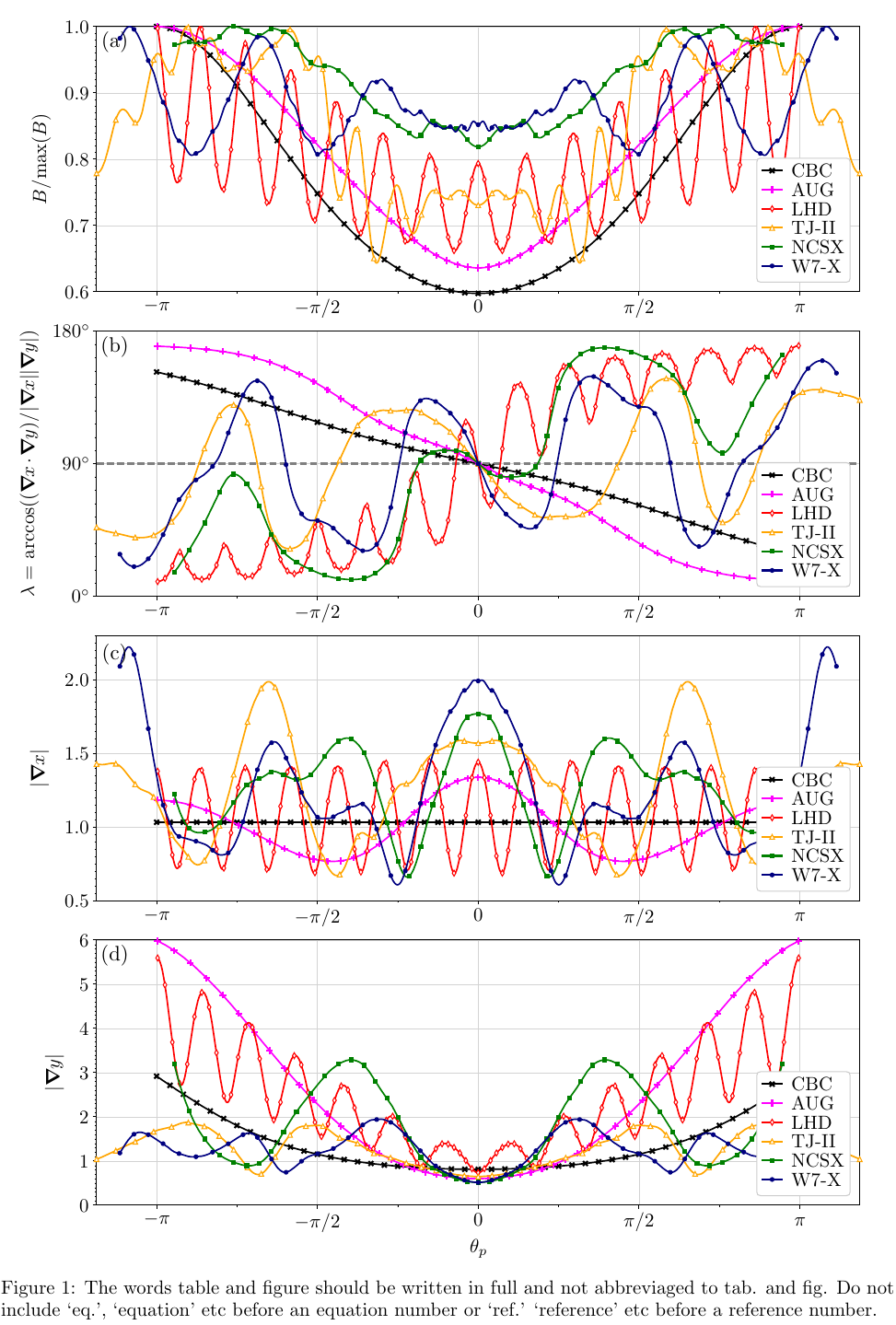} \vspace{-4.5mm}
	\caption{Geometric quantities along the chosen field line: (a) magnetic field strength (b) angle of the perpendicular flux-tube cross-section, (c) magnitude of the perpendicular wavenumber for $k_x \rho_i = 0.0$ and $k_y \rho_i = 1.0$ and (d) magnitude of the perpendicular wavenumber for $k_x \rho_i = 1.0$ and $k_y \rho_i = 0.0$. The markers represent the parallel grid points used in the nonlinear simulations in section \ref{sec:nonlinear}. Note that while $z=\zeta$ is generally used to describe gyrokinetics in stellarators, since the flux tube is typically extended up to one poloidal turn, the geometric quantities are often depicted as a function of the poloidal \texttt{PEST} angle $\theta_p$ instead. \vspace{-3mm}}
	\label{fig:fieldlineAngle}
\end{figure}

The simulation domain---i.e.~the flux tube---is centered around the $\alpha_0=0$ field line, which has its midpoint at $(\zeta, \theta) = (0,0)$. The magnetic field strength along this considered field line is shown in figure \ref{fig:fieldlineAngle}a. It can be seen that tokamaks typically consists of a single magnetic well, while stellarators have a much more complex magnetic field structure, often containing multiple inner magnetic wells. Specifically, the magnetic field strength along a field line in LHD resembles a tokamak with a very strong local ripple. 	Minima of the magnetic field will trap particles which have a relatively low parallel velocity. In the case of a singular magnetic well, the particles which are trapped are those for which $v_\perp/v_\parallel > \sqrt{B_\text{min}/B_\text{max}}$ with $B_\text{max}/B_\text{min}$ the mirror ratio (see table \ref{tab:geometrySixDevices}). Therefore, the relative depth of the magnetic well plays an important role on trapped-electron modes which are driven unstable by the trapped electron population.

\FloatBarrier

Since field-aligned coordinates are employed, the cross-section of the flux tube is twisted into a parallelogram along the field line as a result of the magnetic shear, which has been discussed in detail in section \ref{sec:fluxtube:shearing}. Sharp changes in the shearing of the flux tube cross-section can effectively shear turbulence apart by stretching turbulent eddies (increasing $k_\perp$) which are subsequently suppressed by finite-Larmor-radius effects (section \ref{sec:FLR}). In figure \ref{fig:fieldlineAngle}b the angle between the $\nablam x$ and $\nablam y$ directions (eq.~\ref{eq:coslambda}) is depicted along the considered field line. The AUG and CBC tokamaks have a very large global magnetic shear (\mbox{table \ref{tab:geometrySixDevices}}). As a result, the flux-tube cross-section becomes extremely sheared at the flux tube ends. Similarly, LHD and NCSX have a large global magnetic shear on  the $r/a=0.7$ flux surface. However, for these devices the local magnetic shear also plays an important role. Finally, for W7-X and TJ-II the global magnetic shear is very small. As a result, the local magnetic shear predominantly determines the flux tube shearing. The finite-Larmor-radius effect (section \ref{sec:FLR}) scales with $k_\perp \rho_s$, leaving large scale ($k_\perp \rho_s\ll 1$) fluctuations unaffected while damping small scale ($k_\perp \rho_s \gg 1$) fluctuations. The perpendicular wavenumber associated with the turbulent fluctuations (eq.~\ref{eq:kperp}) not only depends on the flux tube shearing ($\boldsymbol{\nabla} x \cdot \boldsymbol{\nabla} y$) but also on the compression of the flux surfaces $(\nablam x)$ and the field lines  $(\nablam y)$. These quantities change along the field line and are depicted in figures \ref{fig:fieldlineAngle}c and \ref{fig:fieldlineAngle}d, respectively.

Magnetic wells and magnetic shear (figure \ref{fig:fieldlineAngle}) can have an important effect on microturbulence. Additionally, the curvature and gradient of the magnetic field play an important role as they give rise to magnetic drifts (eq.~\ref{eq:MagneticDriftsIntro}). In particular, toroidal ITG modes \cite{plunk2014collisionless} are localized in regions of bad (destabilizing) curvature, for which $k_y T'_i/\psi'_t(\mathbf{B} \times \nablammm B) \cdot \mathbf{k}_\perp > 0$ (eq.~\ref{eq:badCurvatureIntro}). In this
work we only consider peaked temperature profiles ($dT_i/dr <0$) and counter-clockwise configurations  ($d\psi_t/dr<0$). Similar to ITG modes, which are destabilized by the local bad curvature, trapped-electron modes \cite{helander2013collisionless} are driven unstable by the bounce-averaged bad curvature. The most unstable modes typically have \mbox{$k_x = 0$} (but not always, as will be shown in section \ref{sec:gammavskxky}) for which the condition of bad curvature reduces to \mbox{$(\mathbf{B} \times \nablammm B) \cdot \boldsymbol{\nabla} y > 0$}. For simplicity, we define,
\begin{equation}
	\label{eq:Kappay}
	\mathcal{K}_y = (a^2B_r/B^3) \,(\mathbf{B} \times \nablammm B) \cdot \boldsymbol{\nabla} y,
\end{equation}
where \mbox{$\mathcal{K}_y>0$} corresponds to regions of bad curvature. Recall that $\nablam y \approx  r_0\nablam\theta_p - r_0 \iota_0 \nablam\zeta - \iota_0 \hat{s}_0 \,\zeta \nablam x$ has a secular dependence on the parallel coordinate $z=\zeta$. Therefore, regions of bad curvature depend on the parallel coordinate $z=\zeta$ along the field line, which may seem counterintuitive.

\begin{figure}[b!]
	\centering
	\includegraphics[trim={0mm 0mm 0mm 0mm}] {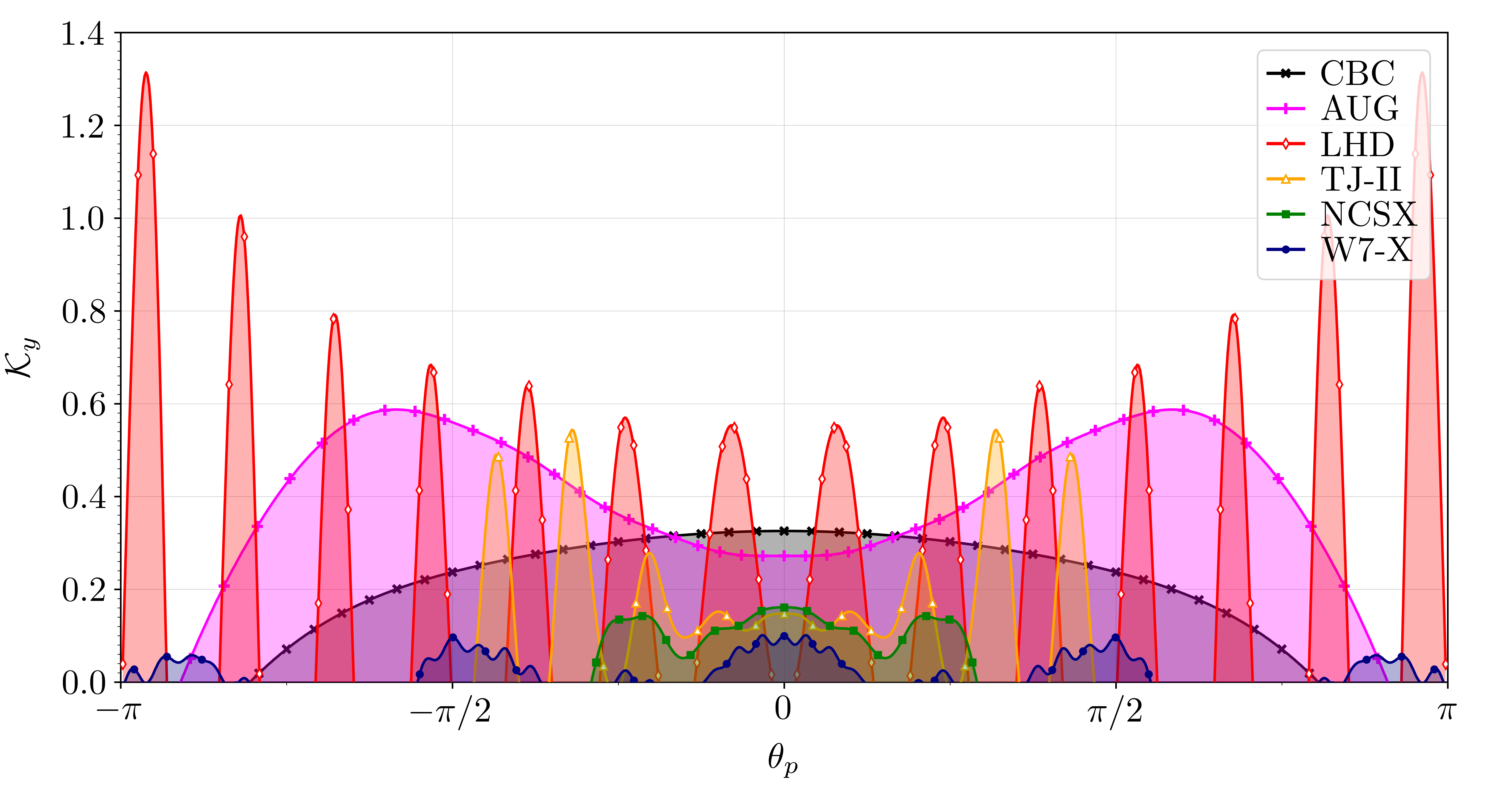}\vspace{-6mm}
	\caption{Bad curvature regions for $k_x = 0.0$ along the considered field lines, defined as $\mathcal{K}_y>0$. \vspace{0mm} }
	\label{fig:fieldlineCurvature}
\end{figure}

\newpage
To study the regions of bad curvature, \mbox{$\mathcal{K}_y>0$} (eq.~\ref{eq:Kappay}) is plotted along the considered magnetic field lines in figure \ref{fig:fieldlineCurvature}, assuming $k_x=0$. In the AUG and CBC tokamaks (black and pink lines), the bad curvature regions are not contained to the outboard side $(\theta_p \in [-\pi/2, \pi/2])$ because these devices have a large positive global magnetic shear on the considered flux surface, specifically $\hat{s} = 1.587$ for AUG and $\hat{s} = 0.796$ for CBC. As a result, along the majority of the field line, the magnetic drifts have a destabilizing effect on perturbations of the plasma. Nonetheless, this destabilizing effect can be counteracted, in part, by the strong stabilizing effect due to the size of the global magnetic shear. Namely, strong magnetic shear is capable of breaking up large-scale turbulent fluctuations, which are subsequently suppressed by FLR effects (section \ref{sec:FLR}).

In contrast, field lines in stellarators generally consist of intermittent regions of good and bad curvature. The distribution function, $\deltafm_s$, associated with a perturbation is generally symmetric in $v_\parallel$, and the majority of the particles have a finite parallel velocity. Therefore, as a perturbation is amplified in a bad curvature region, the charged particles associated with this perturbation will stream along magnetic field lines, and the perturbation effectively propagates in the parallel direction, i.e.~along the field lines, both in the clockwise and counter-clockwise direction. As the perturbation propagates along the field line it will be intermittently damped and amplified as it moves through regions of good and bad curvature. Therefore, devices with very narrow regions of bad curvature, or which have more good curvature than bad curvature along the field lines, should be more stable to ITG and TEM instabilities.

\begin{figure}[b!]
	\centering
	\includegraphics[width=\linewidth, trim={0mm  0mm 0mm -1mm}] {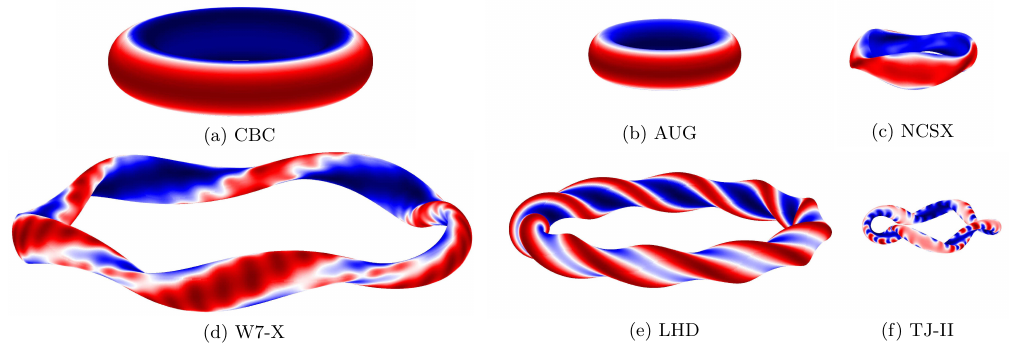}
	\caption{Bad curvature regions (red) and good curvature regions (blue) on the flux surface $r/a=0.7$, defined as $\mathcal{K}_y^{\hat{s}=0} > 0$ and $\mathcal{K}_y^{\hat{s}=0} < 0$, respectively, considering $\hat{s} = 0$ and $k_x=0$. \vspace{8mm}}
	\label{fig:fluxsurfacesbadcurvature}
\end{figure}

The bad curvature regions depend on the parallel location $z=\zeta$ if the global magnetic shear is finite, i.e.~if $\hat{s}\neq 0$ on the considered flux surface. It is possible to illustrate the magnitude of the bad curvature $(\mathcal{K}_y)$ on a specific flux surface by considering $\hat{s} = 0$, which makes $\mathcal{K}_y^{\hat{s}=0}$ periodic in $\zeta$. The magnitude of $\mathcal{K}_y^{\hat{s}=0}$ on the considered flux surfaces is shown in \mbox{figure \ref{fig:fluxsurfacesbadcurvature}.} Neglecting the effect of global magnetic shear, the bad curvature region (red) is confined to the outboard side ($\theta \in [-\pi/2, \pi/2]$) for the CBC and AUG tokamaks, and approximately in NCSX. As a result, for $\hat{s} = 0$ the bad curvature regions overlap with the minima of the magnetic field, which makes trapped-electron modes particularly unstable. In contrast, in W7-X, LHD and TJ-II the bad curvature regions spiral around the torus, similarly to the minima of the magnetic field \mbox{(figure \ref{fig:fluxsurfaces}).} Since they spiral in a similar manner, the bad curvature regions and the magnetic wells tend to overlap in CBC, AUG, NCSX, LHD, TJ-II and the standard configuration of W7-X, when neglecting the effect of the global magnetic shear. It is possible to modify the configurations to reduce the overlap between bad curvature regions and magnetic wells, which should have a stabilizing effect on trapped-electron modes. This overlap will be discussed \mbox{in detail in the next section.}

\newpage
\section{Localization of microinstabilities}
\label{sec:localization}

The background density and temperature gradients can drive a large number of microinstabilities in fusion plasmas. Specifically, the ion-scale heat transport is predominately attributed to turbulence excited by ion-temperature-gradient-driven (ITG) and trapped-electron (TEM) instabilities. These instabilities are destabilized or stabilized along the magnetic field lines depending on geometric characteristics such as magnetic wells (minima of $B$ in figure \ref{fig:fieldlineAngle}a), shearing of the flux-tube cross-section ($\lambda$ in figure \ref{fig:fieldlineAngle}b) and magnetic curvature and magnetic field inhomogeneity which give rise to magnetic drifts that can drive perturbations unstable in the so-called bad-curvature regions ($\mathcal{K}_y>0$ for $k_x=0$ in figure \ref{fig:fieldlineCurvature}). These geometric quantities have been discussed in detail in the previous sections.
Recall that toroidal ITG modes are localized in regions of bad (destabilizing) curvature, whereas  trapped-electron modes are driven unstable by the bounce-averaged bad curvature. Moreover, TEMs reside in regions where electrons are trapped, which is inside the magnetic wells shown in figure \ref{fig:fieldlineAngle}a. Both the TEM and ITG mode can thus be stabilized by reducing the connection length between regions of good and bad curvature. Moreover, the separation of regions of trapping \mbox{and bad curvature has a stabilizing effect on TEMs.}

\begin{figure}[b!]
	\centering
	\vspace{-4mm}
	\includegraphics[trim={0mm 2mm 0mm 0mm}] {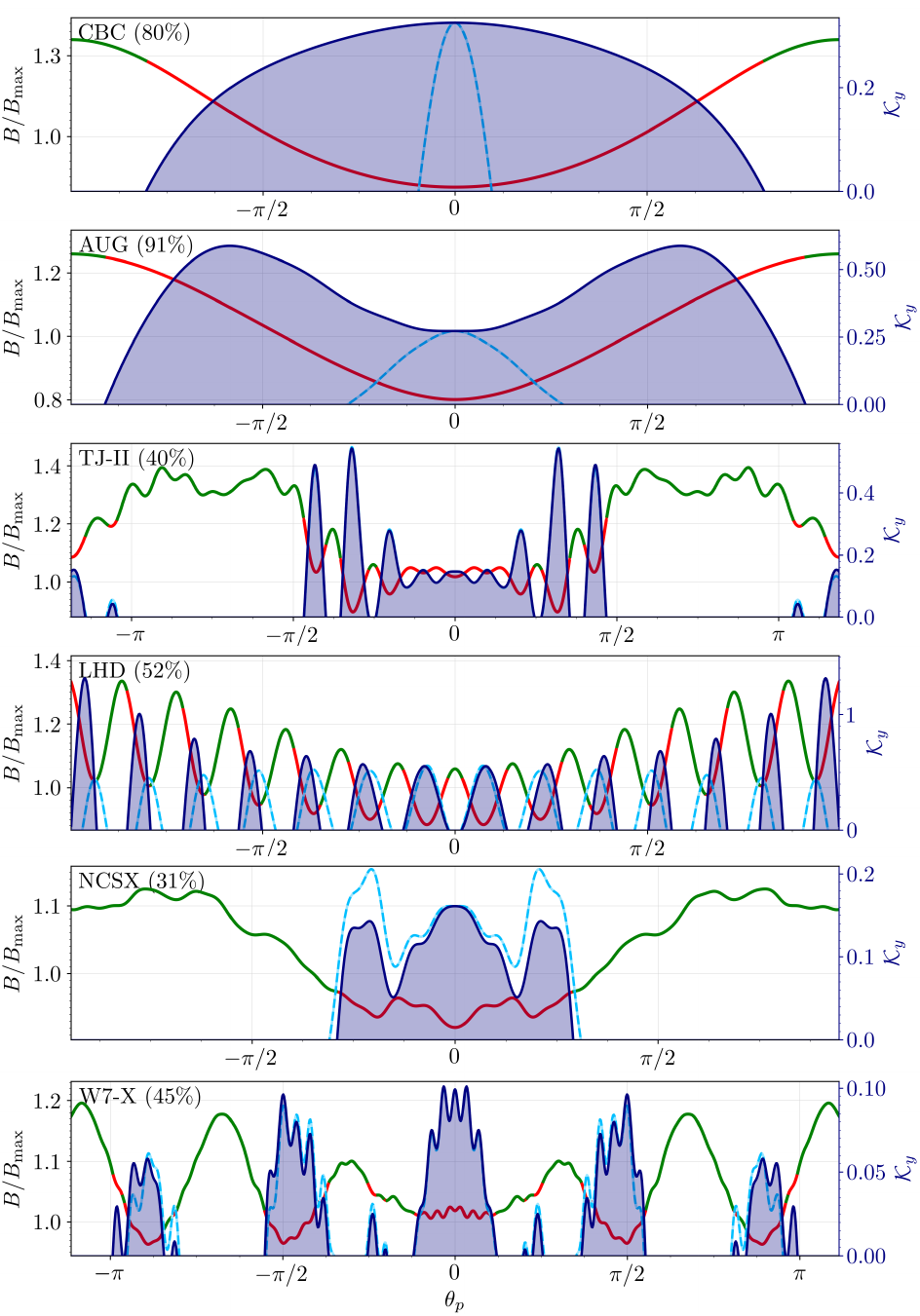}
	\vspace{-1mm}
	\caption{Magnetic field strength (green-red lines) and bad curvature regions ($\mathcal{K}_y\!>\!0$, blue shaded regions) considering $k_x=0$, where the $B$-line is colored red if it has bad curvature, and green if it has good curvature. The bad curvature regions in the absence of global magnetic shear are depicted by the light blue dashed lines. The percentage indicates the amount of the field line which is covered by bad curvature.}
	\vspace{-4mm}
	\label{fig:trappingcurvature}
\end{figure}

Figure \ref{fig:trappingcurvature} shows the regions of particle trapping (red-green lines) and bad curvature (blue shaded areas) along the chosen magnetic field lines, for modes with $k_x=0$, in order to discuss their overlap. Moreover, the bad curvature in the absence of magnetic shear, $\mathcal{K}_y^{\hat{s}=0}$ is depicted by a light blue dashed line to illustrate the effect of global magnetic shear on the bad curvature regions. In devices with positive global magnetic shear (AUG and CBC) the regions of bad curvature become larger. In contrast, the regions of bad curvature are decreased in devices with negative global magnetic shear, which can be observed for NCSX. Moreover, the global magnetic shear can shift the regions of bad curvature away from the magnetic wells, for points along the field further away from $z=0$, which is observed in LHD. Additionally, in LHD, the magnitude of the bad curvature  increases significantly towards the flux tube ends due to the large global magnetic shear.
In TJ-II the  global magnetic shear is so small that the effect on the bad curvature is barely noticeable when a single poloidal turn is considered.  It is important to point out that the effect of the global magnetic shear on the bad curvature regions also depends on $\nablam x$ since $\nablam y \approx  r_0\nablam\theta_p - r_0 \iota_0 \nablam\zeta - \iota_0 \hat{s}_0 \zeta \nablam x$. Therefore, if the flux surface compression $(\nablam x)$ changes drastically along the field line, it can amplify or counteract the stabilizing or destabilizing effect of the global magnetic shear.

For CBC, \mbox{TJ-II} and NCSX, the location with the deepest magnetic well has the most pronounced bad curvature (figure \ref{fig:trappingcurvature}). On the other hand, in AUG and LHD these locations do not coincide. Here, the most pronounced bad curvature is found further along the field line, for $\zeta \neq 0$, because of the $\iota \hat{s} \zeta \boldsymbol{\nabla} \hspace{-0.2mm}x$ term inside $\mathcal{K}_y$. Finally, in W7-X, the central well of the chosen field line corresponds to the most pronounced bad curvature, while the deepest wells are found at \mbox{$z = \pm 1.41$ (and $z = \pm 2.51$)} where the curvature is comparable to the center.
Since bad curvature ($\mathcal{K}_y\!>\!0$) is destabilizing and good curvature ($\mathcal{K}_y\!<0$) is stabilizing, it is important to quantify the amount of bad curvature that occurs along the field line for ITGs, while the amount of bad curvature inside the magnetic wells is decisive for TEMs. It can be seen that the field lines of the tokamaks consist almost entirely of bad curvature, since  $\hat{s} = 0.796$ in CBC and $\hat{s} = 1.587$ in AUG, which increases the region of bad curvature through the $\iota \hat{s} \zeta \boldsymbol{\nabla} \hspace{-0.2mm}x$ term in $\mathcal{K}_y$, whereas the bad curvature region would be confined to the outboard side ($\theta \in [-90^\circ,90^\circ]$) if $\hat{s}=0$.
In contrast, only 30-50\% of the field lines in the stellarators are covered by bad curvature. In particular, in NCSX and \mbox{W7-X} the magnetic wells have a considerable amount of good curvature, while in TJ-II the majority of its large magnetic well has bad curvature. Finally, in LHD the two inner magnetic wells consists almost entirely of bad curvature, whereas for the outer magnetic wells, the regions of bad curvature have been shifted with respect to the regions of trapping because of the \mbox{large global magnetic shear ($\hat{s} = -1.230$).}

\newpage
In this analysis, regions of bad curvature are shown for Fourier modes with $k_x=0$, which most likely corresponds to modes that are excited near the center of the field line, since their ballooning angle is $\zeta = 0$. For modes that are excited further along the field line, a finite $k_x$ (or a non-zero ballooning angle) must be considered. Typically, field lines are centered around a point on the magnetic flux surface where instabilities are most strongly driven, that is, where the magnetic drift amplifies perturbations and where the most trapped particles are found, allowing both ITG and TEM instabilities to be strongly excited. In tokamaks, this location is generally found at the outboard midplane $(\theta=0)$. However, as explained in section \ref{sec:ITG}, very large magnetic drifts can have a stabilizing effect on ITG modes, as they weaken the wave-particle resonance. Therefore, it is important to include sufficient $k_x$ modes to ensure that the ballooning angle covers the entire region of bad curvature. On the other hand, in stellarators it is less straightforward to identify the location where microinstabilities will be strongest, as there are often multiple magnetic wells with varying strengths of the magnetic drift. For this reason, the center of the field line is typically chosen to correspond to a location where instabilities are expected to be driven strongly. To ensure microinstabilities can be excited in any of the magnetic wells or regions of bad curvature, it is essential to take into account sufficient $k_x$ modes to allow excitation at any point along the field line. In other words, the ballooning angle should cover the entire field line for all \mbox{relevant $k_y$ values.}

\section{Linear stability analysis}
\label{sec:linear}

Linear gyrokinetic simulations are performed for the six considered devices in order to discuss the microinstabilities that are driven unstable in the presence of various density and temperature gradients.
To start, in section \ref{sec:gammavskxky}, scans along $(k_x,k_y)$ are performed considering $a/L_{T_i}=3.0$, $a/L_{T_e}=0.0$ and various density gradients to confirm that the most unstable mode is generally found for $k_x=0.0$. Next, detailed scans for $0.0 \leq a/L_n \leq 4.0$ are conducted to study the evolution of the growth rate with the density gradient in section \ref{sec:linearlineallkys}, considering \mbox{$k_x=0.0$} and limiting the binormal wavenumber to ion scales, $k_y\rho_i\leq 2.0$. The most unstable mode is identified and discussed in section \ref{sec:mostunstable}, for which the contributions of the ion temperature gradient and the density gradient are isolated in section \ref{sec:linearlineITGTEM}. A vanishing electron temperature gradient is considered in the simulations to avoid the excitation of the ETG mode, which is isomorphic to the ITG mode, hence it is driven at electron scales, \mbox{$k_{y,\text{ETG}} \approx (\rho_i/\rho_e)\,k_{y,\text{ITG}}$}, and it has much larger growth rates, $\gamma_\text{ETG} \approx (\rho_i/\rho_e) \gamma_\text{ITG}$. Including electron scales is beyond the scope of this work. Nonetheless, the effect of $a/L_{T_e}=3.0$ on the ion-scale instabilities is discussed in \mbox{section \ref{sec:linearlineITGTEM}}.
Finally, in section \ref{sec:gammavsfprimtiprim} growth-rate maps are constructed for the six considered devices considering a wide range of gradients with $a/L_n \in [0.0, 8.0]$ and $a/L_{T_i} \in [0.0, 8.0]$.  It is important to note that this work is limited to the analysis of the fastest-growing mode as \texttt{stella} is an initial value solver. Nonetheless, it has been shown that subdominant modes can play a crucial role on the nonlinear dynamics of the turbulence \cite{faber2018stellarator, mckinney2019comparison}, which is beyond the scope of this work.

\begin{figure}[!b]
	\centering
	\vspace{-1mm}
	\begin{subfigure}{0.5\linewidth}
		\centering
		\includegraphics[trim={0mm 10mm 80mm 1mm},clip] {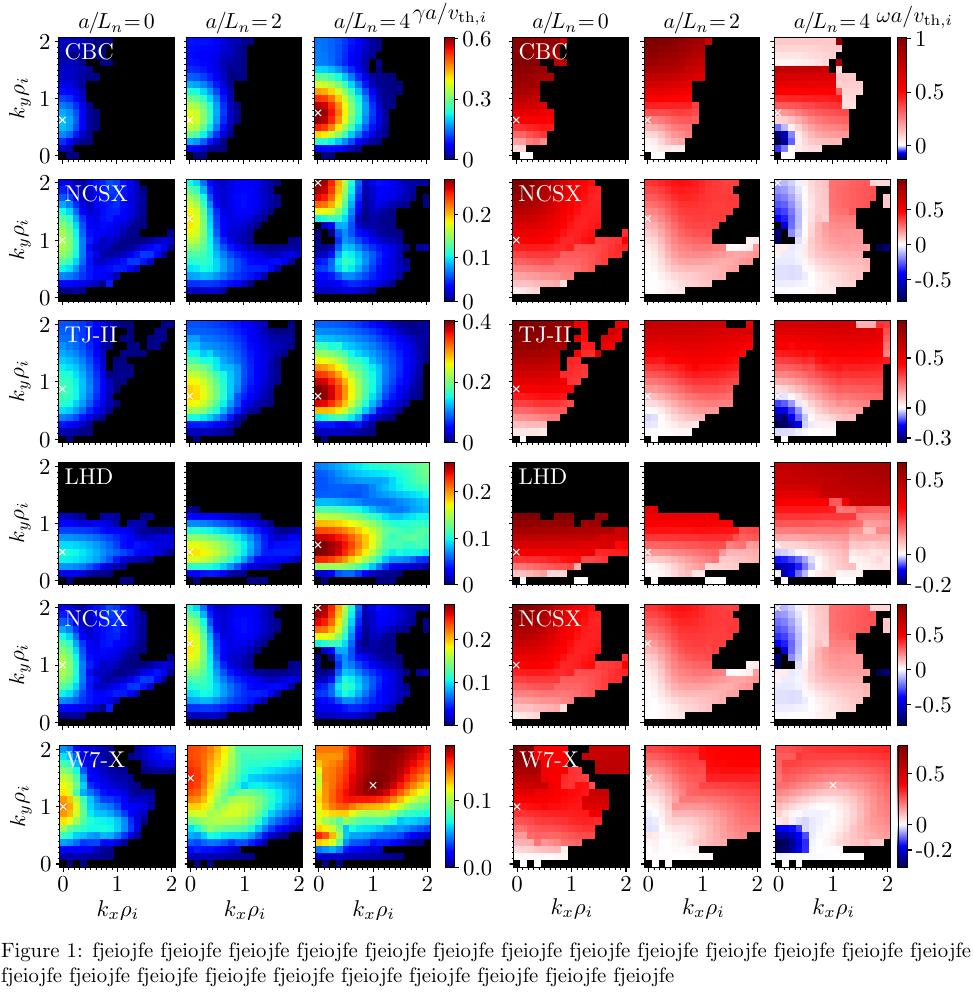}
		\caption{Normalized growth rate $\gamma a/v_{\rm th,i}$}
	\end{subfigure}%
	\begin{subfigure}{0.5\linewidth}
		\centering
		\includegraphics[trim={85mm 10mm 0mm 1mm},clip] {CHAP5_HEATTRANSPORT/Figures/Fig5.7_GammaOmegaVsKxKy.pdf}
		\caption{Normalized frequency $\omega a/v_{\rm th,i}$}
	\end{subfigure}%
		\caption{Normalized growth rate (left) and normalized frequency (right) as a function of $(k_x\rho_i, k_y\rho_i)$ for $a/L_{T_i}=3.0$, $a/L_{T_e}=0.0$ and $a/L_n=0.0$ (left columns), $a/L_n = 2.0$ (middle columns) and $a/L_n=4.0$ (right columns). The most unstable mode is highlighted with a white cross.  \vspace{0mm}}
	\label{fig:gamma_kxky}
\end{figure}

\subsection{Growth rate as a function of $k_x\rho_i$ and $k_y\rho_i$}
\label{sec:gammavskxky}

Generally, linear studies of gyrokinetic simulations are performed considering a vanishing radial wavenumber, i.e.~$k_x=0.0$. This simplification is justified when the most unstable mode peaks at a location along the flux tube where the flux-tube cross-section is perpendicular, or if it peaks symmetrically around such a location. Therefore, most gyrokinetic codes ensure that the center of the flux tube has a perpendicular cross-section, i.e.~$\boldsymbol{\nabla} x \cdot \boldsymbol{\nabla} y = 0$ at $z=0$. Moreover, the center is typically chosen to correspond to the location with the deepest magnetic well, and the most pronounced bad curvature, as discussed in section \ref{sec:localization}. With these choices, the most unstable mode most likely peaks at the center, where it would be characterized by having $k_x = 0.0$, whereas if it would be localized further along the flux tube, at a location where the cross-section has become twisted, it is typically characterized by $k_x \neq 0.0$, due to the FLR effects (see section \ref{sec:FLR}).
In order to verify whether the most unstable modes have $k_x=0.0$, the normalized growth rates $\gamma a/v_{\rm th,i}$ of the linear instabilities are shown in figure \ref{fig:gamma_kxky}a as a function of $(k_x, k_y)$ for $a/L_{T_i}=3.0$, $a/L_{T_e}=0.0$ and considering a density gradient of $a/L_n=0.0$ (left column), $a/L_n=2.0$ (middle column) and $a/L_n=4.0$ (right column). The corresponding normalized frequencies are shown in \mbox{figure \ref{fig:gamma_kxky}b}. The perpendicular wavenumbers are restricted to ion scales, specifically, $0 \leq k_x\rho_i \leq 2.0$ and $0 \leq k_y\rho_i \leq 2.0$, where we limit the scans to positive $k_x$-modes since the field lines are stellarator-symmetric, hence the $k_x$-modes and the $-k_x$-modes have the same growth rates and frequencies. Moreover, only positive $k_y$-modes are considered since negative $k_y$-modes can be recovered using the reality condition, i.e.~$\hat{\varphi}(k_x,k_y,z) = \hat{\varphi}^*(-k_x,-k_y,z)$, with $\hat{\varphi}^*_\mathbf{k}$ the complex conjugate of $\hat{\varphi}_\mathbf{k}$. The length of the flux tube is restricted to one poloidal turn \mbox{in this section. }

\begin{figure}[h!]
	\centering
	\vspace{2mm}
	\includegraphics[clip,trim={0mm 11mm 0mm 3mm}] {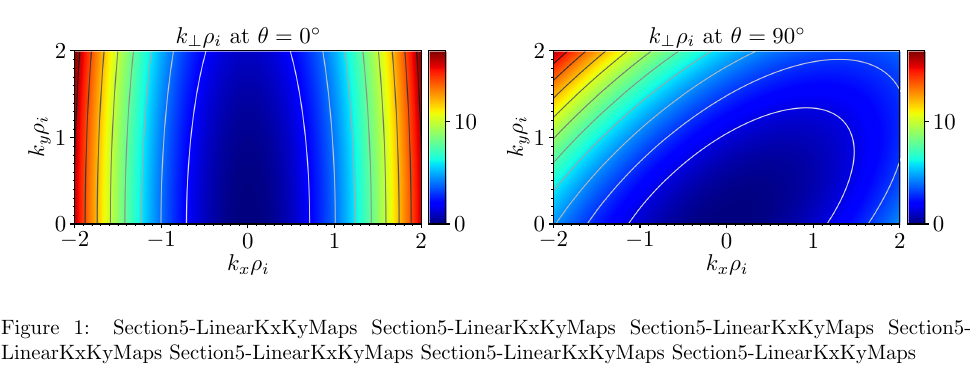}
	\vspace{-11mm}
	\caption{Magnitude of the perpendicular wavenumber in W7-X 	for the $\alpha_0=0$ field line centered at $(\theta, \zeta)=(0,0)$ at the parallel location $\theta=0^\circ$ (left) and $\theta=90^\circ$ (right).} \vspace{2mm}
	\label{fig:kperpcontours}
\end{figure}

For the three considered density gradients, the most unstable mode, which is highlighted by a white cross in figure \ref{fig:gamma_kxky}, is indeed found at $k_x=0.0$ in CBC, AUG, NCSX, LHD and TJ-II.
For \mbox{W7-X}, when considering low density gradients, the most unstable mode is also found at $k_x\rho_i = 0.0$. However, when the density gradient is increased to $a/L_n=4.0$, the most unstable mode is instead found at $k_x\rho_i = 1.0$ and $k_y\rho_i = 1.375$.  Specifically, for $a/L_n=4.0$, the growth rate at \mbox{$k_x\rho_i=0.0$} and $k_y\rho_i=0.5$ is 13\% smaller than at $k_x\rho_i = 1.0$ and $k_y\rho_i=1.375$. As explained in \mbox{section \ref{sec:localization}}, the most unstable mode is most likely excited at the center for W7-X, but it can also be excited in the magnetic wells at $z = \pm 1.41$ (figure \ref{fig:trappingcurvature}) due to the depth of the magnetic well and the strength of the bad curvature at this location. At $z = \pm 1.41$ the cross-section has become sheared and the most unstable mode corresponds to the $(k_x,k_y)$-mode for which the FLR damping is the smallest, which is inside the inner contour shown on the right-hand-side of figure \ref{fig:kperpcontours}.

The normalized frequencies $\omega a/v_{\rm th,i}$ are shown in figure \ref{fig:gamma_kxky}b. From this point onward, we use $\omega$ to denote the oscillation frequency, i.e.~the real part of the complex frequency. Negative frequencies correspond to modes propagating in the electron diamagnetic direction, while those with positive frequencies propagate in the ion diamagnetic direction. For flat density profiles (left column), all considered $(k_x, k_y)$ modes exhibit positive frequencies, indicating that instabilities driven by the ion temperature gradient propagate in the ion diamagnetic direction. When the density gradient is increased to $a/L_n = 2.0$ (middle column), slightly negative frequencies emerge around $k_x \rho_i = 0.0$ in both TJ-II and W7-X. At a stronger density gradient of $a/L_n = 4.0$ (right column), modes with small perpendicular scales (i.e.~small $k_\perp \rho_i$) exhibit negative frequencies (or $\omega \approx 0$) across all devices.  Nevertheless, most $(k_x,k_y)$ modes continue to be characterized by positive frequencies. It is challenging to determine the specific nature of the underlying instabilities based on the frequency alone, as various modes---including TEMs, hybrid TEM-ITG modes, and passing-particle-driven universal instabilities---can propagate in both diamagnetic directions  \cite{kammerer2008exceptional, helander2013collisionless, proll2014trapped,  plunk2017collisionless, helander2015universal, costello2023universal, proll2013collisionless}.

\begin{figure}[h!]
	\centering
	\vspace{3mm}
	\includegraphics[clip,trim={0mm 15mm 0mm 0mm}] {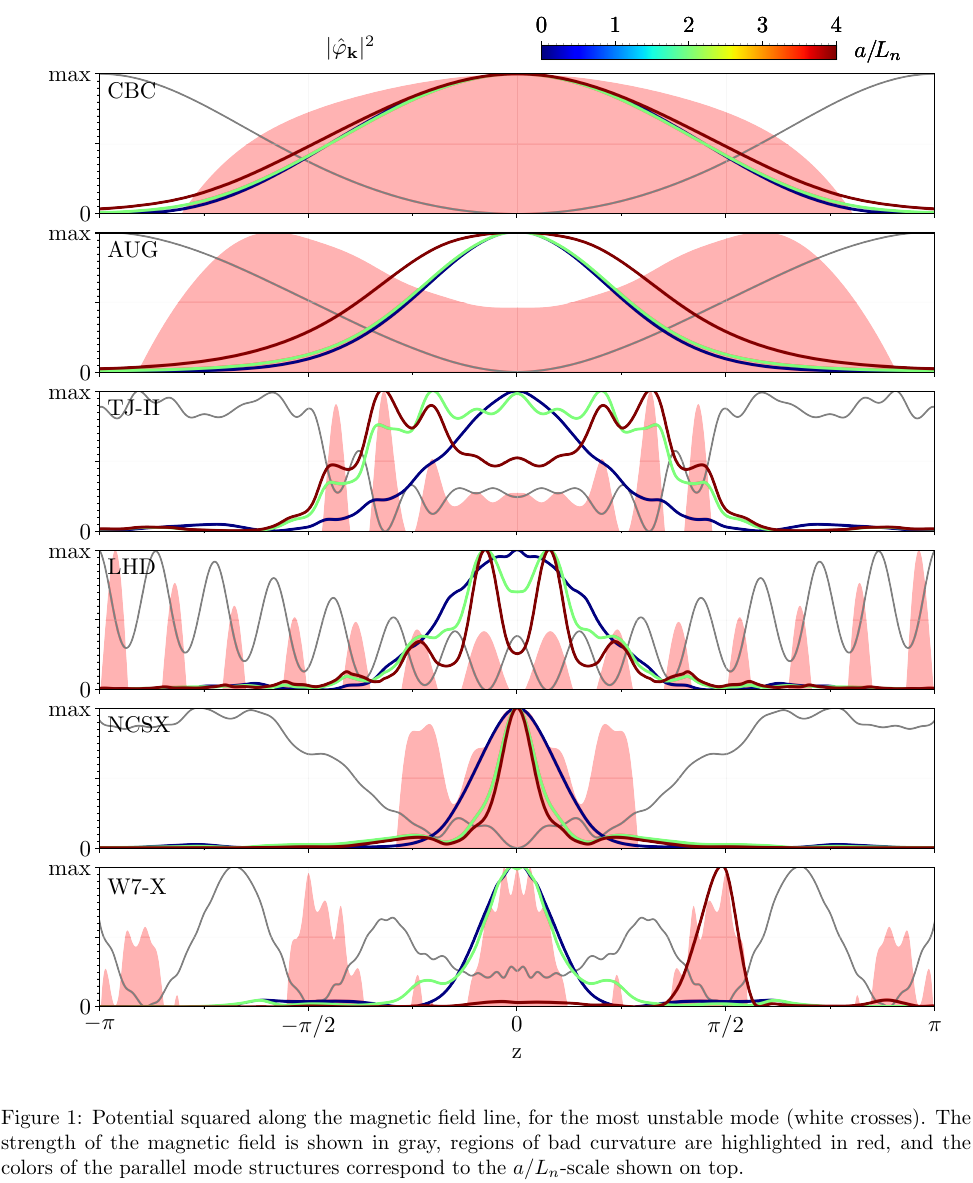} \vspace{-12mm}
	\caption{Potential squared along the magnetic field line, for the most unstable mode (white crosses) of figure \ref{fig:gamma_kxky}. The strength of the magnetic field is shown in gray, regions of bad curvature are highlighted in red, and the colors of the parallel mode structures correspond to the $a/L_n$-scale shown on top. \vspace{-2mm}}
	\label{fig:linear_phi_vs_z}
\end{figure}

Nonetheless, certain features of the driving mechanism can be inferred by examining the electrostatic potential of the mode along the magnetic field line. This is illustrated in figure \ref{fig:linear_phi_vs_z} by plotting the parallel mode structure of the most unstable mode, corresponding to the white crosses in figures \ref{fig:gamma_kxky}. First of all, from the parallel mode structure it can be inferred whether the most unstable mode is characterized by $k_x=0.0$ or not.
In CBC, AUG and NCSX the most unstable mode peaks at the center of the flux tube where the cross-section is perpendicular, hence it has $k_x=0.0$. On the other hand, in TJ-II and LHD, for $a/L_n = \{2.0,4.0\}$, the most unstable mode peaks symmetrically around the center of the flux tube, in the two magnetic wells at $z = \pm 0.23$ for LHD, and in the magnetic wells at $z = \pm 0.54$ and $z = \pm 0.83$ for TJ-II, also corresponding to $k_x=0.0$. Finally, in W7-X, the most unstable mode has $k_x\rho_i \neq 0.0$ for $a/L_n=4.0$, for which the mode peaks in the well at $z =  1.41$. On the other hand, it is clear that the electrostatic potential of the modes tends to peak in regions of bad curvature. Additionally, as the density gradient increases, the potential becomes more pronounced within the magnetic wells, as trapped electrons start playing a more significant role in driving the instability. Consequently, at higher density gradients, specific details of the structure of the magnetic field strength along the field line can be observed in the parallel mode structure of the modes.

\newpage
To conclude, if the cross-section at the center of the flux tube is chosen to be perpendicular and if it corresponds to the location with the strongest instability drives, then the most unstable mode is generally found for $k_x\rho_i=0.0$. However, in some cases, the most unstable mode is characterized by $k_x\rho_i \neq 0.0$, because the mode peaks further along the flux tube, at a location where the cross-section has become sheared. It is important to note that in nonlinear simulations, all the fastest-growing $(k_x,k_y)$-modes shown in \mbox{figure \ref{fig:gamma_kxky}}, as well as subdominant modes (which are not computed in this work), are coupled and will influence the nonlinear saturation \mbox{of the turbulence \cite{hatch2011role}.}

\subsection{Growth rate as a function of $a/L_n$ and $k_y\rho_i$}
\label{sec:linearlineallkys}

\begin{figure}[!b]
	\centering
	\includegraphics[clip,trim={0mm 18mm 0mm 0mm}] {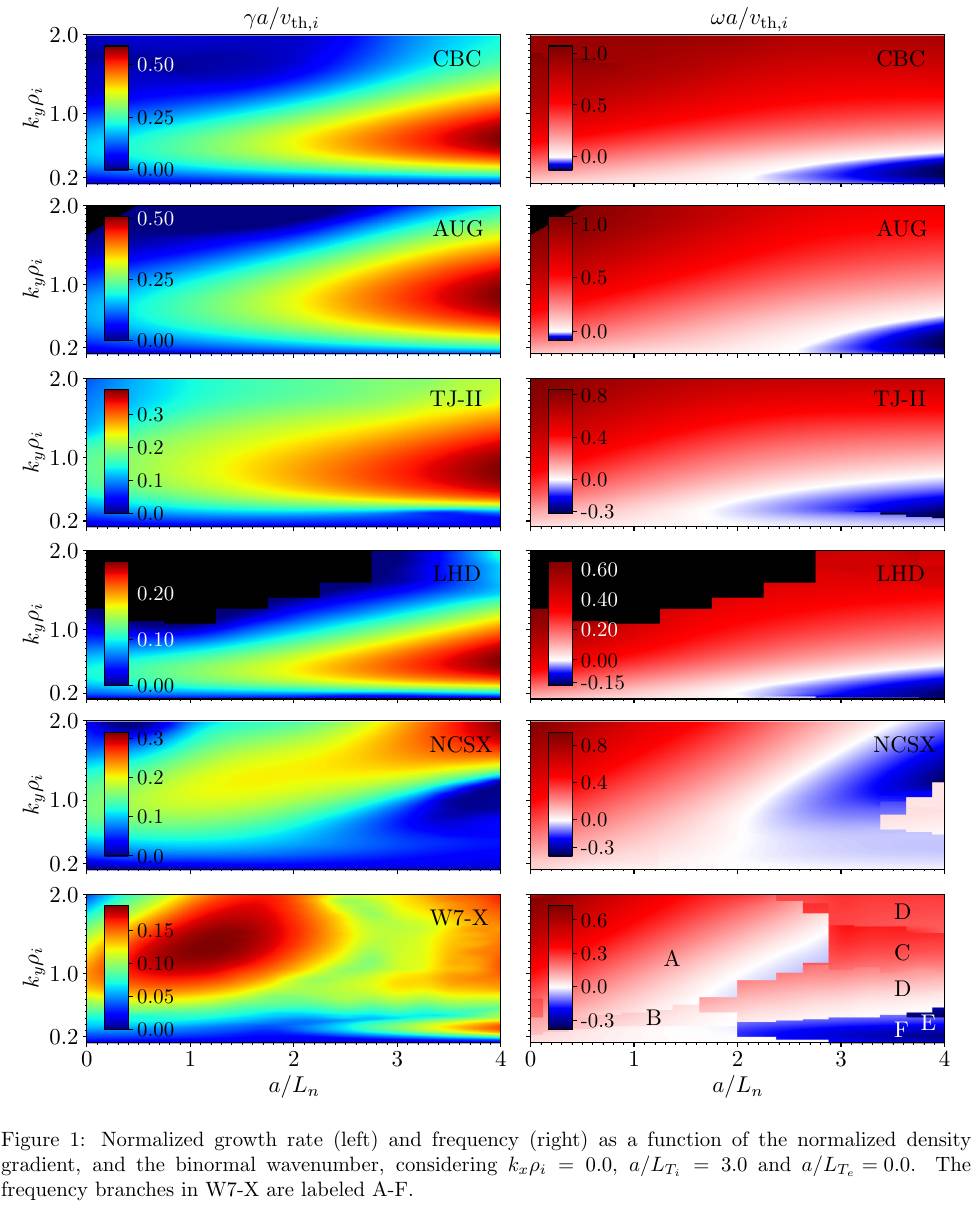}
	\vspace{-6mm}
	\caption{Normalized growth rate (left) and frequency (right) as a function of the normalized density gradient and the binormal wavenumber, considering $k_x\rho_i=0.0$,  $a/L_{T_i}=3.0$ and \mbox{$a/L_{T_e}=0.0$}. The frequency branches in W7-X are labeled A-F. }
	\label{fig:gamma_map_line}
\end{figure}

The linear analysis is continued considering \mbox{$k_x\rho_i=0.0$}, while scanning $k_y\rho_i$ in $[0.125, 2.0]$ to investigate the perpendicular scale of the modes. We simultaneously scan $a/L_n$ in $[0.0,4.0]$ to study the evolution of the instabilities as the density gradient is increased. An ion temperature gradient of \mbox{$a/L_{T_i}=3.0$} and a vanishing electron temperature gradient ($a/L_{T_e}=0.0$) are considered. The flux tube is extended up to 3 poloidal turns in order to capture extended modes. The growth rates and frequencies of the modes are shown in figure \ref{fig:gamma_map_line}.

In CBC, AUG, TJ-II and LHD, the linear instabilities become increasingly more unstable with increasing density gradients. Considering the range of scanned density gradients, it is expected that distinct analytical instabilities are present in these devices. Specifically, ITG modes should be dominant at \mbox{$a/L_n=0.0$} and density-gradient-driven instabilities such as TEMs, iTEMs and universal instabilities should be present at $a/L_n=4.0$, where it is likely that they are mixed with the ITG mode due to the presence of the strong ion temperature gradient of $a/L_{T_i} = 3.0$. It can be seen that CBC, AUG, \mbox{TJ-II} and LHD have continuous frequency maps, except for very small $k_y\rho_i$ in LHD and TJ-II. This continuous change of the frequency is accompanied by a continuous change of the parallel mode structure of these modes, which is discussed in detail for the most unstable mode along $k_y\rho_i$ in the next section (figure \ref{fig:linear_phi_vs_z2}). These continuous frequency branches indicate that the modes are transforming into one another through a continuous change of the density \mbox{gradient \cite{kammerer2008exceptional}}. As a result, one should refrain from labeling modes traveling in the ion diamagnetic direction as ITG modes, and those traveling in the electron diamagnetic direction as TEM modes, since in the presence of both $a/L_n$ and $a/L_{T_i}$ the modes are most \mbox{likely mixed to a certain degree.}

In contrast, the growth rate and frequency spectra of NCSX and W7-X are very different. Specifically, in NCSX, the growth rate also increases with increasing density gradients, however, the most unstable mode is found for \mbox{larger $k_y\rho_i$} at $a/L_n=4$, in contrast to the other five devices. According to mixing length arguments, where \mbox{$Q_i \propto \gamma/k_y^2$}, one could expect the heat transport to be reduced as the density gradient increases  because the perpendicular scale of the modes becomes smaller,  although it has been demonstrated that reliable quasi-linear models must account for all subdominant unstable modes  \cite{pueschel2016stellarator}.  Moreover, note that a second instability with positive frequencies shows up at large density gradients.
On the other hand, the growth rate spectrum of W7-X consists of multiple different peaks, which correspond to six distinct frequency branches, characterized by having different parallel mode structures (shown for the most unstable mode along $k_y\rho_i$ in figure \ref{fig:linear_phi_vs_z2}). For the highest density gradient, $a/L_n=4.0$, four distinct instability branches co-exist within  \mbox{$k_y\rho_i \leq 2.0$}.  The different branches have been labeled (A) to (F).  In W7-X, the growth rate thus initially increases with the density gradient until $a/L_n = 1.25$, after which the growth rate decreases up to $a/L_n = 2.75$, due to the stabilization of instability (A) by increasing density gradients---which is most likely an ITG mode---without driving the other instabilities particularly unstable yet. As the density gradient is increased further, the instabilities corresponding to frequency branches (C) to (F) appear, which are driven unstable by the density gradient. Note that instability (F) has large perpendicular wavelengths, hence it could be \mbox{particularly detrimental to the heat transport. }

In summary, CBC, AUG, TJ-II and LHD have  continuous frequency spectra, where the modes undergo a smooth transition from one instability into another as the density gradient is increased, whereas NCSX and W7-X are characterized by abrupt transitions between different frequency branches.
Hence, it is clear that in most cases, a frequency branch does not necessarily correspond to a distinct analytical instability such as the ITG or TEM, and the identification of the instabilities should be done very carefully or be avoided altogether.

\subsection{Most unstable mode as a function of $a/L_n$}
\label{sec:mostunstable}

\begin{figure}[h!]
	\centering
	\vspace{2mm}
	\includegraphics[trim={0mm 0mm 0mm 6mm}, clip] {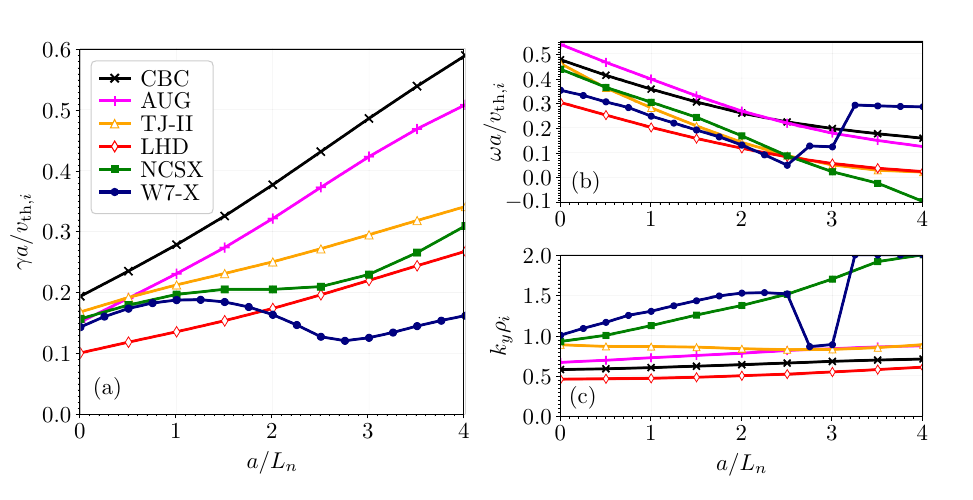} \vspace{-10mm}
	\caption{The (a) normalized growth rate, (b) normalized frequency, and (c) normalized binormal wavenumber of the most unstable mode as a function of the normalized density gradient, $a/L_n$, considering $k_x = 0.0$, $a/L_{T_i}=3.0$ and  $a/L_{T_e}=0.0$. }
	\vspace{2mm}
	\label{fig:gamma_vs_aLn}
\end{figure}

Next, the growth rate, frequency and wavenumber of the most unstable modes of figure \ref{fig:gamma_map_line} are plotted as a function of the density gradient in figure \ref{fig:gamma_vs_aLn}, which is a linear analysis commonly performed in gyrokinetic studies. Recall that \mbox{we consider  $a/L_{T_i}=3.0$}, \mbox{$a/L_{T_e} = 0.0$, $k_x\rho_i=0.0$} and \mbox{$k_y\rho_i \leq 2.0$}.
For CBC, AUG, TJ-II and LHD, the growth rate of the most unstable mode increases approximately linearly with the density gradient (figure \ref{fig:gamma_vs_aLn}a), while there is no significant change in the perpendicular scales of the modes (figure \ref{fig:gamma_vs_aLn}c).
In NCSX, the growth rate also increases as the density gradient is increased, however, it has a weaker dependence \mbox{on $a/L_n$}. Additionally, the perpendicular scales of the modes become smaller.  In contrast, in W7-X, the growth rates are reduced by 36\% from $a/L_n=1.25$ to $a/L_n=2.75$. A decrease in the perpendicular scales of the modes is also observed in W7-X  in figure \ref{fig:gamma_vs_aLn}c, however, this is misleading since \mbox{figure \ref{fig:gamma_map_line}} indicates that multiple instabilities co-exist at high density gradients, some of which are characterized by having large perpendicular scales which can be particularly detrimental to heat transport. Therefore, one has to be careful when interpreting linear results based only on the most unstable $(k_x,k_y)$-mode.
Nonetheless, the linear analysis suggest that a reduction of the heat transport with increasing density gradients is expected for NCSX and W7-X based on mixing length arguments, i.e.~$Q_i \propto \gamma/k_y^2$, while an increase is expected for CBC, AUG, TJ-II and LHD.

\begin{figure}[b!]
	\centering
	\vspace{0mm}
	\includegraphics[clip,trim={0mm 14mm 0mm 0mm}] {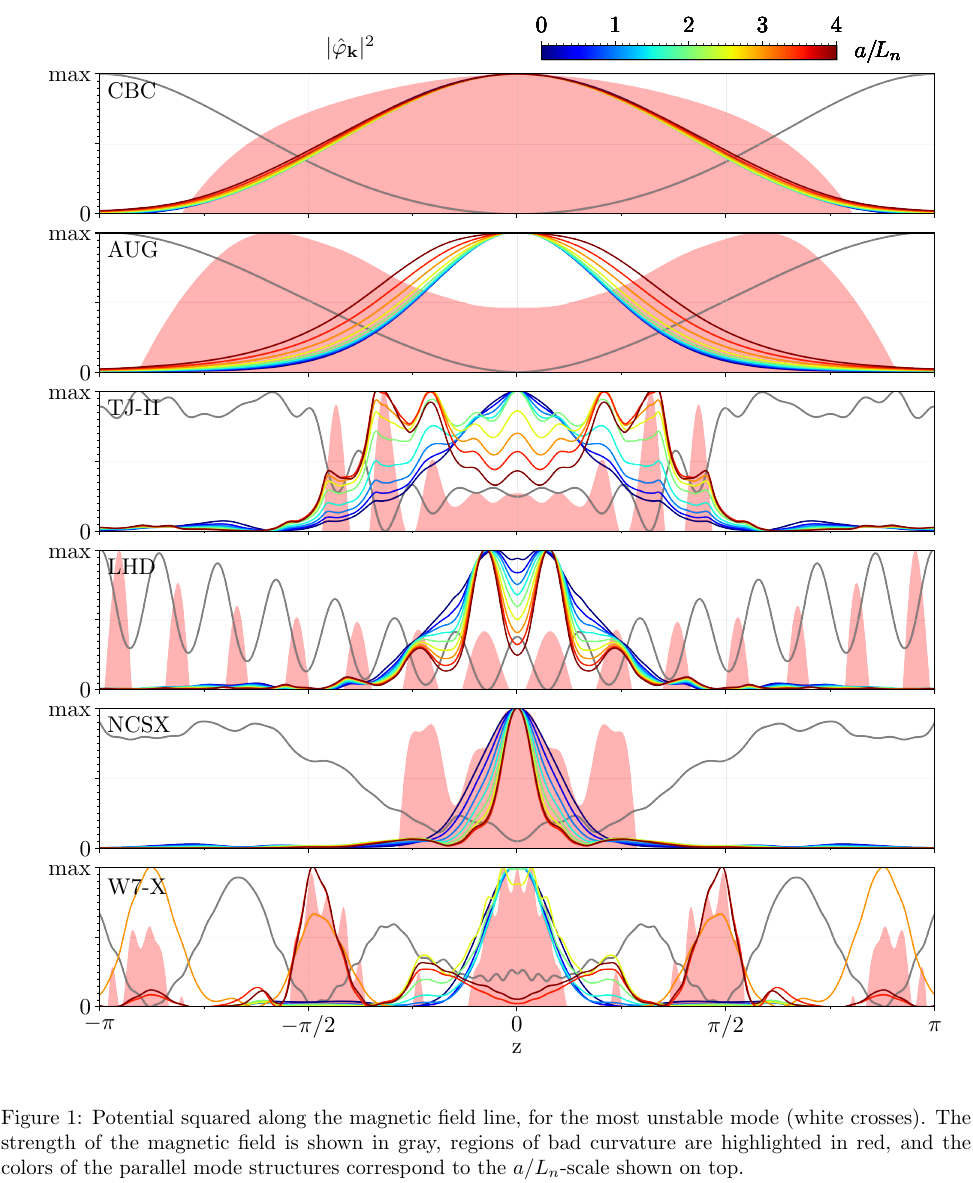}
	\vspace{-13mm}
	\caption{Potential squared along the magnetic field line, for the modes in figure \ref{fig:gamma_vs_aLn}. The strength of the magnetic field is shown in gray, regions of bad curvature are highlighted in red, and the colors of the parallel mode structures correspond to the $a/L_n$-scale shown on top.}
	\vspace{4mm}
	\label{fig:linear_phi_vs_z2}
\end{figure}

In order to explain the differences in the linear behavior between these devices, $|\hat{\varphi}_\mathbf{k}|^2$ is plotted along the field line in figure \ref{fig:linear_phi_vs_z2}, for the modes from \mbox{figure \ref{fig:gamma_vs_aLn}}. These parallel mode structures give insight into the geometric characteristics that drive the instability. In every device, except for LHD, the modes peak at the center of the field line for small density gradients, where the ITG mode is believed to be dominant, which is driven unstable in bad curvature regions \mbox{(figure \ref{fig:fieldlineCurvature})}. This is mainly due to the FLR damping, which scales with $k_\perp\rho_i$ for the ions, and has a minimum at the center of the flux tube for \mbox{$k_x=0.0$} \mbox{(figure \ref{fig:fieldlineAngle}d)}. On the other hand, as the density gradient is increased, the modes become localized inside the deepest magnetic wells, since TEMs start contributing significantly.
Specifically, CBC and AUG have a single magnetic well consisting of mostly bad curvature, hence the modes have the same parallel mode structure regardless of the density gradient. On the other hand, in LHD, the modes become more localized inside the magnetic wells as the density gradient is increased since trapped electrons start driving the instabilities.
Note that the modes are more localized (the mode structures are more narrow) in AUG ($\hat{s}=1.6$) and LHD \mbox{($\hat{s}=1.2$)} compared to CBC ($\hat{s}=0.8$), due to the higher global shear and thus stronger FLR damping towards the ends of the flux tube. NCSX shows improved confinement of the modes, despite the moderate global shear ($\hat{s}=0.7$) which is likely due to the very strong local magnetic shear \mbox{(figure \ref{fig:fieldlineAngle}b)}. This limits the modes to a region which is even smaller than the already quite narrow region of bad curvature. In contrast, the mode structures in \mbox{TJ-II} are quite broad, since the shear is low and there is a large central well which consists almost entirely of bad curvature.
\mbox{Finally, in W7-X}, distinct mode structures occur, corresponding to distinct frequency branches (figure \ref{fig:gamma_vs_aLn}). For low density gradients, the modes peak at the center where  the FLR damping is the weakest and the bad curvature is the largest \mbox{(figures \ref{fig:fieldlineAngle}d and \ref{fig:fieldlineCurvature})}. Whereas for $a/L_n=3.0$ the mode peaks predominately inside the wells at $z= \pm 2.51$ while a smaller contribution comes from the wells $z= \pm 1.41$. Finally, for $a/L_n \geq 3.5$ the modes peak in the wells at $z= \pm 1.41$.

The magnetic geometry is correlated to the magnitude of the growth rates of the instabilities. Specifically, in the devices where the mode structures are confined to narrow regions of the field line, such as in LHD, NCSX and W7-X, the growth rates are the smallest (figure \ref{fig:gamma_vs_aLn}). Additionally, the high local magnetic shear in NCSX, and the  intermittent regions of bad and good curvature in the central well of W7-X have a stabilizing effect.  TJ-II has the largest growth rates of the stellarators due to its low shear and broad magnetic well, which is covered almost entirely by bad curvature. Finally, the tokamaks have the highest growth rates since most of the field line is driving the modes unstable, where AUG performs slightly better, most likely \mbox{due to its higher shear.}

\subsection{Separate contributions to the growth rate from the ion temperature gradient and the density gradient}
\label{sec:linearlineITGTEM}

\begin{figure}[b!]
	\centering
	\vspace{-3mm}
	\includegraphics[trim={0mm 0mm 0mm 4mm}] {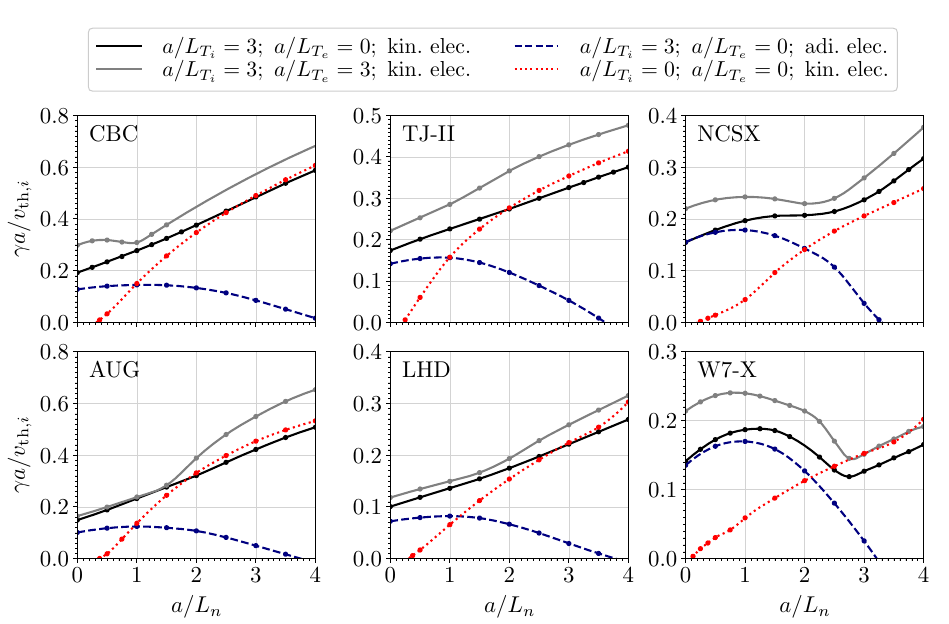}
	\vspace{-2mm}
	\caption{Normalized growth rate as a function of the normalized density gradient with kinetic electrons, $a/L_{T_i}=3$ and a vanishing electron temperature gradient (black solid lines),  and with $a/L_{T_e}=3$ (gray solid lines). Ion-temperature-gradient-driven modes are isolated by treating the electrons as adiabatic (blue dashed lines) and density-gradient-driven modes are isolated by setting $a/L_{T_i}=0$ (red dotted lines).}
	\vspace{-5mm}
	\label{fig:linear_teprim}
\end{figure}

In most devices, we are unable to distinguish between different instability types, since the transitions between them are smooth (section \ref{sec:linearlineallkys}). Nonetheless, it is possible to isolate specific instabilities based on their instability drive by modifying the simulations accordingly, which is shown in figure \ref{fig:linear_teprim}. The original simulations considered in figure \ref{fig:gamma_vs_aLn} are reproduced with black lines. Firstly, we consider the case of adiabatic electrons to isolate the instabilities which are driven by the ion temperature gradient and which rely solely on the ion dynamics (blue dashed lines).  Next, we impose vanishing temperature gradients while treating the electrons kinetically, to isolate modes which are driven by the density gradient and require electron dynamics (red dotted lines).

The instabilities driven by the ions (blue dashed lines) show an initial increase in their growth rates with the density gradient until \mbox{$a/L_n = 1.0$}, after which the growth rates decrease, and eventually the instability is suppressed by the density gradient. The plasma parameters and behavior of these modes suggest that this instability corresponds to the toroidal ITG \mbox{mode \cite{plunk2014collisionless}}.  The density gradient for which these modes are suppressed is the highest in CBC, AUG, TJ-II and LHD, and the lowest in NCSX and \mbox{W7-X}.  Moreover, note that for $a/L_n=0.0$, including the electron dynamics \mbox{(blue lines $\rightarrow$ black lines)} has a destabilizing effect in every device, except in NCSX and W7-X, where it has no effect on the growth rate.

The red dotted line in figure \ref{fig:linear_teprim} represents the contribution of the density-gradient-driven modes, which are most likely classical electron-driven trapped-electron modes (TEM) \cite{helander2013collisionless} but they could also be ion-driven trapped-electron modes (iTEM) \cite{plunk2017collisionless} or passing-particle-driven universal instabilities \cite{costello2023universal, podavini2023electrostatic}, however these instabilities are typically subdominant to TEMs. A minimum density gradient is required in order to drive these modes unstable, after which the growth rate increases with the density gradient. Note that for high density gradients in W7-X, TJ-II and AUG, the linear growth rates are reduced by the presence of an ion temperature gradient (red lines $\rightarrow$ black lines), in contrast to NCSX, where the ion temperature gradient has a destabilizing effect.

In CBC, AUG, TJ-II and LHD the growth rates of the modes in the presence of both the ion temperature gradient and kinetic electrons (black lines) increase monotonously with the density gradient. In contrast, in \mbox{W7-X}, and to lesser extent in NCSX, distinct instabilities are observed in the frequency spectra (figure \ref{fig:gamma_map_line}), and the growth rates of the original simulations (black lines) resembles the sum of the growth rates of the ion-temperature-gradient-driven modes (blue lines) and the density-gradient-driven modes \mbox{(red lines)}.
The presence of a clear separation between these two contributions seems to be related to the density gradient above which the growth rates driven solely by the density-gradient-driven modes exceeds those driven by the ion-temperature-gradient-driven modes, i.e.~where the red line crosses the blue line. This occurs at around $a/L_n \approx 1.0$ for CBC, AUG,  TJ-II and LHD, whereas it occurs around around $a/L_n \approx 2.0$ for NCSX and W7-X. Hence, in NCSX and W7-X, there is a clear separation between these two instabilities, and there is a region of density gradients for which the  ion-temperature-gradient-driven modes are stabilized by the density gradient, without driving the density-gradient-driven modes \mbox{particularly unstable yet.}

Finally, the effect of adding a non-vanishing electron temperature gradient, set to \mbox{$a/L_{T_e} = a/L_{T_i}$} with $a/L_{T_i} = 3.0$, is represented by the gray lines in figure \ref{fig:linear_teprim}. The presence of an electron temperature gradient increases the growth rates in every device, hence it has a destabilizing effect on the ion-scale instabilities. Note that in CBC the ETG mode is dominant for $a/L_n \leq 0.75$, since the most unstable mode  is found at the end of the scanned wavenumber range, at $k_y\rho_i = 2.0$, and it propagates in the electron diamagnetic direction (not shown here). Increasing the scanned $k_y\rho_i$-values shows that this mode belongs to an ETG branch which peaks at electron scales.

\begin{figure}[!b]
	\centering
	\vspace{-3mm}
	\includegraphics[trim={0mm 10mm 0mm 2mm},clip] {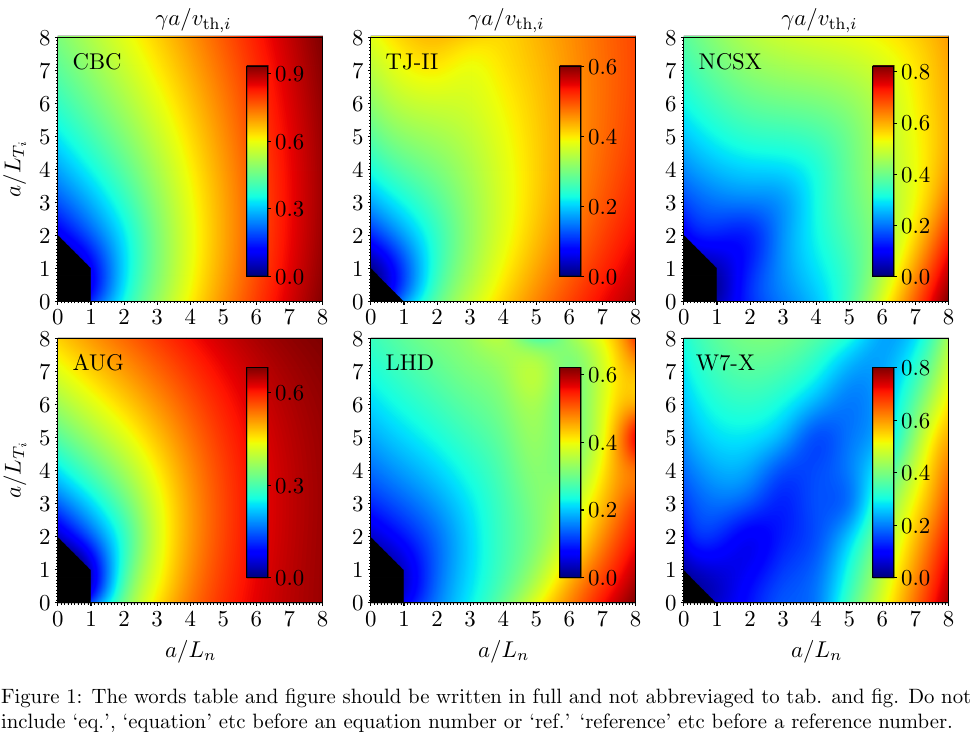} \vspace{-9mm}
	\caption{\mbox{Normalized growth rate of the most unstable} mode for $k_x\rho_i=0.0$ and $k_y\rho_i \leq 2.0$, as a function of the \mbox{normalized density gradient} and the normalized ion temperature gradient, considering $a/L_{T_e}=0.0$. \vspace{-5mm} }
	\label{fig:gamma_maps}
\end{figure}

\subsection{Growth rate as a function of $a/L_n$ and $a/L_{T_i}$}
\label{sec:gammavsfprimtiprim}
\vspace{-1mm}

\begin{figure}[b!]
	\centering
	\vspace{-4mm}
	\includegraphics[trim={0mm 10mm 0mm 2mm}, clip] {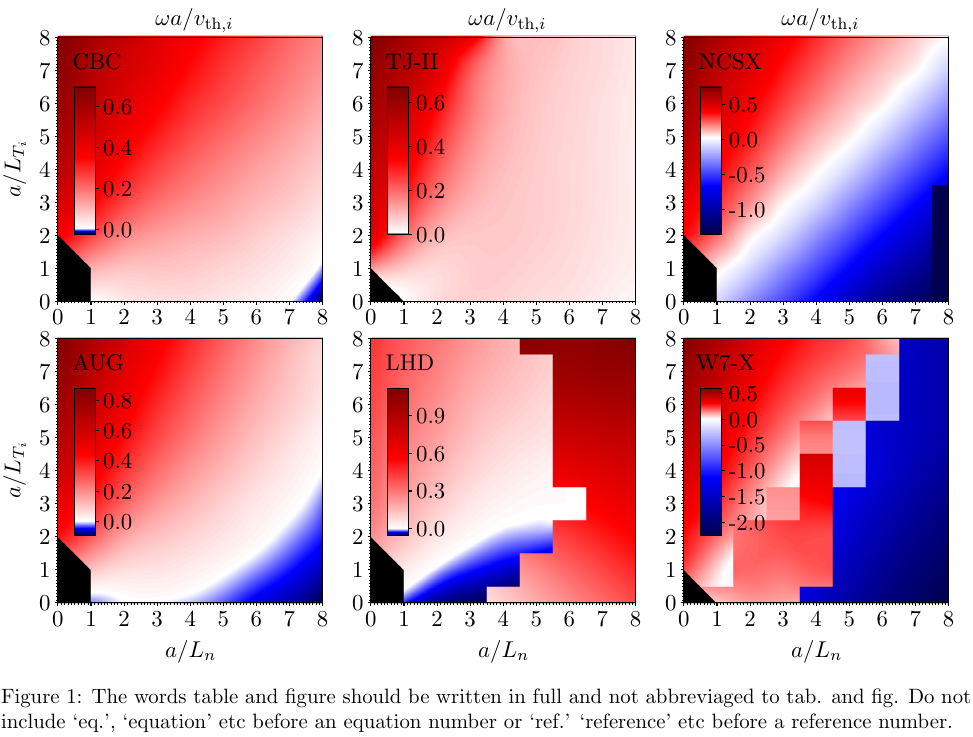} \vspace{-9mm}
	\caption{Normalized frequency of the most unstable mode for $k_x\rho_i=0.0$ and $k_y\rho_i \leq 2.0$, as a function of the normalized density gradient and the normalized ion temperature gradient, considering $a/L_{T_e}=0.0$. \vspace{-5mm}}
	\label{fig:omega_maps}
\end{figure}

\begin{figure*}[!b]
	\centering
	\vspace{-2mm}
	\includegraphics[trim={0mm 0mm 0mm -3mm}] {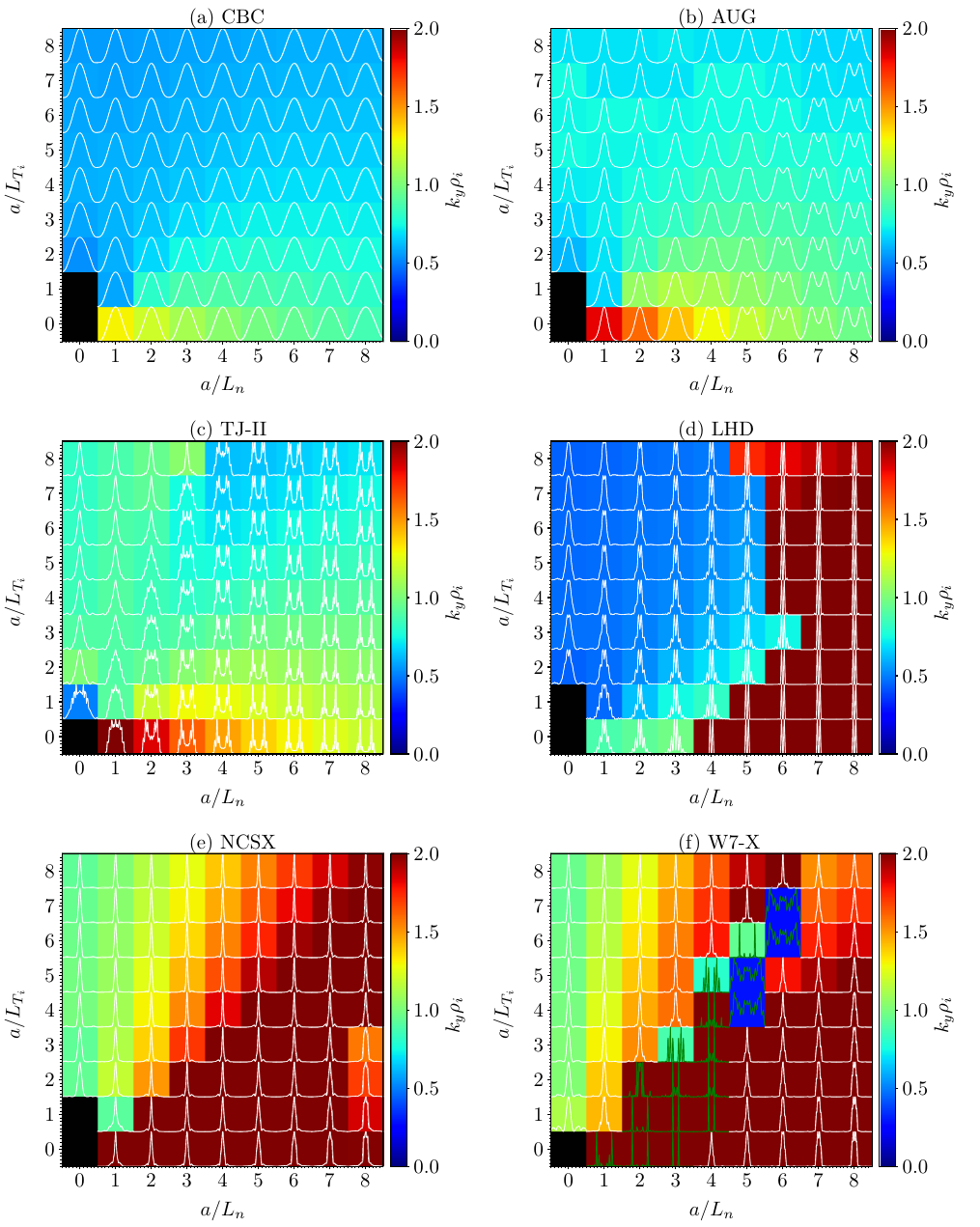}
	\vspace{-7mm}
	\caption{Binormal wavenumber, $k_y\rho_i$,  of the most unstable mode for $k_x\rho_i=0.0$ and $k_y\rho_i \leq 2.0$, as a function of the normalized density gradient and the normalized ion temperature gradient, considering a vanishing electron temperature gradient. On top of each $k_y\rho_i(a/L_n, a/L_{T_i})$ tile, the parallel mode structure of the most unstable mode is shown, by plotting the potential squared along the field line, plotted in white if it is confined to the first poloidal turn, or in green if the mode extends up to three poloidal turns. \vspace{-2mm}}
	\label{fig:ky_maps}
\end{figure*}

The linear analysis is extended to a wide range of plasma parameters, varying the gradients from flat density and ion temperature profiles ($a/L_n=0.0$ and $a/L_{T_i}=0.0$) up to extremely peaked profiles ($a/L_n=8.0$ and $a/L_{T_i}=8.0$), scanned with a step of $ a/L_n=1.0$ and $a/L_{T_i}=1.0$, while considering a vanishing electron temperature gradient \mbox{($a/L_{T_e}=0.0$)}. The growth rates, frequencies and binormal wavenumber of the most unstable modes, found for $k_x\rho_i=0.0$ and $k_y\rho_i \leq 2.0$, are shown in figures \ref{fig:gamma_maps}, \ref{fig:omega_maps} and \ref{fig:ky_maps}, respectively. These maps contribute to the collection of growth-rate maps for various configurations presented in Ref.~\cite{alcuson2020suppression}. Note that the black regions indicate that no unstable modes were found within the considered CPU time. Resolving these modes, which lie near marginality, would require a more extensive study.  Moreover, it is important to acknowledge that for most devices, the gradients that are considered in this section are far beyond the experimentally relevant values at \mbox{$r/a=0.7$}. Therefore, one needs to be careful when drawing conclusions based on such extreme scans.

It is clear that \mbox{W7-X} is the only device which exhibits a prominent \mbox{valley \cite{alcuson2020suppression}}, i.e.~a region along the diagonal \mbox{$a/L_n \approx a/L_{T_i}$} which is characterized by having significantly reduced growth rates \mbox{(figure \ref{fig:gamma_maps})}. However, note that for the specific choice of \mbox{$a/L_{T_i}=3.0$} and $a/L_n \leq 4.0$, the reduction of the growth rate in W7-X is very modest \mbox{(figure \ref{fig:gamma_vs_aLn})}. As explained in the previous section, this valley in W7-X is formed by the stabilization of the ion-temperature-gradient-driven modes by increasing density gradients, without driving the density-gradient-driven modes particularly unstable yet.
Similarly, a small reduction of the growth rate is observed along \mbox{$a/L_n \approx a/L_{T_i}$} in NCSX, while for the other four devices, no reduced growth rates are observed along the diagonal.

In figure \ref{fig:omega_maps}, the frequency of the most unstable mode is depicted. Each frequency branch---whose modes have a similar parallel mode structure and occur at a similar binormal wavenumber (figure \ref{fig:ky_maps})---has been interpolated separately. In CBC, AUG, \mbox{TJ-II}, and NCSX the most unstable mode changes continuously from a pure ion-temperature-gradient-driven mode (top left) to a pure density-gradient-driven mode (bottom right), as can be seen by the continuous change of the frequency throughout all scanned gradients---except for the jump to a new frequency branch at \mbox{$a/L_n=8.0$} for NCSX. On the other hand, LHD exhibits a similar frequency map as the other devices for low density gradients, while for high density gradients, the most unstable mode belongs to a different frequency branch, which is characterized by a different parallel mode structure (figure \ref{fig:ky_maps}).
In contrast, many distinct instabilities live inside the growth-rate valley of \mbox{W7-X}.

The binormal wavenumber of the most unstable mode is shown in figure \ref{fig:ky_maps}. In NCSX and  W7-X, the instabilities have smaller perpendicular scales (larger $k_y\rho_i$) as the density gradient is increased. In contrast, in CBC, AUG, TJ-II and LHD, the most unstable mode retains large perpendicular scales when increasing the density gradient. Note that for \mbox{W7-X}, NCSX and LHD, the most unstable modes in the bottom right corners, are found at the end of our scanned range of $k_y\rho_i$ values. Therefore, the growth rates and frequencies shown in figures  \ref{fig:gamma_maps} and \ref{fig:omega_maps} do not correspond to a peak in the $\gamma(k_y\rho_i)$ spectra. This is the reason why the frequency spectrum of NCSX has such negative frequencies in the bottom right corner of figure \ref{fig:omega_maps}. If one would increase the scanned $k_y\rho_i$ values, the most unstable mode would be found further along the frequency branch, for higher, and possibly positive, frequencies. Nonetheless, we restricted the linear studies to  $k_y\rho_i\leq 2.0$ to remain on ion scales.

On top of each $k_y\rho_i(a/L_n, a/L_{T_i})$ tile, the parallel mode structure of the most unstable mode is shown, by plotting the potential squared along the field line, plotted in white if it is confined to the first poloidal turn, or in green if the mode extends up to three poloidal turns. It can be seen that in CBC and AUG, the binormal wavenumber of the most unstable mode and its parallel mode structure change continuously throughout the scanned gradients. Similarly, in \mbox{TJ-II} and NCSX the binormal wavenumber and parallel mode structure change continuously for most of the scanned gradients, however, for some adjoining tiles, the mode changes rather strongly. Nonetheless by going from any tile A to tile B by first passing through other adjoining tiles, the changes are continuous. When looking at the $\gamma(k_y\rho_i)$ spectra (not shown here) this corresponds to a single peak in $\gamma(k_y\rho_i)$ which slowly splits up into two peaks as the gradients are modified. Finally, for LHD and W7-X clear jumps between distinct modes can be observed. Specifically, in W7-X, extended modes with both large and small perpendicular scales show up along $a/L_n \approx a/L_{T_i}$.

In summary, the linear analysis at $a/L_{T_i}=3.0$ reveals a modest reduction of the growth rates with increasing density gradients in \mbox{W7-X} and the absence of large-scale modes  at high density gradients in NCSX \mbox{(figure \ref{fig:gamma_map_line})}, indicating a potential reduction of nonlinear heat transport with increased density gradients.  In contrast, in CBC, AUG, TJ-II and LHD the linear observations point towards an increase of the heat transport with increasing density gradients, since the growth rates increase with $a/L_n$, and the perpendicular scale of the turbulence remains large.
Moreover, analyzing the geometric characteristics of the chosen field line, it is evident that the modes in NCSX and \mbox{W7-X} are confined to narrow regions along the field lines, due to the high local magnetic shear in NCSX and due to the narrow magnetic wells and regions of bad curvature in W7-X (figure \ref{fig:linear_phi_vs_z2}).

\vspace{-1mm}
\section{Nonlinear analysis}
\label{sec:nonlinear}

In the second part of this chapter, nonlinear simulations are performed to investigate the effect of density gradients on turbulence. Section \ref{sec:heat} provides a detailed discussion on the ion heat transport, while section \ref{sec:nonlinearITGTEM} isolates the contributions from the ion-temperature-gradient-driven modes and the density-gradient-driven modes. Next, section \ref{sec:nonlinearWithALTe} examines the electron heat flux, addressing the influence of a non-vanishing electron temperature gradient. Finally, more insight is given into the characteristics of the instabilities that drive the turbulence in section \ref{sec:distributon}, by identifying the contribution of the trapped and passing particles to the turbulent distribution function. The resolution used for these simulations is provided in table \ref{tab:resolutionSixDevices}, and extensive resolution checks are performed in Appendix \mbox{\ref{sec:resolution}}.

\begin{table}[h!]
	\centering
	\vspace{10mm}
	\caption{\label{tab:Qi} Time-averaged ion heat fluxes, $Q_i/Q_{\text{gB},i}$, considering $a/L_{T_i} = 3.0$ and $a/L_{T_e} = 0.0$, corresponding to the data plotted in figure \ref{fig:Qi_vs_aLn}, as well as the black curves in figure \ref{fig:qi_contributions}.} \vspace{-2.5mm}
	\lineup
	\begin{tabular}{@{}ccccccc}
		\specialrule{.12em}{.1em}{.1em} \\[-4mm]
		& CBC & AUG & TJ-II & LHD &  NCSX &  W7-X \\[0.5mm] \hline\\[-4mm]
		\hspace{24mm} & \hspace{18.5mm}  & \hspace{18.5mm} &
		\hspace{18.5mm} & \hspace{18.5mm} & \hspace{18.5mm} & \hspace{18.5mm} \\[-3.5mm]
		$a/L_n = 0$ & 19.2 & 12.1 & 23.3 & 5.6 & 9.0 & 9.1 \\[0.5mm]
		$a/L_n = 1$ & 26.5 & 19.0 & 10.6 & 6.9 & 8.8 & 6.4\\[0.5mm]
		$a/L_n = 2$ & 37.2 & 30.4 & 9.3 & 6.1 & 3.6 & 1.4\\[0.5mm]
		$a/L_n = 3$ & 57.4 & 48.8 & 7.0 & 6.1 & 0.9 & 1.5\\[0.5mm]
		$a/L_n = 4$ & 53.4 & 44.5 & 5.4 & 7.5 & 1.1 & 4.7 \\[0.5mm]
		\specialrule{.12em}{.1em}{.1em}
	\end{tabular}
	\vspace{-1.5mm}
\end{table}

\subsection{Turbulent heat transport}\vspace{-1mm}
\label{sec:heat}

\begin{figure}[!b]
	\centering
	\includegraphics[trim={0mm 2mm 0mm 3mm}] {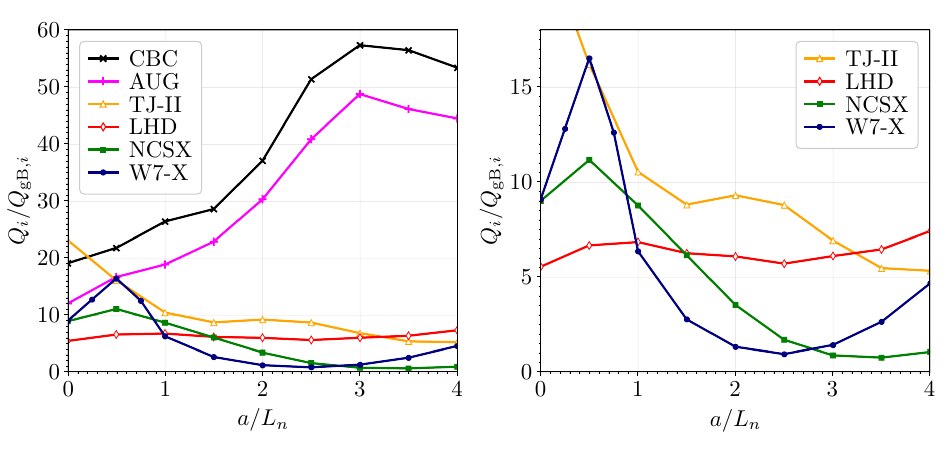}  \vspace{-5mm}
	\caption{Ion heat flux as a function of the normalized density gradient, considering $a/L_{T_i}=3.0$ and $a/L_{T_e}=0.0$. The right-hand side represents a zoomed view of the plot shown on the left.}
	\label{fig:Qi_vs_aLn}
\end{figure}

In the following sections we will discuss the ion and electron heat fluxes, $Q_i$ and $Q_e$, as defined by equation (\ref{eq:Q}). These fluxes are normalized with respect to the ion gyro-Bohm flux, given by \mbox{$Q_{\rm gB,i}= n_i\,T_i\,v_{{\rm th},i}\,(\rho_i/a)^2$}.  The evolution of the ion heat flux as a function of the normalized density gradient, considering a fixed ion temperature gradient of $a/L_{T_i}=3.0$ and a vanishing electron temperature gradient ($a/L_{T_e}=0.0$), is shown in figure \ref{fig:Qi_vs_aLn}, and summarized in table \ref{tab:Qi}.

For the tokamak configurations, it can be seen on the left-hand-side of figure \ref{fig:Qi_vs_aLn} that the density gradient gives rise to a significant increase of the ion heat transport up to \mbox{$a/L_n = 3.0$}, after which it decreases slightly.
In order to discuss the stellarators, a zoom of the this plot is shown on the right-hand-side of \mbox{figure \ref{fig:Qi_vs_aLn}}. For LHD, the ion heat flux is approximately independent of the density gradient. On the other hand, in \mbox{TJ-II} the ion heat flux is reduced significantly for $a/L_n \leq 1.5$ and \mbox{$a/L_n\geq2.5$}, whereas it remains approximately constant within \mbox{$1.5 \leq a/L_n \leq 2.5$}.
In contrast, in NCSX and W7-X, a very strong reduction of the ion heat flux is observed across a wide range of density gradients. Initially, the ion heat flux increases for $a/L_n\leq 0.5$, after which there is a reduction of the ion heat flux by an order of magnitude in both devices. Note that the ion heat flux starts increasing again for $a/L_n\geq 2.5$ in W7-X  and for $a/L_n \geq 3.5$ in NCSX.

Comparing the nonlinear ion heat fluxes with the linear growth rates in figure \ref{fig:gamma_vs_aLn}a, it is clear that the reduction of the ion heat flux in W7-X is an order of magnitude stronger than the reduction of the growth rates. Moreover, the ion heat flux in NCSX is reduced very strongly, while the growth rates do not indicate any reduction. For LHD and \mbox{TJ-II} the growth rates increase linearly with the density gradient, while their ion heat flux is either constant or gets reduced.  Even when taking into account the perpendicular scale of the modes  observed in the linear simulations (figure \ref{fig:gamma_vs_aLn}c), the reduction of the ion heat flux in NCSX and \mbox{W7-X} is much larger than the reduction of $\gamma/k_y^2$. Therefore, it is clear that there is no correlation between the nonlinear ion heat fluxes and the linear growth rates of the fastest-growing modes for the stellarators.
The fastest-growing mode is thus not necessarily representative of the nonlinear dynamics of the turbulence. Instead, nonlinear saturation mechanisms play a crucial role on the heat transport within these devices.  Determining these underlying mechanisms, which includes the role of subdominant modes through the nonlinear transfer of energy from unstable to damped modes \cite{hatch2011role, terry2006role, hegna2018theory, terry2018saturation} and the effects of zonal flow shearing and energy cascades to dissipative scales \cite{diamond2005zonal, 2015-Plunk-Understanding, terry2000suppression}, is beyond the scope of this work. Nonetheless, NCSX and W7-X are the only two stellarators for which the linear analysis of the fastest-growing mode points towards a reduction of the nonlinear heat transport, albeit the magnitude of the reduction was significantly underestimated. On the other hand, for the tokamaks, the increase of the ion heat flux is in agreement \mbox{with the increase of the growth rates for $a/L_n \leq 3.0$.}

\begin{figure}[h!]
	\centering
	\vspace{-1mm}
	\includegraphics[trim={0mm 3mm 0mm -1mm}] {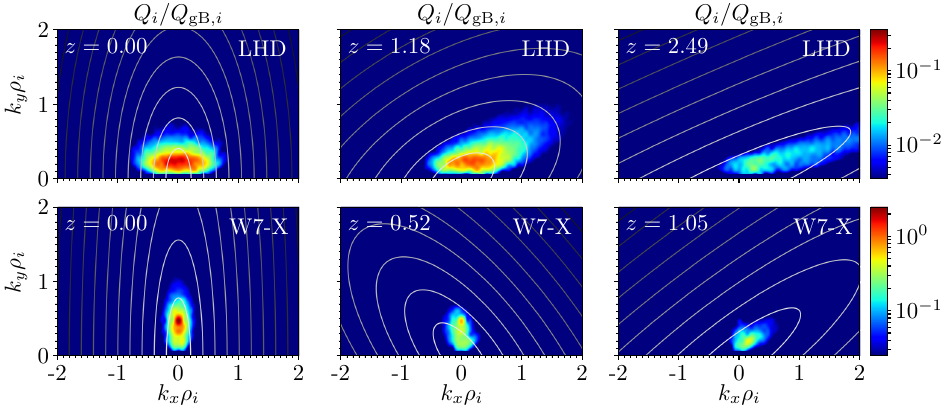}
	\vspace{-1mm}
	\caption{Contributions of the individual $(k_x,k_y)$-modes to the ion heat flux of LHD considering \mbox{$a/L_n=4.0$}, $a/L_{T_i}=3.0$ and $a/L_{T_e}=0.0$ (top row) and of W7-X considering $a/L_n=0.0$, $a/L_{T_i}=3.0$ and $a/L_{T_e}=0.0$ (bottom row), for different points along the field line.  The contour lines represent the magnitude of the perpendicular wavenumber $k_\perp\rho_i$ (white is minimum, black is maximum). \vspace{0mm}}
	\label{fig:kxkyspectraLHD}
\end{figure}

The contributions of the specific $(k_x,k_y)$-modes to the ion heat flux of LHD and W7-X are shown in figure \ref{fig:kxkyspectraLHD} for three different locations along the field line, in order to illustrate the effect of the local magnetic shear and FLR damping. As can be seen in \mbox{figure \ref{fig:qfluxvsz}}, the largest contributions to the heat flux tend to originate from the center of the field line. At $z=0$ the cross-section of the flux tube is perpendicular and the minimum perpendicular wavenumber  occurs for $k_x=0.0$ (see section \ref{sec:localization}). As one moves along the field line, the cross-section becomes sheared \mbox{(figure \ref{fig:fieldlineAngle}b)}, and the turbulent eddies become very twisted and are subsequently suppressed through FLR effects, leading to a reduction in heat losses towards the ends of the flux tube, as can be seen in figure \ref{fig:kxkyspectraLHD}. Note that the modes that contribute the most to the heat flux are those for which the perpendicular wavenumber $k_\perp\rho_i$ is minimum---corresponding to the modes within the inner contour line---where the FLR damping is the weakest. Seeing that the heat flux typically peaks at the center of the field line, the largest contributions thus arise from the $k_x=0.0$ modes. For this reason, the bad curvature regions are shown for $k_x=0.0$ in \mbox{figure \ref{fig:qfluxvsz}}. Nonetheless, other $k_x$-modes can also contribute significantly to the heat flux, for which the \mbox{bad curvature regions are located elsewhere.}

\begin{figure}[h!]
	\centering
	\vspace{3mm}
	\includegraphics[trim={0mm 3mm 0mm 0mm},clip] {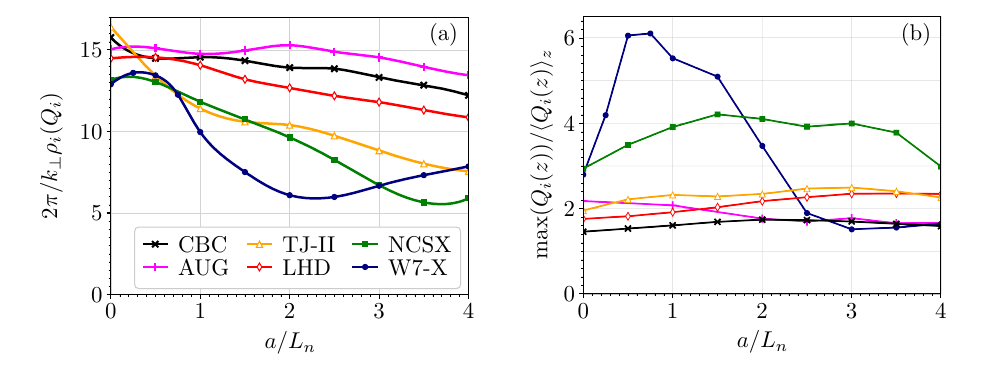}
	\vspace{-7mm}
	\caption{(a) Dominant perpendicular scale of the fluctuations that contribute to the ion heat flux and (b) ratio of the maximum ion heat flux along the field line with respect to the field-line averaged ion heat flux as a function of the normalized density gradient, $a/L_n$, considering $a/L_{T_i}=3.0$ and $a/L_{T_e}=0.0$. \vspace{1mm}}
	\label{fig:dominantkperp}
\end{figure}

\begin{figure}[b!]
	\centering
	\vspace{0mm}
	\includegraphics[trim={0mm 8mm 0mm 1mm}] {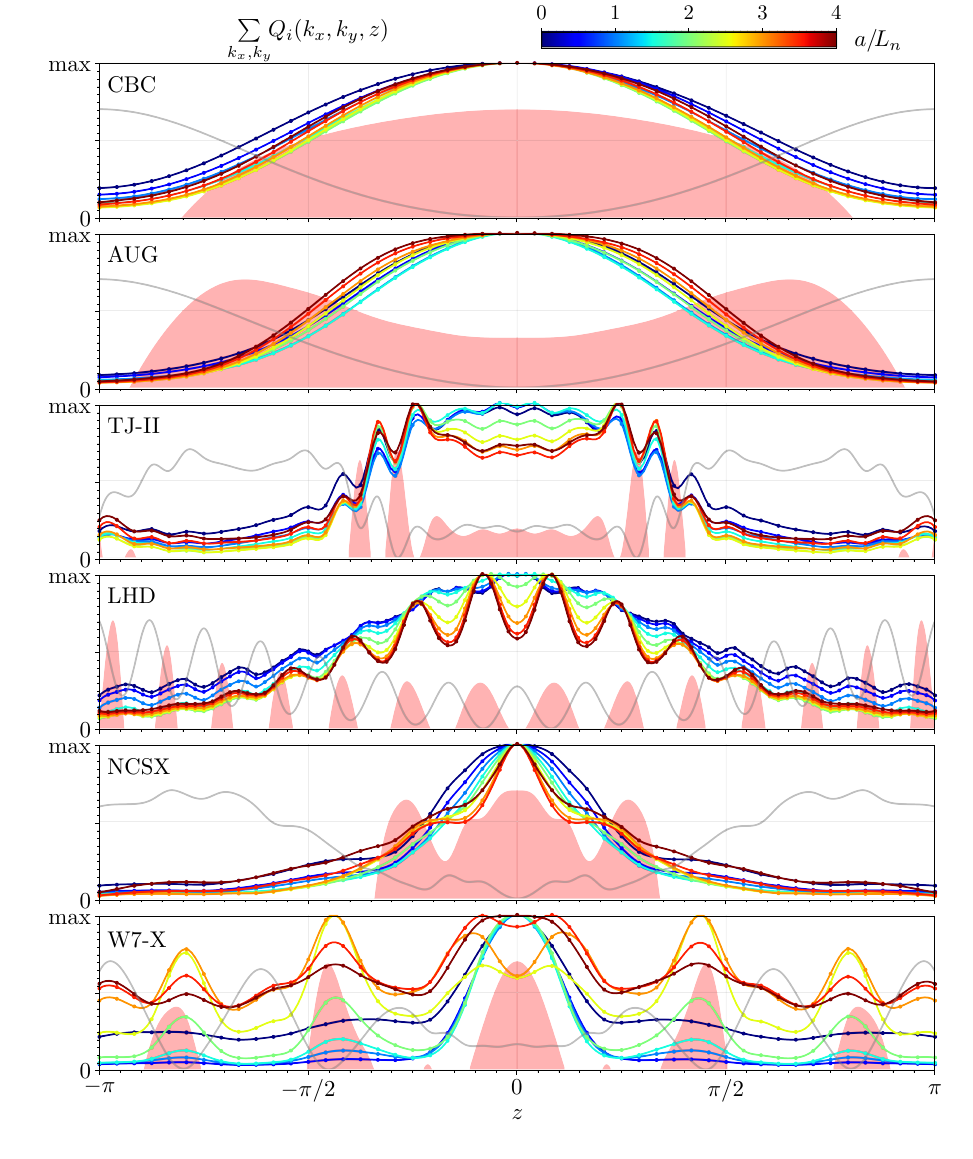} \vspace{-5mm}
	\caption{Contributions of the individual $z$-points to the ion heat flux. The strength of the magnetic field is shown in gray, regions of bad curvature considering $k_x=0.0$ are highlighted in red, and the colors of the curves correspond to the $a/L_n$-scale shown on top. \vspace{0mm} }
	\label{fig:qfluxvsz}
\end{figure}

The contributions $Q_i(k_x,k_y,z)$ to the ion heat flux are transformed to $Q_i(k_\perp\rho_i)$, and the $k_\perp\rho_i$ for which $Q_i(k_\perp\rho_i)$ is maximum is considered the dominant scale of the perpendicular fluctuations of the turbulence, which is shown in figure \ref{fig:dominantkperp}a as $2\pi/k_\perp\rho_i(Q_i^\textrm{max})$. \mbox{For W7-X}, NCSX, and to lesser extent for TJ-II, the perpendicular fluctuations become significantly smaller as the density gradient increases. Moreover, the trend of $2\pi/k_\perp\rho_i$ for W7-X, NCSX and TJ-II is very similar to the trend of their ion heat fluxes (figure \ref{fig:Qi_vs_aLn}), hence the heat losses are reduced partly because the scale of the turbulence becomes smaller. In contrast, for CBC and AUG the scale of the turbulence is not correlated to the trend of the ion heat flux, while in LHD both the ion heat flux and the scale of the fluctuations are approximately independent of the density gradient.

The contribution to the ion heat flux coming from specific points along the field line is visualized in figure \ref{fig:qfluxvsz}. In every device, except for W7-X, the localization of the ion heat losses does not change significantly when increasing the density gradient, besides the increased contribution coming from the magnetic wells.
Specifically, for \mbox{$a/L_n=0.0$}, the ion heat losses are the largest at the center of the field line ($z=0$). As explained in section \ref{sec:localization}, the ITG mode, which is believed to be dominant for \mbox{$a/L_n=0.0$}, generally drives turbulence in locations which have the most pronounced bad curvature---unless the magnitude of the curvature becomes too large which can have a stabilizing effect. The most pronounced bad curvature typically corresponds to the center of the field line (figure \ref{fig:fieldlineCurvature}). For CBC, AUG and NCSX the center also corresponds to the deepest magnetic well, hence the heat losses remain dominant at $z=0$ as the density gradient is increased, and trapped-electron modes start playing an important role. Whereas for W7-X, LHD and TJ-II the heat losses at high density gradients predominantly arise from locations further along the field line, corresponding to the deepest magnetic wells.

Moreover, for W7-X the distribution of the ion heat losses along the field line changes multiple times as the density gradient is increased. For $a/L_n = 0.0$, the heat losses arise predominately from the center of the field line, however, a substantial portion of the heat losses also arise uniformly from the  entire field line. When a finite density gradient is introduced \mbox{$(a/L_n = \{0.5,1.0,1.5\})$},  the uniform contribution disappears and the heat flux is generated almost entirely at the center. For \mbox{$a/L_n=2.0$}, which is close to the minimum ion heat flux (figure \ref{fig:Qi_vs_aLn}), heat losses start arising from the trapping regions, indicating that trapped-electron modes are beginning to become unstable. As the density gradient is increased further $(a/L_n = \{2.5,3.0\})$, heat is lost inside each magnetic well, however, the heat losses in the central well are localized predominantly on the sides of the magnetic well. For the highest density gradients that are considered $(a/L_n = \{3.5,4.0\})$, the preference of the magnetic wells is diminished and instead the majority of heat losses arise uniformly from the entire field line, suggesting a significant involvement of passing electrons.

To quantify how localized the heat losses are, the maximum ion heat flux along the field line is compared to the field-line averaged heat flux in \mbox{figure \ref{fig:dominantkperp}b}. Note that it has been demonstrated that heat losses depend on the field-line label \cite{sanchez2021gyrokinetic}, hence full-flux-surface simulations would be required to accurately discuss the localization. For CBC, AUG, TJ-II and LHD, the maximum heat flux along the field line is around twice as big as the field-line averaged heat flux, regardless of the density gradient, because the heat losses remain localized at the center of the field line. Similarly, in NCSX the localization does not change with the density gradient, however, the heat losses are much more localized. This is likely due to the fact that the bad curvature region, as well as the deepest magnetic well, are rather narrow, and the strong local magnetic shear \mbox{(figure \ref{fig:fieldlineAngle}b)} damps the fluctuations arising from the two outer peaks of bad curvature. On the other hand, in \mbox{W7-X}, there is a uniform loss of heat along the entire field line, as well as a strong contribution at $z=0$ for $a/L_n=0.0$, after which the heat losses become strongly localized to the center of the field line as the density gradient is increased. The heat losses remain strongly localized until $a/L_n = 1.5$ after which a uniform contribution along the entire field line emerges, indicating that an instability driven by passing electrons is likely becoming unstable for density \mbox{gradients above $a/L_n \geq 2.0$.}

\newpage
\subsection{Contributions to $Q_i$ from the ion temperature and density gradient}
\label{sec:nonlinearITGTEM}
\vspace{0.5mm}

\begin{figure}[b!]
	\centering
	\vspace{-1mm}
	\includegraphics[trim={0mm 0mm 0mm 2mm}] {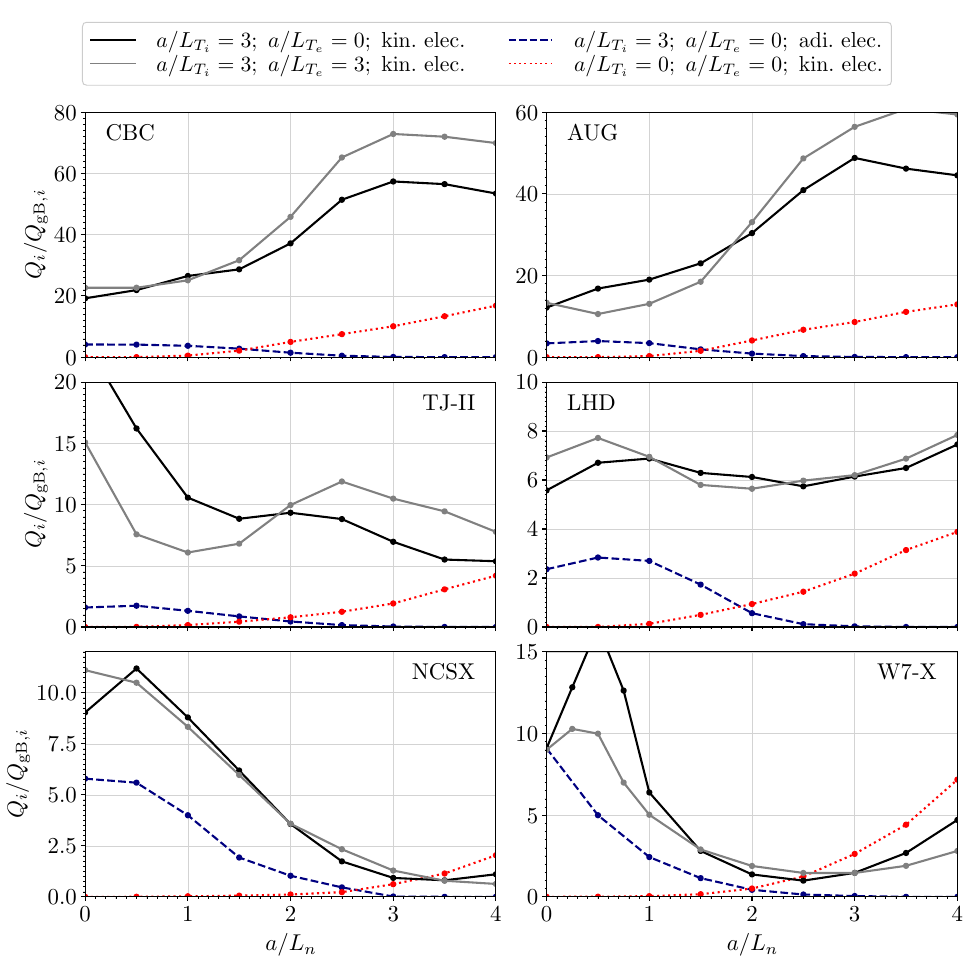}
	\vspace{-10mm}
	\caption{Normalized ion heat flux as a function of the normalized density gradient with kinetic electrons for \mbox{$a/L_{T_e}=0.0$} (black) and  $a/L_{T_e}=3.0$ (gray). Ion-temperature-gradient driven modes are isolated by treating the electrons as adiabatic (blue) and density-gradient-driven modes are isolated by setting \mbox{$a/L_{T_i} = 0.0$} (red), both considering \mbox{$a/L_{T_e}=0.0$}. \vspace{-5mm} }
	\label{fig:qi_contributions}
\end{figure}

\begin{figure}[b!]
	\centering
	\vspace{-2mm}
	\includegraphics[trim={0mm 0mm 0mm 2mm},clip] {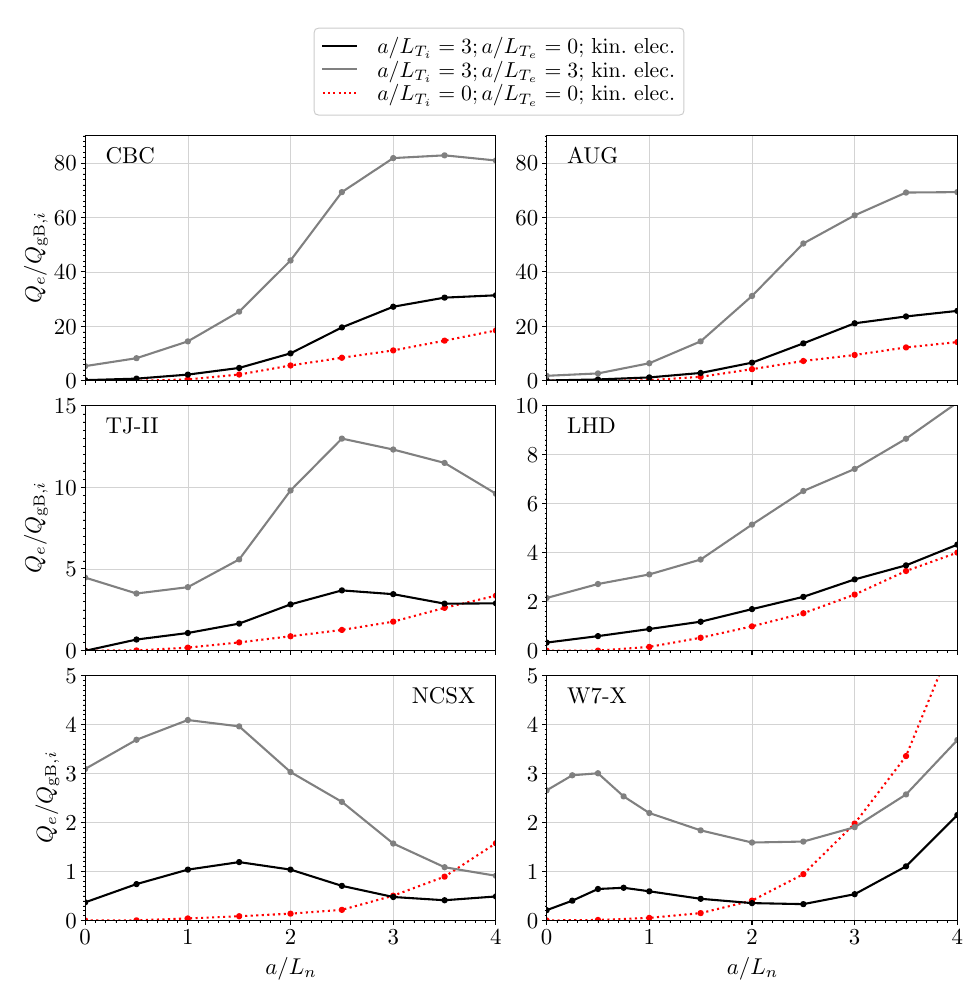}
	\vspace{-10mm}
	\caption{Normalized electron heat flux as a \mbox{function of the} density gradient for \mbox{$a/L_{T_e}=0.0$} (black) and  $a/L_{T_e}=3.0$ (gray). The density-gradient-driven modes are isolated by \mbox{setting $a/L_{T_i} = a/L_{T_e} = 0.0$ (red).} \vspace{-3mm} }
	\label{fig:qe_contributions}
\end{figure}

To gain more insights into the trends of the heat fluxes, the contributions of the ion-temperature-gradient-driven modes (blue dashed lines) and the density-gradient-driven modes (red dotted lines) to the ion and electron heat flux are isolated and shown in figures \ref{fig:qi_contributions} and \ref{fig:qe_contributions}, respectively. The simulations containing both the ion temperature gradient and kinetic electrons, shown in figure \ref{fig:Qi_vs_aLn}, are represented here by the black lines. The ion-temperature-gradient-driven modes are isolated by treating the electrons adiabatically, whereas the density-gradient-driven modes are isolated by considering vanishing temperature gradients.

The ion-temperature-gradient-driven modes (blue dashed lines in figure \ref{fig:qi_contributions}) are stabilized with increasing density gradients, in the same way as in the linear simulations (figure \ref{fig:linear_teprim}). Under these conditions the toroidal ITG mode is believed to be dominant. Note that at $a/L_n = 0.0$ the largest ion heat flux is driven in W7-X and NCSX, and the lowest in LHD and TJ-II (see table \ref{tab:QiITG}).
In \mbox{TJ-II}, including kinetic electrons in the simulations (blue lines $\rightarrow$ black lines) drastically increases the ion heat flux at $a/L_n=0.0$, while a significant increase in the ion heat flux is observed in CBC and AUG, a moderate increase in LHD and NCSX and finally in W7-X the electron dynamics do not effect the ion heat flux at $a/L_n=0.0$.

\begin{table}[t!]
	\centering
	\vspace{2mm}
	\caption{\label{tab:QiITG} Ion heat fluxes, $Q_i/Q_{\text{gB},i}$, considering $a/L_{T_i} = 3.0$ and adiabatic electrons to isolate ion-temperature-gradient driven modes, corresponding to the blue dashed lines in figure \ref{fig:qi_contributions}.} \vspace{-2.5mm}
	\lineup
	\begin{tabular}{@{}ccccccc}
		\specialrule{.12em}{.1em}{.1em} \\[-4mm]
		& CBC & AUG & TJ-II & LHD &  NCSX &  W7-X \\[0.6mm] \hline\\[-4mm]
		\hspace{24mm} & \hspace{18.5mm}  & \hspace{18.5mm} &
		\hspace{18.5mm} & \hspace{18.5mm} & \hspace{18.5mm} & \hspace{18.5mm} \\[-3.5mm]
		$a/L_n = 0$ & 4.1 & 3.4 & 1.6 & 2.4 & 5.8 & 9.1 \\[0.6mm]
		$a/L_n = 1$ & 3.7 & 3.4 & 1.3 & 2.7 & 4.0 & 2.4 \\[0.6mm]
		$a/L_n = 2$ & 1.4 & 0.9 & 0.4 & 0.6 & 1.0 & 0.4 \\[0.6mm]
		$a/L_n = 3$ & 0.1 & 0.0 & 0.0 & 0.0 & 0.0 & 0.0 \\[0.6mm]
		$a/L_n = 4$ & 0.0 & 0.0 & 0.0 & 0.0 & 0.0 & 0.0 \\[0.6mm]
		\specialrule{.12em}{.1em}{.1em}
	\end{tabular}
	\vspace{-1mm}
\end{table}

\begin{table}[t!]
	\centering
	\vspace{2mm}
	\caption{\label{tab:QiTEM} Ion heat fluxes, $Q_i/Q_{\text{gB},i}$, considering $a/L_{T_i} = 0.0$, $a/L_{T_e} = 0.0$ and kinetic electrons to isolate density-gradient-driven modes, corresponding to the red dotted lines in figure \ref{fig:qi_contributions}.} \vspace{-2.5mm}
	\lineup
	\begin{tabular}{@{}ccccccc}
		\specialrule{.12em}{.1em}{.1em} \\[-4mm]
		& CBC & AUG & TJ-II & LHD &  NCSX &  W7-X \\[0.6mm] \hline\\[-4mm]
		\hspace{24mm} & \hspace{18.5mm}  & \hspace{18.5mm} &
		\hspace{18.5mm} & \hspace{18.5mm} & \hspace{18.5mm} & \hspace{18.5mm} \\[-3.5mm]
		$a/L_n = 0$ & 0.0  & 0.0 & 0.0 & 0.0 & 0.0 & 0.0 \\[0.6mm]
		$a/L_n = 1$ & 0.5  & 0.3 & 0.2 & 0.1 & 0.0 & 0.0 \\[0.6mm]
		$a/L_n = 2$ & 5.0  & 4.1 & 0.8 & 0.9 & 0.1 & 0.5 \\[0.6mm]
		$a/L_n = 3$ & 10.1 & 8.6 & 1.9 & 2.2 & 0.6 & 2.6 \\[0.6mm]
		$a/L_n = 4$ & 16.8 & 12.9 & 4.2 & 3.9 & 2.0 & 7.2 \\[0.6mm]
		\specialrule{.12em}{.1em}{.1em}
	\end{tabular}
	\vspace{-1mm}
\end{table}

The ion heat flux driven by the density-gradient-driven modes (red dotted lines in figure \ref{fig:qi_contributions}) increases with increasing density gradients, which is in line with the linear results (figure \ref{fig:linear_teprim}). In this case, the turbulence is believed to be primarily driven by trapped-electron modes. At $a/L_n = 4.0$, the lowest ion heat fluxes are found in the stellarators, as shown in table \ref{tab:QiTEM}, with NCSX having the lowest of all. As expected, the largest ion heat fluxes are found in the tokamaks, as here the majority of trapped electrons are located in the bad curvature region. Moreover, in CBC and AUG the ion heat flux is increased significantly by the ion temperature gradient at $a/L_n=4.0$ (red line $\rightarrow$ black line), whereas in LHD it is increased moderately. Finally, in TJ-II, NCSX and W7-X, the effect of the ion temperature gradient is very small, moreover, it is capable of reducing the ion heat flux driven by the density-gradient-driven modes.

When including both kinetic electrons and the ion temperature gradient at the same time (black lines), the ion heat flux can resemble a direct sum of the two contributions, or it is enhanced strongly. The latter is true for CBC, AUG, TJ-II and LHD, indicating that the instabilities are destabilized significantly by the free energy in the ion temperature gradient, and by the electron dynamics. In contrast, in W7-X and NCSX, the ion heat flux from the complete simulations (black lines) resembles the direct sum of the separate effects, where NCSX benefits particularly well from the weak density-gradient-driven modes (red dotted lines).
Therefore, for W7-X and NCSX there is a region of density gradients for which the ion-temperature-gradient-driven modes are being suppressed and the density-gradient-driven modes are not driving a significant amount of turbulence yet, which is qualitatively in line with the linear observations \mbox{(figure \ref{fig:linear_teprim})}.
Moreover, from the nonlinear simulations we can conclude that the reduction of the ion heat flux by the density gradient is also related to the fact that the ion-temperature-gradient-driven modes (blue dashed lines) in NCSX are only moderately destabilized by the electron dynamics, while they are not destabilized by kinetic electrons in W7-X, in contrast to the other devices. Additionally, the density-gradient-driven modes in NCSX and \mbox{W7-X} are slightly stabilized in the presence of a finite ion temperature gradient, while they are significantly destabilized in \mbox{the other devices.  }

\newpage
\subsection{Effect of the electron temperature gradient on $Q_i$ and $Q_e$}
\label{sec:nonlinearWithALTe}

The effect of adding a non-vanishing electron temperature gradient on the heat transport is shown in figures \ref{fig:qi_contributions} and \ref{fig:qe_contributions}, respectively, by comparing the black lines $(a/L_{T_e}=0.0)$ with the gray lines $(a/L_{T_e}=3.0)$. The values of the ion and electron heat fluxes in the presence of a finite electron temperature gradient are summarized in tables \ref{tab:QiTEPRIM} and \ref{tab:QeTEPRIM}, respectively. It can be seen that the electron temperature gradient has a relatively small effect on the ion heat losses \mbox{(figure \ref{fig:qi_contributions})}.  Specifically, in AUG and TJ-II, the presence of an electron temperature gradient  decreases the ion heat flux moderately for $a/L_n\leq 2.0$ and increases it for $a/L_n\geq 2.0$, a similar increase in the ion heat flux is observed for CBC. In W7-X, the electron temperature gradient has a moderate stabilizing effect on the ion heat loses in the presence of small density gradients, reducing the ion heat flux by 40\% for $a/L_n = 0.5$. Finally, in LHD and NCSX the effect is minimal.

In contrast, the electron temperature gradient increases the drive for the electron heat transport strongly, resulting in a significant increase of the electron heat flux, which becomes of the order of the ion heat flux. Note that the electron heat flux in the absence of an electron temperature gradient (black lines in \mbox{figure \ref{fig:qe_contributions}}) is much smaller than the ion heat flux (black lines in \mbox{figure \ref{fig:qi_contributions}}), especially at low density gradients. This difference in magnitude is expected in the presence of a steep ion temperature gradient of $a/L_{T_i} = 3.0$.
Specifically, in CBC and AUG, both the electron heat flux and the ion heat flux increase with the density gradient. For LHD, the electron heat flux increases roughly linearly with the density gradient, similar as its linear growth rates. In TJ-II the electron heat flux initially decreases, after which it increases strongly with the density gradient for $a/L_{T_e}=3.0$, comparable to the evolution of its ion heat flux. Finally, for both NCSX and W7-X, the electron heat flux undergoes a similar reduction with increasing density gradients as the ion heat flux, albeit it is less pronounced.

\begin{table}[t!]
	\centering
	\vspace{2mm}
	\caption{\label{tab:QiTEPRIM} Ion heat fluxes, $Q_i/Q_{\text{gB},i}$, considering $a/L_{T_i} = 3.0$ and $a/L_{T_e} = 3.0$, corresponding to the gray curves in figure \ref{fig:qi_contributions}, in order to study the effect of a finite electron temperature gradient.} \vspace{-2mm}
	\lineup
	\begin{tabular}{@{}ccccccc}
		\specialrule{.12em}{.1em}{.1em} \\[-4mm]
		& CBC & AUG & TJ-II & LHD &  NCSX &  W7-X \\[1mm] \hline\\[-4mm]
		\hspace{24mm} & \hspace{18.5mm}  & \hspace{18.5mm} &
		\hspace{18.5mm} & \hspace{18.5mm} & \hspace{18.5mm} & \hspace{18.5mm} \\[-3.5mm]
		$a/L_n = 0$ & 22.7 & 13.3 & 15.1 & 6.9 & 11.1 & 9.0 \\[1mm]
		$a/L_n = 1$ & 25.1 & 13.1 & \wzero6.1  & 6.9 & \wzero8.3 & 5.0 \\[1mm]
		$a/L_n = 2$ & 45.8 & 33.1 & 10.0 & 5.6 & \wzero3.6 & 1.9 \\[1mm]
		$a/L_n = 3$ & 72.9 & 56.4 & 10.5 & 6.2 & \wzero1.3 & 1.5 \\[1mm]
		$a/L_n = 4$ & 69.9 & 59.4 & \wzero7.8  & 7.8 & \wzero0.6 & 2.8 \\[1mm]
		\specialrule{.12em}{.1em}{.1em}
	\end{tabular}
	\vspace{-0.9mm}
\end{table}

\begin{table}[b!]
	\centering
	\vspace{2mm}
	\caption{\label{tab:Qe} Electron heat fluxes, $Q_e/Q_{\text{gB},i}$, for $a/L_{T_i} = 3.0$ and a vanishing electron temperature gradient ($a/L_{T_e} = 0.0$), corresponding to the black curves in figure \ref{fig:qe_contributions}.} \vspace{-2mm}
	\lineup
	\begin{tabular}{@{}ccccccc}
		\specialrule{.12em}{.1em}{.1em} \\[-4mm]
		& CBC & AUG & TJ-II & LHD &  NCSX &  W7-X \\[1mm] \hline\\[-4mm]
		\hspace{24mm} & \hspace{18.5mm}  & \hspace{18.5mm} &
		\hspace{18.5mm} & \hspace{18.5mm} & \hspace{18.5mm} & \hspace{18.5mm} \\[-3.5mm]
		$a/L_n = 0$ & \wzero0.2 & \wzero0.1 & 0.0 & 0.3 & 0.4 & 0.2 \\[1mm]
		$a/L_n = 1$ & \wzero2.3 & \wzero1.2 & 1.1 & 0.9 & 1.0 & 0.6 \\[1mm]
		$a/L_n = 2$ & 10.1 & \wzero6.6 & 2.8 & 1.7 & 1.0 & 0.4 \\[1mm]
		$a/L_n = 3$ & 27.2 & 21.1 & 3.5 & 2.9 & 0.5 & 0.5 \\[1mm]
		$a/L_n = 4$ & 31.4 & 25.7 & 2.9 & 4.3 & 0.5 & 2.2 \\[1mm]
		\specialrule{.12em}{.1em}{.1em}
	\end{tabular}
	\vspace{-0.9mm}
\end{table}

\begin{table}[b!]
	\centering
	\vspace{2mm}
	\caption{\label{tab:QeTEPRIM} Electron heat fluxes, $Q_e/Q_{\text{gB},i}$, for $a/L_{T_i} = 3.0$ and $a/L_{T_e} = 3.0$, corresponding to the gray curves in figure \ref{fig:qe_contributions}, in order to study the effect of a finite electron temperature gradient.}\vspace{-2mm}
	\lineup
	\begin{tabular}{@{}ccccccc}
		\specialrule{.12em}{.1em}{.1em} \\[-4mm]
		& CBC & AUG & TJ-II & LHD &  NCSX &  W7-X \\[1mm] \hline\\[-4mm]
		\hspace{24mm} & \hspace{18.5mm}  & \hspace{18.5mm} &
		\hspace{18.5mm} & \hspace{18.5mm} & \hspace{18.5mm} & \hspace{18.5mm} \\[-3.5mm]
		$a/L_n = 0$ & \wzero5.4  & \wzero1.8  & \wzero4.5  & \wzero2.1  & 3.1 & 2.7 \\[1mm]
		$a/L_n = 1$ & 14.5 & \wzero6.4  & \wzero3.9  & \wzero3.1  & 4.1 & 2.2 \\[1mm]
		$a/L_n = 2$ & 44.2 & 31.1 & \wzero9.8  & \wzero5.1  & 3.0 & 1.6 \\[1mm]
		$a/L_n = 3$ & 81.8 & 60.8 & 12.3 & \wzero7.4  & 1.6 & 1.9 \\[1mm]
		$a/L_n = 4$ & 80.9 & 69.3 & \wzero9.6  & 10.1 & 0.9 & 3.7 \\[1mm]
		\specialrule{.12em}{.1em}{.1em}
	\end{tabular}
	\vspace{-0.9mm}
\end{table}

Recall that these simulations are performed at ion-Larmor scales, therefore the contribution to the heat transport that arises from the electron-Larmor scales is missing. Nonetheless, it has been shown that when the ion temperature gradient is sufficiently above the critical ion temperature gradient, the ion and electron heat flux driven at ion-Larmor scales gives an accurate picture of the heat transport in tokamaks \cite{howard2016multi, howard2015multi}. Whether or not this holds for stellarators needs to be investigated and is beyond the scope of this work.

\subsection{Contribution of trapped and passing particles}
\label{sec:distributon}

\begin{figure}[h!]
	\vspace{-3.5mm}
	\includegraphics{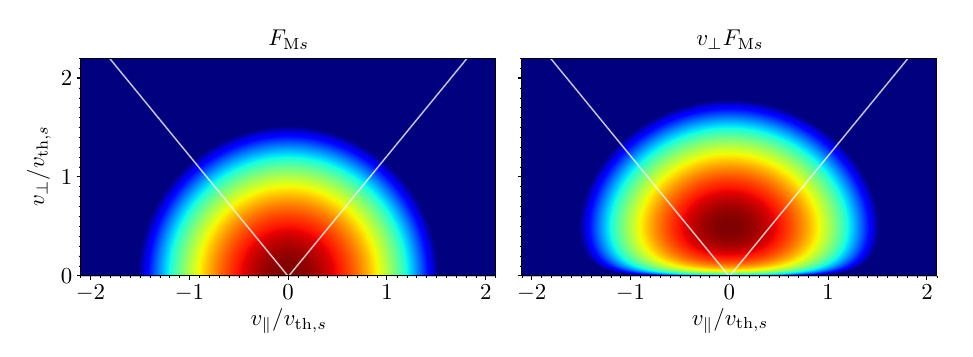}
	\vspace{-10.5mm}
	\caption{Maxwellian distribution in velocity space with a trapping cone corresponding to \mbox{$B/B_\textrm{max}=0.75$}. The Maxwellian distribution is rescaled with $v_\perp$ on the right-hand-side.}
	\vspace{0mm}
	\label{fig:hmuvpa_max}
\end{figure}

In order to gain a better understanding of the microinstabilities that drive the turbulence, the turbulent part of the distribution function is examined in detail in this section.
Let us denote by $\hat{g}_{\mathbf{k},s}$ the Fourier components of the gyro-averaged turbulent distribution function and let
\begin{equation}
	\hat{h}_{\mathbf{k},s} = \hat{g}_{\mathbf{k},s} +  \frac{Z_s e}{T_s} \, \hat{\varphi}_\mathbf{k}\, F_{\text{M}s} \, J_0(k_\perp\rho_s)
	\label{eq:g}
\end{equation}
be its non-adiabatic component (eq.~\ref{eq:deltaf}). In figure \ref{fig:hmuvpa_max} a Maxwellian distribution is shown as a function of the parallel and perpendicular velocities. The distribution is rescaled by a factor $v_\perp$ on the right-hand-side of figure \ref{fig:hmuvpa_max}, because $v_\perp$ corresponds to the Jacobian of the transformation from Euclidean velocity coordinates, $(v_x,v_y,v_z)$, to guiding-center velocity coordinates, $(v_\perp, v_\parallel, \vartheta)$.\!

Particles are trapped if $\smash{v_\parallel^2 + v_{\!\perp}^2 < 2\mu B_\text{max}}$, with \mbox{$v_{\!\perp}^2 = 2\mu B(z)$} and $B_\text{max}$ is the maximum value of $B(z)$ along the field line. This trapping condition is visualized in figure \ref{fig:hmuvpa_max} by the white lines---commonly known as the trapping cone. Passing particles (outside the cone) are capable of exploring the entire field line, while trapped particles (inside the cone) are confined to magnetic wells. For a Maxwellian distribution, the fluctuations are isotropic in velocity space, and the trapped particle fraction is given by \mbox{$f_\text{trapped}(z) = \sqrt{1-B(z)/B_\textrm{max}}$}. In figure \ref{fig:hmuvpa_max}, we have $f_\text{trapped}=0.5$, therefore, half of the particles are trapped in this scenario.

\begin{figure}[b!]
	\vspace{-3mm}
	\includegraphics{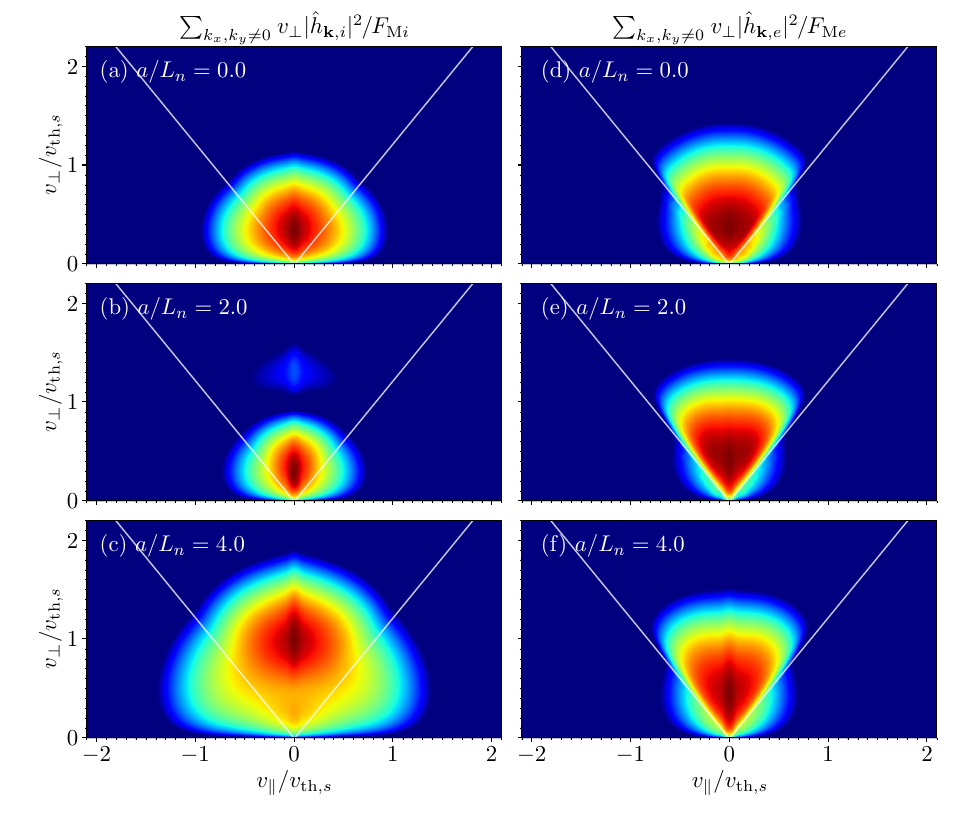}
	\vspace{-10mm}
	\caption{Non-adiabatic part of the distribution function at \mbox{$z=0$} for ions (left columns) and electrons (right columns) in CBC considering \mbox{$a/L_{T_i}=3.0$} and $a/L_{T_e}=0.0$, for $a/L_n=0.0$ (top row),  $a/L_n=2.0$ (middle row) and  $a/L_n=4.0$ (bottom row).  }
	\vspace{-5mm}
	\label{fig:hmuvpa_CBC}
\end{figure}

\begin{figure}[h!]
	\vspace{0mm}
	\includegraphics{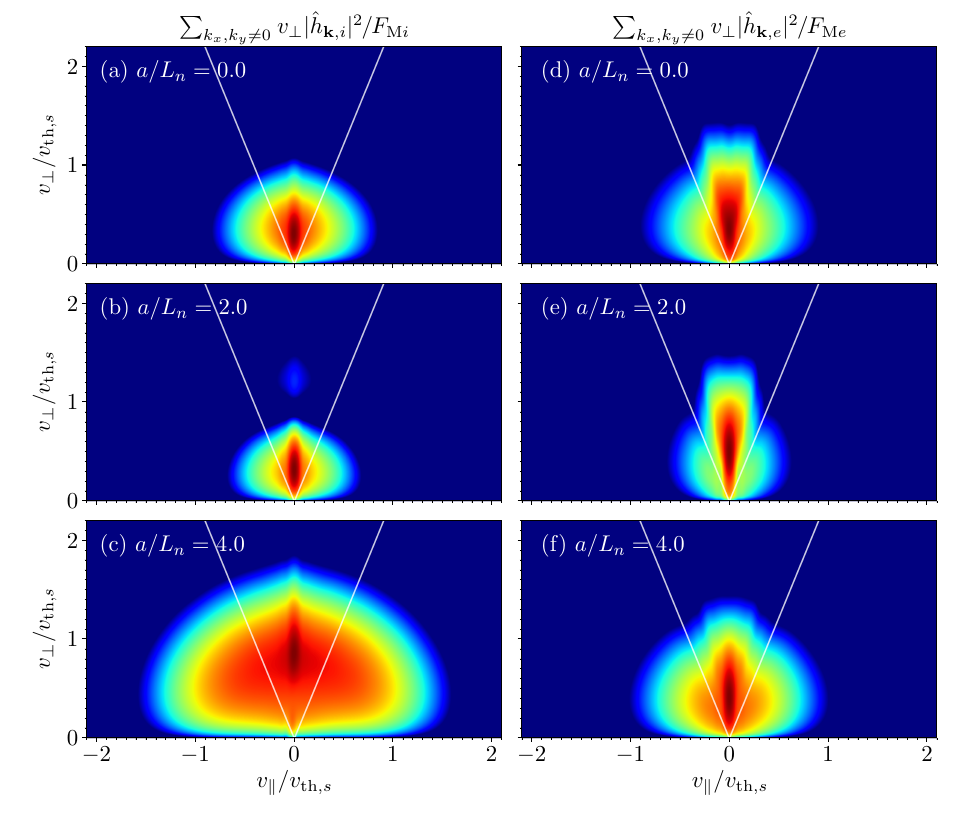}
	\vspace{-10mm}
	\caption{Non-adiabatic part of the distribution function at \mbox{$z=0$} for ions (left columns) and electrons (right columns) in W7-X considering \mbox{$a/L_{T_i}=3.0$} and $a/L_{T_e}=0.0$, for $a/L_n=0.0$ (top row),  $a/L_n=2.0$ (middle row) and  $a/L_n=4.0$ (bottom row).  }
	\vspace{-2.5mm}
	\label{fig:hmuvpa_W7X}
\end{figure}

Note that a large fraction of trapped particles does not necessarily imply that trapping plays an important role on the turbulence,  it may simply indicate that the magnetic well is relatively deep. Only if plots of \smash{$\hat{h}_{\mathbf{k},s} (v_\parallel,v_\perp)$} exhibit a clear deviation from isotropy, can we expect that trapping could play an important role. Moreover, the non-adiabatic part \smash{$\hat{h}_{\mathbf{k},s}$} has to be sufficiently large compared to the total distribution \smash{$\hat{g}_{\mathbf{k},s}$} to have a notable influence on the turbulence. Therefore, it is not sufficient to only consider the fraction of trapped particles to discuss the nature of the turbulence, instead, one has to carefully examine the distribution in velocity space while simultaneously taking into account the relative size of $\hat{h}_{\mathbf{k},s}$, which is discussed in the following paragraphs.

In figures \ref{fig:hmuvpa_CBC} and \ref{fig:hmuvpa_W7X}, the non-zonal contributions to the non-adiabatic part of the distribution function for the ions and electrons, i.e.~$\smash{\sum_{k_x, k_y\neq 0}|\hat{h}_{\mathbf{k},s}|^2}$, are shown for CBC and \mbox{W7-X}, respectively, at $z=0$, considering \mbox{$a/L_{T_i}=3.0$} and $a/L_{T_e}=0.0$. Note that the $k_y=0$ modes do contain a non-zonal contribution, arising from their dependence on $z$, which, for simplicity, we neglect here. It is clear that for electrons trapping can play a dominant role, e.g.~in CBC the distribution function is clearly localized within the trapping region (right column of figure  \ref{fig:hmuvpa_CBC}) and the electrons are strongly trapped for all density gradients. On the other hand, in W7-X the electrons are fairly trapped at $a/L_n=0.0$, they become very trapped at $a/L_n=2.0$ and finally they are moderately trapped at $a/L_n=4.0$ (right column of figure  \ref{fig:hmuvpa_W7X}). Note that in stellarators particles can also be trapped with respect to a local maximum of the magnetic field, which results in a more narrow trapping cone. This is observed for W7-X (right column of figure  \ref{fig:hmuvpa_W7X}), which has a shallow magnetic well at the center of the field line (figure \ref{fig:fieldlineAngle}a). On the other hand, for CBC and W7-X, ions are more strongly trapped than in the case of an isotropic distribution (figure \ref{fig:hmuvpa_max}), which can be seen in the left columns of figures \ref{fig:hmuvpa_CBC} and \ref{fig:hmuvpa_W7X}. Except for $a/L_n=4.0$ in W7-X, where the ion distribution is almost isotropic. Note however that trapping does not necessarily play a notable role on the ion dynamics, it is possible that the instability is simply predominantly driven by ions that have small parallel velocities. This could explain why the ion distribution is rather smooth across the trapping-passing boundary, while for electrons the distribution changes \mbox{strongly across the trapping-passing boundary.}

\begin{figure}[h!]
	\centering
	\includegraphics[trim={0mm 0mm 0mm 2mm}, clip] {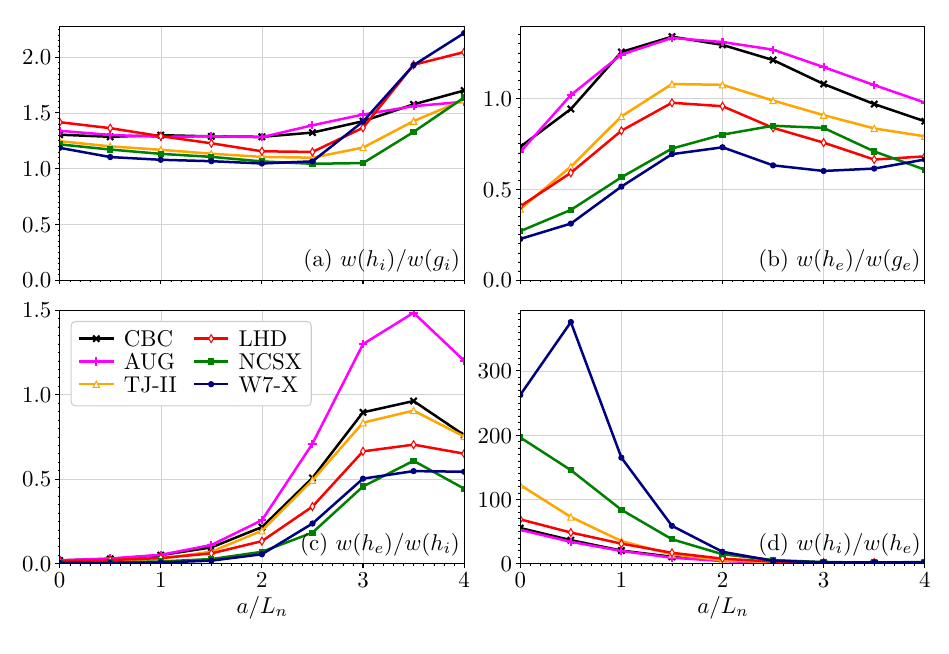}  \vspace{-8mm}
	\caption{Ratio of the non-adiabatic part of the distribution function to the gyro-averaged distributed function (a) for ions and (b) for electrons. Fraction of the non-adiabatic part of the distribution function of (c) the electrons to the ions and (d) of the ions to the electrons, considering $a/L_{T_i}=3.0$ and $a/L_{T_e}=0.0$.}  \vspace{2mm}
	\label{fig:ratios}
\end{figure}

To quantify the degree of non-adiabaticity and the importance of trapped particles we are going to introduce new diagnostics in the following paragraphs. Let us define
\begin{equation}
	\label{eq:w}
	w(h_s, z) = \int d^3\mathbf{v} \!\! \sum\limits_{k_x, k_y \neq 0}  \frac{T_s}{2} \frac{|\hat{h}_{\mathbf{k},s}|^2}{F_{\text{M}s}} .
\end{equation}
Given that the role of trapped particles could be significant in $h_s$, but, at the same time, the non-adiabatic part $h_s$ could be very small compared to the distribution $g_s$ in some situations, we quantify the degree of non-adiabaticity as $w(h_i)/w(g_i)$ and $w(h_e)/w(g_e)$, which are shown in figures \ref{fig:ratios}a and \ref{fig:ratios}b, respectively. Note that $w$ (with dimensions of energy density) is reminiscent to one of the pieces of the free energy (see equation (74) in \cite{schekochihin2009astrophysical}). This quantity is evaluated at $z=0$ for CBC, AUG, NCSX and \mbox{W7-X}, and at $z=\pm 0.23$ for LHD and $z = \pm 0.54$ for \mbox{TJ-II} since these are the locations where the potential fluctuations peak. The parallel mode structures of the potential are not shown here, nonetheless, they are similar to the localization of the heat losses shown in figure \ref{fig:qfluxvsz}, with the exception that the potential fluctuations always peak at the center for W7-X, and in the deepest wells for LHD and \mbox{TJ-II}, regardless of the density gradient. The parameter $z$ is henceforward dropped in $w(h_s, z)$ for simplicity.

For the ions $w(h_i)>w(g_i)$ for all density gradients in every device (figure \ref{fig:ratios}a), which means that the potential counter-acts the ion dynamics. Moreover, the ratio is around unity for small density gradients, and it increases substantially at large density gradients, hence the ion dynamics are almost entirely non-adiabatic. On the other hand, for small density gradients, the contribution of the electrons is much smaller than that of the ions, i.e., \mbox{$w(h_e) \ll w(h_i)$} (figure \ref{fig:ratios}c), moreover, the non-adiabatic part of the electrons is rather small, $w(h_e) < w(g_e)$ \mbox{(figure \ref{fig:ratios}b)}.  Therefore, for small density gradients, the electron dynamics do not play a significant role on the turbulence. This is expected because a vanishing electron temperature gradient is considered here, hence when the density gradient is small, the electrons experience little to no drive, while the ions are strongly driven by the steep ion temperature gradient of $a/L_{T_i} = 3.0$.
As the density gradient is increased from \mbox{$a/L_n=0.0$} to $a/L_n=1.5$, the ITG mode is damped, and while the electron dynamics remain negligible ($w(h_e) \ll w(h_i)$), the non-adiabatic part of the electron distribution function (figure \ref{fig:ratios}b) increases with the density gradient. When the density gradient is increased further, $a/L_n\geq2.0$, the electron dynamics start playing an important role since $w(h_e)/w(h_i)$ becomes relatively large, reaching values around 0.5 for W7-X and NCSX, up to 1.5 in AUG (figure \ref{fig:ratios}c). Moreover, the non-adiabatic part of the distribution function of the electrons is relatively large at $a/L_n=2.0$, with  $w(h_e) > w(g_e)$ in  CBC and AUG, $w(h_e) \approx w(g_e)$ in TJ-II and LHD, while $w(h_e) < w(g_e)$ in W7-X and NCSX. After $a/L_n=2.0$, $w(h_e)/w(g_e)$ decreases slowly with the density gradient in CBC, AUG, TJ-II and LHD, whereas in NCSX this decrease starts after $a/L_n=3.0$ and in W7-X the decrease starts at $a/L_n=2.0$, however $w(h_e)/w(g_e)$ starts increasing again after $a/L_n=3.0$.

\begin{figure}[b!]
	\centering
	\includegraphics[trim={0mm 5mm 0mm 15mm}] {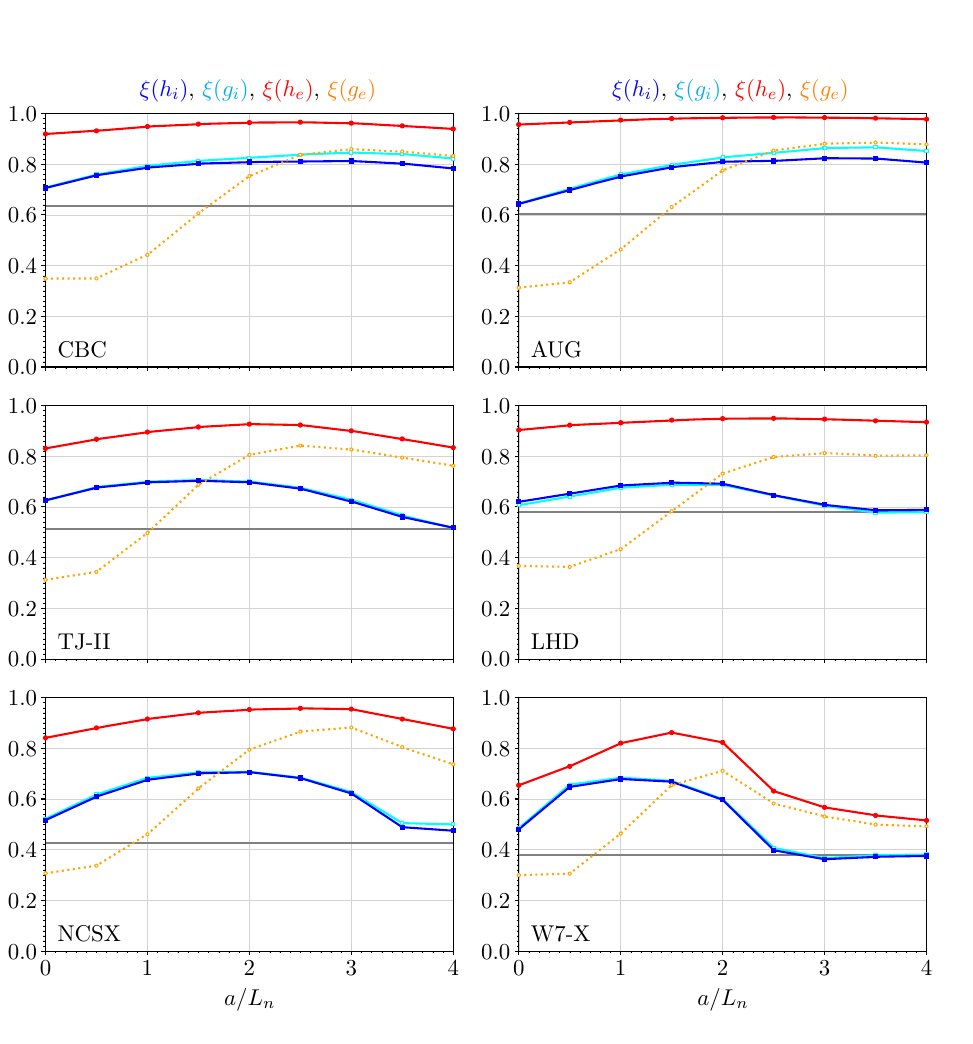}
	\vspace{-7mm}
	\caption{ Fraction of trapped particles considering $h_i$ (blue), $g_i$ (cyan), $h_e$ (red) and $g_e$ (orange), for $a/L_{T_i}=3$,  $a/L_{T_e}=0$, evaluated at the center of the field line $(z=0)$ for CBC, AUG, NCSX and W7-X and at $z=\pm 0.23$ for LHD and  $z = \pm 0.54$ for \mbox{TJ-II}. The trapped particle fraction $f_\text{trapped} = \sqrt{1-B(z)/B_\textrm{max}}$ is visualized by the horizontal gray lines. \vspace{-4mm}}
	\label{fig:trappedratios}
\end{figure}

To quantify whether the turbulence is driven by trapped or passing particles, we define a measure of the particle trapping at each point $z$ as,
\begin{equation}
	\label{eq:xi}
	\xi(h_s, z) = \frac{1}{w(h_s, z)} \int  \! d^3\mathbf{v} \ H_\text{trap} \!\!\! \sum\limits_{k_x, k_y \neq 0}  \frac{T_s}{2} \frac{|\hat{h}_{\mathbf{k},s}|^2}{F_{\text{M}s}} ,
\end{equation}
with $\smash{H_\text{trap} = H(2\mu (B_\text{max}\!-\!B(z)) - v_\parallel^2)}$ the Heaviside function. This measure is evaluated at the same $z$-locations considered in the previous paragraph, hence $z$ is dropped in $\xi (h_s,z)$ for simplicity. It is plotted in figure \ref{fig:trappedratios} considering the non-adiabatic part of the distribution function for the ions ($h_i$ in blue) and the electrons ($h_e$ in red), as well as for the full turbulent distribution function of the ions ($g_i$ in cyan) and electrons ($g_e$ in orange).

The quantity $\xi$ needs to be compared to the fraction of trapped particles that would occur if the fluctuations were isotropic in velocity space, which corresponds to the trapped particle fraction \mbox{$f_\text{trapped} = \sqrt{1-B(z)/B_\textrm{max}}$}, represented by the gray lines in figure \ref{fig:trappedratios}. The trapped particle fraction is 63\% in CBC, 60\% in AUG, 51\% in TJ-II, 58\% in LHD, 43\% in NCSX and 38\% in \mbox{W7-X} considering the $\alpha=0$ field line at $r/a=0.7$.
For small density gradients, the majority of the ions are trapped, i.e.~$\xi(h_i) > f_\text{trapped}$ (blue lines in figure \ref{fig:trappedratios}).   As the density gradient is increased, $\xi(h_i)$ approaches $f_\text{trapped}$  in the stellarators (blue lines $\rightarrow$ gray lines), because the velocity distribution becomes isotropic. In contracst, in CBC and AUG $\xi(h_i)$ \mbox{remains high at $a/L_n=4.0$.}

On the other hand, the majority of the electrons which are driving the non-adiabatic part of the distribution function (red lines) are trapped regardless of the density gradient, i.e., $\xi(h_e) > f_\text{trapped}$. With the exception of W7-X, which will be discussed separately in the next section. The non-adiabatic part $h_e$ contains the electrons which are driven unstable by the geometric characteristics, whereas the distribution function $g_e$ also contains the contribution of the electrons that follow the perturbed electrostatic potential $\varphi$. Therefore, the geometric characteristics, such as the magnetic wells and regions of bad curvature, mostly excite trapped electrons, hence $\xi(h_e)$ is close to unity (red lines).
Finally, looking at the full distribution function of the electrons (orange lines), it is clear that for flat density profiles the electrons are mostly passing, $\xi(g_e) < f_\text{trapped}$. This is because the electron dynamics do not play a notable role at small density gradients (figure \ref{fig:ratios}) and hence the majority of the electrons are passing since they respond to the electrostatic potential determined by the ion dynamics. As the density gradient is increased, and the electron dynamics start playing a significant role, the majority of the electrons become trapped, $\xi(g_e) > f_\text{trapped}$, in CBC, AUG, TJ-II, LHD and NCSX where the trapped-electron mode is believed to become dominant.

\subsection{The case of W7-X: trapped-electron modes and passing-particle-driven universal instabilities}

The amount of trapped and passing particles that are driving the turbulence in W7-X is very different. Specifically, $\xi(h_e)$ is significantly lower in \mbox{W7-X} compared to the other five devices (red lines in \mbox{figure \ref{fig:trappedratios}}), and it depends strongly on the density gradient.  In particular, for small density gradients, the particles that are driving the instability become increasingly trapped as the role of the electron dynamics increases (figures \ref{fig:ratios}b \mbox{and \ref{fig:ratios}c}), similarly as in the other devices.
On the other hand, for \mbox{$a/L_n \geq 2.5$}, $\xi(h_e)$  decreases significantly, indicating that the turbulence is driven by both trapped and passing electrons. Therefore, in \mbox{W7-X, the} density-gradient-driven modes likely consist of trapped-electron modes, as well as another instability that is driven by passing electrons, which is most likely the passing-particle-driven \mbox{universal instability \cite{costello2023universal, podavini2023electrostatic}}. This is supported by figure \ref{fig:hmuvpa_W7X}f which shows that the non-adiabatic part of the distribution function consists of a trapped particle contribution coming from the central magnetic well (since the cone is very narrow) with the addition of an isotropic contribution which is likely originating from the passing-particle-driven universal instability. Moreover, this instability explains the large amount of heat loss across the entire field line for \mbox{$a/L_n \geq 2.5$} in W7-X (figure \ref{fig:qfluxvsz}), since passing particles explore the entire geometry. Comparing with the results in \cite{costello2023universal} this instability likely corresponds to instability (F) in the linear simulations shown in figure \ref{fig:gamma_map_line}.
Therefore, it is likely that in W7-X the trapped-electron mode co-exists with the passing-particle-driven universal instability at high density gradients. In the other five devices, the passing-particle-driven universal instability is probably absent, or subdominant to the trapped-electron mode. The detailed analysis of the distribution function thus gives valuable insights into the microinstabilities which \mbox{are driving the turbulence.}

 \vspace{-1mm}
\section{Conclusions}
\label{sec:conclusions}

The effect of the density gradient on the turbulent heat transport is investigated by carrying out a comprehensive comparative study of different devices, including the  W7-X, LHD, TJ-II and NCSX stellarators, which represent helias, heliotron, heliac and quasi-axisymmetric configurations, respectively, as well as the ASDEX Upgrade (AUG) tokamak and the tokamak geometry of the Cyclone Base Case (CBC). Both linear and nonlinear simulations covering a wide range of parameters are performed, examining the effect of density gradients, as well as the presence of electrons with and without a temperature gradient. Moreover, a detailed study of the geometric characteristics---magnetic wells, curvature and shear---that drive the instabilities is conducted, and the contributions of trapped and passing particles to the turbulence  are identified.

A detailed study of the linear growth rates along $(k_y\rho_i, a/L_n)$ reveals that many distinct instability branches co-exist at ion-scales in \mbox{W7-X}, while only two branches co-exist in NCSX, which can have an important effect on the nonlinear dynamics, hence one has to be careful when studying only the most unstable mode along $k_y\rho_i$. On the other hand, in the other devices a continuous transition between different microinstabilities has been observed for most of the considered density gradients, ion temperature gradients, and wavenumbers, hence the identification of the instabilities should be done very carefully or be avoided altogether. It has been shown that the growth rates increase with increasing density gradients in CBC, AUG, TJ-II and LHD, and to lesser extent in NCSX, while a moderate decrease of the growth rate is found between $a/L_n = 1.25$ and $a/L_n = 2.75$ in \mbox{W7-X}. Moreover, in NCSX and W7-X the growth rates of the density-gradient-driven modes exceeds those driven by the ion-temperature-gradient-driven modes around \mbox{$a/L_n \approx 2.0$}, whereas this occurs around $a/L_n \approx 1.0$ for CBC, AUG, TJ-II and LHD. Therefore, in NCSX and W7-X there is a region of density gradients for which the ion-temperature-gradient-driven modes are stabilized, without driving the density-gradient-driven modes particularly unstable yet. It is important to note that these linear studies have only taken into account the fastest-growing mode, even though subdominant modes have been shown to have a significant impact on the nonlinear state \cite{faber2018stellarator, mckinney2019comparison}, which is beyond the scope of this work.

The analysis is continued by performing nonlinear simulations covering a wide range of density gradients, $0.0 \leq a/L_n \leq 4.0$, considering a fixed \mbox{ion temperature} gradient of $a/L_{T_i}=3.0$ and taken into account both vanishing and finite electron temperature gradients. In NCSX and \mbox{W7-X}, a strong reduction of the ion heat flux with increasing density gradients is observed. This reduction is also evident in TJ-II for \mbox{$a/L_n \leq 1.0$}, whereas in LHD, the ion heat flux is nearly unaffected by the density gradient.  In contrast, in the tokamaks, the ion heat flux increases strongly with the density gradient. Notably, the ion heat fluxes for the stellarators do not correlate with the growth rates of the fastest-growing modes obtained in the linear analysis. This indicates that underlying nonlinear saturation mechanisms, such as the nonlinear transfer of energy from unstable to damped modes \cite{hatch2011role, terry2006role, hegna2018theory, terry2018saturation} and the effects of zonal flow shearing and energy cascades to dissipative scales \cite{terry2000suppression, diamond2005zonal, 2015-Plunk-Understanding}, play a crucial role on the turbulence.
Nonetheless, the strong reduction of the ion heat flux in \mbox{W7-X} and NCSX is partly attributed  to the stabilization of the ion-temperature-gradient-driven modes by the density gradient, without significantly destabilizing the \mbox{density-gradient-driven modes.} Moreover, NCSX and W7-X benefit from the fact that the ion-temperature-gradient-driven modes are only moderately or not at all destabilized by kinetic electrons, while the density-gradient-driven modes are slightly stabilized in the presence of an ion temperature gradient. This is in contrast to the other four devices where these modes are significantly destabilized by kinetic electrons and ion temperature gradients. Furthermore, in NCSX, W7-X and TJ-II the reduction of the ion heat flux can be attributed in part to the fact that the size of the perpendicular fluctuations diminishes as the density gradient increases.
Lastly, more light is shed on the nature of the microinstabilities by investigating the contributions of the trapped and passing particles to the turbulence. This reveals that trapped-electron modes are dominant at high density gradients in CBC, AUG, TJ-II, LHD and NCSX, while the passing-particle-driven universal instability drives a large part of the turbulence in W7-X.

It should be noted that the heat fluxes obtained in this chapter are generally higher than those typically observed in experiments. In tokamaks, it is well established that density peaking alone cannot account for the enhanced performance of pellet-fueled discharges, and it is essential to consider additional stabilizing mechanisms, such as $E\times B$ shearing, reversed magnetic shear and high-collisionality \cite{hugon1992shear, smeulders1995survey, maget1999drift, baylor2000improved, romanelli2004effects}. In contrast, in stellarators, density peaking alone may explain the reduction in turbulence observed in devices like W7-X, NCSX, and TJ-II. This is due to the favorable properties of density-gradient-driven modes in these devices, along with enhanced nonlinear saturation mechanisms. Nonetheless, other stabilization mechanisms, such as the effects of ambipolar electric fields \cite{xanthopoulos2020turbulence, fu2021global} and collisions \cite{kauffmann2012including} are likely required to fully align with experimental observations.

\chapter{Turbulent particle transport in stellarators and its role in preventing core particle depletion}
\label{chap:particles}
\vspace{-2mm}

It has long been known that, in the core of stellarator plasmas, neoclassical theory predicts an outward particle flux that is unmatched by the typically very small central \mbox{particle source \cite{Maassberg_ppcf_1999}.} This imbalance would result in strongly hollow density profiles and core particle depletion, posing a serious challenge to the feasibility of the stellarator as a fusion reactor concept. However, experimental observations contradict these neoclassical predictions.  In particular, in \mbox{Wendelstein 7-X} (W7-X)---the first large, optimized stellarator---density profiles are generally flat or only weakly peaked \cite{Wolf_nf_2017}, suggesting that neoclassical theory alone cannot account for the observed profiles, since an inward contribution to the particle flux is missing in the core. In this chapter, an extensive number of nonlinear gyrokinetic simulations are performed with the gyrokinetic code \mbox{\texttt{stella} \cite{stella}} to demonstrate that turbulence provides this \mbox{missing inward contribution.}

The results presented in this chapter closely follow those recently published in Ref.~\cite{thienpondt2023prevention} as well as some complementary results presented in Ref.~\cite{alonso2024density}, as indicated in the ``Published and submitted content'' declaration on page~\pageref{page:content}.
First, the magnetic configurations examined in this chapter are discussed in section \ref{sec:MC}. Next, section \ref{sec:neo} demonstrates through analytical and numerical arguments the prediction of a very strong inward neoclassical particle flux in the core of large stellarators. Despite this robust neoclassical prediction, flat or weakly peaked density profiles have generally been measured in e.g.~ECRH plasmas of \mbox{W7-X \cite{Wolf_nf_2017}}. This disagreement between neoclassical theory and experiments indicates that a significant inward contribution to the particle flux has not yet been identified. Therefore, in section \ref{sec:parameterscans} extensive parameter scans are performed for the W7-X and TJ-II stellarators, demonstrating that the turbulence resulting from the onset of the ion and electron temperature gradients produces an inward pinch in broad regions of parameter space. This behavior is not unique to these devices; as shown in section \ref{sec:heatparticles}, similar inward pinches are observed in a range of other stellarator and tokamak configurations.
Additionally, the phase shift between the density and potential fluctuations is analyzed in \mbox{section \ref{sec:phaseshift}}, as it is an experimentally accessible quantity that provides insight into the direction of the particle flux.
Finally, a specific W7-X discharge is studied in section \ref{sec:w7xprofiles}, performing turbulence simulations throughout the plasma radius. The resulting profile of the turbulent particle flux is compared with the shortfall in the flux inferred from a careful particle balance analysis that includes estimates of the particle source and computations of the neoclassical particle flux. It is found that the sign of the missing contribution to the particle flux needed to sustain the experimental density profile agrees with that of the calculated turbulent particle flux, not only at the core \mbox{but also at the edge of the plasma.}

\vspace{-2mm}
\section{Magnetic configurations}
\label{sec:MC}
\vspace{-1mm}

In section \ref{sec:parameterscans} extensive parameter scans are performed for W7-X and TJ-II. The standard EIM-003 and 100\_44\_64 configurations are considered for W7-X and TJ-II, respectively, which have been previously examined in chapter \ref{chap:heat}. However, in contrast to the previous simulations, which are performed towards the edge of the plasma at $r/a=0.7$, the scans in this chapter are performed towards the plasma core at $r/a=0.25$. On the other hand, in section \ref{sec:w7xprofiles} a specific discharge (\#20180920.017) of the W7-X stellarator is examined. While this discharge is also performed using the standard configuration of W7-X, there are small differences with the W7-X configuration examined previously, as this discharge was performed using the standard EIM-169 configuration of W7-X. Nonetheless, the differences between these two standard \mbox{configurations are minimal.}

The resolutions used for the simulations performed in this chapter are given in table \ref{tab:resolutionParticleFlux}. Note that it is important to extend $k_{y,\text{max}}\rho_i$ sufficiently in order to accurately capture the particle flux. A small amount of hyper-viscosity, with D = 0.1, was applied in the simulations presented in this chapter. The form of the hyper-viscosity used in stella is given by equation (53) in \cite{stella}.

\begin{table}[h!]
	\centering
	\vspace{1mm}
	\caption{\label{tab:resolutionParticleFlux} Resolution used for the nonlinear simulations of chapter \ref{chap:particles}. Approximately 1 poloidal turn is used to ensure that $\Delta k_x = \Delta k_y$. The exact flux tube length is given in table \ref{tab:geometryParticleFlux}. The Fourier grid corresponds to a computational domain of $L_x = L_y = 94\rho_i$ with $N_x = 91$ and $N_y = 271$ grid \mbox{points in real space.}} \vspace{-2.5mm}
	\lineup

	\begin{tabular}{@{}ccccccccccc}
		\specialrule{.12em}{.1em}{.1em} \\[-4mm]
		& $N_{\rm pol}$ & $N_z$ & $k_x\rho_i$ & $k_y\rho_i$ & $N_{k_x}$ & $N_{k_y}$ & $v_\parallel/v_{{\rm th},s}$ & $v_\perp/v_{{\rm th},s}$ & \!$N_{v_\parallel}$\! & \!$N_{v_\perp}$\!  \\[1mm] \hline\\[-4mm]
		\hspace{0mm} & \hspace{8mm} & \hspace{8mm} & \hspace{8mm} & \hspace{8mm} & \hspace{8mm} & \hspace{8mm} & \hspace{17mm} & \hspace{17mm} & \hspace{8mm} & \hspace{8mm} \\[-4mm]
		& $\sim$1 & 49 & \![0.067, 2.00]\! & \![0.067, 6.00]\! & 61 & 91 & \![$-$3.0, 3.0]\! & \![0.0, 3.0]\! & 48 & 12  \\[1mm]
		\specialrule{.12em}{.1em}{.1em}
	\end{tabular}
	\vspace{-1mm}
\end{table}

\subsection{W7-X (EIM-003) and TJ-II (100\_44\_64) configurations}

\begin{figure}[h!]
	\centering
	\vspace{-3mm}
	\begin{subfigure}{0.5\linewidth}
		\includegraphics[width=0.95\linewidth, trim={180mm 95mm 170mm 93mm}, clip] {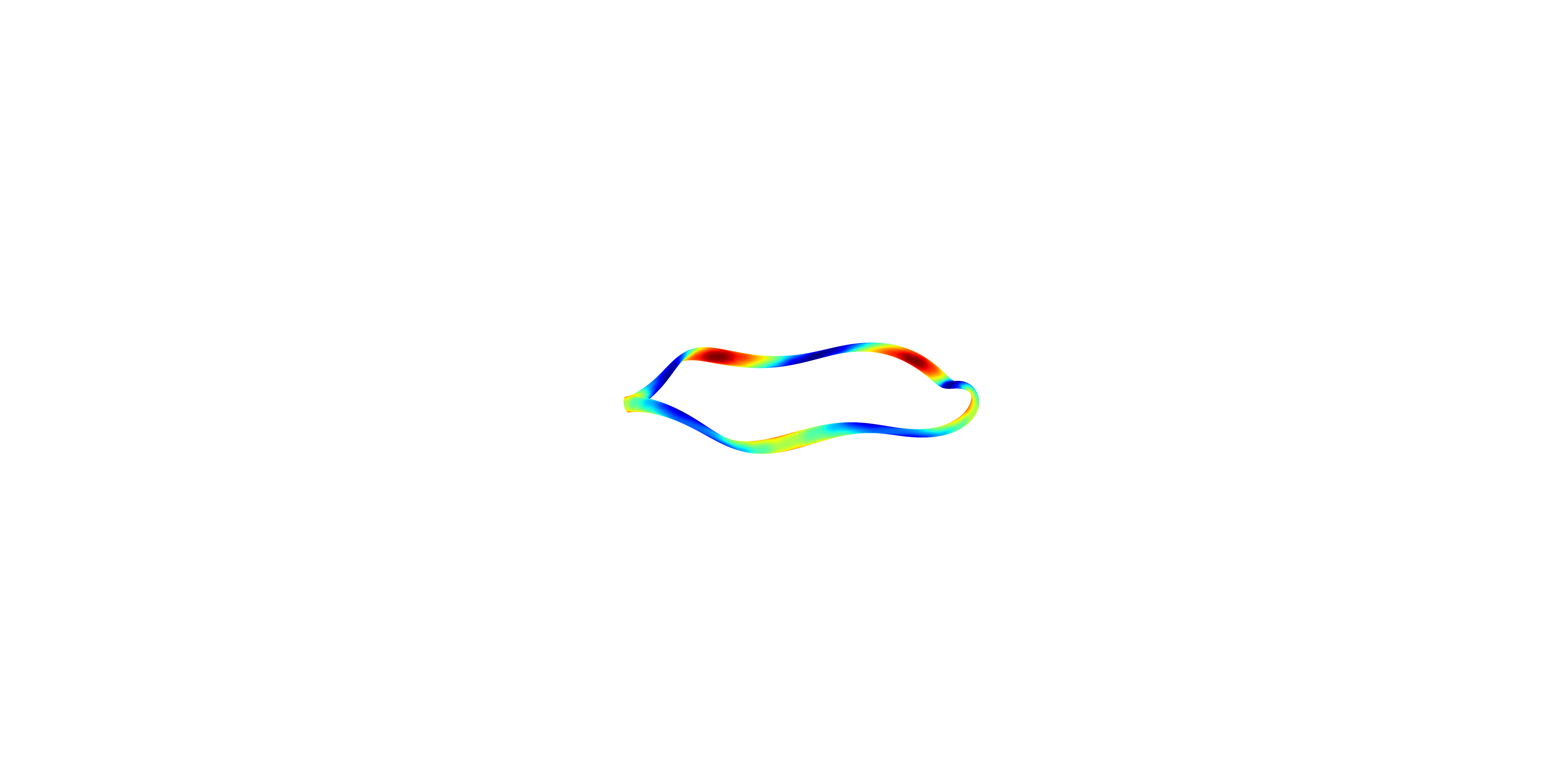}
		\caption{W7-X (EIM-003)}
	\end{subfigure}%
	\begin{subfigure}{0.5\linewidth}
		\includegraphics[width=0.95\linewidth, trim={180mm 95mm 170mm 93mm}, clip] {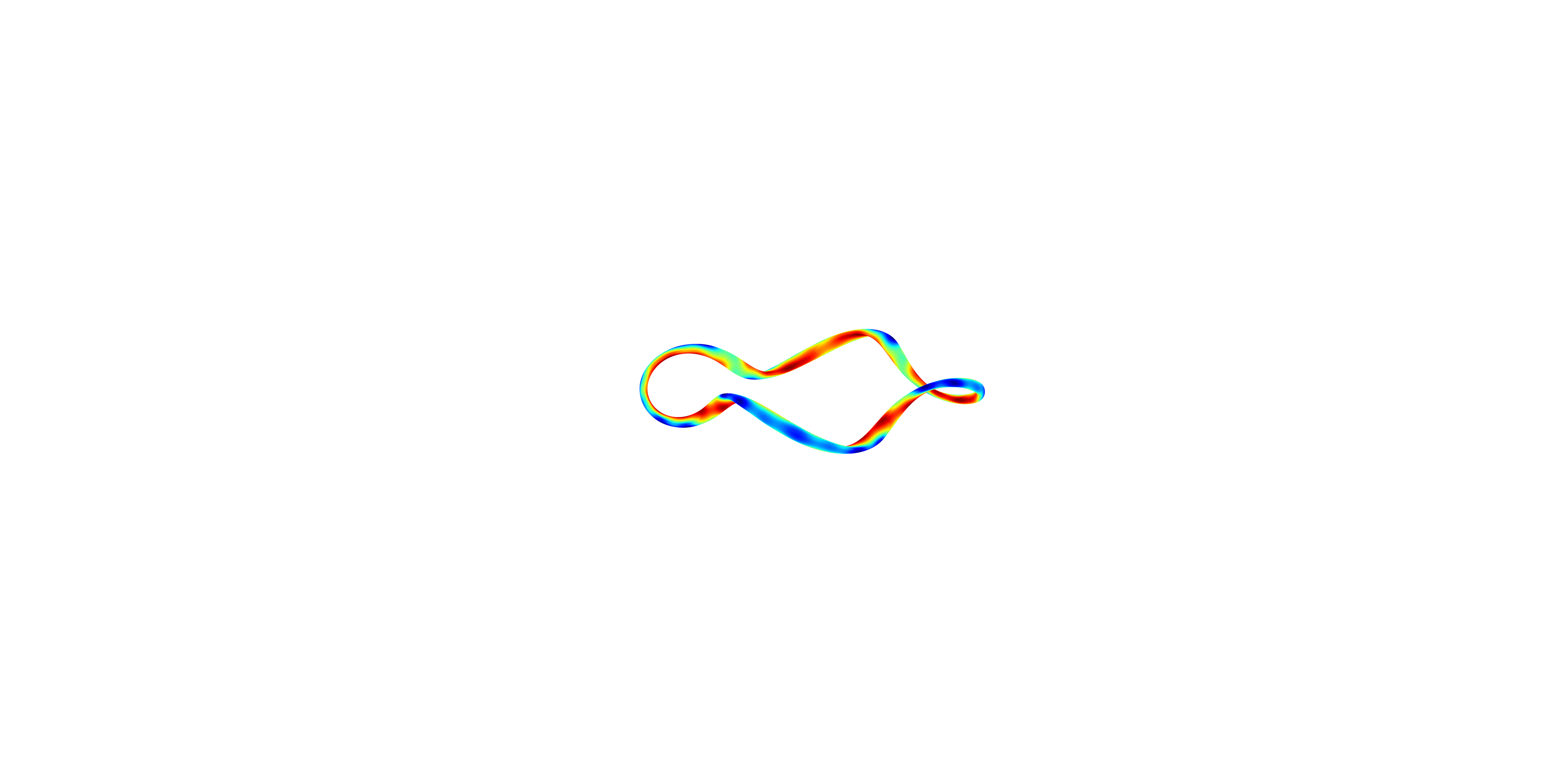}
		\caption{TJ-II (100\_44\_64)}
	\end{subfigure}
	\vspace{-6mm}
	\caption{Magnetic field strength---minimum is blue, maximum is red---on the flux surface at $r/a = 0.25$ for (a) the standard EIM-003 configuration of W7-X and (b) the standard 100\_44\_64 configuration of TJ-II. }
	\vspace{-2mm}
	\label{fig:FFSW7XTJII}
\end{figure}

\begin{figure}[h!]
	\centering
	\vspace{0mm}
	\includegraphics[width=\linewidth, trim={0mm  1mm 0mm 2mm}] {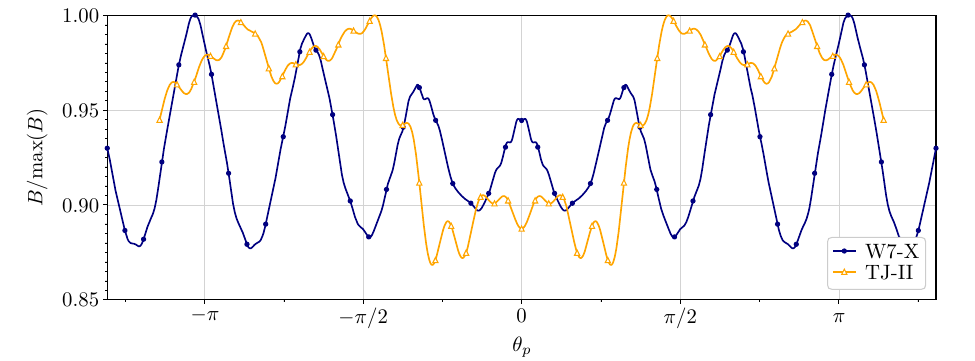}  \vspace{-6mm}
	\caption{Magnetic field strength along the $\alpha_0=0$ field line for the standard EIM-003 configuration of the W7-X stellarator (blue) and the standard 100\_44\_64 configuration of the TJ-II stellarator (gold). \vspace{-6mm}}
	\label{fig:BmagW7XTJII}
\end{figure}

The parameter scans presented in section \ref{sec:parameterscans} are conducted for the standard EIM-003 configuration of W7-X and the standard 100\_44\_64 configuration of TJ-II, towards the plasma core at $r/a=0.25$. The corresponding flux surfaces are depicted in figure \ref{fig:FFSW7XTJII}.
The simulations are performed around the $\alpha_0=0$ field line centered at $(\theta,\zeta)=(0,0)$. The magnetic field strength along this field line is illustrated in figure \ref{fig:BmagW7XTJII}. The markers represent the parallel grid points considered in the nonlinear simulations in section \ref{sec:parameterscans}. The magnetic field strength differs slightly from that shown in figure \ref{fig:fieldlineAngle}a, as these field line are located closer to the magnetic axis. An overview of the geometric characteristics of these field lines is given in table \ref{tab:geometryParticleFlux2}.

\begin{table}[h!]
	\vspace{2mm}
	\caption{\label{tab:geometryParticleFlux2} Geometric quantities for the standard EIM-003 configuration of W7-X and the standard 100\_44\_64 configuration of TJ-II at $r/a=0.25$ considered in the nonlinear simulations performed in section \ref{sec:parameterscans}. } \vspace{-3mm}
	\lineup
	\centering
	\begin{tabular}{@{}cccccccc}
		\specialrule{.12em}{.1em}{.1em} &&&&&&& \\[-4mm]
		& $\0a$\,(m) & $\0R_0$\,(m)  & $\hat{s}_0$ &$\iota_0$ & $B_\text{max}/B_\text{min}$ & $N_{f}$ & $N_{\rm pol}$\ \\[1mm] \hline &&&&&& \\[-4mm]
		\hspace{15mm} & \hspace{15.5mm}  & \hspace{15.5mm} &
		\hspace{17mm} & \hspace{17mm} & \hspace{15.5mm} & \hspace{15.5mm} & \hspace{15.5mm} \\[-3.5mm]
		W7-X & \m 0.51 & \m 5.51 & -0.007 &  0.859 & 1.14 & 5 & 1.31 \\[0.5mm]
		TJ-II & \m 0.19 & \m 1.50 & -0.006 & -1.557 & 1.15 & 4 & 1.14 \\[0.5mm]
		\specialrule{.12em}{.1em}{.1em}
	\end{tabular}
	\vspace{0mm}
\end{table}
\FloatBarrier

\begin{table}[t!]
	\vspace{2mm}
	\caption{\label{tab:geometryParticleFlux} Geometric quantities for the standard EIM-169 configuration of W7-X, evaluated at different radial locations, considered in the nonlinear simulations performed in section \ref{sec:w7xprofiles}. } \vspace{-3mm}
	\lineup
	\centering
	\begin{tabular}{@{}cccccccc}
		\specialrule{.12em}{.1em}{.1em} &&&&&&& \\[-4mm]
		$r/a$ & $\0a$\,(m) & $\0R_0$\,(m)  & $\hat{s}_0$ &$\iota_0$ & $B_\text{max}/B_\text{min}$ & $N_{f}$ & $N_{\rm pol}$\ \\[1mm] \hline &&&&&& \\[-4mm]
		\hspace{15mm} & \hspace{15.5mm}  & \hspace{15.5mm} &
		\hspace{17mm} & \hspace{17mm} & \hspace{15.5mm} & \hspace{15.5mm} & \hspace{15.5mm} \\[-3.5mm]
		0.2 & \m 0.52 & \m 5.52 & -0.003 & 0.862 & 1.11 & \05 & 0.97 \\[0.5mm]
		0.3 & \m 0.52 & \m 5.52 & -0.010 & 0.864 & 1.13 & \05 & 0.98 \\[0.5mm]
		0.4 & \m 0.52 & \m 5.52 & -0.021 & 0.867 & 1.15 & \05 & 1.00 \\[0.5mm]
		0.5 & \m 0.52 & \m 5.52 & -0.041 & 0.873 & 1.17 & \05 & 1.03 \\[0.5mm]
		0.6& \m 0.52 & \m 5.52 & -0.074 & 0.882 & 1.20 & \05 & 1.06 \\[0.5mm]
		0.7 & \m 0.52 & \m 5.52 & -0.124 & 0.895  & 1.23 & \05 & 1.11 \\[0.5mm]
		\specialrule{.12em}{.1em}{.1em}
	\end{tabular}
	\vspace{0mm}
\end{table}
\FloatBarrier

\subsection{Standard EIM-169 configuration of W7-X}
\label{sec:geo:particleflux}

\begin{figure}[h!]
	\centering
	\vspace{4mm}
	\includegraphics[width=\linewidth, trim={0mm  1mm 0mm 2mm}] {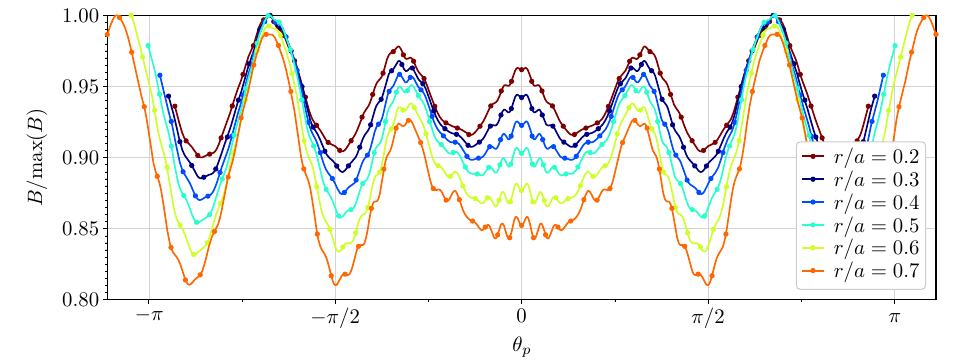}  \vspace{-5mm}
	\caption{Magnetic field strength along the $\alpha_0=0$ field line for the standard EIM-169 configuration of the W7-X stellarator, considering different radial locations between $r/a=0.2$ and $r/a=0.7$.}
	\label{fig:BmagW7XVsRho}
\end{figure}

In section \ref{sec:w7xprofiles} a specific discharge (\#20180920.017) of the W7-X stellarator is examined. This discharge has been performed using the standard EIM-169 configuration of W7-X, which is very similar to the standard EIM-003 configuration discussed in the previous section. The particle transport is examined from $r/a=0.2$ up to $r/a=0.7$. Simulations deeper inside the plasma, for $r/a<0.2$, are challenging due to convergence issues of the \texttt{VMEC} magnetic equilibria towards the magnetic axis. At the other end, for $r/a>0.7$, steep plasma gradients emerge, requiring specialized codes to accurately model the edge physics. The magnetic field strength along the $\alpha_0=0$ field lines centered at $(\theta,\zeta)=(0,0)$ is depicted in figure \ref{fig:BmagW7XVsRho} for radial locations between $r/a=0.2$ and $r/a=0.7$. Finally, the geometric quantities along the field lines for each radial location are summarized in table \ref{tab:geometryParticleFlux}. The length of the field line is chosen to be approximately one poloidal turn, to ensure that $\Delta k_x = \Delta k_y$ when using the stellarator-symmetric boundary conditions.

\section{Neoclassical particle transport in large stellarators}
\label{sec:neo}

Neoclassical transport---associated with the inhomogeneity of the magnetic field and particle collisions---has traditionally handicapped the performance of stellarators. For this reason, its minimization is one of the main targets in stellarator optimization. W7-X  has demonstrated that large stellarators with optimized neoclassical transport can be designed \cite{Grieger_fst_1992} and built with high accuracy \cite{Pedersen_nature_2016} and that this has been fundamental for achieving triple product record values \cite{Beidler_nature_2021}.
However, even in optimized stellarators, neoclassical transport may play a major role in the plasma core \cite{Dinklage_nf_2013}. In this region, high temperatures are achieved, causing strong neoclassical particle and energy fluxes due to their unfavourable scaling with the temperature in low collisionality plasma regimes. For example, in the $1/\nu$ regime, the neoclassical particle and energy fluxes scale, respectively, as $\smash{T_s^{7/2}}$ and $\smash{T_s^{9/2}}$, with $T_s$ the temperature of species $s$ (see e.g.~Ref.~\cite{Beidler_nature_2021}). In what follows, we only deal with plasmas consisting of singly charged ions and electrons, so that $s$ can take the values $s = i$ and $s=e$, respectively.

Stellarator plasmas with core particle transport dominated by neoclassical processes have been predicted to present problems of core particle depletion \cite{Maassberg_ppcf_1999, Beidler_ppcf_2018}.
If the core particle source is negligible, as is the case in large stellarators with peripheral fueling, the steady-state particle balance equation for each of the species reads $\Gamma_s=\Gamma^{\mathrm{neo}}_s+\Gamma^{\mathrm{turb}}_s=0$. Here, $\Gamma_s$ is the flux-surface-averaged radial particle flux, and it is assumed to be the sum of a neoclassical ($\Gamma^{\mathrm{neo}}_s$) and a turbulent contribution ($\Gamma^{\mathrm{turb}}_s$). Assuming $\Gamma^{\mathrm{turb}}_s=0$, the above steady-state condition reduces to $\Gamma^{\mathrm{neo}}_i=\Gamma^{\mathrm{neo}}_e=0$. This relation, which additionally implies that no net radial current exists in the plasma, i.e.~that the particle fluxes are ambipolar, imposes a constraint on the plasma profiles. The neoclassical particle flux can be written as,
\begin{equation}
	\frac{{\Gamma_s^{\mathrm{neo}}}}{n_s}= - L_{11}^s\left(\frac{1}{n_s}\frac{\text{d}n_{s}}{\text{d}r}- \frac{Z_seE_r}{T_s}+\delta_{12}^s\frac{1}{T_s}\frac{\text{d}T_{s}}{\text{d}r}\right),
	\label{EQ_GNC}
\end{equation}
where $r\in[0,a]$ is the effective minor radius, $a$ is the minor radius, $E_r$ is the radial electric field, $n_s$ is the density and $Z_se$ is the electric charge ($Z_i = 1$ for ions and $Z_e = -1$ for electrons). Quasineutrality implies $n_i=n_e$, and its radial derivative implies $\text{d}n_i/\text{d}r = \text{d}n_e/\text{d}r$. Equation (\ref{EQ_GNC}) is valid at low collisionality and has been discussed in detail in Ref.~\cite{Beidler_nature_2021} and references therein. The neoclassical transport coefficients $L_{11}^s$ and $\delta_{12}^s$ generally depend on the magnetic configuration and plasma parameters, including $E_r$. However, if the electrons and main ions are assumed to be, respectively, in the neoclassical $1/\nu$ and $\sqrt{\nu}$ collisionality regimes---the relevant regimes in the core of reactor-relevant stellarator plasmas---we can approximate $\delta_{12}^e=7/2$ and $\delta_{12} ^i=5/4$. After imposing $\Gamma^{\mathrm{neo}}_i=\Gamma^{\mathrm{neo}}_e=0$, we arrive at a constraint on the plasma \mbox{profiles \cite{Beidler_ppcf_2018}. Namely,}
\begin{equation}
	\frac{1}{n_e}\frac{\text{d}n_{e}}{\text{d}r}=
	-\frac{7/2}{T_e+T_i}\frac{\text{d}T_{e}}{\text{d}r} -\frac{5/4}{T_e+T_i}\frac{\text{d}T_{i}}{\text{d}r}\,.
	\label{EQ_DN}
\end{equation}\vskip-0.0cm

Reactor-relevant plasmas are expected to display peaked temperature profiles, i.e.~$\text{d}T_{s}/\text{d}r<0$, with $T_e\approx T_i$. Since $\delta_{12}^s>0$, this automatically leads to a hollow density profile, i.e.~$\text{d}n_{e}/\text{d}r>0$. Moreover, $\delta_{12}^s>1$ because the so-called thermodiffusion coefficient $L_{11}^s\delta_{12}^s$ is relatively large. As a result, the hollowness can become large enough to make the pressure gradient positive.

\begin{figure}[h!]
	\centering
	\includegraphics[trim={0mm 0mm 0mm 0mm},clip] {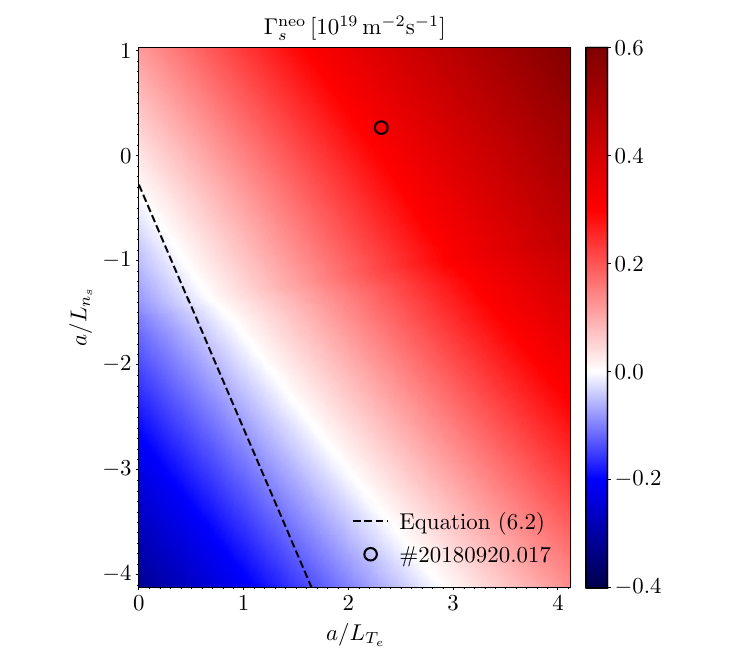}
	\vspace{-3mm}
	\caption{Neoclassical particle flux as a function of $a/L_{T_e}$ and $a/L_{n_s}$ for the standard EIM-169 configuration of W7-X. Here $r/a=0.25$, $n_e\!=\!n_i\!=6.3 \times 10^{19}\,$m$^{-3}$, $T_e=2.2\,$keV, $T_i=1.1\,$keV and $a/L_{T_i}=0.81$. The circle corresponds to the experimental discharge at $r/a=0.25$ analyzed in section \ref{sec:w7xprofiles}. The dashed line shows the combination of $a/L_{T_e}$ and $a/L_{n_s}$ that gives $\Gamma^{\mathrm{neo}}_s=0$ in equation (\ref{EQ_DN}). }
	\vspace{0mm}
	\label{fig:knosos_2d_scan}
\end{figure}

The above analytical argument illustrates the generality of the mechanism of core depletion by neoclassical transport. Of course, more accurate relations between the stationary density and temperature gradients can be obtained by means of numerical simulations. Figure \ref{fig:knosos_2d_scan} shows the neoclassical radial particle flux, $\Gamma_s^{\mathrm{neo}}$, obtained with the neoclassical code \texttt{KNOSOS} \cite{Velasco_nf_2021}, over a wide range of values for the density and electron temperature gradients. Recall that peaked density profiles have $a/L_{n_s} >0$ while hollow density profiles have $a/L_{n_s} < 0$. For the rest of the parameters,
typical values from W7-X hydrogen plasmas sustained by Electron Cyclotron Resonance Heating (ECRH) are employed (see the caption of figure \ref{fig:knosos_2d_scan} and Ref.~\cite{Beurskens_nf_2021}), and $E_r$ is set by the ambipolarity of the neoclassical fluxes.
Outward fluxes ($\Gamma_s^\text{neo} > 0$) are obtained in most of the represented area, driven by $a/L_{T_e}$, indicating the importance of thermodiffusion.

The contour $\Gamma_s^{\mathrm{neo}}=0$ (white line) in figure \ref{fig:knosos_2d_scan} predicts a slightly hollow stationary density profile \mbox{($a/L_{n_s} < 0$)} even in the presence of a practically flat electron temperature profile \mbox{($a/L_{T_e}=0$)}. While for the larger values of $a/L_{T_e}$ obtained in W7-X, a distinctive hollow density profile should be measured if particle transport were described by neoclassical theory only.  The dashed line in figure \ref{fig:knosos_2d_scan} shows the combination of $a/L_{T_e}$ and $a/L_{n_s}$ that gives $\Gamma^{\mathrm{neo}}_s=0$ in equation (\ref{EQ_DN}), which is a good approximation of the $\Gamma_s^{\mathrm{neo}}=0$ contour obtained by \texttt{KNOSOS}.

\begin{figure}[b!]
	\centering
	\vspace{-4mm}
	\includegraphics[trim={0mm 2mm 0mm 1mm},clip] {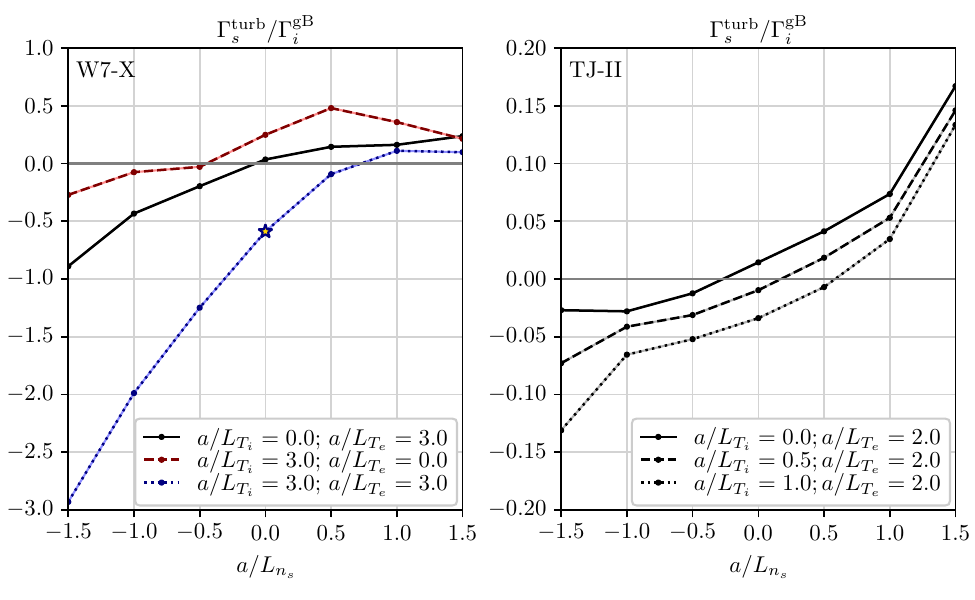} \vspace{-7.5mm}
	\caption{Turbulent radial particle flux as a function of the normalized density gradient for the standard EIM-003 configuration of W7-X (left) and the standard 100\_44\_64 configuration of TJ-II (right). The reference point $\left\{T_e/T_i, a/L_{T_i}, a/L_{T_e}, a/L_{n_s}\right\}=\left\{1, 3, 3, 0\right\}$ for W7-X is highlighted with a golden star.}
	\vspace{-3mm}
	\label{fig:scansW7XTJII}
\end{figure}

\section{Parametric dependence of the turbulent particle flux}
\label{sec:parameterscans}

Let us write the radial turbulent particle flux as the sum of a diffusive and a convective contribution,
\begin{equation}
	\label{eq:DifConv}
	\frac{\Gamma_s^\text{turb}}{n_s} = -D \frac{1}{n_s}\frac{\text{d}n_s}{\text{d}r} + V\,,
\end{equation}
with $D$ the diffusion coefficient (which, in general, depends on the density gradient) and $V$ the convection velocity. As $D$ is always positive, the diffusive term adds a contribution to the flux with opposite direction to that of the density gradient. In other words, the diffusive flux is inwardly or outwardly directed depending on whether the density gradient is hollow or peaked, respectively. On the other hand, $V$ can be both positive or negative, supporting the formation of hollow or peaked density profiles, respectively. More specifically, the sign of $V$ for a flat density profile ($a/L_{n_s}=0$) is of	particular interest, as it predicts whether a hollow (if $V>0$) or peaked (if $V<0$) density profile will develop. In order to gain insight into this matter, and prior to the comparison between numerical and W7-X experimental results, we provide a comprehensive numerical characterization of the turbulent particle flux, with emphasis on the case with a flat density profile, i.e.~$a/L_{n_s}=0$.

Gyrokinetic simulations are performed which are nonlinear, collisionless, electrostatic and account for kinetic ions and electrons. The calculations are performed towards the inner plasma core, at $r/a\!=\!0.25$, for the standard EIM-003 configuration of W7-X and the standard 100\_44\_64 configuration of TJ-II. The influence of the density gradient on the turbulent particle transport is depicted in figure \ref{fig:scansW7XTJII}. In W7-X, the effect of adding temperature gradients of $a/L_{T_i}=3.0$ and $a/L_{T_e}=3.0$ are investigated. It can be observed that the turbulence driven solely by the electron temperature gradient (black line) or solely by the ion temperature gradient (red line) cannot lead to peaked density profiles, as convection is outwards ($V>0$ or $\Gamma_s^\text{turb}>0$) for the flat density profile cases ($a/L_{n_s}=0$). In contrast, when both temperature gradients coexist, depicted by the blue line in figure \ref{fig:scansW7XTJII}, turbulence gives rise to a significant inward convection $(V<0)$, i.e.~$\Gamma_s^\text{turb}/\Gamma_i^\text{gB} \approx -0.6$, at a vanishing density gradient. Moreover, the normalized density gradient at which the particle flux is zero is $a/L_{n_s} \approx 0.7$ when both temperature gradients coexist (blue line). In other words, if all transport were due to turbulence (and for fixed values of the other parameters used in this plot), the equilibrium density profile would be clearly peaked.

On the other hand, for the TJ-II stellarator, the turbulent particle flux is scanned on the right-hand-side of figure \ref{fig:scansW7XTJII} considering a fixed electron temperature gradient of $a/L_{T_e}=2.0$ while the ion temperature gradient is scanned from $a/L_{T_i}=0.0$ to $a/L_{T_i}=1.0$. These choices correspond to the experimental plasma profiles observed in the plasma core of TJ-II which are characterized by moderate electron temperature gradients and vanishing to small ion temperature gradients. For a flat density profile ($a/L_{n_s}=0$), an outward turbulent particle flux is observed if only an electron temperature gradient is considered (black solid line). As the ion temperature gradient increases, the turbulent particle flux turns inwards (black dashed and dotted lines) when considering $a/L_{n_s}=0$.

Moreover, in both W7-X and TJ-II, inward turbulent particle fluxes are observed for very hollow density profiles ($a/L_{n_s}=-1.5$), while outward particle fluxes are obtained in the presence of very peaked density profiles ($a/L_{n_s}=1.5$). This behavior is driven by the diffusive component of the particle flux, as described by equation \ref{eq:DifConv}, which overshadows the convective contribution at large density gradients. Since diffusion acts to flatten density profiles, it results in inward fluxes for hollow density profiles and outward fluxes for peaked density profiles.

\begin{figure}[b!]
	\centering
	\vspace{-5mm}
	\includegraphics[trim={0mm 3mm 0mm 0mm},clip] {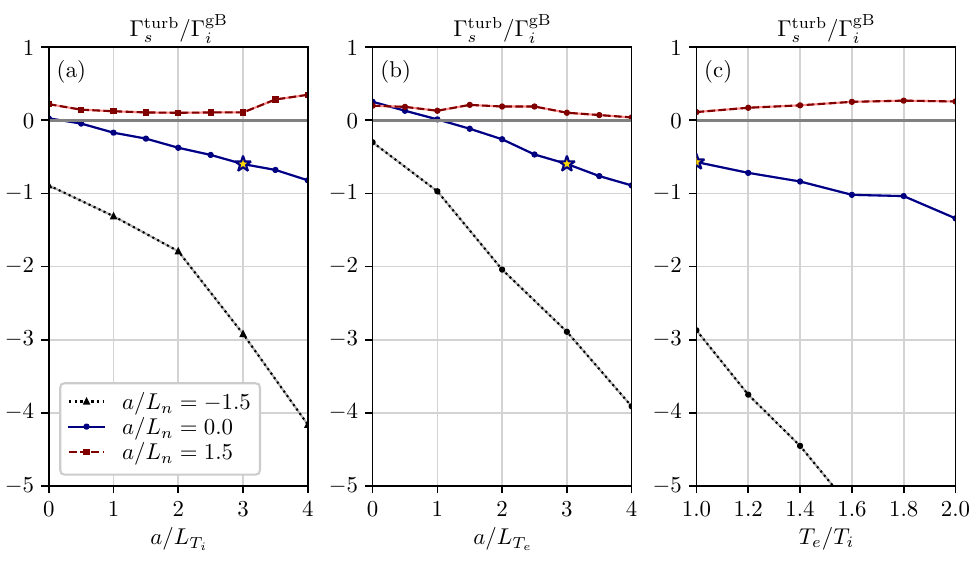} \vspace{-7mm}
	\caption{Turbulent radial particle flux  as a function of the (a) normalized ion temperature gradient, (b) normalized electron temperature gradient, and (c) the electron to ion temperature ratio for the standard EIM-003 configuration of W7-X, performed towards the plasma core at $r/a=0.25$.  The reference point in parameter space $\left\{T_e/T_i, a/L_{T_i}, a/L_{T_e}, a/L_{n_s}\right\}=\left\{1, 3, 3, 0\right\}$ is highlighted with a golden star.}
	\vspace{0mm}
	\label{fig:scansW7X}
\end{figure}

To study the effect of the temperature profiles on the turbulent particle transport in more detail, a parameter scan has been carried out for W7-X, around the reference point in parameter space $\left\{T_e/T_i, a/L_{T_i}, a/L_{T_e}, a/L_{n_s}\right\}=\left\{1, 3, 3, 0\right\}$ which has been highlighted by a golden star in figures  \ref{fig:scansW7XTJII} and \ref{fig:scansW7X}. The dependence of the turbulent particle flux on the following quantities is depicted: the normalized ion temperature gradient $a/L_{T_i}$ in figure \ref{fig:scansW7X}a; the normalized electron temperature gradient $a/L_{T_e}$ in figure \ref{fig:scansW7X}b; and the  electron to ion temperature ratio $T_e/T_i$ in \mbox{figure \ref{fig:scansW7X}c.} Such exhaustive scans cover both reactor-relevant conditions, where ions and electrons are expected to be thermalized and have comparable values of density and temperature gradients, and conditions corresponding to the first W7-X campaigns, where the temperature and its gradient are noticeably larger for electrons than for ions in the core.

First, the case of a flat density profile is discussed, corresponding to the blue lines in \mbox{figure \ref{fig:scansW7X}}. From figure \ref{fig:scansW7X}a it can be observed that the ion temperature gradient, $a/L_{T_i}$, need not be large or comparable to $a/L_{T_e}$ in order to turn the particle flux inwards. The negative convection starts to manifest from values as modest as $a/L_{T_i}\sim 0.25$ when considering $a/L_{T_e}=3$, and its magnitude increases roughly linearly with the size of $a/L_{T_i}$.
Similarly, figure \ref{fig:scansW7X}b shows that, although the case with only $a/L_{T_i}=3$ (and $a/L_{T_e}=0$) drives positive flux, adding an electron temperature gradient as moderate as $a/L_{T_e}\gtrsim 1.0$, turns the particle flux inward. Furthermore, figure \ref{fig:scansW7X}c shows that a larger electron to ion temperature ratio enhances the particle pinch.
In contrast, when considering a peaked density profile (red lines), the particle flux remains outward regardless of the scanned temperature gradients and temperature ratios, because the diffusive contribution to the particle flux is dominant. Similarly, for hollow density profiles (black lines) the particle flux is always directed inward for the parameters considered in these scans. Moreover, larger temperature gradients increase the magnitude of the particle flux significantly for $a/L_n=-1.5$ (black line) while the particle flux remains mostly unaffected by the temperature profiles \mbox{for $a/L_n=-1.5$ (red line).}

In summary, the existence of an inward turbulent particle flux is very robust and is found over broad regions of the scanned parameter space, whenever the temperature gradient of ions and electrons are both finite and not too small.
\FloatBarrier

\section{Universality of the turbulent inward pinch}
\label{sec:heatparticles}

\begin{figure}[h!]
	\vspace{-2mm}
	\includegraphics[width=0.95\linewidth]{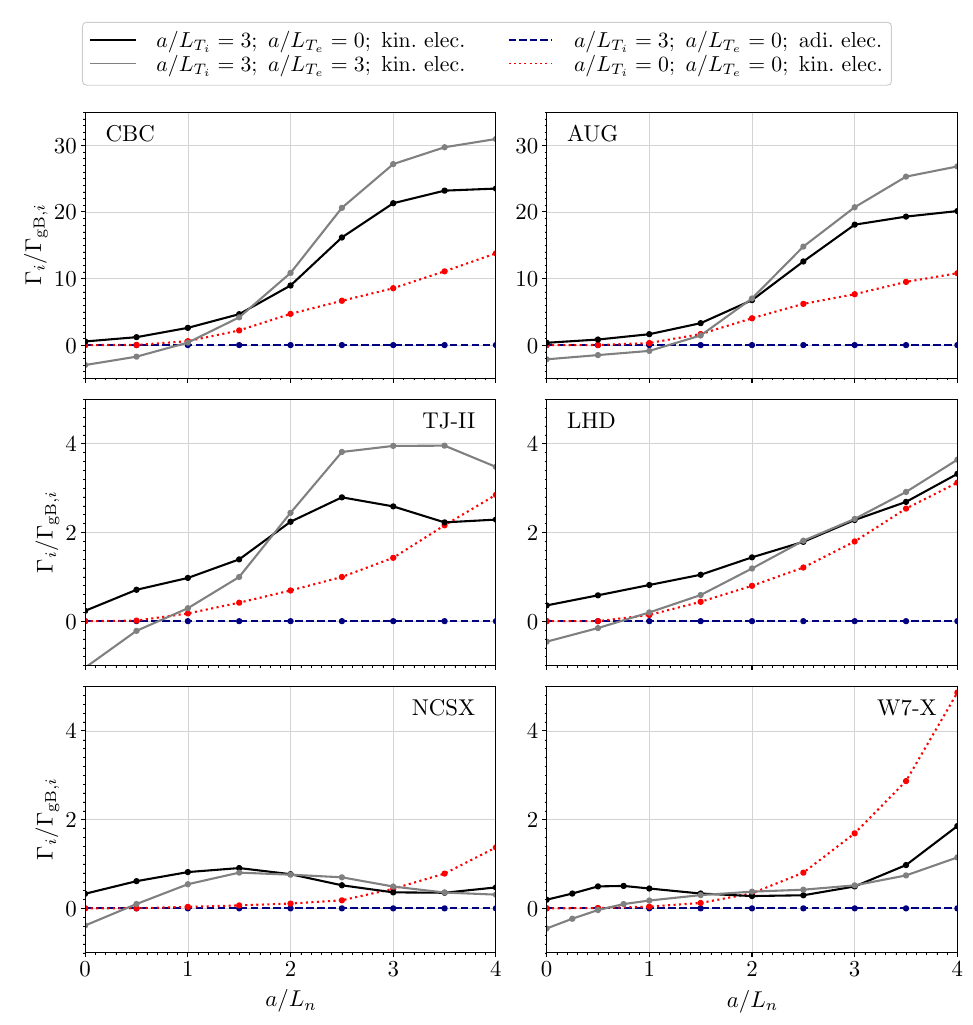}
	\vspace{-4mm}
	\caption{Normalized turbulent particle flux, $\Gamma_i/\Gamma_{\rm gB,i}$, as a function of the normalized density gradient considering kinetic electrons with \mbox{$a/L_{T_e}=0.0$} (black) and  $a/L_{T_e}=3.0$ (gray). Ion-temperature-gradient driven modes are isolated by treating the electrons as adiabatic (blue) and density-gradient-driven modes are isolated by setting $a/L_{T_i} = 0.0$ (red), both considering \mbox{$a/L_{T_e}=0.0$}. }
	\vspace{-4mm}
	\label{fig:pflux}
\end{figure}

The inward turbulent particle pinch observed in W7-X and TJ-II is not unique to these devices. It appears to be a robust and universal feature, largely independent of magnetic geometry as it is also observed in other stellarators, such as LHD and NCSX, as well as in tokamaks like AUG and CBC. This is demonstrated in figure \ref{fig:pflux} where the turbulent particle flux is depicted as a function of the normalized density gradient for the six fusion devices analyzed in \mbox{chapter \ref{chap:heat}}. The data correspond to the simulations presented in figures \ref{fig:qi_contributions} and \ref{fig:qe_contributions}, considering a finite ion temperature gradient of $a/L_{T_i}=3.0$. Interestingly, the particle transport closely mirrors the trend observed for the electron heat flux. It is important to note that when electrons are treated adiabatically (blue dashed lines), there is no radial particle flux. For cases with a flat electron temperature profile (black and red lines), the particle flux is consistently positive, indicating outward particle transport that degrades plasma confinement. In contrast, introducing a finite electron temperature gradient (gray lines) leads to an inward particle flux at low density \mbox{gradients, thereby improving confinement.}

\section{Phase shift between density and potential fluctuations}
\label{sec:phaseshift}

\begin{figure}[h!]
	\centering
	\vspace{4mm}
	\includegraphics[]{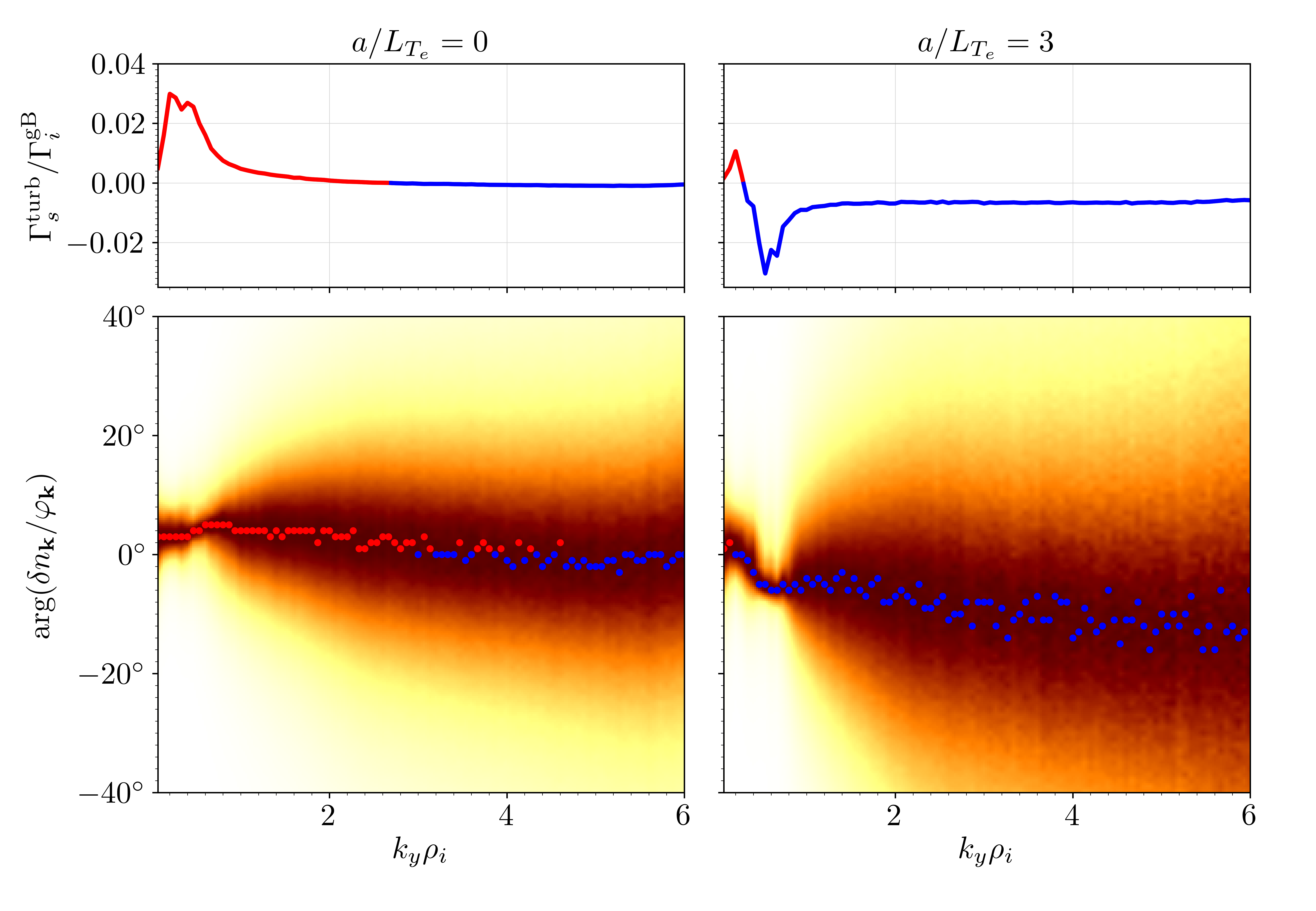}
	\vspace{-11.5mm}
	\caption{Spectra of the particle flux (top) and phase difference between the density ($\delta n_\mathbf{k}$) and electrostatic potential ($\varphi_\mathbf{k}$) fluctuations (bottom), with $a/L_{T_e}=0$ (left) and $a/L_{T_e}=3$ (right) for the standard EIM-003 configuration of W7-X considering $a/L_{T_i}=3.0$ and $a/L_n=0.0$. The dominant phase difference for each $k_y\rho_i$ is denoted by a dot. Positive values are highlighted in red and negative \mbox{values in blue.} }
	\vspace{-2mm}
	\label{fig:phaseshift}
\end{figure}

\begin{figure}[h!]
	\centering
	\includegraphics[]{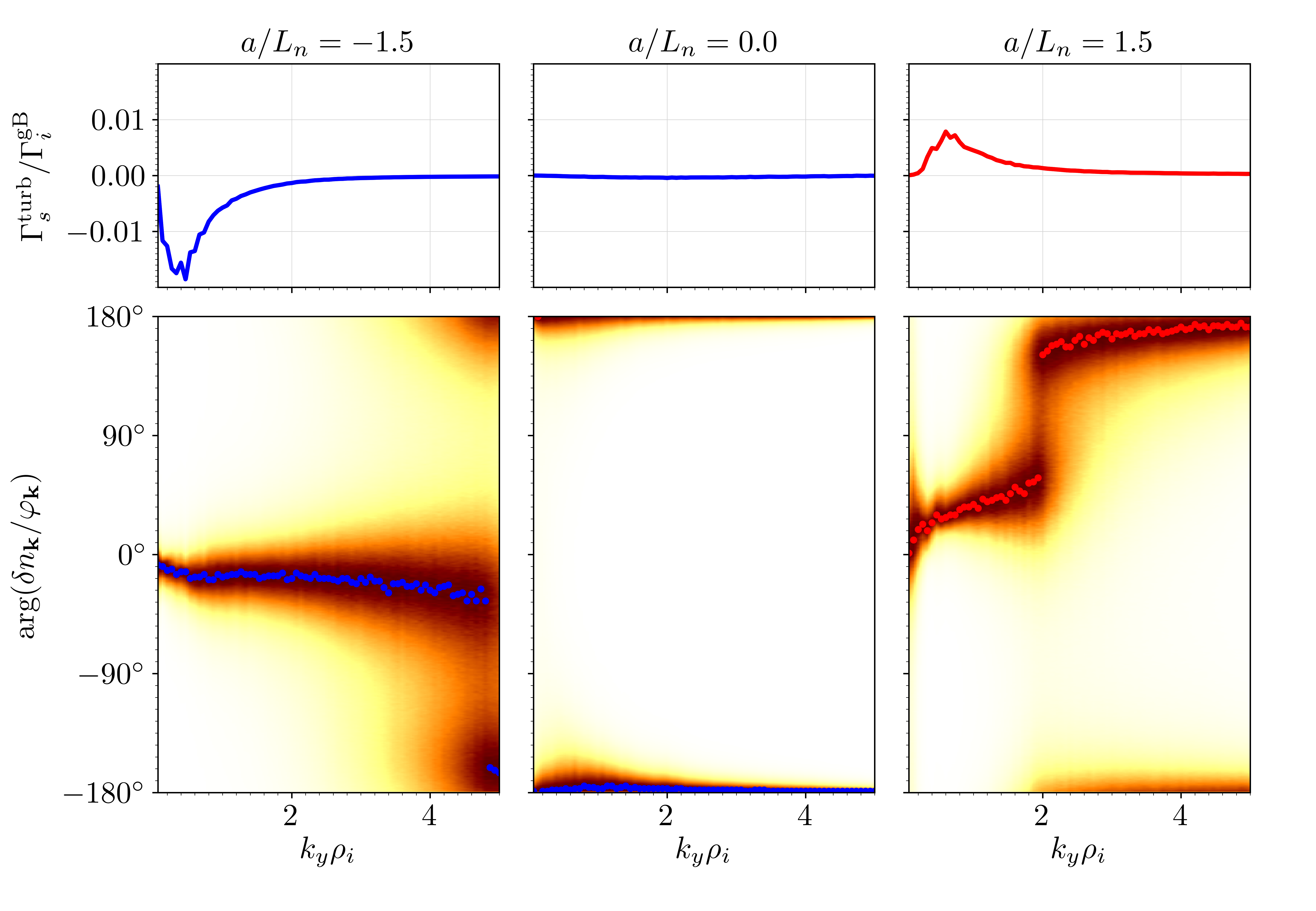}
	\vspace{-11.5mm}
	\caption{Spectra of the particle flux (top) and phase difference between the density ($\delta n_\mathbf{k}$) and electrostatic potential ($\varphi_\mathbf{k}$) fluctuations (bottom), with $a/L_{n}=-1.5$ (left), $a/L_{n}=0.0$ (middle) and $a/L_{n}=1.5$ (right) for TJ-II considering $a/L_{T_i}=0.5$ and $a/L_{T_e}=2.0$. The dominant phase difference for each $k_y\rho_i$ is denoted by a dot. Positive values are highlighted in red and negative \mbox{values in blue.}  }
	\vspace{2mm}
	\label{fig:phaseshiftTJII}
\end{figure}

Particle fluxes are driven by density and electrostatic potential fluctuations that are out of phase. For example, if the electrons are treated adiabatically, the density and potential would be in phase and no particle fluxes would be driven. The sign of the particle flux is thus correlated with the sign of the phase shift between the density and potential fluctuations, which is an experimentally accessible quantity. Figures \ref{fig:phaseshift} and \ref{fig:phaseshiftTJII} show the spectra of the turbulent particle flux (top row) and of the phase difference between density fluctuations, $\delta n_\mathbf{k}$, and electrostatic potential fluctuations, $\varphi_\mathbf{k}$, (bottom row) along the normalized binormal wavenumber $k_y\rho_i$, \mbox{for W7-X and TJ-II, respectively}.

For W7-X, figure \ref{fig:phaseshift} depicts the cases for $a/L_{T_e}=0$ (left column) and \mbox{$a/L_{T_e}=3$ (right column)}, considering $a/L_{T_i}=3$ and $a/L_{n}=0$. Inward particle flux contributions occur when the density fluctuations lag behind on the potential fluctuations (negative values of the phase difference), which is what predominantly happens at almost all represented scales for the case with $a/L_{T_e}=3$. In contrast, the phase difference is essentially positive at the scales with largest contributions to the total particle flux when $a/L_{T_e}=0$. Specifically, the phase shift is around $4^\circ$ for $a/L_{T_e}=0$ and $-6^\circ$ for $a/L_{T_e}=3$ at the $k_y\rho_i$ scales where the contributions to the particle flux are the largest. On the other hand, for TJ-II, various density gradient are depicted in figure \ref{fig:phaseshiftTJII}, namely $a/L_{n}=-1.5$ (left), $a/L_{n}=0.0$ (middle) and $a/L_{n}=1.5$ (right), considering $a/L_{T_i}=0.5$ and $a/L_{T_e}=2.0$. For hollow density profiles ($a/L_{n}=-1.5$) a phase shift of approximately $-20^\circ$ is observed around $k_y\rho_i \approx 0.5$, for peaked density profiles  ($a/L_{n}=1.5$) the phase shift is around $45^\circ$ and finally for flat density profiles the density and potential fluctuations are almost perfectly out of phase. These gyrokinetic predictions may be validated through experimental measurements.

\FloatBarrier

\section{Particle balance analysis for a  W7-X plasma discharge}
\label{sec:w7xprofiles}

Finally, a specific W7-X discharge ($\#20180920.017$) is studied by performing turbulence simulations throughout the plasma radius, and comparing the obtained turbulent particle flux with the flux inferred from a careful particle balance analysis that includes estimates of the particle source and computations of the neoclassical particle flux. To determine the radial particle flux in an experiment one needs to compute the neutral ionization source in the confined plasma. In current fusion experiments, the recycling neutrals---i.e. the plasma that is neutralized at the divertor plates and returns to the main plasma---are generally the main source of particles. In Ref.~\cite{Kremeyer_nf_2022} it was concluded that both the main chamber and divertor targets contribute to the fueling of the plasma core in W7-X ECRH plasmas. Once the neutral hydrogen atoms reach the confined plasma they are transported towards the core by charge exchange reactions with \mbox{the plasma ions, while they ionize by electron impact.}

\begin{figure}[h!]
	\includegraphics{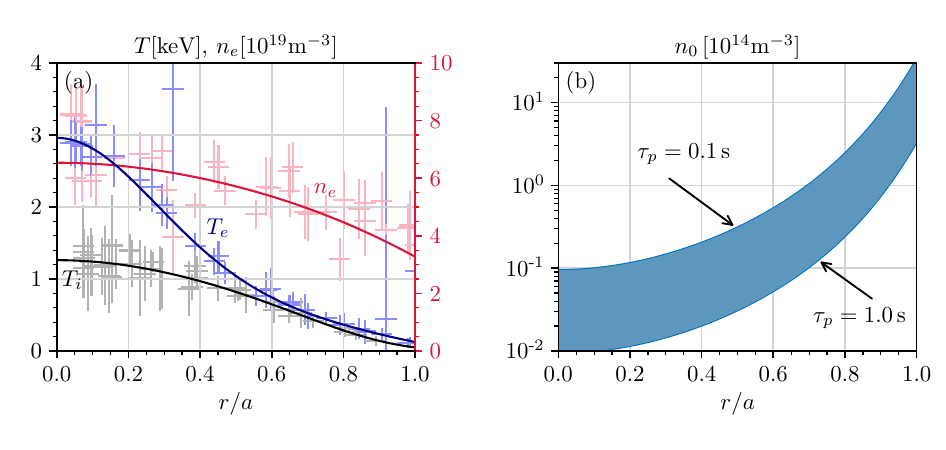} \vspace{-7mm}
	\caption{Plasma profiles (a) and estimated neutral densities (b) for discharge $\#20180920.017$. }
	\vspace{-2mm}
	\label{fig:profiles}
\end{figure}

The radial profile of the neutral density, $n_0(r)$, and the associated ionization source are estimated using a short-mean-free-path, one-dimensional neutral transport model given by
\begin{equation}\label{eq:n0}
	\frac{1}{r}\frac{\text{d}}{\text{d}r}r D_\text{CX}\left(\frac{\text{d} n_0}{\text{d}r} + \frac{1}{T_0}\frac{\text{d}T_0}{\text{d}r}n_0\right) = \nu_\text{ion}n_0
\end{equation}
(see e.g.~Ref.~\cite{Hazeltine_nf_1992}), where $D_\text{CX} = T_0/m_0\nu_\text{CX}$ is the charge exchange diffusivity coefficient, $T_0$ and $m_0$ are the neutral temperature and mass and $\nu_\text{ion}$ and $\nu_\text{CX}$ are the temperature- and density-dependent ionization and charge exchange frequencies. Neutrals and plasma ions are assumed to be isothermal, $T_0(r) = T_i(r)$, due to frequent charge exchange reactions between them. Figure \ref{fig:profiles}b shows the neutral profile obtained with this model for a set of W7-X \mbox{plasma profiles (figure \ref{fig:profiles}a).}

\vspace{-1mm}
\vspace{0.5cm}

Equation (\ref{eq:n0}) determines the neutral profile up to a multiplicative constant that can be determined from the global particle confinement time $\tau_p$. In Ref.~\cite{Kremeyer_nf_2022}, a confinement time of $\tau_p=0.258$\,s was estimated for a similar discharge using a single-reservoir plasma particle balance. The neutral transport model (\ref{eq:n0}) has recently been compared with neutral density measurements in W7-X (see figure 14 in Ref.~\cite{Beurskens_nf_2021}), showing varying degrees of agreement for $\tau_p$ between 0.15 and 1.0\,s. In figure \ref{fig:profiles}b, $\tau_p$ is varied from 0.1\,s to 1.0\,s to account for uncertainties in this number.

\begin{figure}[b!]
	\vspace{-4.5mm}
	\includegraphics{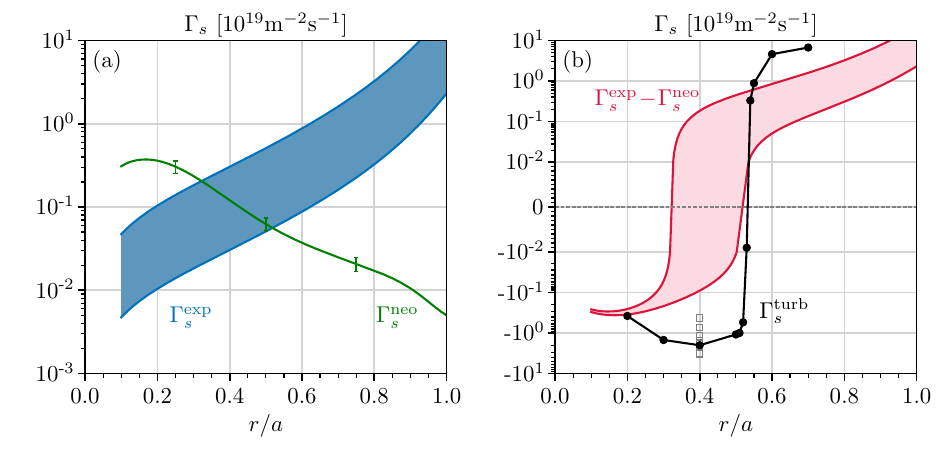} \vspace{-3.5mm}
	\caption{$\Gamma_s^\text{exp}$, $\Gamma_s^\text{neo}$ and $\Gamma_s^\text{turb}$ for discharge $\#20180920.017$. The green error bars (figure \ref{fig:gamma}a) and the gray squares (figure \ref{fig:gamma}b) correspond to $\pm10\%$ variations of the plasma gradients.} \vspace{-4mm}
	\label{fig:gamma}
\end{figure}

Let us denote by $\Gamma_s^{\rm exp}$ the flux profiles calculated, for this range of values of $\tau_p$, by integrating the neutral ionization source $\nu_\text{ion}n_0$ over the plasma radius. The neutral density $n_0$ is obtained from equation (\ref{eq:n0}) using the experimental profiles shown in figure \ref{fig:profiles}a.  The results for $\Gamma_s^{\rm exp}$ are displayed in figure \ref{fig:gamma}a. There is a region in the core (the width of which depends on the \mbox{assumed $\tau_p$)} where the estimated central particle sources are too small to sustain the outward neoclassical radial flux $\Gamma_s^{\rm neo}$ calculated with \texttt{DKES}~\cite{Hirshman1986}. One is led to conclude that an additional particle flux must exist which moves particles towards the magnetic axis in that region. Further outside, on the contrary, the additional flux must be outward directed and account for essentially all the particle flux, since the neoclassical contribution falls 1 to 3 orders of magnitude short with respect to $\Gamma_s^{\rm exp}$. A similar shortfall is observed in other stellarators, such as HSX \cite{HSX1, HSX2}.

\pagebreak
The turbulent particle flux, $\Gamma_s^\text{turb}$, is modeled with \texttt{stella} and represented in figure \ref{fig:gamma}b by the black line. The plasma parameters and resulting particle fluxes depicted in figures \ref{fig:profiles} and \ref{fig:gamma} are summarized in table \ref{tab:Gammai}. The size of $\Gamma_s^\text{turb}$ agrees with $\Gamma_s^\text{exp} - \Gamma_s^\text{neo}$ reasonably well, taking into account, among other factors, the uncertainties in the plasma profiles and the simplicity of the neutral transport model employed. Remarkably, the sign (direction) of the turbulent particle flux is found to match the sign of the discrepancy between the experimental and neoclassical particle fluxes across the entire plasma radius. Specifically, the simulations predict a sign reversal of the turbulent flux halfway through the plasma cross-section. This reversal explains why neoclassical theory overestimates the particle flux in the core of the plasma and underestimates it near the edge. The effect of the uncertainties of the plasma profiles on the turbulent particle flux is quantified in the next section. In addition, in the future, it would be desirable to carry out this kind of analysis using full-flux-surface or \mbox{radially global gyrokinetic codes.}

\section{Uncertainties on the plasma profiles and gradients}
\vspace{-1mm}
\label{sec:appendixb}

\begin{figure}[h!]
	\centering
	\vspace{0mm}
	\includegraphics[trim={0mm 0mm 0mm 0mm},clip]  {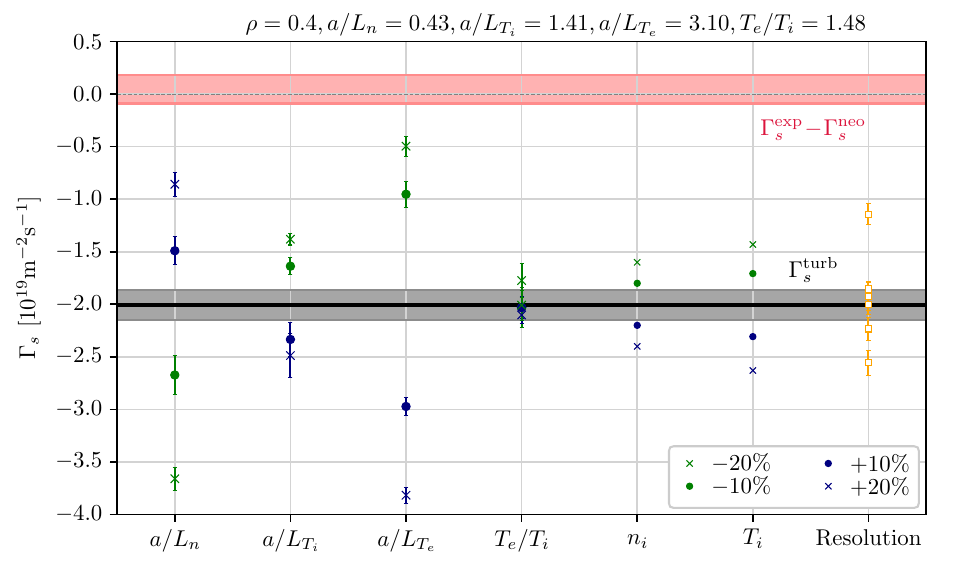}
	\vspace{-8mm}
	\caption{Effect on the turbulent particle flux of a 10\% and 20\% change of the plasma parameters at $r/a=0.4$.  The values of $\Gamma_s^\text{exp}-\Gamma_s^\text{neo}$ and $\Gamma_s^\text{turb}$ shown in figure 5 are displayed in red and black, respectively. The error bar on $\Gamma_s^\text{exp}-\Gamma_s^\text{neo}$ represents the uncertainty on $\tau_p$ while the error bar on $\Gamma_s^\text{turb}$ represents the standard mean deviation of the saturated state of the time trace of the simulation. The final column of data shows the effect on the turbulent particle flux of doubling the resolution parameters 1 by 1.}
	\vspace{0mm}
	\label{fig:uncertainty}
\end{figure}

To quantify the effect of the uncertainty on the plasma profiles (figure \ref{fig:profiles}a), gyrokinetic simulations are performed with a 10\% (or 20\%) change of the plasma parameters at $r/a=0.4$, which is shown in figure \ref{fig:uncertainty}. 	Since \texttt{stella} performs simulations with dimensionless parameters, the uncertainties on the plasma density $n_i=n_e$ and the ion temperature $T_i$ are visualized by denormalizing the particle fluxes accordingly.
Figure \ref{fig:uncertainty} suggests that at $r/a=0.4$ higher density gradients and lower temperature gradients would improve the agreement between $\Gamma_s^\text{exp}-\Gamma_s^\text{neo}$ and $\Gamma_s^\text{turb}$. Similarly, lower densities and temperatures would improve agreement, although their influence is smaller compared to the effect of changing the gradients.

\begin{table}[h!]
	\centering
	\vspace{2mm}
	\caption{\label{tab:Gammai} Overview of the plasma profiles and gradients depicted in figure \ref{fig:profiles}, and the experimental ($\Gamma_s^\text{exp}$), neoclassical ($\Gamma_s^\text{neo}$) and turbulent ($\Gamma_s^\text{turb}$) particle fluxes shown in figure \ref{fig:gamma}. For the experimental particle flux, the average value is shown for the confinement times of $\tau_p=0.1\,$s and $\tau_p=1.0\,$s. Temperatures are given in keV, densities in 10$^{19}$m$^{-3}$ and particle fluxes in 10$^{19}$m$^{-2}$s$^{-1}$.} \vspace{-2.5mm}
	\lineup
	\begin{tabular}{@{}cccccccccc}
		\specialrule{.12em}{.1em}{.1em} \hspace*{12mm} & \hspace*{12mm} & \hspace*{12mm} & \hspace*{12mm} & \hspace*{12mm} & \hspace*{12mm} & \hspace*{12mm} & \hspace*{13mm} & \hspace*{13mm} & \hspace*{13mm} \\[-4mm]
		$r/a$ & $T_e/T_i$ & $T_i$ & $n$ & $a/L_{T_i}$ & $a/L_{T_e}$ & $a/L_{n}$ & $\Gamma_s^\text{exp}$ & $\Gamma_s^\text{neo}$ & $\Gamma_s^\text{turb}$   \\[0.5mm] \hline &&&&&&&&&\\[-4mm]
		&&&&&&&&& \\[-3.5mm]
		0.2 & 2.05 & 1.18 & 6.40 & 0.63 & 1.93 & 0.21 & 0.06 & 0.36 & $-$0.38 \\[0.5mm]
		0.3 & 1.76 & 1.09 & 6.23 & 0.99 & 2.64 & 0.32 & 0.10 & 0.24 & $-$1.49 \\[0.5mm]
		0.4 & 1.48 & 0.97 & 6.00 & 1.41 & 3.10 & 0.43 & 0.17 & 0.12 & $-$2.01 \\[0.5mm]
		0.5 & 1.27 & 0.82 & 5.72 & 1.97 & 3.33 & 0.56 & 0.28 & 0.06 & $-$1.09 \\[0.5mm]
		0.6 & 1.15 & 0.65 & 5.37 & 2.77 & 3.39 & 0.71 & 0.48 & 0.04 & \wminus 4.60 \\[0.5mm]
		0.7 & 1.14 & 0.46 & 4.95 & 3.98 & 3.49 & 0.90 & 0.88 & 0.02 & \wminus 6.70 \\[0.5mm]
		\specialrule{.12em}{.1em}{.1em}
	\end{tabular}
	\vspace{4mm}
\end{table}

\section{Conclusions}
\label{sec:conclusions}

Whereas neoclassical theory predicts strongly hollow density profiles in the plasma core of large stellarators, measured density profiles in Wendelstein 7-X are flat or weakly peaked, indicating that theoretical particle flux calculations are missing an inward contribution. In this chapter, we have shown that the discrepancy between theory and experiment disappears when the turbulent component of the particle flux is included in the calculations: turbulence driven by both the ion and electron temperature gradients provides the missing inward particle flux needed to sustain the experimentally measured density profiles. Moreover, we have performed the first one-dimensional particle balance analysis in a Wendelstein 7-X discharge where the sign of the turbulent particle flux agrees, over the entire plasma radius, with the sign of the difference between the experimental and the neoclassical particle fluxes. In particular, the simulations predict a sign change of the turbulent flux, accounting for the fact that the neoclassical flux is too large at the core and too small at the edge of the plasma to explain the experimental particle flux. Moreover, it is shown that the existence of an inward turbulent particle flux is very robust and is found over broad regions of the scanned parameter space, independent of the magnetic geometry.

\part{Summary and Discussion}
\chapter{Conclusions and Future work}

\vspace{-2mm}
Understanding and ultimately mitigating turbulent transport is central to the success of magnetic confinement fusion as a sustainable energy source. This dissertation has addressed key aspects of turbulence-driven transport in fusion plasmas, focusing on the influence of the density gradient on heat transport and examining turbulent particle transport within the core of fusion devices. Through extensive numerical simulations using the gyrokinetic code \texttt{stella}, and by examining a wide range of fusion-relevant magnetic geometries and plasma gradients, this work provides detailed insights into how magnetic geometry and plasma profiles affect turbulent transport, offering both new physical understanding and comparative benchmarks across a range of devices.

The first major part of the thesis explored turbulent heat transport under the influence of varying density gradients. Previous studies, both theoretical and experimental, had hinted at the potential for steep density profiles to suppress heat transport by stabilizing temperature-gradient-driven modes. However, these effects had not been systematically explored across different magnetic configurations.
By conducting linear and nonlinear simulations spanning a wide parameter space in multiple devices---including the Wendelstein 7-X (W7-X) stellarator, the Large Helical Device (LHD), the TJ-II stellarator and the National Compact Stellarator Experiment (NCSX), which represent helias, heliotron, heliac and quasi-axisymmetric configurations, respectively, as well as the Axially Symmetric Divertor Experiment or ASDEX Upgrade (AUG) tokamak and the tokamak geometry of the Cyclone Base Case (CBC)---a comprehensive picture emerged. Namely, steep density gradients can reduce the ion heat flux, particularly in devices like W7-X and NCSX, due to the stabilization of ion-temperature-gradient-driven modes without significantly destabilizing density-gradient-driven modes. This effect was not captured by linear growth rates of the fastest-growing modes alone, highlighting the importance of nonlinear saturation mechanisms, such as energy transfer to damped modes, zonal flow shearing, and energy cascades to dissipative scales.
Furthermore, the thesis dissected the nature of microinstabilities across devices, identifying that trapped-electron modes (TEMs) dominate in LHD, TJ-II, NCSX, AUG and CBC, and highlighting the unique contribution of universal instabilities driven by passing particles in W7-X. These insights help bridge the gap between observed turbulent behavior and theoretical modeling, revealing why certain stellarators might exhibit naturally improved confinement properties under density peaking.

The second major part of the thesis turned to the question of particle transport in stellarators, with particular focus on the discrepancy between neoclassical predictions and experimental measurements in the plasma core. Neoclassical theory anticipates strong outward particle fluxes leading to hollow density profiles. However, experiments, for example in W7-X, consistently measure flat or weakly peaked profiles. To investigate this inconsistency, nonlinear gyrokinetic simulations were carried out to compute the turbulent particle flux in W7-X and TJ-II.
These simulations reveal a robust inward turbulent particle flux driven by both ion and electron temperature gradients, sufficient to reconcile the neoclassical prediction with the experimental profile. In a detailed case study of a W7-X discharge, a one-dimensional particle balance analysis was performed, showing that the sign of the turbulent particle flux agreed throughout the plasma radius with the sign of the discrepancy between experimental and neoclassical fluxes. These inward fluxes were found to be robust across parameter scans and magnetic field configurations, suggesting that turbulence plays a constructive role in sustaining experimentally favorable density profiles. This dissertation therefore demonstrates turbulence-driven particle retention in the core of a stellarator, and provides compelling evidence that turbulence is not merely a source of degradation in confinement but can under \mbox{certain conditions act constructively to sustain desirable profiles.}

Taken together, the results presented in this thesis reinforce several important conclusions. First of all, density gradients play a dual role in turbulence: while stabilizing temperature-gradient-driven modes, they may also drive turbulence via trapped-electron modes and universal instabilities. The balance between these effects depends strongly on the magnetic geometry and is therefore highly device-specific. Secondly, it has been shown that nonlinear simulations are essential for understanding turbulence. Linear growth rates, particularly of only the fastest-growing modes, do not reliably correlate with transport levels, especially in complex geometries such as those of stellarators where multiple instability branches coexist or interact. Thirdly, stellarator geometry can naturally support improved confinement via favorable linear and nonlinear turbulence properties. Devices like W7-X and NCSX show reduced heat transport under density peaking, potentially explaining experimental observations without invoking additional stabilization mechanisms. Finally, turbulence can be a beneficial mechanism for particle confinement in stellarators, countering neoclassical losses and helping to sustain experimentally observed density profiles. This overturns the traditional view of turbulence as purely deleterious and suggests new pathways for optimization.

Despite the novel insights and contributions, several limitations inherent to the modeling approach constrain the generality of the conclusions. Linear simulations considered only the fastest-growing mode, neglecting subdominant instabilities that can substantially influence the nonlinear dynamics and saturation \cite{faber2018stellarator, mckinney2019comparison, hatch2011role, terry2006role, hegna2018theory, terry2018saturation, terry2000suppression, diamond2005zonal, 2015-Plunk-Understanding}. Moreover, this study adopted the flux-tube approximation, which restricts the analysis to a narrow radial domain and neglects full-flux-surface and global effects \cite{cole2020nonlinear, mishchenko2022gyrokinetic, singh2022global, wilms2023full, mishchenko2023global, wilms2024global}.
Additionally, collisions were not included in the simulations. While low-collisionality plasmas were assumed, collisional effects can influence both instability thresholds and nonlinear saturation mechanisms \cite{von2024linearised, kauffmann2010global, sanchez2019validation}. Electromagnetic effects were also excluded. While electrostatic turbulence dominates in many core plasma regimes, electromagnetic modes can become significant at higher beta and in more realistic reactor scenarios \cite{mulholland2024finite,  ishizawa2014electromagnetic, aleynikova2018kinetic, zocco2018strongly, mckinney2021kinetic, aleynikova2022influence}. Finally, the simulations did not incorporate self-consistent equilibrium evolution or feedback from the turbulent transport to the background profiles, limiting the ability to \mbox{predict time-evolving behavior or profile stiffness \cite{navarro2023first, rodriguez2024enhancing}}.

Looking forward, the insights gained from this work lay the groundwork for several future directions. The role of subdominant modes in nonlinear saturation and transport should be explored more systematically. Including full-flux-surface effects, global effects, finite collisionality, and electromagnetic effects in future simulations will help bridge remaining gaps between theory and experiment. In addition, multi‑species plasma effects remain relatively underexplored, despite its potential to significantly impact turbulent heat transport \cite{garcia2024reduction}. Validation studies against experimental measurements are also essential, particularly in devices like W7-X where detailed profile and fluctuation data are increasingly available, to test model predictions and refine simulation assumptions. Initial efforts in this direction have already begun, including comparisons of Doppler reflectometry measurements \cite{gonzalez2024electrostatic} and zonal flow structures \cite{carralero2025first} with \texttt{stella} simulations for W7-X.  Finally, one could explore turbulence optimization strategies in stellarator designs, leveraging insights from this thesis to propose magnetic geometries that suppress deleterious modes.

Ultimately, this dissertation advances our understanding of how turbulent transport can be shaped by plasma profiles and magnetic geometry. The key insight---that turbulence can both be suppressed and used constructively---opens new avenues for magnetic confinement optimization. In doing so, it brings us closer to designing fusion reactors with intrinsically enhanced confinement and more predictable, controllable turbulence, key steps on the path to sustainable fusion energy.

\addtocontents{toc}{\protect\newpage}
\part{Appendices}
\appendix

\chapter{Expressions using different conventions for $\alpha$ and $\psi$}
\label{chap:app:tok_stell}

The tokamak geometry of CBC is defined by specifying a set of Miller parameters that are used to construct a local solution to the Grad--Shafranov equation \cite{miller1998noncircular}, whereas the magnetic equilibria of AUG, W7-X, LHD, TJ-II and NCSX are generated by the three-dimensional MHD equilibium code \texttt{VMEC} \cite{hirshman1983steepest}. Depending on whether a set of Miller parameters or a \texttt{VMEC} equilibrium  are used to construct the magnetic geometry, \texttt{stella} uses a set of coordinates which are preferred in the tokamak literature or the stellarator literature, respectively. Specifically, the choices $\alpha_q = \zetaB - q \thetaB$ and $\psi=\psi_p$ are implemented when defining the magnetic equilibrium through the Miller formalism, with $(\theta_B,\zeta_B)$ a set of Boozer coordinates \cite{boozer1981plasma}. On the other hand, $\alpha_\iota =  \theta_p  - \iota \zeta$ and $\psi=-\psi_t$ are employed when a \texttt{VMEC} magnetic equilibrium is read, with $(\theta_p,\zeta)$ the so-called \texttt{PEST} \mbox{coordinates \cite{grimm1983ideal}}. In the main manuscript, the stellarator conventions are used, corresponding to the coordinate system implemented in \texttt{stella} for \texttt{VMEC} equilibria. In this Appendix, the equivalent expressions for the tokamak conventions are listed, which are used in \texttt{stella} when defining the magnetic geometry based on a set of Miller parameters. To differentiate between the two conventions, a subindex ``$q$'' is used for conventions which are popular in the tokamak literature, while a subindex ``$\iota$'' is used for conventions preferred in the stellarator literature.

\section{Clebsch form of $\mathbf{B}$ using different conventions for $\alpha$ and $\psi$}
\label{app:clebsch}

The \textit{Clebsch form}\index{Clebsch form} of the magnetic field, which is derived in section 6.1 of Ref.~\cite{haeseleer2012flux}, is given by,
\begin{equation}
	\label{app:generalClebsch}
	\mathbf{B} = \nablam \psi \times \nablam \big( \psi'_t \theta_s  - \psi'_p \zeta_s \big),
\end{equation}
where primes denote derivatives with respect to $\psi$, and  $(\psi,\theta_s,\zeta_s)$ is a set of straight-field-line flux coordinates, as defined in section \ref{sec:straight-field-line-coords}. The $(\psi'_t \theta_s  - \psi'_p \zeta_s )$-term is simplified by defining a specific field-line label $\alpha$. The two most straightforward definitions are $\alpha_q = \zeta_s - q \theta_s$ and $\alpha_\iota =  \theta_s  - \iota \zeta_s$. In \texttt{stella}, Boozer coordinates \cite{boozer1981plasma}  are used when defining the magnetic geometry based on a set of Miller parameters, hence $\alpha_q = \zetaB - q \thetaB$ while \texttt{PEST} coordinates \cite{grimm1983ideal} are used when reading a \texttt{VMEC} equilibrium, i.e.~$\alpha_\iota =  \theta_p  - \iota \zeta$. For these choices of the field-line label, we obtain,
\begin{equation}
	\label{app:Clebsch}
	\mathbf{B} = \Big( \frac{d\psi_t}{d\psi} \Big)  \nablam \psi \times \nablam (\theta_p  - \iota \zeta) = \Big(\!- \frac{d\psi_p}{d\psi} \Big) \,\nablam \psi \times \nablam (\zetaB - q \thetaB).
\end{equation}
Generally, the choice $\psi=\psi_p$ is made in the tokamak literature, which allows us to write the magnetic field as $\mathbf{B} = \nablam \alpha_q \times \nablam \psi_p$, while for stellarators it is common to use $\psi=\psi_t$ which gives $\mathbf{B} = \nablam \psi_t \times \nablam \alpha_\iota$. These are the Clebsch expressions commonly found in the gyrokinetic literature.
In the paper by Barnes et al.~\cite{stella} describing the \texttt{stella} code, $\psi=\psi_p$ is implemented when defining the magnetic equilibrium through the Miller formalism while $\psi=-\psi_t$ is employed when a \texttt{VMEC} magnetic equilibrium is read. These choices make it possible to derive the gyrokinetic equations using \mbox{$\mathbf{B} =  \nablam \alpha \times \nablam \psi$}. However, while the choice $\psi=-\psi_t$ ensures that $\psi$ points radially outwards for counter-clockwise magnetic fields (where the toroidal magnetic field points counter-clockwise when viewed from above), for clockwise magnetic fields, $\psi$ will point radially inwards which is counterintuitive, since $\psi$ is typically used as a radial coordinate. Therefore, as part of this PhD work, \texttt{stella} has been modified so that the definition of the flux-surface label can be set to $\psi=r$ by the user, to ensure that $\psi$ always points radially outward.

\section{Standard twist-and-shift parallel boundary condition using tokamak conventions}
\label{app:twistandshift}

The standard twist-and-shift boundary condition is generally derived using the tokamak conventions for the field-line label and parallel coordinate, i.e.~$\alpha = \zeta_s - q(\psi)\,\theta_s$ and $z=\theta_s$. In this Appendix, the derivation of the twist-and-shift boundary conditions, summarized in section \ref{sec:twistandshift}, is repeated, using the tokamak conventions instead of the stellarator conventions. The starting point is the Fourier expansion in straight-field-line coordinates obtained by substituting $\alpha = \zeta_s - q(\psi)\,\theta_s$ and $q(\psi) \approx q_0 + (dq/d\psi)\, (\psi-\psi_0)$ into equation (\ref{eq:phiFourier}) which gives,
\begin{align}
	\label{app:varphikpsikalpha3}
	\varphi(\psi,\alpha,z,t)
	& = \sum_{k_\psi,k_\alpha} \hat{\varphi}_{k_\psi,k_\alpha} (z, t) \, \exp \Bigg( \ir \Big( k_\psi -  k_\alpha \frac{dq}{d\psi} \theta_s \Big) (\psi-\psi_0) +  \ir k_\alpha (\zeta_s - q_0\theta_s ) \Bigg).
\end{align}
Fluctuating quantities are assumed to be periodic in $z=\theta_s$ with a period $2\pi N$ while holding $(\psi,\zeta_s)$ fixed, with $N$ an integer number. This implies that when a particular $(k'_\psi,k'_\alpha)$ Fourier mode gets to the end of the $z$-domain $(\theta_s = \pi N)$, the mode connects exactly to another Fourier mode $(k_\psi,k_\alpha)$ at the beginning of the the $z$-domain $(\theta_s = -\pi N)$. Substituting this condition \mbox{into eq.~(\ref{app:varphikpsikalpha3}) gives, }
\begin{equation}
	\label{app:parallelBC}
	\begin{split}
		\hat{\varphi}_{k_\psi,k_\alpha}(-\pi N,t) \,  \exp  \! \Bigg(\ir \Big(k_\psi - k_\alpha \frac{dq}{d\psi} (-\pi N) \Big) (\psi-\psi_0) + \ir k_\alpha (\zeta_s - q_0(-\pi N)) \Bigg) \\
		=
		\hat{\varphi}_{k'_\psi,k'_\alpha}(\pi N,t) \,  \exp \! \Bigg(\ir \Big(k'_\psi - k'_\alpha \frac{dq}{d\psi} \, (\pi N)  \Big) (\psi-\psi_0)  + \ir k'_\alpha \Big(\zeta_s - q_0 (\pi N)  \Big) \! \Bigg)
	\end{split}
\end{equation}
For this equation to hold for all field lines, the factors in front of $\zeta_s$ should match. Therefore, the binormal wavenumber remains unchanged, i.e.~$k'_\alpha=k_\alpha$. Moreover, for equality to hold for all flux surfaces, the factors in front of $\psi$ should match, thus we must have,
\begin{equation}
	\label{app:kpsi}
	\delta k_\psi = k'_\psi - k_\psi = k_\alpha \, \frac{dq}{d\psi} \, 2\pi N.
\end{equation}
The wavenumbers $(k_\psi, k_\alpha)$ in the Fourier expansions (eqs.~\ref{eq:gFourier} and \ref{eq:phiFourier}) are discretized, specifically, \mbox{$k_\psi=2\pi j/L_\psi$} and $k_\alpha = 2\pi k/L_\alpha$, where $j$ and $k$ are integers. To ensure that $k'_\psi$ belongs to the wavenumber grid, and assuming that the $k_\psi$ grid consists of integer multiples of some \mbox{$k_{\psi,\text{min}} = 2\pi/L_\psi$}, then the difference $k'_\psi-k_\psi$ should be an integer multiple of $k_{\psi,\text{min}}$, i.e.%
\begin{equation}
	\delta k_\psi = k'_\psi - k_\psi  =  J\, k_{\psi,\text{min}}
\end{equation}
In particular, equation (\ref{app:kpsi}) must be true for the smallest nonzero $k_{\alpha,\text{min}}=2\pi/L_\alpha$, which gives,
\begin{equation}
	\label{app:kalphamin}
	k_{\alpha,\text{min}} \, \frac{dq}{d\psi} \, 2\pi N = J\, k_{\psi,\text{min}}
\end{equation}
Using this quantization, the parallel periodicity condition imposed in equation (\ref{app:parallelBC}) gives,
\begin{align}
	\label{app:bcphaseshift}
	\hat{\varphi}_{k_\psi,k_\alpha}(-\pi N, t) \, \exp \Big( \ir  k_\alpha q_0  2\pi N \Big) = \hat{\varphi}_{k_\psi+\delta k_\psi,k_\alpha}(\pi N, t),
\end{align}
with $C_{k_\alpha} = \exp ( i  k_\alpha q_0  2\pi N )$ a phase-factor which recenters the flux tube around $\alpha_0 + q_0  2\pi N$ as a perturbation is connected across the $z$-boundary. The phase factor is equal to one for a rational surface, i.e.~if $q_0N$ is an integer and $L_\alpha=2\pi$. This means that on a rational surface, field lines connect exactly back onto themselves after $q_0N$ toroidal turns.
This phase factor is important in full-flux-surface simulations, to ensure that field lines are connected properly. In contrast, in flux-tube simulations the binormal wavenumber $k_\alpha$ is assumed to be very small. As a result, the phase factor $C_{k_\alpha}$ is typically taken to be 1 in the flux-tube approximation,
\begin{equation}
	C_{k_\alpha} = \exp ( \ir  k_\alpha q_0  2\pi N ) \approx 1.
\end{equation}

\chapter{Visual derivation of the standard twist-and-shift boundary condition}

\label{app:visualTwistAndShift}

\vspace{-3mm}
A particle which is streaming along a specific field line and exits the flux tube at $z=L_z/2$, needs to be confined to the same field line as it reenters the flux tube at $z=-L_z/2$. This is enforced through the \textit{parallel boundary conditions}, which connects a certain $(k_\psi,k_\alpha)$-mode at the end of the flux tube, to another Fourier mode $(k_\psi+\delta k_\psi,k_\alpha)$ at the beginning of the flux tube. Here, the equivalence of the twist-and-shift parallel boundary condition (section \ref{sec:twistandshift}) is explained in real space \cite{ball2020eliminating}.
As illustrated in figure \ref{fig:MagneticShear}, field lines starting at the same location, e.g.~at $(\theta_s, \zeta_s)=(0,0)$, on different flux surfaces, end at a different toroidal angle $\zeta_s$ after covering a distance $z=\theta_s$. Using \mbox{equation (\ref{eq:alphaq})}, the distance traversed by a field line along the toroidal angle is given by,
\begin{align}
	\label{eq:zetas}
	\zeta_s(\alpha,\theta_s,\psi) & = \alpha + q_0 \theta_s + \theta_s (dq/d\psi)  (\psi - \psi_0).
\end{align}
First of all, as a field line passes through the end of the $z$-domain at $z=L_z/2$, it needs to reconnect to the same field line at $z=-L_z/2$, which is characterized by a different field-line label. For example, a field line starting at $\theta_s=-\pi$, which extends up to one poloidal turn, does not end up at the same physical location at $\theta_s = \pi$, unless $q$ is an integer (rational surface). Instead, it will connect to the $\alpha' = \alpha + 2\pi q_0$ field line at $\theta_s=-\pi$. When a perturbation exits the flux tube domain it thus reenters in an identical copy of the flux tube, centered around a \mbox{different $\alpha$}. The shift in $\alpha$ corresponds to the shift in $\zeta_s$ on the $\psi_0$ flux surface, i.e.
\begin{align}
	\Delta \zeta_\text{fixed-$\psi_0$}(\theta_s) & =
	\zeta(\alpha,0,\psi_0) -
	\zeta(\alpha,L_z,\psi_0) = q_0L_z.
\end{align}
Using the twist-and-shift parallel boundary condition, the flux tube is recentered around a different field line $\alpha' = \alpha + q_0L_z$ as it crosses the $z$-boundary through the phase factor $C_{k_\alpha} = \exp ( - i  k_\alpha q_0  L_z )$ in equation (\ref{app:bcphaseshift}).
Field lines that are located on flux surfaces with $\psi \neq \psi_0$ have a different tilt since $q(\psi) \approx q_0 + (dq/d\psi)\, (\psi-\psi_0)$. As a result, the cross-section is sheared into a parallelogram as the flux tube moves in the parallel direction, as illustrated in figure \ref{fig:parallelogramFourier}. The shearing of the cross-section is quantified by the difference in $\zeta_s(\alpha,\theta,\psi)$ between two matching field lines located on the $\psi+L_\psi/2$ and $\psi-L_\psi/2$ flux surfaces. The shift between matching field lines is,
\begin{align}
	\Delta \zeta_\text{fixed-$\theta_s$}(\theta_s) & =
	\zeta(\alpha,\theta_s,\psi_0+L_\psi/2) -
	\zeta(\alpha,\theta_s,\psi_0-L_\psi/2) =  \theta_s L_\psi (dq/d\psi).
\end{align}
At $\theta_s=0$, the cross-section is taken to be perpendicular, therefore, the shift in $\zeta_s(\alpha,\theta,\psi)$ between matching field lines is zero, $\Delta \zeta_\text{fixed-$\theta_s$}(0) = 0$. This alignment is lost as one moves along the flux tube.  Between the flux tube ends, located at $\theta_s=\pm L_z/2$, the total shift is
\begin{align}
	\Delta \zeta_{z_\pm} = \Delta \zeta_\text{fixed-$\theta_s$}(L_z/2) - \Delta \zeta_\text{fixed-$\theta_s$}(-L_z/2)  & =  L_z L_\psi (dq/d\psi).
\end{align}

\begin{figure}[h!]
	\centering
	\includegraphics{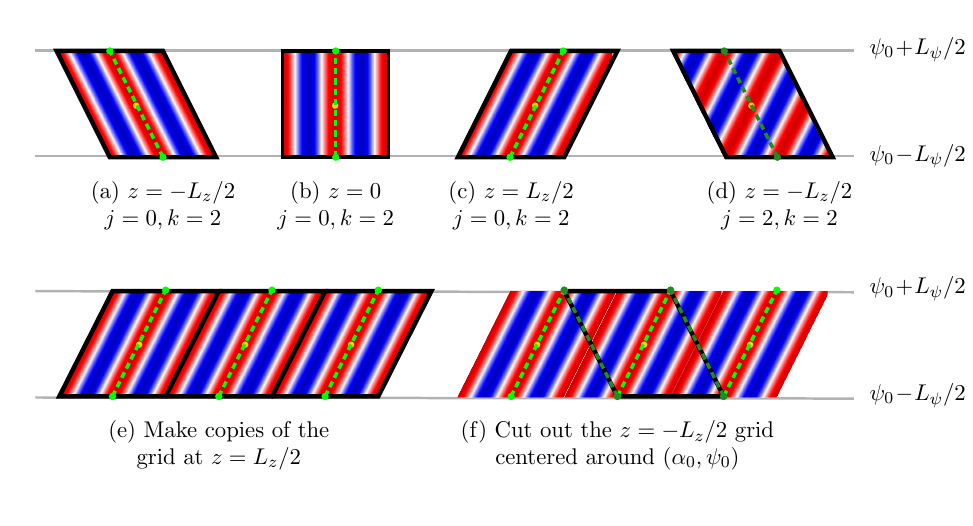}
	\captionsetup{singlelinecheck=false,  margin=0.0cm} \vspace{-15mm}
	\caption{A perturbation of the electrostatic potential is introduced at $z=0$, corresponding to the $k_\psi=0$ and $k_\alpha = 2 \, k_{\alpha,\text{min}}$ mode, with $k_{\alpha,\text{min}}=2\pi/L_\alpha$. The perturbation becomes increasingly sheared  along the flux tube due to the global magnetic shear. At the end of the flux tube $(z=L_z/2)$ the sheared cross-section is shifted onto the cross-section at the beginning of the flux tube $(z=-L_z/2)$ which introduces a shift in the radial wavenumber of the mode. In this example $L_\psi=L_\alpha$ and $dq/d\psi=-1/L_z$ hence $[k_\psi]_{z_-} = [k_\psi]_{z_+} + k_\alpha$.}  \vspace{1mm}
	\label{fig:parallelogramFourierBC}
\end{figure}

\noindent
Therefore, to connect the tilted cross-section at $z=L_z/2$ to the cross-section at $z=-L_z/2$ which has an opposing tilt, the twisted cross-section at the end of the flux tube is shifted onto the twisted cross-section at the beginning of the flux tube, hence the name twist-and-shift, which is illustrated in figure \ref{fig:parallelogramFourierBC}. An initial perturbation of the electrostatic potential is introduced in the flux tube, corresponding to the $(k_\psi,k_\alpha) = (0, 2 \,k_{\alpha,\text{min}})$ Fourier mode. The Fourier mode becomes sheared towards the flux tube ends (figures \ref{fig:parallelogramFourierBC}a and \ref{fig:parallelogramFourierBC}c). As the perturbation exits the flux tube at $z=L_z/2$ it needs to reenter through the twisted cross-section at $z=-L_z/2$. This is achieved by placing multiple copies of the perturbation at \mbox{$z=L_z/2$} next to each other \mbox{(figure \ref{fig:parallelogramFourierBC}e)} and then cutting out the twisted cross-section corresponding to $z=-L_z/2$ (\mbox{figure \ref{fig:parallelogramFourierBC}f}), centered around $(\alpha_0,\psi_0)$. Using this twist-and-shift boundary condition, the perturbation has reentered the flux tube at $z=-L_z/2$ which is shown in figure \ref{fig:parallelogramFourierBC}d. It is clear that in $(\psi,\alpha)$-coordinates, the perturbation has undergone a shift in the radial wavenumber $k_\psi$ as it crossed the $z$-boundary (eq.~\ref{app:kpsi}). Specifically, in figure \ref{fig:parallelogramFourierBC}, the $(k_\psi,k_\alpha) = (0, 2 \, k_{\alpha,\text{min}})$  Fourier mode connects onto the $(k_\psi,k_\alpha) = (2\, k_{\psi,\text{min}}, 2 \, k_{\alpha,\text{min}})$ mode as it crosses the $z$-boundary.

\vspace{-6mm}
\FloatBarrier

\begin{figure}[h!]
	\centering
	\includegraphics[trim={0mm 0mm 0mm 40mm},clip] {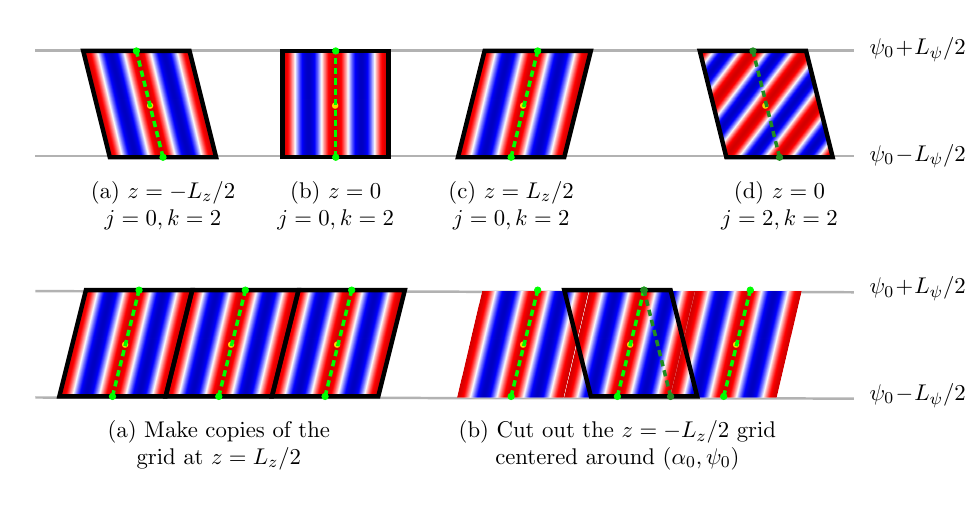}
	\captionsetup{singlelinecheck=false,  margin=0.0cm} \vspace{-15mm}
	\caption{A similar example as in figure \ref{fig:parallelogramFourierBC}, with half the global magnetic shear $dq/d\psi=-1/(2L_z)$. Taking $L_\psi=L_\alpha$, condition (\ref{app:kalphamin}) is not fulfilled since it gives $J=1/2$ which is not an integer. As a result, matching field lines are no longer matching as the perturbation crosses the $z$-boundary.}  \vspace{-3mm}
	\label{fig:parallelogramFourierBCBad}
\end{figure}

\noindent
It is important that field lines are shifted back onto their matching pairs, otherwise particles which are streaming along specific field lines would be transported across field lines as they cross the $z$-boundary. In figure \ref{fig:parallelogramFourierBC}, the light green dashed line connects two matching field lines within the flux tube. As the perturbation crosses the $z$-boundary (figure \ref{fig:parallelogramFourierBC}f) the cross-section at $z=-L_z/2$, again connects the originally matching field lines. To illustrate this in more detail, a counter example is shown in figure \ref{fig:parallelogramFourierBCBad}. In this example, $dq/d\psi=-1/(2L_z)$ and $L_\psi=L_\alpha$ which means that condition (\ref{app:kalphamin}) is not fulfilled since it gives $J=1/2$ which is not an integer. As a result, if the twist-and-shift boundary condition would be applied, the field lines are no longer matching as they cross the $z$-boundary. This can be seen in figure \ref{fig:parallelogramFourierBCBad}b where the dark green line, which connects matching field lines at $z=-L_z/2$, no longer connects the field lines that were matching before the perturbation crossed the $z$-boundary, highlighted by the light green lines.

\vspace{-3mm}
\FloatBarrier

\noindent
It is clear that the total shift in $\zeta_s(\alpha,\theta,\psi)$ thus needs to equal an integer multiple of $L_\zeta$,
\begin{equation}
	\label{eq:deltaz}
	\Delta \zeta_{z_\pm}  =  L_z L_\psi (dq/d\psi)  = J L_\zeta =  J L_\alpha ,
\end{equation}
with $J$ a nonzero integer. To see that $L_\zeta  = L_\alpha$, evaluate equation (\ref{eq:zetas}) at fixed $\theta_s$ and fixed $\psi$,
\begin{align}
	L_\zeta & = \zeta(\alpha_0+L_\alpha/2,\theta_s,\psi) -
	\zeta(\alpha_0-L_\alpha/2,\theta_s,\psi) = L_\alpha.
\end{align}
Ensuring particles remain confined to the same field line as they pass through the $z$-boundary thus imposes a quantization on the domain's perpendicular aspect ratio,
\begin{equation}
	\label{eq:LpsiLalpha}
	\frac{L_\psi}{L_\alpha} = \frac{J}{L_z} \frac{d\psi}{dq}
\end{equation}
This is equivalent to condition (\ref{app:kalphamin}) derived for the standard twist-and-shift boundary condition.

\chapter{Default input variables in \texttt{stella}}

\label{app:defaultInputVariables}

The \deltaf-gyrokinetic code \texttt{stella} is open source and available on \href{https://github.com/stellaGK}{https://github.com/stellaGK}.
The default \texttt{stella} input variables of the ``v0.6'' release of the code, found at  \href{https://github.com/stellaGK/stella/releases/tag/v0.6}{https://github.com/ stellaGK/stella/releases/tag/v0.6}, are summarized in this section. Moreover, the default input variables are directly available through the python diagnostics package \texttt{stellapy}. This Python package has been developed as a part of this PhD work to diagnose the output from \texttt{stella}. In \texttt{stella/stellapy/data/stella/load\_defaultInputParameters.py} the input parameters and their default values, used in \texttt{stella}, are defined. They can be accessed through the dictionary \texttt{input\_parameters["namelist"]["variable"]}. For example, the default number of $z$-points is given by \texttt{input\_parameters["zgrid\_parameters"]["nzed"]}.

Moreover, as a part of this PhD work, new options have been implemented in stella. These are not available in \texttt{stella v0.6}, nonetheless they are already available on the master branch of \texttt{stella} as of September 26, 2024. The master branch is currently undergoing a major clean-up by Hanne Thienpondt and Georgia Acton, to make \texttt{stella} more readable, more robust and user-friendly. Better documentation is being written and automatic tests have been added to Github to ensure that the main physics implemented in the code remain correct. Moreover, as a part of this clean-up many variables and namelists are being renamed and variables are being relocated to different namelists. Therefore, no documentation can be written on the new namelists and variables at the moment of writing this thesis. Nonetheless, the new \texttt{stella} options, which have been implemented as a part of this PhD work, are documented here and will be referred to as belonging to \mbox{\texttt{stella vPhD}}. These options are available on the master \mbox{branch on September 26, 2024.}

To define the physics and numerical parameters used to run a \texttt{stella} simulation, an input file is written. The following
namelists are available:

\begin{itemize}
	\item Define the discrete coordinate grids $(k_x,k_y,z,\mu,v_\parallel)$ (section \ref{sec:defaultinputs:grids}) \vspace{-2mm}
	\begin{itemize}
		\item \texttt{kt\_grids\_knobs} \vspace{-1mm}
		\item \texttt{kt\_grids\_box\_parameters} \vspace{-1mm}
		\item \texttt{kt\_grids\_range\_parameters} \vspace{-1mm}
		\item \texttt{zgrid\_parameters} \vspace{-1mm}
		\item \texttt{vpamu\_grids\_parameters} \vspace{-1mm}
	\end{itemize}

	\item Define the magnetic geometry (section \ref{sec:defaultinputs:geometry}) \vspace{-2mm}
	\begin{itemize}
		\item \texttt{geo\_knobs} \vspace{-1mm}
		\item \texttt{vmec\_parameters} \vspace{-1mm}
		\item \texttt{millergeo\_parameters} \vspace{-1mm}
	\end{itemize}

	\item Define the kinetic species present in the plasma (section \ref{sec:defaultinputs:species}) \vspace{-2mm}
	\begin{itemize}
		\item \texttt{species\_knobs} \vspace{-1mm}
		\item \texttt{species\_parameters} \vspace{-1mm}
		\item \texttt{euterpe\_parameters} \vspace{-1mm}
	\end{itemize}

	\item Introduce an initial perturbation of the electrostatic potential (section \ref{sec:defaultinputs:fields}) \vspace{-2mm}
	\begin{itemize}
		\item \texttt{init\_g\_knobs} \vspace{-1mm}
	\end{itemize}

	\item Physics parameters (section \ref{sec:defaultinputs:phsyics}) \vspace{-2mm}
	\begin{itemize}
		\item \texttt{parameters} \vspace{-1mm}
		\item \texttt{physics\_flags} \vspace{-1mm}
	\end{itemize}

	\item Numerical parameters such as numerical schemes, parallelization, ...  (section \ref{sec:defaultinputs:numerical}) \vspace{-2mm}
	\begin{itemize}
		\item \texttt{knobs} \vspace{-1mm}
		\item \texttt{layouts\_knobs} \vspace{-1mm}
		\item \texttt{time\_advance\_knobs} \vspace{-1mm}
	\end{itemize}

	\item Diagnose the potential and distribution function (section \ref{sec:defaultinputs:diagnostics}) \vspace{-2mm}
	\begin{itemize}
		\item \texttt{stella\_diagnostics\_knobs} \vspace{-1mm}
	\end{itemize}

	\item Add dissipation in the form of hyper-dissipation or collisions (section \ref{sec:defaultinputs:dissipation}) \vspace{-2mm}
	\begin{itemize}
		\item \texttt{hyper} \vspace{-1mm}
		\item \texttt{collisions\_fp} \vspace{-1mm}
		\item \texttt{collisions\_dougherty} \vspace{-1mm}
		\item \texttt{dissipation} \vspace{-1mm}
	\end{itemize}

	\item Run radially global \texttt{stella} simulations (section \ref{sec:defaultinputs:radiallyglobal}) \vspace{-2mm}
	\begin{itemize}
	\item \texttt{multibox\_parameters} \vspace{-1mm}
	\end{itemize}

	\item Add neoclassical terms to the gyrokinetic equation (section \ref{sec:defaultinputs:neoclassical}) \vspace{-2mm}
	\begin{itemize}
		\item \texttt{sfincs\_input} \vspace{-1mm}
		\item \texttt{sources} \vspace{-1mm}
		\item \texttt{neoclassical\_input} \vspace{-1mm}
	\end{itemize}
\end{itemize}

 \section{Input variables for coordinate grids}
 \label{sec:defaultinputs:grids}

 \vspace{1mm}
 \begin{lstlisting}[language=Fortran,firstnumber=1]
! Select the option to define the (kx,ky) grid:
!    - A range of modes ["default", "range"]
!    - A box of modes ["box", "annulus", "nonlinear"] 	
&kt_grids_knobs
  grid_option = "default"
/
 \end{lstlisting}

 \vspace{1mm}
 \begin{lstlisting}[language=Fortran,firstnumber=1]
! This namelist is only read if grid_option = ["default", "range"]
&kt_grids_range_parameters

  ! For the kx grid, include nakx modes ranging from akx_min to akx_max
  ! For the ky grid, include naky modes ranging from aky_min to aky_max
  nakx = 1.0
  naky = 1.0
  akx_min = 0.0
  akx_max = -1.0
  aky_min = 0.0
  aky_max = 0.0

  ! Alternatively, the akx_min and akx_max values can be set using 
  !    kx = theta0*ky*shat for grid_option = "Miller" and q_as_x=.false.
  !    kx = theta0*ky for grid_option = "Miller" and q_as_x=.true.
  theta0_min = 0.0
  theta0_max = -1.0

  ! Select the option to define spacing along the ky grid:
  !    - Linear spacing ["default", "linear"]
  !    - Exponential spacing ["exponential"] 	
  kyspacing_option = "default"
/
 \end{lstlisting}

 \vspace{1mm}
 \begin{lstlisting}[language=Fortran,firstnumber=1]
! This namelist is only read if grid_option = ["box", "annulus", "nonlinear"]
&kt_grids_box_parameters

  ! Select the number of (x,y) grid points in real space which corresponds to
  !     naky = (ny - 1) / 3 + 1
  !     nakx = 2 * ((nx - 1) / 3) + 1
  nx = 1
  ny = 1

  ! The factor J in the parallel boundary condition can be set manually.
  ! If it remains unset, stella will select an appropriate value.
  jtwist = -1.0
  jtwistfac = 1.0

  ! The box size Ly in real space is set through:
  !     y0 = Ly/(2*pi*rho_ref) = Lalpha/(2*pi*rho_ref)(dy/dalpha) 
  ! In a full-flux-surface simulation rhostar determines y0 since L_alpha=2*pi 
  !     y0 = Ly/(2*pi*rho_ref) = Lalpha/(2*pi*rho_ref)(dy/dalpha) 
  !        = (1/rho_ref)(r0) = (r0/a)*(a/rho_ref) = rhotor/rhostar
  x0 = -1.0
  y0 = -1.0

  ! Other options
  centered_in_rho = .true.
  periodic_variation = .false.
  randomize_phase_shift = .false.
  phase_shift_angle = 0.0
/
 \end{lstlisting}

 \vspace{1mm}
 \begin{lstlisting}[language=Fortran,firstnumber=1]
&zgrid_parameters

  ! Select the parallel boundary condition:
  !    - Each mode is unconnected ["default", "zero", "unconnected"]
  !    - Self-periodic boundary condition ["self-periodic", "periodic"]
  !    - Twist-and-shift boundary condition ["linked"]
  !    - Stellarator-symmetric boundary condition ["stellarator"]
  boundary_option = "default"

  ! For <geo_option> = "miller" by default z = theta_Boozer
  ! For <geo_option> = "vmec" by default z = zeta_PEST
  ! The arc length can be used as the parallel coordinate instead:
  zed_equal_arc = .false.

  ! Number of grid points in the parallel direction is given by:
  !     nzgrid = nzed / 2 + (nperiod - 1) * nzed
  nzed = 24
  nperiod = 1

  ! Number of flux tubes for radially global simulations
  ntubes = 1

  ! If boundary_option="linked" but |shat|<shat_zero
  ! If boundary_option="stellarator" but |grad_x_grad_y_end|<|grad_x_grad_y_zero|
  ! Then use self-periodic boundary conditions instead
  shat_zero = 1e-05
  grad_x_grad_y_zero = 1e-05

  ! For the stellarator-symmetric boundary condition, the ratio dkx/dky can be set 
  ! manually, stella will then look for an optimal length Lz where this is valid.
  dkx_over_dky = -1.0
/
 \end{lstlisting}

 \vspace{1mm}
 \begin{lstlisting}[language=Fortran,firstnumber=1]
&vpamu_grids_parameters

  ! Set the (nmu, nvpa) velocity grid points, with nvpa = 2*nvgrid
  nvgrid = 24
  nmu = 12

  ! The velocity grid is limited to |vpa|<vpa_max and |vperp|<vperp_max
  vpa_max = 3.0
  vperp_max = 3.0

  ! By default the mu grid is set through a Gauss-Laguerre quadrature
  ! However, a mu grid with an equal step size can be chosen instead
  equally_spaced_mu_grid = .false.

  ! Options for the collision operator
  conservative_wgts_vpa = .false.
/
\end{lstlisting}

\section{Input variables for the magnetic geometry}
\label{sec:defaultinputs:geometry}

\vspace{1mm}
\begin{lstlisting}[language=Fortran,firstnumber=01]
! Define geometry parameters
&geo_knobs

  ! Select the option to define the magnetic geometry:
  !    - Based on a set of Miller parameters ["local", "default", "miller"]
  !    - Read from a VMEC equilibrium ["vmec"] 	
  !    - Read from an input file ["input.profiles"]
  geo_option = "local"
  
  ! Use a constant B(z) 
  set_bmag_const = .false.
  
  ! Use psi = q and x = (dpsi_p/dr)(1/Bref)(q-q0) in radially global simulations
  q_as_x = radial_variation
  
  ! The geometry can be read from a text file
  geo_file = "input.geometry"
  overwrite_bmag = .false.
  overwrite_b_dot_grad_zeta = .false.
  overwrite_gds2 = .false.
  overwrite_gds21 = .false.
  overwrite_gds22 = .false.
  overwrite_gds23 = .false.
  overwrite_gds24 = .false.
  overwrite_gbdrift = .false.
  overwrite_cvdrift = .false.
  overwrite_gbdrift0 = .false.
/
\end{lstlisting}

\vspace{1mm}
\begin{lstlisting}[language=Fortran,firstnumber=01]
! Define parameters related to Miller geometry 
! This namelist is only read if <geo_option>  = ["local", "default", "miller"]
&millergeo_parameters
  rhoc = 0.5
  rmaj = 3.0
  shift = 0.0
  qinp = 1.4
  shat = 0.8
  kappa = 0.0
  kapprim = 0.0
  tri = 0.0
  triprim = 0.0
  rgeo = 3.0
  betaprim = 0.0
  betadbprim = 0.0
  d2qdr2 = 0.0
  d2psidr2 = 0.0
  nzed_local = 128.0
  read_profile_variation = .false.
  write_profile_variation = .false.
/
\end{lstlisting}

\vspace{1mm}
\begin{lstlisting}[language=Fortran,firstnumber=01]
! Define parameters related to VMEC equilibria 
! This namelist is only read if <geo_option>  = "vmec"
&vmec_parameters
  alpha0 = 0.0
  zeta_center = 0.0
  nfield_periods = -1.0
  torflux = 0.6354167
  zgrid_refinement_factor = 1.0
  surface_option = 0.0
  radial_coordinate = "sgn(psi_t)psi_t"
  verbose = .true.
  vmec_filename = "equilibria/wout_w7x_standardconfig.nc"
  n_tolerated_test_arrays_inconsistencies = 0.0 
/
\end{lstlisting}

\vspace{2mm}
\noindent
In \texttt{stella vPhD} a new flag has been introduced to ensure that the perpendicular cross-section of the flux tube is rectangular at the midpoint of the flux tube as opposed to being sheared into a parallelogram. In other words, this flags ensures that the $\nablam x$ and $\nablam y$ directions are perpendicular at $z=0$. If the magnetic geometry is defined through a set of Miller parameters, then $\nablam x \cdot \nablam y = 0$ is always guaranteed at $z = \theta_s = 0$. On the other hand, if the magnetic geometry is read from a \texttt{VMEC} equilibrium, then the value of $\nablam x \cdot \nablam y$ at the midpoint of the flux tube, $z = \zeta_s = 0$, is not necessarily zero. Specifically, the cross-section is inherently rectangular for the flux tube centered around the $\alpha=0$ field line which has its midpoint at $(\theta_p,\zeta) = (0,0)$. However, if the flux tube is centered around the $\alpha = \iota \pi/N_\text{nfp}$ field line with its midpoint at $(\theta_p,\zeta) = (0,-\pi/N_\text{nfp})$, which is a stellarator-symmetric field line, then $\nablam x \cdot \nablam y$ is not guaranteed to be zero at the midpoint $(\theta_p,\zeta) = (0,-\pi/N_\text{nfp})$. To ensure that the $\nablam x$ and $\nablam y$ directions are perpendicular at the midpoint of the flux tube, the field-line label $\alpha = \theta_p - \iota \zeta$ has to be redefined as $\alpha = \theta_p - \iota (\zeta - \zeta_\text{center})$. This is achieved by setting \texttt{rectangular\_cross\_section = .true.} in the \texttt{vmec\_parameters} namelist.

\vspace{1mm}
\begin{lstlisting}[language=Fortran,firstnumber=01]
&vmec_parameters
  rectangular_cross_section = .false.
/
\end{lstlisting}

\section{Input variables for the kinetic species}
\label{sec:defaultinputs:species}

\begin{lstlisting}[language=Fortran,firstnumber=1]
! Define the number of kinetic species included in the simulation.
! The species_option can be ["default", "stella", "input.profiles", "euterpe"]
! with "default" being equal to choosing "stella". 
! In the species_parameters_X namelists, the species type can be:
!    - ions ["ion", "default"]
!    - electrons ["electrons", "e"]
! If no electron species are present, stella adds an adiabatic electron response 
! and the electron density and temperature are set through <tite> and <nine> in 
! the parameters namelist. If electrons are present, <tite> and <nine> are ignored
&species_knobs
  nspec = 2.0
  species_option = "stella"
  read_profile_variation = .false.
  write_profile_variation = .false.
  ecoll_zeff = .false.
/

! Define the parameters of the first kinetic species if nspec>=1
&species_parameters_1
  z = 1.0
  mass = 1.0
  dens = 1.0
  temp = 1.0
  tprim = -999.9
  fprim = -999.9
  d2ndr2 = 0.0
  d2tdr2 = 0.0
  bess_fac = 1.0
  type = "default"
/

! Define the parameters of the second kinetic species if nspec>=2
&species_parameters_2
  z = 1.0
  mass = 1.0
  dens = 1.0
  temp = 1.0
  tprim = -999.9
  fprim = -999.9
  d2ndr2 = 0.0
  d2tdr2 = 0.0
  bess_fac = 1.0
  type = "default"
/
\end{lstlisting}

\vspace{2mm}
\noindent
The species data can be read directly from an output file of the \texttt{EUTERPE} code \cite{mishchenko2022gyrokinetic, mishchenko2023global, jost2001global, kleiber2024euterpe}. Here the kinetic ion species will be read, as well as the kinetic electron species, unless Boltzmann electrons are used, in which case \texttt{tite} and \texttt{nine} have to be set in the \texttt{parameters} namelist.

\vspace{1mm}
\begin{lstlisting}[language=Fortran,firstnumber=01]
&euterpe_parameters
  nradii = 1000.0
  data_file = "euterpe.dat"
/
\end{lstlisting}

\section{Input variables to initialize the electrostatic potential}
\label{sec:defaultinputs:fields}

There are multiple options to initialize the perturbed electrostatic potential. Whereas the quasineutrality condition is used to consequently initialize the distribution function of the guiding centers.

\vspace{1mm}
\begin{lstlisting}[language=Fortran,firstnumber=01]
&init_g_knobs

  ! Select the initialization option 
  ! ["default", "noise", "many", "kpar", "rh", "remap"]
  ginit_option = "default"
  
  ! Used for all initialization options
  scale_to_phiinit = .false.
  phiinit = 1.0
  
  ! Used when a simulation is restarted
  scale = 1.0
  tstart = 0.0
  restart_file = "dummy.nc"
  restart_dir = "./"
  read_many = .true.
  
  ! Used for ginit_option = "default", "noise" and  "kpar"
  chop_side = .false.
  left = .true.
  
  ! Used for ginit_option = "default" and "kpar"
  den0 = 1.0
  upar0 = 0.0
  width0 = -3.5
  
  ! Used for ginit_option = "default"
  oddparity = .false.
  
  ! Used for ginit_option = "noise"
  zf_init = 1.0
  
  ! Used for ginit_option = "kpar"
  tpar0 = 0.0
  tperp0 = 0.0
  imfac = 0.0
  refac = 1.0
  den1 = 0.0
  upar1 = 0.0
  tpar1 = 0.0
  tperp1 = 0.0
  den2 = 0.0
  upar2 = 0.0
  tpar2 = 0.0
  tperp2 = 0.0
  
  ! Used for ginit_option = "rh"
  kxmax = 1e+100
  kxmin = 0.0
/
\end{lstlisting}

\section{Input variables for the physics parameters}
\label{sec:defaultinputs:phsyics}

\vspace{1mm}
\begin{lstlisting}[language=Fortran,firstnumber=1]
&physics_flags

  ! If an adiabatic (Boltzmann) species is included select how it is treated 
  !    Adiabatic electrons ["field-line-average-term"]
  !    Adiabatic ions ["no-field-line-average-term", "default"]
  adiabatic_option = "default"
  
  ! Include different terms in the gyrokinetic equation
  include_parallel_nonlinearity = .false.
  include_parallel_streaming = .true.
  include_mirror = .true.
  nonlinear = .false.
  
  ! By default stella runs local flux-tube simulations
  ! This flag turns on the full-flux-surface (FFS) version
  full_flux_surface = .false.
  
  ! As a check, each field line can be taken to be identical in FFS
  const_alpha_geo = .false.
  
  ! Radially global stella simulations
  radial_variation = .false.
  include_pressure_variation = .true.
  include_geometric_variation = .true.
  
/
\end{lstlisting}

\vspace{1mm}
\begin{lstlisting}[language=Fortran,firstnumber=1]
&parameters
	
  ! If Boltzmann/adiabatic electrons are included, then their
  ! temperature and density are set by tite and nine
  tite = 1.0
  nine = 1.0

  ! Set the collision frequency of each species 
  vnew_ref = -1.0  
  zeff = 1.0

  ! Used for electromagnetic version of stella
  ! For electrostatic stella, beta = 0
  beta = 0.0

  ! Define rhostar = rho_ref/a = m_ref*vt_ref/(e*B_ref*a_ref) = 1/irhostar
  rhostar = -1.0  
  irhostar = -1.0

  ! Include an equilibrium flow shear	
  g_exb = 0.0
  g_exbfac = 1.0
  omprimfac = 1.0
/
\end{lstlisting}

\section{Input variables for the numerical schemes}
\label{sec:defaultinputs:numerical}

\vspace{1mm}
\begin{lstlisting}[language=Fortran,firstnumber=1]
&time_advance_knobs

  ! Select the Runge-Kutta method
  !     Second order Runge-Kutta ["rk2"]
  !     Third order Runge-Kutta ["rk3", "default"]
  !     Fourth order Runge-Kutta ["rk4"]
  explicit_option = "default"
  
  ! Modify the frequencies in the gyrokinetic equation
  xdriftknob = 1.0
  ydriftknob = 1.0
  wstarknob = 1.0
  
  ! Modify the time-advance scheme
  flip_flop = .false.
/
\end{lstlisting}

\vspace{1mm}
\begin{lstlisting}[language=Fortran,firstnumber=1]
&knobs	

  ! For an electrostatic simulation set fphi=1, fapar=0 and fbpar=0
  fphi = 1.0
  fapar = 1.0
  fbpar = -1.0

  ! stella runs until t*v_{th,i}/a=tend or until istep=nstep
  tend = -1.0
  nstep = -1

  ! The time step of the code (code_dt) needs to  be smaller than cfl_dt, which is 
  ! the CFL constraint on the time step size. Therefore, code_dt needs to stay 
  ! within [cfl_dt*cfl_cushion_lower, cfl_dt*cfl_cushion_upper]. The code_dt 
  ! can thus be increased if cfl_dt increases, however, it never increases code_dt 
  ! above delt_max (which is set to delt by default). Exit stella if code_dt < 
  ! delt_min (i.e. when the potential blows up). By default stella stays a factor 
  ! of 2 under the CFL condition, otherwise the potential tends to blow up. By 
  ! default, if code_dt > cfl_dt/2 or code_dt < cfl_dt/100000, set code_dt to 
  ! cfl_dt/4. The default cfl_cushion_lower is very low to not trigger it.
  cfl_cushion_upper = 0.5       
  cfl_cushion_middle = 0.25     
  cfl_cushion_lower = 0.00001   
  delt_max = -1
  delt_min = 1.e-10

  ! Set the available cpu time in seconds, 5 minutes before this time,
  ! stella will make a clean exit. If this is not set, the simulation is 
  ! sometimes ended abruptly and the netcdf file can become corrupted
  avail_cpu_time = 1.e10

  ! Select the option to perform the LU decomposition of the response matrix
  !     No parallelisation (single core is used) ["default"]
  !     Parallelisation on jtwist nodes ["local"]
  !     Parallelisation on all cores ["global"]
  lu_option = "default"

  ! If the random-number-generator (rng) seed is fixed, then the initialization 
  ! of the fields with ginit_option = "noise" uses the same seed
  ! If rng_seed<0, then the current time is used as the seed
  rng_seed = -1 

  ! Numerical schemes
  stream_implicit = .true.
  mirror_implicit = .true.
  drifts_implicit = .false.
  driftkinetic_implicit = .false.
  maxwellian_inside_zed_derivative = .false.
  mirror_semi_lagrange = .true.
  mirror_linear_interp = .false.
  stream_matrix_inversion = .false.
  use_deltaphi_for_response_matrix = .false.
  maxwellian_normalization = .false.
  delt_option = "default"
  zed_upwind = 0.02
  vpa_upwind = 0.02
  time_upwind = 0.02
  ky_solve_radial = 0
  ky_solve_real = .false.
  mat_gen = .false.
  mat_read = .false. 

  ! Parallelisation along (kx,ky,z) grid points or (mu,vpa,s) grid points
  fields_kxkyz = .false.

/
\end{lstlisting}

\vspace{2mm}
\noindent
Parallelization of the code.

\vspace{1mm}
\begin{lstlisting}[language=Fortran,firstnumber=01]
&layouts_knobs
  xyzs_layout = "yxzs"
  vms_layout = "vms"
/
\end{lstlisting}

\section{Input variables for the diagnostics}
\label{sec:defaultinputs:diagnostics}

The diagnostics namelist for \texttt{stella v0.6} is given by:

\vspace{1mm}
\begin{lstlisting}[language=Fortran,firstnumber=01]
&diagnostics
  nwrite = 50.0
  navg = 50.0
  nsave = -1.0
  save_for_restart = .false.
  flux_norm = .true.
  nc_mult = 1.0
  write_omega = .false.
  write_phi_vs_time = .false.
  write_gvmus = .false.
  write_gzvs = .false.
  write_kspectra = .false.
  write_moments = .false.
  write_radial_fluxes = radial_variation
  write_radial_moments = radial_variation
  write_fluxes_kxkyz = .false.
  flux_norm = .true.
  nc_mult = 1.0
/
\end{lstlisting}

\vspace{2mm}
\noindent
Note that some of the novel diagnostics, developed as part of this PhD work are not present in \texttt{stella v0.6}. Nonetheless they are already available on the master branch of \texttt{stella} and they will be released in the upcoming \texttt{stella} version.
Moreover, the diagnostics namelist, available on the master branch on September 26, 2024, contains a lot of name changes, since an effort has been made as a part of this PhD work to clean the \texttt{stella} code and make it more readable and user-friendly. The diagnostics namelist on the master branch on September 26, 2024, is given by,

\vspace{1mm}
\begin{lstlisting}[language=Fortran,firstnumber=01]
&diagnostics
  nwrite = 50.0
  navg = 50.0
  nsave = -1.0
  autostop = .false.
  save_for_restart = .false.
  flux_norm = .true.
  nc_mult = 1.0
  write_phi2_vs_time = .true.
  write_apar2_vs_time = .true.
  write_bpar2_vs_time = .true.
  write_fluxes_vs_time = .true.
  write_phi_vs_kxkyz = .false.
  write_g2_vs_vpamus = .false.
  write_g2_vs_zvpas = .false.
  write_g2_vs_zmus = .false.
  write_g2_vs_kxkyzs = .false.
  write_g2_vs_zvpamus = .false.
  write_distribution_g = .true.
  write_distribution_h = .false.
  write_distribution_f = .false.
  write_omega_vs_kxky = .true.
  write_omega_avg_vs_kxky = .false.
  write_phi2_vs_kxky = .false.
  write_moments = .false.
  write_radial_fluxes = .false.
  write_radial_moments = .false.
  write_fluxes_kxkyz = .false.
  write_fluxes_kxky = .false.
  write_all = .false.
/
\end{lstlisting}

\section{Input variables for dissipation}
\label{sec:defaultinputs:dissipation}

The collision operator has been included in \texttt{stella} by von Boetticher in Ref.~\cite{von2024linearised}, where the details on its implementation can be found, and benchmarks with the gyrokinetic code \texttt{GS2} are performed. The collision operator is similar to the one implemented in \texttt{GS2} described in Refs.~\cite{abel2008linearized} and \cite{barnes2009linearized}. In this work only hyper-dissipation is used.

\vspace{1mm}
\begin{lstlisting}[language=Fortran,firstnumber=01]
&dissipation
  include_collisions = .false.
  collisions_implicit = .true.
  collision_model = "dougherty"
  hyper_dissipation = .false.
/
&collisions_dougherty
  momentum_conservation = .true.
  energy_conservation = .true.
  vpa_operator = .true.
  mu_operator = .true.
/
&collisions_fp
  testpart = .true.
  fieldpart = .false.
  lmax = 1.0
  jmax = 1.0
  nvel_local = 512.0
  interspec = .true.
  intraspec = .true.
  iiknob = 1.0
  ieknob = 1.0
  eeknob = 1.0
  eiknob = 1.0
  eiediffknob = 1.0
  eideflknob = 1.0
  deflknob = 1.0
  eimassr_approx = .false.
  advfield_coll = .true.
  spitzer_problem = .false.
  density_conservation = .false.
  density_conservation_field = .false.
  density_conservation_tp = .false.
  exact_conservation = .false.
  exact_conservation_tp = .false.
  vpa_operator = .true.
  mu_operator = .true.
  cfac = 1.0
  cfac2 = 1.0
  nuxfac = 1.0
  i1fac = 1.0
  i2fac = 0.0
  no_j1l1 = .true.
  no_j1l2 = .false.
  no_j0l2 = .false.
/
&hyper
  d_hyper = 0.05
  d_zed = 0.05
  d_vpa = 0.05
  hyp_zed = .false.
  hyp_vpa = .false.
  use_physical_ksqr = .true.
  scale_to_outboard = .false.
/
\end{lstlisting}

\section{Input variables for radially global simulations}
\label{sec:defaultinputs:radiallyglobal}

The flux-tube version of \texttt{stella} has been extended by St-Onge to allow for radially global simulations. Details on its implementation can be found in Ref.~\cite{st2022novel}.

\vspace{1mm}
\begin{lstlisting}[language=Fortran,firstnumber=01]
&multibox_parameters
  boundary_size = 4.0
  krook_size = 0.0
  smooth_zfs = .false.
  zf_option = "default"
  lr_debug_option = "default"
  krook_option = "default"
  rk_step = .false.
  nu_krook_mb = 0.0
  mb_debug_step = -1.0
  krook_exponent = 0.0
  comm_at_init = .false.
  phi_bound = 0.0
  phi_pow = 0.0
  krook_efold = 3.0
  use_dirichlet_bc = .false.
/
\end{lstlisting}

\section{Input variables for neoclassical fluxes and electric field}
\label{sec:defaultinputs:neoclassical}

The \texttt{stella} code is integrated with the stellarator neoclassical code \texttt{SFINCS} (Stellarator Fokker-Planck Iterative Neoclassical Conservative Solver) developed by Landreman. The \texttt{SFINCS} code is described in Ref.~\cite{landreman2014comparison}. The \texttt{SFINCS} code is run from within \texttt{stella} to calculate the radial electric field, which is then used in \texttt{stella}.

\vspace{1mm}
\begin{lstlisting}[language=Fortran,firstnumber=01]
&neoclassical_input

  ! The only supported option is ["sfincs"]
  neo_option = "sfincs"
  
  ! Include neoclassical terms in the gyrokinetic equation
  include_neoclassical_terms = .false.
  
  ! Settings for radial derivatives
  nradii = 5.0
  drho = 0.01
/
\end{lstlisting}

\vspace{1mm}
\begin{lstlisting}[language=Fortran,firstnumber=01]
&sfincs_input
  read_sfincs_output_from_file = .false.
  nproc_sfincs = 1
  irad_min = -nradii / 2; irad_max = nradii / 2
  calculate_radial_electric_field = .true.
  includeXDotTerm = .true.
  includeElectricFieldTermInXiDot = .true.
  magneticDriftScheme = 0
  includePhi1 = .true.
  includePhi1InKineticEquation = .false.
  geometryScheme = 1
  VMECRadialOption = 0
  equilibriumFile = 'wout_161s1.nc'
  coordinateSystem = 3
  inputRadialCoordinate = 3
  inputRadialCoordinateForGradients = 3
  aHat = 1.0
  psiAHat = geo_surf%psitor_lcfs
  Delta = -1.0
  nu_n = -1.0
  dPhiHatdrN = -9999.9
  Er_window = 0.3
  nxi = 48
  nx = 12
  Ntheta = 65
  Nzeta = 1
/
\end{lstlisting}

\vspace{1mm}
\begin{lstlisting}[language=Fortran,firstnumber=01]
&sources
  source_option = "none"
  nu_krook = 0.05
  tcorr_source = 0.02
  ikxmax_source = 1.0
  krook_odd = .true.
  exclude_boundary_regions = .false.
  tcorr_source_qn = 0.0
  exclude_boundary_regions_qn = .false.
  from_zero = .true.
  conserve_momentum = .false.
  conserve_density = .false.
/
\end{lstlisting}

\chapter{Most important \texttt{stella} variables}

\section{Setting the discrete $(k_x, k_y, z, v_\parallel, \mu)$ grids}
\label{app:stella:grids}

To numerically evolve a perturbation of the electrostatic field within a flux tube, the $(k_x, k_y, z, v_\parallel, \mu)$ dimensions must be discretized, and a balance has to be found between the simulation accuracy and the computational cost. The number of grid points, the step size between the grid points and the minimum (or maximum) values are set in the input file given to the \texttt{stella} code.

\subsection[Discrete $z$-grid]{Discrete $\bm{z}$-grid}

Before discretizing the $z$-grid, the length of the flux tube must be specified. If the magnetic geometry is defined by a set of Miller parameters (\texttt{geo\_option\,=\,"Miller"}), then
\begin{equation}
	\texttt{nperiod} = (N+1)/2
\end{equation}
specifies the length of the flux tube, with $N$ the number of poloidal turns which has to be an odd integer. For example, to simulate 3 poloidal turns, \texttt{nperiod = 2} should be set in the \texttt{zgrid\_parameters} namelist. The length of the selected field line is then given by,
\begin{equation}
	L_z = L_{\theta_s} = 2\pi N = 2\pi\,(2*\texttt{nperiod} - 1).
\end{equation}
On the other hand,  if a \texttt{VMEC} equilibrium is read (\texttt{geo\_option\,=\,"VMEC"}), then the length of the flux tube is set through \texttt{nfield\_periods} in the \texttt{vmec\_parameters} namelist. Typically $N$ poloidal turns are simulated for a stellarator, for which
\begin{equation}
	\texttt{nfield\_periods} = qN\,N_\text{nfp} = P\, N_\text{nfp},
\end{equation}
with $P$ the number of toroidal turns and $N_\text{nfp}$ the number of field periods within a single toroidal turn of the device. Therefore, the length of the selected field line is given by,
\begin{equation}
	\label{eq:Lzeta}
	L_z = L_\zeta = 2\pi P = 2\pi \,\, \texttt{nfield\_periods}/N_\text{nfp}.
\end{equation}
Once the length of the flux tube has been set, the $z$-grid can be discretized. The number of grid points in the parallel direction is set through the input variable \texttt{nzed} in the \texttt{zgrid\_parameters} namelist. The variable \texttt{nzed} corresponds to the number of $z$ points within a single poloidal turn if \texttt{geo\_option\,=\,"Miller"}, while it corresponds to the number of $z$ points along the entire field line if \texttt{geo\_option\,=\,"VMEC"}. Moreover,  \texttt{nzed} must be an even integer, since the $z=0$ point, corresponding to the midpoint of the flux tube, is added by \texttt{stella}. Using \texttt{nzed}, the number of positive (or negative) $z$ points is calculated as,
\begin{align}
	\texttt{nzgrid} & = \texttt{nzed}/2 + (\texttt{nperiod}-1)*\texttt{nzed}.
\end{align}
As a result, the total number of grid points $N_z$ along the parallel direction is given by,
\begin{align}
	\label{eq:nzeta}
	\texttt{nzeta} & = 2*\texttt{nzgrid} + 1 = N_z.
\end{align}
For \texttt{geo\_option\,=\,"Miller"} the parallel coordinate corresponds to the poloidal Boozer angle, \mbox{$z=\thetaB$},  by default, whereas for \texttt{geo\_option\,=\,"VMEC"} the parallel coordinate is taken to be the conventional cylindrical toroidal angle, $z=\zeta$, by default. Nonetheless, the arc length $\ell$ of the field line can be used as the parallel coordinate by setting $\texttt{zed\_equal\_arc\,=\,.true.}$ in the input file. For $z=\ell$ the parallel gradient, $\hatb \cdot \tilde{\nablam} \tilde{B}$ in equation (\ref{eq:NormalizedGyrokineticEquation}) is constant along the field line, which improves the numerical resolution along $z$. Especially when including kinetic electrons in the simulation, setting $z=\ell$ tends to avoid numerical instabilities. Regardless of the choice of the parallel coordinate, the normalized $\tilde{z}$ grid ranges from $-N\pi$ to $N\pi$ if \texttt{geo\_option\,=\,"Miller"}, whereas it ranges from $-\pi$ to $\pi$ if \texttt{geo\_option\,=\,"VMEC"}, since $\texttt{nperiod} = 1$ in this case.

\subsection{Discrete $(k_x, k_y)$ grid}
\label{eq:discretekxkygrid}

The $(k_x,k_y)$-dimensions are discretized by specifying the number of modes and the step size in the \texttt{kt\_grids\_knobs}, \texttt{kt\_grids\_box\_parameters} and \texttt{kt\_grids\_range\_parameters} namelists in the input file (appendix \ref{sec:defaultinputs:grids}). The number of $(k_x,k_y)$ grid points in reciprocal space are defined by the number of $(x,y)$ grid points in real space, through,
\begin{align}
	N_{k_x} & = 2\,(N_x - 1)/3 + 1, \\[1mm]
	N_{k_y} & = (N_y - 1)/3 + 1,
\end{align}
where $\texttt{nx} = N_x$ and $\texttt{ny} = N_y$ are specified in the input file.
The number of grid points in spectral coordinates, $\texttt{nakx} = N_{k_x}$ and $\texttt{naky} = N_{k_y}$, is only 2/3 of those in real space, to ensure that the $(N_{k_x},N_{k_y})$ Fourier modes in the convolution (equation \ref{eq:NormalizedNonlinearTerm}) are free of aliasing \cite{isliker2004tutorial}. Moreover, negative binormal wavenumbers are not evolved, therefore, $N_{k_y}$ is only 1/3 of $N_y$. For example, if $N_x = N_y = 91$ grid points are selected in the input file, then $N_{k_x} = 61$ and $N_{k_y} = 31$ spectral grid points are evolved in the code. To perform the Fourier transformation in equation (\ref{eq:NormalizedNonlinearTerm}), the $(k_x,k_y)$ vectors of length $(N_{k_x},N_{k_y})$ are extended to vectors of length $(N_{x}, N_{y})$. The negative values of $k_y$ are recovered using the reality condition, which states,
\begin{equation}
	\hat{\varphi}(k_x,k_y,z) = \hat{\varphi}^*(-k_x, -k_y, z),
\end{equation}
with $\hat{\varphi}^*_\mathbf{k}$ the complex conjugate of $\hat{\varphi}_\mathbf{k}$. The remaining 1/3 of the vector is padded with zeros to avoid aliasing errors when performing the convolution in equation (\ref{eq:NormalizedNonlinearTerm}). Enough $k_x$ and $k_y$ modes have to be kept to be able to resolve fluctuations up to a desired value of $k_\perp\rho_i$, above which the Fourier coefficients $\hat{\varphi}(k_x,k_y,z,t)$ are assumed to vanish due to the finite-Larmor-radius effects (section \ref{sec:FLR}). Specifically, the largest wavenumbers included in the Fourier expansions are,
\begin{align}
	k_{x, \text{max}} & = (N_{k_x}-1)/2 \ k_{x, \text{min}},
	\\[1mm]
	k_{y, \text{max}} & = (N_{k_y}-1) \ k_{y, \text{min}}.
\end{align}
The smallest wavenumbers are defined through the perpendicular box size of the flux tube, i.e.
\begin{equation}
	\label{eq:y0}
	k_{x,\text{min}} = \frac{2\pi}{L_x} = \Big(\frac{d\psi}{dx}\Big) \frac{2\pi}{L_\psi}
	\qquad\text{and}\qquad
	k_{y,\text{min}} = \frac{2\pi}{L_y} = \Big(\frac{d\alpha}{dy}\Big) \frac{2\pi}{L_\alpha}.
\end{equation}
In particular, the minimum binormal wavenumber, $k_{y,\text{min}}$, is specified explicitly by the user by setting the normalized binormal box width of the flux tube, through,
\begin{equation}
	\texttt{y0} = \frac{L_y}{2\pi \rho_r} = \frac{1}{k_{y,\text{min}}\rho_r} = \frac{1}{\texttt{dky}}
\end{equation}
For example, setting $\texttt{y0} = 15$ yields a flux tube with a width of $94.2\,\rho_r$ in the binormal direction, corresponding to $k_{y,\text{min}}\rho_r = 0.067$. Seeing that $\rho_r$ typically corresponds to the ion Larmor radius, the box would be $94.2$ ion Larmor radii wide. On the other hand, the radial box size $L_x$ follows from the parallel boundary conditions. If the $(k_x,k_y)$-modes remain unconnected at the flux tube ends (\texttt{boundary\_option = "unconnected"}), if they are connected onto themselves (\texttt{boundary\_option = "self-periodic"}), or if stellarator-symmetric boundary conditions are used with \mbox{$[\nablam \psi \cdot  \nablam \alpha]_{z_\pm}\! = 0$} (\texttt{boundary\_option = "stellarator"}), then the radial box width can be chosen freely. If $L_x$ remains undefined then $k_{x,\text{min}} = k_{y,\text{min}}$ is used, otherwise it is set through,
\begin{equation}
	\texttt{x0} = \frac{L_x}{2\pi \rho_r} = \frac{1}{k_{x,\text{min}}\rho_r} = \frac{1}{\texttt{dkx}}
\end{equation}
Alternatively, if the twist-and-shift parallel boundary condition (section \ref{sec:twistandshift}) is selected through \texttt{boundary\_option = "linked"}, then $L_x$ must satisfy equation (\ref{eq:kalphamin}). When the magnetic geometry is defined by a set of Miller parameters, $N=\texttt{nperiod}$ specifies the number of poloidal turns that are considered. In local $(x,y)$-coordinates, using $\alpha_q=\zeta_s - q \theta_s$ and $z=\theta_s$, the twist-and-shift boundary condition defines the smallest radial wavenumber which is included in the simulation,
\begin{equation}
	\texttt{dkx} = k_{x,\text{min}}\rho_r =  \frac{2\pi N}{J}   \frac{dq}{d\psi} \,   \Big(\frac{dy}{d\alpha} \frac{d\psi}{dx}\Big) \, k_{y,\text{min}}\rho_r,
\end{equation}
where  $J = \texttt{jtwist}$ can be set in the input file. If $J$ is not set by the user, an optimal value is set by \texttt{stella}, to ensure that $k_{y,\text{min}}$ is as close as possible to $k_{x,\text{min}}$. On the other hand,  if a \texttt{VMEC} equilibrium is read, the length of the flux tube is set through \texttt{nfield\_periods}, which determines the number of field periods that are simulated. Typically $N$ poloidal turns are simulated for a stellarator, for which $\texttt{nfield\_periods} = qN\,N_\text{nfp} = P\, N_\text{nfp}$, with $P$ the number of toroidal turns and $N_\text{nfp}$ the number of field periods within a single toroidal turn of the device. Using \mbox{$\alpha_\iota=\theta_s - \iota \zeta_s$} and $z=\zeta_s$, the twist-and-shift boundary condition (eq.~\ref{eq:kalphamin}) defines $k_{x,\text{min}}\rho_r$ through,
\begin{equation}
   \texttt{dkx} =  k_{x,\text{min}}\rho_r = \frac{2\pi P}{J} \, \frac{d\iota}{d\psi} \, \Big(\frac{dy}{d\alpha} \frac{d\psi}{dx}\Big) \, k_{y,\text{min}}\rho_r,
\end{equation}
Finally, if stellarator-symmetric boundary conditions are used with $[\nablam \psi \cdot  \nablam \alpha]_{z_\pm}\! \neq 0$, then the smallest radial wavenumber included in the simulation is given by equation (\ref{eq:kalphaminlocal}), which yields,
\begin{equation}
	\texttt{dkx} =  k_{x,\text{min}}\rho_r = \frac{2}{J}  \frac{[\nablam x \cdot  \nablam y]_{z_-} }{|\nablam x|^2}  \, k_{y,\text{min}}\rho_r.
\end{equation}
To summarize, by setting \texttt{ny} and \texttt{y0} in the input file, the grid of binormal wavenumbers $k_y$ is completely defined. On the other hand, the radial wavenumbers $k_x$ depend on the specified \texttt{nx} as well as on a combination of \texttt{x0}, \texttt{jtwist}, \texttt{boundary\_option}, \texttt{nperiod} and \texttt{nfield\_periods}.

\subsection{Discrete $(v_\parallel, \mu)$ grid}

The parallel velocity and magnetic moment are discretized using the variables in the \texttt{vpamu\_grids\_ parameters} namelist (appendix \ref{sec:defaultinputs:grids}). Specifically, the number of grid points with positive velocities, and the maximum normalized parallel velocity, are specified in the input file as,
\begin{align}
	\texttt{nvgrid} & = N^+_{v_\parallel} = N_{v_\parallel}/2, \\[1mm]
	\texttt{vpa\_max} & = \tilde{v}_{\parallel, \text{max}}  = v_{\parallel, \text{max}}/v_{\text{th},s}.
\end{align}
The cutoff value, $\tilde{v}_{\parallel, \text{max}}$ should be large enough to ensure that the velocity integral (eq.~\ref{eq:NormalizedQuasiNeutrality}) is negligible for $\tilde{v}_\parallel > \tilde{v}_{\parallel, \text{max}}$.   The $\tilde{v}_\parallel$ grid points are equally spaced, therefore, the total number of grid points and the step size along the parallel velocity grid are given by,
\begin{align}
	\texttt{nvpa} & = N_{v_\parallel}, \\[1mm]
	\texttt{dvpa} & = \Delta \tilde{v}_{\parallel} = 2 \tilde{v}_{\parallel, \text{max}}/(N_{v_\parallel}-1).
\end{align}
The number of parallel velocities is always even to ensure that $v_\parallel = 0$ is not included in the grid, which simplifies the numerical implementation of the gyrokinetic equation. As a result, the minimum normalized parallel velocity is $\tilde{v}_{\parallel, \text{min}} = \Delta \tilde{v}_{\parallel}/2$. Similarly, the grid of magnetic moments $\mu = v_\perp^2/2B$, is defined by specifying the number of grid points and the maximum normalized perpendicular velocity in the input file, which are defined as,
\begin{align}
	\texttt{nmu} & = N_{\mu}, \\[1mm]
	\texttt{vperp\_max} & = \tilde{v}_{\perp, \text{max}} = v_{\perp, \text{max}}/v_{\text{th},s}.
\end{align}
If the grid is chosen to be equally spaced (\texttt{equally\_spaced\_mu\_grid\,=\,.true.}) then the \mbox{step size is,}
\begin{align}
	\texttt{dmu} & =  \Delta \tilde{\mu} = \tilde{\mu}_\text{max}/(N_{\mu}-0.5).
\end{align}
with $\tilde{\mu}_\text{max} = \tilde{v}^2_{\perp, \text{max}} / (2\, \text{max}(\tilde{B}))$ the maximum normalized magnetic moment, which should be large enough to ensure that the velocity integral is negligible for $\tilde{\mu} > \tilde{\mu}_\text{max}$. Moreover, the $\mu = 0$ is not included in the grid to simplify the implementation, which means that the  minimum normalized magnetic moment is given by $\tilde{\mu}_\text{min} = \Delta \tilde{\mu}/2$, if the $\mu$ grid is equally spaced. Alternatively, the $\mu$ grid points can be set according to a Gauss-Laguerre quadrature \cite{stella, abramowitz1968handbook}, which states that,
\begin{equation}
	\int\limits_{0}^{\infty} e^{-x} \, f(x) \, dx \approx \sum_{i=1}^{N_x} w_i \, f(x_i)
	\qquad \text{with} \qquad
	w_i = \frac{x_i}{(n+1)^2 [L_{n+1}(x_i)]^2}.
\end{equation}
where $x_i$ is the $i^\text{th}$ root of Laguerre polynomial $L_n(x)$ and $w_i$ are the weights. These $x_i$ grid points are rescaled so that the maximum value of $x_i$ corresponds to $\tilde{\mu}_\text{max} = \tilde{v}^2_{\perp, \text{max}} / (2\, \text{min}(\tilde{B}))$. Moreover, using $f(x) = g(x) \,e^x$ we can rewrite the integral with respect to the magnetic moment as,
\begin{equation}
	\int\limits_{0}^{\infty} g(\tilde{\mu}) \, d\tilde{\mu} = \frac{\tilde{\mu}_\text{max}}{x_\text{max}} \int\limits_{0}^{\infty} e^{-x} \, g(x) \, e^x \, dx \approx \frac{\tilde{\mu}_\text{max}}{x_\text{max}} \, \sum_{i=1}^{N_x} \, w_i \, g(x_i) \, e^{x_i}.
\end{equation}
This ensures that the step size is very small close to $\mu=0$, while it becomes bigger towards $\tilde{\mu}_\text{max}$. The velocity integral in equation (\ref{eq:NormalizedQuasiNeutrality}) is then given by,
\begin{equation}
	\frac{2\tilde{B}}{\sqrt{\pi}} \int\limits_{-\infty}^\infty \!\! d\tilde{v}_\parallel \int\limits_0^\infty \! d\tilde{\mu} =
	\int\limits_{-\infty}^\infty \! \Bigg(\frac{d\tilde{v}_\parallel}{\sqrt{\pi}}\Bigg) \int\limits_0^\infty \! \Big(2\tilde{B} \, d\tilde{\mu}\Big)
	\approx \!\! \sum_{\texttt{i}=-\texttt{nvgrid}}^\texttt{nvgrid} \!\!\!\!\! \texttt{wgts\_vpa[i]} \, \sum_{\texttt{j}=1}^\texttt{nmu} \, \texttt{wgts\_mu[j]}
\end{equation}
The integration weights for the parallel velocity are constructed according to the composite Simpson's rule, while the 3/8 Simpson's rule is used at the outer boundaries. The integration weights for the inner grid points are given by \mbox{$\texttt{wgts\_vpa[i]} = \Delta \tilde{v}_{\parallel}/\sqrt{\pi}$}. The weights for the magnetic moments are given by $\texttt{wgts\_mu[i]} = 2 \tilde{B} \Delta \tilde{\mu}$  if the $\tilde{\mu}$ grid is evenly spaced and \mbox{$\texttt{wgts\_mu[i]} = 2 \tilde{B}\, w_i\, \tilde{\mu}_\text{max}/x_\text{max}$} if a Gauss-Laguerre quadrature is used. Note that it is important to correctly apply the integration weights in velocity space to diagnose the outputs from \texttt{stella}.

\section{Magnetic geometry within a flux tube}
\label{sec:geometryicQuantities}

There are eight independent geometrical quantities appearing in the gyrokinetic equation (\ref{eq:NormalizedGyrokineticEquation}). Therefore, the quantities needed to specify the magnetic geometry in \texttt{stella} are the following:

\noindent
\begin{minipage}{0.4\textwidth}
	\begin{align}
		\label{eq:stella:geometricquantitiesfirst}
		\texttt{bmag} & = B/B_r \\[3mm]
		\texttt{gradpar} & = a\,\nablam_\parallel z \\[3mm]
		\texttt{gds2} & = |\nablam y|^2 \\[3mm]
		\texttt{gds21} & = \hat{s} \,\nablam x \cdot \nablam y \\[3mm]
		\texttt{gds22} & = \hat{s}^2 \, |\nablam x|^2
	\end{align}
\end{minipage}%
\begin{minipage}{0.1\textwidth}
	\textcolor{white}{filler}
\end{minipage}%
\begin{minipage}{0.5\textwidth}
	\begin{align}
		\texttt{gbdrift} & = \frac{2aB_r}{B^3} \, \mathbf{B} \times \nablammm B \cdot \nablam y \\[2mm]
		\texttt{gbdrift0} & = \hat{s} \, \frac{2aB_r}{B^3}\,  \mathbf{B} \times \nablammm B \cdot \nablam x \\[2mm]
		\texttt{cvdrift} & = \frac{2aB_r}{B^2} \, \mathbf{B} \times \boldsymbol{\kappa} \cdot \nablam y \\[2mm]
		\texttt{cvdrift0} & = \hat{s} \, \frac{2aB_r}{B^2}\,  \mathbf{B} \times \boldsymbol{\kappa} \cdot \nablam x
		\label{eq:stella:geometricquantitieslast}
	\end{align}
\end{minipage}

\vspace{6mm}
\noindent
with $\boldsymbol{\kappa} = \hatb \cdot \nablam \hatb$ the curvature vector, and $\mathbf{B} \times \boldsymbol{\kappa} \cdot \nablam x = \hatb \times \nablammm B \cdot \nablam x$ for any static ideal MHD equilibrium.
The quantities above are functions of the parallel coordinate $z$ and are calculated explicitly in \texttt{stella} based on a \texttt{VMEC} equilibrium or for a set of Miller parameters.
The Jacobian from Cartesian coordinates to the $(y,x,z)$ \texttt{stella} coordinates, with $\mathbf{B} = \psi_t' \nablam \psi \times \nablam \alpha$ is given by,
\begin{align}
	\label{eq:jac}
	\frac{1}{\mathcal{J}} & = (\boldsymbol{\nabla}y \times \boldsymbol{\nabla} x) \cdot \boldsymbol{\nabla} z = \Big(\frac{dy}{d\alpha}\Big) \Big( \frac{dx}{d\psi} \Big) \, (\boldsymbol{\nabla}\alpha \times \boldsymbol{\nabla} \psi) \cdot \boldsymbol{\nabla} z \\[2mm]
	& = -\,\frac{1}{\psi_t'} \Big(\frac{dy}{d\alpha}\Big) \Big( \frac{dx}{d\psi} \Big) \, \mathbf{B} \cdot \boldsymbol{\nabla} z
	= -\,\frac{1}{\psi_t'} \Big(\frac{\rho_r}{a}\Big) \frac{d\tilde{x}}{d\tilde{\psi}} \Big(\frac{\rho_r}{a}\Big) \frac{d\tilde{y}}{d\tilde{\alpha}} \, \tilde{\mathbf{B}} \cdot \tilde{\nablam} \tilde{z}
\end{align}
Within the \texttt{stella} code, the Jacobian is computed as,
\begin{equation}
	\texttt{jacob} = - \texttt{clebsch\_factor} / (\texttt{dydalpha * dxdpsi * b\_dot\_grad\_z * bmag}).
\end{equation}
Note that the $(y,x,z)$ coordinate system used in \texttt{stella} is left-handed.
The normalized derivatives of the local perpendicular coordinates $(x,y)$ with respect to $(\psi,\alpha)$ are assumed to be of the order of the gyrokinetic ordering parameter $\rho_{*s} = {\rho_s}/{L}$ (eq.~\ref{eq:GyrokineticOrderings}). Therefore, the derivatives are multiplied with $\rho_{*} = \rho_r/L$ with $L=a$ in \texttt{stella} to ensure that they are of order unity, i.e.
\begin{align}
	\texttt{dxdpsi}
	= \Big(\frac{\rho_r}{a}\Big) \frac{d\tilde{x}}{d\tilde{\psi}},
	\\[2mm]
	\texttt{dydalpha} =  \Big(\frac{\rho_r}{a}\Big) \frac{d\tilde{y}}{d\tilde{\alpha}}.
\end{align}

\section{Magnetic drift frequency and diamagnetic frequency}

The normalized \textit{magnetic drift frequency} $\omega_{d,\mathbf{k},s}$  (eq.~\ref{eq:NormalizedOmegaD}) in the gyrokinetic equation can be split up into the curvature drift and $\nablammm B$ drift. The normalized curvature drift frequency \mbox{is calculated as,}
\begin{equation}
	\begin{split}
		\omega^\text{curvature}_{d,\mathbf{k},s} &
		= \frac{\tilde{T}_s }{Z_s \tilde{B}} \, \hatb \times \tilde{v}_\parallel^2 \tilde{\boldsymbol{\kappa}} \cdot (k_x\rho_r\boldsymbol{\nabla} x + k_y\rho_r \boldsymbol{\nabla} y )\\
		& =  \tilde{v}_\parallel^2 \, \frac{\tilde{T}_s}{Z_s \tilde{B}}\, \Big( \frac{a}{B} \Big)  \Big( k_x\rho_r  \, (\mathbf{B} \times  \boldsymbol{\kappa}  \cdot \boldsymbol{\nabla} x)  +  k_y\rho_r \, (\mathbf{B} \times  \boldsymbol{\kappa}  \cdot \boldsymbol{\nabla} y) \Big)   \\
		& =   \tilde{v}_\parallel^2 \, \frac{1}{2} \frac{\tilde{T}_s}{Z_s}  \Big( \frac{k_x\rho_r}{\textcolor{black}{\hat{s}}}  \, \texttt{xdriftknob}*\texttt{cvdrift0}  +  k_y\rho_r \, \texttt{ydriftknob}*\texttt{cvdrift} \Big),
	\end{split}
\end{equation}
where \texttt{xdriftknob} and \texttt{ydriftknob} are set to 1 by default. These are input variables that have been added to the expression, in order to modify the strength of the drift. On the other hand, the normalized $\nablammm B$ drift frequency is computed as,
\begin{equation}
	\begin{split}
		\omega^{\boldsymbol{\nabla} \hspace{-1pt}B\text{-drift}}_{d,\mathbf{k},s} & = \frac{\tilde{T}_s}{Z_s \tilde{B}} \, \hatb \times \tilde{\mu} \, \boldsymbol{\tilde{\nabla}} \hspace{-1pt}\tilde{B} \cdot (k_x\rho_r\boldsymbol{\nabla} x + k_y\rho_r \boldsymbol{\nabla} y ) \\
		& = \tilde{\mu}\, \frac{\tilde{T}_s}{Z_s \tilde{B}} \, \Big( \frac{a}{B_r B} \Big)  \, \Big( k_x\rho_r \, (\mathbf{B} \times  \boldsymbol{\nabla} \hspace{-1pt} B  \cdot \boldsymbol{\nabla} x)  +  k_y\rho_r \, (\mathbf{B} \times  \boldsymbol{\nabla} \hspace{-1pt} B \cdot \boldsymbol{\nabla} y) \Big)     \\
		& = \tilde{\mu}_s \tilde{B} \, \frac{1}{2} \frac{\tilde{T}_s }{Z_s} \, \Big( \frac{k_x\rho_r}{\textcolor{black}{\hat{s}}} \, \texttt{xdriftknob}*\texttt{gbdrift0}  +  k_y\rho_r \, \texttt{ydriftknob}*\texttt{gbdrift} \Big).
	\end{split}
\end{equation}
The $\langle \mathbf{v}_\text{E} \rangle_\mathbf{R} \cdot \nablam|_\mathcal{E} F_{\text{M}s}$ term in the gyrokinetic equation (eq.~\ref{eq:GyrokineticEquation}) includes the combined effect of the $\mathbf{E} \times \mathbf{B}$ drift and the plasma gradients. In the flux-tube approximation this is equivalent to the $ i\omega_{*,\mathbf{k},s} J_0(a_{\mathbf{k},s}) \tilde{\varphi}_\mathbf{k}$ term in  equation (\ref{eq:NormalizedGyrokineticEquation}), where the normalized \textit{diamagnetic frequency} is given by,
\begin{align}
	\omega_{*,\mathbf{k},s} & = - \frac{1}{\psi_t'} \, \frac{k_y\rho_r}{2} aB_r \frac{dy}{d\alpha} \frac{e^{-\tilde{v}^2_s}}{F_\text{Ms}} \frac{dF_\text{Ms}}{d\psi}
\end{align}
The radial derivative of the Maxwellian distribution function yields,
\begin{equation}
	\label{eq:GradientOfF0s}
	\partial_\psi F_\text{Ms} = \Bigg[ \frac{n'_s}{n_s} + \frac{T'_s}{T_s} \, \Big( \frac{m_s}{T_s} (v_\parallel^2 / 2 + \mu B ) - \frac{3}{2} \Big) \Bigg]\, F_\text{Ms}.
\end{equation}
Substituting this into the  diamagnetic frequency gives,
\begin{align}
	\label{app:eq:omegastar}
		\omega_{*,\mathbf{k},s} &
		= - \frac{1}{\psi_t'}  \, \frac{k_y\rho_r}{2} aB_r \frac{dy}{d\alpha}  \Big[ \frac{1}{n_s} \frac{dn_s}{d\psi} + \frac{1}{T_s} \frac{dT_s}{d\psi} \, \Big( \frac{m_s}{2T_s} (v_\parallel^2  + 2 \mu B ) - \frac{3}{2} \Big) \Big] e^{-\tilde{v}^2_s}\notag \\[2mm]
		&= - \frac{1}{\psi_t'}  \, \frac{k_y\rho_r}{2} aB_r \Big(\rho_r \frac{d\tilde{y}}{d\alpha} \Big) \Big(\frac{\rho_r}{a^2B_r} \frac{d\tilde{x}}{d\tilde{\psi}} \Big) \Big[ \frac{1}{n_s} \frac{dn_s}{dx} + \frac{1}{T_s} \frac{dT_s}{dx} \, \Big( \tilde{v}_\parallel^2  + 2 \tilde{\mu} \tilde{B}  - \frac{3}{2} \Big) \Big] e^{-\tilde{v}^2_s} \notag\\[2mm]
		&= - \frac{1}{\psi_t'}  \, \frac{k_y\rho_r}{2} \Big(\frac{\rho_r}{a} \frac{d\tilde{y}}{d\alpha} \Big) \Big(\frac{\rho_r}{a} \frac{d\tilde{x}}{d\tilde{\psi}} \Big) \Big[ \frac{a}{n_s} \frac{dn_s}{dx} + \frac{a}{T_s} \frac{dT_s}{dx} \, \Big( \tilde{v}_\parallel^2  + 2 \tilde{\mu} \tilde{B}  - \frac{3}{2} \Big) \Big] e^{-\tilde{v}^2_s}  \\[2mm]
		&= -\frac{\texttt{wstarknob}}{\texttt{clebsch\_factor}}  \, \frac{k_y\rho_r}{2} \, \texttt{dydalpha} *  \texttt{dxdXcoord} \Big( \texttt{fprim[is]} + \texttt{tprim[is]} * (\texttt{vpa[iv]}^2  \notag\\[0mm]
		& + \texttt{vperp2[iz,imu]} - 1.5)
		\,*\, \texttt{maxwell\_vpa[iv,is]}* \texttt{maxwell\_mu[iz,imu,is]}  \Big) \notag
\end{align}
where $\texttt{(iz,imu,is)}$ refers to the discrete grid points along $(\tilde{z}, \tilde{\mu}, \tilde{v}_\parallel)$, and \texttt{wstarknob} is set to 1 by default and can be used in the input file to modify the contribution of the diamagnetic frequency.

\chapter{Input files for \texttt{stella}}

\label{app:usedinputfiles}

\section{Input files for linear simulations}
\label{app:usedinputfileslinear}

Two example input files are shown here, which reproduce the results from figure \ref{fig:gamma_vs_aLn} for W7-X and CBC considering $a/L_n=1.0$, $a/L_{T_i}=3.0$ and $a/L_{T_e}=0.0$. Setting $k_y\rho_i=1.3$ within these input files corresponds to the most unstable mode for $k_x=0.0$ and $k_y\rho_i \leq 2.0$ in W7-X, which has $\gamma a/v_{\text{th},i} = 0.187$ and $\omega a/v_{\text{th},i} = 0.247$. On the other hand, for CBC, the most unstable mode is found for $k_y\rho_i=0.6$, characterized by $\gamma a/v_{\text{th},i} = 0.278$ and $\omega a/v_{\text{th},i} = 0.353$. The main input file is the same for each device, and is given by,

\vspace{1mm}
\begin{lstlisting}[language=Fortran,firstnumber=1]
!include geometry.in
!include grids.in
!include species.in

&physics_flags
  nonlinear = .false.
  adiabatic_option = "field-line-average-term"
/

&knobs
  fphi = 1.0
  fapar = 0.0
  fbpar = 0.0
  delt = 0.05
  tend = 500
  LU_option = "local"
/

&stella_diagnostics_knobs
  nwrite = 10
  write_omega = .true.
/
\end{lstlisting}

\vspace{3mm}
\noindent
The difference between the devices is the magnetic geometry, which is set in ``geometry.in":

\vspace{1mm}
\begin{lstlisting}[language=Fortran,firstnumber=1]
! The "geometry.in" file considering a VMEC equilibrium of the standard W7-X 
! configuration. The chosen magnetic field line is located at r/a=0.7, 
! corresponding to torflux = (r/a)^2 = psit/psit_LCFS = 0.49 and extends up to 
! 3 poloidal turns, hence nfield_periods = Npol*Nnfp/iota = 3*5/0.894 = 16.778.
&geo_knobs
  geo_option = "vmec"
/

&vmec_parameters
  vmec_filename = "wout_w7xr003.nc"
  torflux = 0.49
  nfield_periods = 16.7777867495387
/

&zgrid_parameters
  nzed = 512
  nperiod = 1
/
\end{lstlisting}

\vspace{1mm}
\begin{lstlisting}[language=Fortran,firstnumber=1]
! The "geometry.in" file considering a set of Miller parameters corresponding 
! to the Cylone Base Case (CBC) benchmark with r/a = 0.7 and Npol = 3
&geo_knobs
  geo_option = "miller"
/

&millergeo_parameters
  nzed_local = 24
  rhoc = 0.7
  shat = 0.796
  qinp = 1.4
  rmaj = 2.77778
  rgeo = 2.77778
  shift = 0.0
  kappa = 1.0
  kapprim = 0.0
  tri = 0.0
  triprim = 0.0
  betaprim = 0.0
  d2qdr2 = 0.0
  d2psidr2 = 0.0
  betadbprim = 0.0
/

&zgrid_parameters
  nzed = 128
  nperiod = 2
/
\end{lstlisting}

\vspace{3mm}
\noindent
All linear simulations use the same resolution in $(\mu,v_\parallel)$, while the Fourier modes $(k_x,k_y)$ are run one by one. The ``grid.in" file is given by,

\vspace{1mm}
\begin{lstlisting}[language=Fortran,firstnumber=1]
! The "grids.in" file	
&kt_grids_knobs
  grid_option = "range"
/

&kt_grids_range_parameters
  nakx = 1 
  naky = 1
  akx_min = 0.0
  akx_max = 0.0
  aky_min = 1.3
  aky_max = 1.3
/

&vpamu_grids_parameters
  nvgrid = 64
  nmu = 12
  vpa_max = 3.0
  vperp_max = 3.0
/
\end{lstlisting}

\vspace{3mm}
\noindent
Finally, unless states otherwise, both kinetic ions and electrons are included in the simulations, with $a/L_n=1.0$, \mbox{$a/L_{T_i}=3.0$} and $a/L_{T_e}=0.0$ in this case. The profile gradients (\texttt{fprim} and \texttt{tprim}) are varied in order to investigate the effect of the density gradients and temperature gradients in chapter \ref{chap:heat}. The ``species.in" file is given by

\vspace{1mm}
\begin{lstlisting}[language=Fortran,firstnumber=1]
! The "species.in" file		
&species_knobs
  nspec = 2
/

&species_parameters_1
  z = 1.0
  mass = 1.0
  dens = 1.0
  temp = 1.0
  tprim = 3.0
  fprim = 1.0
  type = "ion"
/

&species_parameters_2
  z = -1.0
  mass = 5.43867E-4
  dens = 1.0
  temp = 1.0
  tprim = 0.0
  fprim = 1.0
  type = "electron"
/
\end{lstlisting}

\vspace{3mm}
\noindent
To perform simulations with a Boltzmann electron response, it is sufficient to set \mbox{\texttt{nspec = 1}}. Consequently only \texttt{species\_parameters\_1} is read, which adds kinetic ions to the simulation. The Boltzmann response is then set through \texttt{adiabatic\_option = "field-line-average-term"}, \mbox{\texttt{nine = 1}} and \texttt{tite = 1}. The \texttt{nine} and \texttt{tite} variables do not need to be included explicitly in the input files, since these are set to their default values in this work.

\section{Input files for nonlinear simulations}
\label{app:usedinputfilesnonlinear}

To perform nonlinear simulations, the nonlinear term has to be included in the gyrokinetic equation by setting \texttt{nonlinear} to true, and a finite hyper dissipation is set to prevent a pile-up of energy at small perpendicular scales due to turbulent cascades. The species parameters are identical to those used for the linear simulations. The main ``input.in'' file is given by,

\vspace{1mm}
\begin{lstlisting}[language=Fortran,firstnumber=1]
!include geometry.in
!include grids.in
!include species.in

&physics_flags
  adiabatic_option = "field-line-average-term"
  nonlinear = .true.
/

&knobs
  fphi = 1.0
  fapar = 0.0
  fbpar = 0.0
  delt = 0.1
  tend = 1000
  LU_option = "local"
/

&stella_diagnostics_knobs
  nwrite = 10
  write_fluxes_kxkyz = .true.
  flux_norm = .false.
/ 

&init_g_knobs
  ginit_option = "noise"
/

&dissipation
  hyper_dissipation = .true.
/

&hyper
  d_hyper = 0.1
/
\end{lstlisting}

\vspace{3mm}
\noindent
For nonlinear simulations, a box of $(k_x,k_y)$ Fourier modes is constructed. The binormal box length $L_y$ is set in the input file through $\texttt{y0} = L_y / (2\pi \rho_r)$ (eq.~\ref{eq:y0}) while the radial box length $L_x$ is set through the parallel boundary conditions. For the \texttt{VMEC} equilibria, the length of the flux tube is chosen so that $k_{x,\text{min}} = k_{y,\text{min}}$. On the other hand, for the tokamak geometry of CBC, $\hat{s}=0.796$ and $J=5$ ensures that $k_{x,\text{min}} = k_{y,\text{min}}$ as well. Moreover, note that \texttt{nx = 91} and \mbox{\texttt{ny = 91}} grid points in real space correspond to \texttt{nakx = 61} and \texttt{naky = 31} Fourier modes in reciprocal space.

\vspace{1mm}
\begin{lstlisting}[language=Fortran,firstnumber=1]
! The "grids.in" file	
&kt_grids_knobs
  grid_option = "box"
/

&kt_grids_box_parameters
  nx = 91
  ny = 91
  y0 = 15
/

&vpamu_grids_parameters
  nmu = 12
  nvgrid = 24
  vpa_max = 3.0
  vperp_max = 3.0
/	
\end{lstlisting}

\vspace{3mm}
\noindent
Seeing that a box of $(k_x,k_y)$ Fourier modes are being evolved, parallel boundary conditions have to be selected in the \texttt{zgrid\_parameters} namelist. For \texttt{VMEC} equilibria the stellarator-symmetric boundary conditions are used (section \ref{sec:StellaratorSymmetricBC}), whereas for the tokamak geometry of CBC, which is based on a set of Miller parameters, the standard  twist-and-shift boundary conditions are used (section \ref{sec:twistandshift}). Moreover, since the geometric quantities can vary very strongly along the parallel coordinate, the arc length is used as the parallel coordinate, which ensures that the simulations are less prone to blow up due to numerical instabilities. For all considered devices \texttt{nzed = 48} is used, except for LHD for which \texttt{nzed = 96} is used instead to capture the rapid variation of its geometric quantities along the field line.

\vspace{1mm}
\begin{lstlisting}[language=Fortran,firstnumber=1]
! The "geometry.in" file considering a VMEC equilibrium of the standard W7-X 
! configuration with nfield_periods = Npol*Nnfp/iota = 1.114*5/0.894 = 6.230.
&vmec_parameters
  vmec_filename = "wout_w7xr003.nc"
  torflux = 0.49
  nfield_periods = 6.230232651695381 
/
&zgrid_parameters
  nzed = 48
  zed_equal_arc = .true.
  boundary_option = "stellarator"
/
&geo_knobs
  geo_option = "vmec"
/
\end{lstlisting}

\chapter{Convergence checks of heat fluxes}

\label{sec:resolution}

\begin{figure}[h!]
	\centering
	\includegraphics[trim={0mm 1mm 0mm 0mm}]{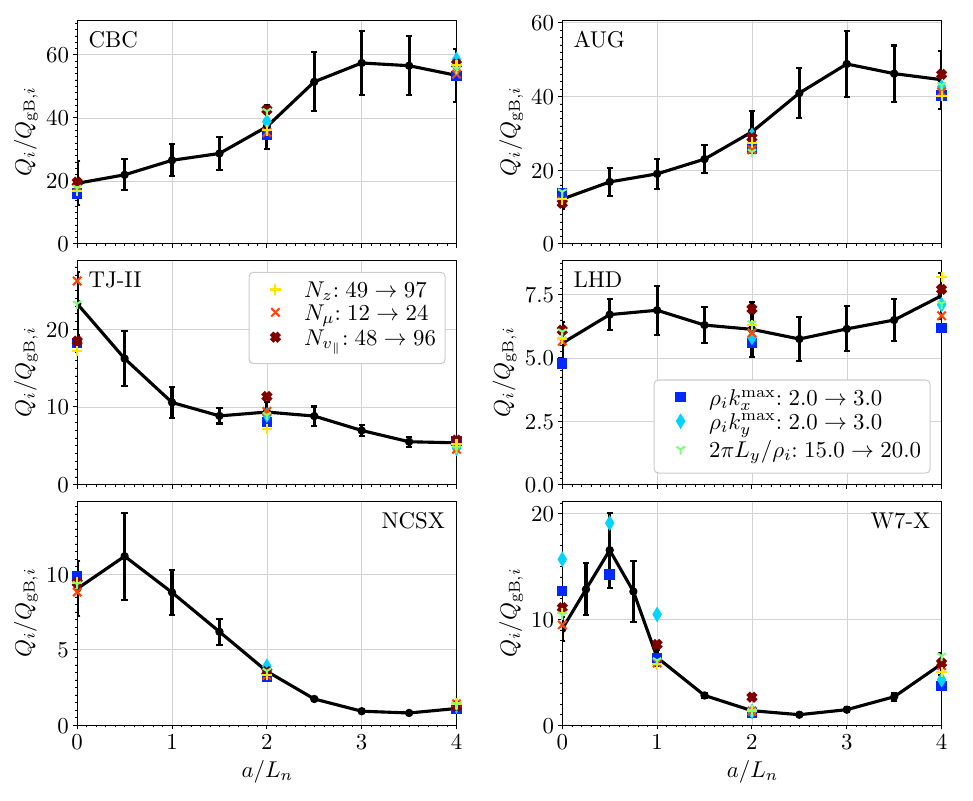}
	\vspace{-9mm}
	\caption{Convergence checks by increasing the nonlinear resolution parameters one by one, considering $a/L_{T_i}=3.0$ and \mbox{$a/L_{T_e}=0.0$}, corresponding to the simulations presented in figure \ref{fig:Qi_vs_aLn}. For LHD, the parallel resolution is increased from $N_z = 97$ to $193$. \vspace{-3mm} }
	\label{fig:resolution}
\end{figure}

\vspace{-6mm}
The nonlinear resolutions have been obtained from very rigorous convergence checks performed in \mbox{W7-X} for different sets of density and ion temperature gradients. The same resolution has been used for the other five devices, checking the validity of the choices by doubling the resolution parameters one by one, and examining the effect on the ion heat flux, which is shown in \mbox{figure \ref{fig:resolution}}. Here the black lines represent the ion heat fluxes presented in figure \ref{fig:Qi_vs_aLn}. First tests for LHD revealed that $N_z=49$ is too low in order to resolve its detailed magnetic field structure (figure \ref{fig:fieldlineAngle}). Therefore, a resolution of $N_z=97$ is used for LHD and it is doubled to $N_z = 193$ \mbox{in figure \ref{fig:resolution}.}

The error bars in figure \ref{fig:resolution} represent the standard mean deviation on the time trace of the simulation, taken between $tv_\text{th,i}/a=500$ and \mbox{$tv_\text{th,i}/a=1000$}, where the heat fluxes have saturated. In the simulations run with higher resolutions (colored markers), an ion heat flux has been obtained which falls within the error bars of the original simulations, hence the resolution used in this work produces reliable heat fluxes. Only in W7-X does the resolution in the perpendicular $(k_x,k_y)$-grid change the ion heat flux beyond the error bars, nonetheless, the change is not significant and it does not influence the conclusions drawn in this chapter.

\newpage
\bibliographystyle{unsrtnat_modified}
\bibliography{./bibliography}
\end{document}